\documentclass[11pt,a4paper]{article}
\pdfoutput=1

\parskip=\baselineskip
\usepackage{jheppub}
\usepackage[utf8]{inputenc}
\usepackage{mathtools} 
\usepackage{enumitem}
\usepackage{amsmath}
\usepackage{amsthm} 
\usepackage{amssymb}
\usepackage{amscd}
\usepackage{amsthm}
\usepackage{comment}
\usepackage{color} 
\usepackage{textcomp}
\usepackage{array} 
\usepackage{graphicx}
\usepackage{adjustbox}
\usepackage{mathrsfs}
\usepackage[samesize]{cancel}
\usepackage{ytableau}
\usepackage{booktabs}
\usepackage{tikz-cd}
\usepackage{relsize}
\usepackage{bbold}

\usepackage{tikz}
\usetikzlibrary{positioning,fit}
\usetikzlibrary{calc,intersections,through,backgrounds}
\usetikzlibrary{calc,arrows,decorations.markings,shapes.arrows,patterns, decorations.pathmorphing}
\usetikzlibrary{positioning,matrix,arrows,decorations.pathmorphing,decorations.pathreplacing}
\usetikzlibrary{arrows,matrix,positioning,shapes,arrows}
\usetikzlibrary{shapes.geometric}
\usetikzlibrary{decorations.text}
\usetikzlibrary{arrows.meta}
\usepackage{circuitikz}

\tikzset{
  ->-/.style={decoration={markings, mark=at position 0.5 with {\arrow{to}}},
              postaction={decorate}},}
\tikzset{
  -<-/.style={decoration={markings, mark=at position 0.5 with {\arrow{to reversed}}},
              postaction={decorate}},}
              
\tikzset{
  pics/torus/.style n args={3}{
    code = {
      \providecolor{pgffillcolor}{rgb}{1,1,1}
      \begin{scope}[
          yscale=cos(#3),
          outer torus/.style = {draw,line width/.expanded={\the\dimexpr2\pgflinewidth+#2*2},line join=round},
          inner torus/.style = {draw=pgffillcolor,line width={#2*2}}
        ]
        \draw[outer torus] circle(#1);\draw[inner torus] circle(#1);
        \draw[outer torus] (180:#1) arc (180:360:#1);\draw[inner torus,line cap=round] (180:#1) arc (180:360:#1);
      \end{scope}
    }
  }
}

\newcommand{\tikznode}[2]{\relax
    \ifmmode%
    \tikz[remember picture,baseline=(#1.base),inner sep=0pt] \node (#1) {$#2$};
    \else
    \tikz[remember picture,baseline=(#1.base),inner sep=0pt] \node (#1) {#2};%
    \fi
}

\newcommand{\largesupset}{\mathrel{\scalebox{1.5}{\ensuremath{\supset}}}}  

\allowdisplaybreaks[2]
\numberwithin{equation}{section}
\graphicspath{{figures/}}

\newcommand{\cA}{\mathcal A}
\newcommand{\cB}{\mathcal B}

\newcommand{\cD}{\mathcal D}
\newcommand{\cF}{\mathcal F}
\newcommand{\cG}{\mathcal G}
\newcommand{\cH}{\mathcal H}
\newcommand{\cJ}{\mathcal J}

\newcommand{\cL}{\mathscr{L}}

\newcommand{\cM}{\Sigma}

\newcommand{\cN}{\mathcal N}
\newcommand{\cP}{\mathcal P}
\newcommand{\cR}{\mathcal R}

\newcommand{\cO}{\mathcal O}

\newcommand{\JJ}{\mathcal{J}}

\newcommand{\mc}{\mathcal}
\newcommand{\mf}{\mathfrak}
\def\tilde{\widetilde}
\def\bar{\overline}

\newcommand{\br}{\overline}

\newcommand{\C}{\mathbb C}

\newcommand{\Z}{\mathbb Z}

\renewcommand{\d}{\mathrm{d}}
\newcommand{\no}{\nonumber}

\def\Tr{{\mathrm{Tr}}}

\def\sh{{\sf h}}

\newcommand{\vac}{\left| \emptyset \right> }
\newcommand{\wbar}{\br{w}} 
\newcommand{\zbar}{\br{z}}
\newcommand{\la}{\lambda}

\newcommand{\R}{\mathbb{R}}

\newcommand{\Lax}{{\mathscr L}} 
\newcommand{\si}{\sigma}
\newcommand{\del}{\partial}

\newcommand{\hgamma}{\hat{\gamma}}
\newcommand{\hbeta}{\hat{\beta}}
\newcommand{\ha}{\hat{a}}
\newcommand{\gm}{\gamma}
\newcommand{\bt}{\beta}
\newcommand{\phibar}{\bar{\phi}}
\newcommand{\alphabar}{\bar{\alpha}}

\newcommand{\U}{\mathrm{U}}
\newcommand{\SU}{\mathrm{SU}}
\newcommand{\GL}{\mathrm{GL}}
\newcommand{\PGL}{\mathrm{PGL}}
\newcommand{\SL}{\mathrm{SL}}

\def\ie{\begin{equation}\begin{aligned}}
\def\fe{\end{aligned}\end{equation}}

\begin{document}

\pagenumbering{Alph}
\begin{titlepage}
\begin{flushright}
\end{flushright}

\vskip 1.5in
\begin{center}
{\bf\Large{Dualities and Discretizations of}} \\
    \bigskip 
    {\bf\Large{Integrable Quantum Field Theories}} \\
    \bigskip 
    {{\bf\Large from 4d Chern-Simons Theory}}
\vskip 1.0cm 
{Meer Ashwinkumar$^{\mathcal{R}_{12}}$, Jun-ichi Sakamoto$^{\mathcal{R}_{13}}$, and Masahito Yamazaki$^{\mathcal{R}_{12}, \mathcal{R}_{23}}$} \vskip 0.05in 
\vskip 0.5cm 
{\small{ \textit{$^{\mathcal{R}_{12}}$Kavli Institute for the Physics and Mathematics of the Universe (WPI),\\ University of Tokyo, Kashiwa, Chiba 277-8583, Japan}
\vskip 0 cm 
\textit{$^{\mathcal{R}_{13}}$INFN sezione di Torino, via Pietro Giuria 1, 10125 Torino, Italy}
\vskip 0 cm 
\textit{$^{\mathcal{R}_{23}}$Trans-Scale Quantum Science Institute, University of Tokyo, Tokyo 113-0033, Japan
}}}

\end{center}


\vskip 0.5in
\baselineskip 16pt

\begin{abstract}

We elucidate the relationship between 2d integrable field theories and 2d integrable lattice models, in the framework of the 4d Chern-Simons theory. The 2d integrable field theory is realized by coupling the 4d theory to multiple 2d surface order defects, each of which is then discretized into 1d defects. We find that the resulting defects can be dualized into Wilson lines, so that the lattice of discretized defects realizes integrable lattice models. Our discretization procedure works systematically for a broad class of integrable models (including trigonometric and elliptic models), and uncovers a rich web of new dualities among integrable field theories. We also study the anomaly-inflow mechanism for the integrable models, which is required for the quantum integrability of field theories. By analyzing the anomalies of chiral defects, we derive a new set of bosonization dualities between generalizations of massless Thirring models and coupled Wess-Zumino-Witten (WZW) models. We study an embedding of our setup into string theory, where the thermodynamic limit of the lattice models is realized by polarizations of D-branes. 
\end{abstract}
\date{June, 2023}

\end{titlepage}
\pagenumbering{arabic} 

\tableofcontents
\newpage

\section{Introduction}
  \label{sec:introduction}

In physics we often invoke two different types of descriptions: discrete and continuous.
When we start with discrete lattice models, we have an emergent description 
in terms of continuous field theories in the Infrared (IR); in the opposite direction, we start with 
a field theory in the continuum and then find discretized lattice models as an Ultraviolet (UV) description.
In general it is a subtle question to understand the precise relation between field theories and their discretizations.

This paper aims to revisit this decades-old problem
in the context of the discretization of integrable field theories,
where the ``exact solvability'' via the existence of infinitely many conserved charges 
makes the analysis much more tractable. Such discussion has a long history,\footnote{In this paper we will discuss lightcone discretization of integrable field theories. 
The topic has a long history and consequently there are many related references, which are too many to be listed here. See e.g.\ \cite{Faddeev:1985qu,Destri:1987ze,Destri:1987hc,Destri:1987ug,Faddeev:1992xa,Faddeev:1993pe,Bazhanov:1995zg,Faddeev:1996iy} for some early references.} and indeed, the discretization of quantum field theories was 
one of the historical motivations for studying integrable lattice models. Since discretized field theories are free from UV divergences, one hopes that such a discussion will be of help in attacking the problem of quantization of integrable field theories, which is in general still an unsolved problem.

We will revisit this problem from the novel perspective of the four-dimensional Chern-Simons (CS) theory \cite{Costello:2013zra,Costello:2017dso,Costello:2018gyb,Costello:2019tri}.

The four-dimensional CS theory has emerged in recent years as a systematic, unifying approach to two-dimensional integrable systems. Its main applications thus far can be divided into two classes, namely, the realization of integrable lattice models/spin chains \cite{Costello:2013zra,Costello:2017dso,Costello:2018gyb} and integrable field theories \cite{Costello:2019tri}. The former theories are constructed via nets of intersecting Wilson line operators, while the latter are obtained as the effective two-dimensional field theory of surface defects coupled to the 4d CS theory. The aim of this work is to explicitly relate the integrable systems in these two classes from the perspective of the 4d CS theory, with emphasis on the case of integrable field theories realized using order surface defects.

As an unexpected byproduct of our analysis, we encounter duality statements among the defects,
and consequently among integrable quantum field theories with different Lagrangians. This originates from
the rigidity of the integrable lattice models---while we can consider a variety of surface defects in the discussions 
of integrable quantum field theories, we often find that their discretizations are described by the Wilson lines associated with the representations of an infinite-dimensional algebra (such as the Yangian). This means that integrable field theories associated with different defects are discretized into the same integrable lattice models, and hence are actually dual to each other.

One of the underlying reasons for the existence of such dualities is the dualities among the defects themselves. We therefore discuss the dualities among the defects more systematically without necessarily resorting to discretizations. As a concrete application,
we discuss the dualities (non-Abelian bosonizations) between free fermion defects, edge mode actions that arise in the study of the 4d CS coupled to disorder defects, and Wess-Zumino-Witten (WZW) defects, and generate an infinite class of new bosonization statements hitherto unknown in the literature.


The following is an outline of the paper. 
In Section \ref{sec:idea} we begin by summarizing the main ideas and highlights of this paper.
In Section \ref{sec:4dCS_review}, we review the construction of classically integrable field theories and integrable lattice models from the 4d CS theory. In Section \ref{sec:disc_FT}
we provide general procedures for discretization of integrable
field theories into integrable lattice models. We also discuss the opposite limit, where the thermodynamic limit of the lattice model reproduces the field theory.
In Sections \ref{sec:disc_defect} and \ref{sec:VA} we analyze more concretely the discretization of various surface defects and discuss the connection of the resulting 1d defects to Wilson lines. In Section \ref{sec:anomaly}
we discuss the cancellation of the anomalies of the surface defects from the anomaly inflow mechanism from the bulk 4d CS theory. 
In Section \ref{sec:duality} we study dualities between surface defects, and consequently for 
integrable field theories.  The general setup for our discretization/thermodynamic limit is briefly summarized in Section \ref{sec:summary}. 
In Section \ref{sec:duality-boson}
we discuss dualities between a quartet of defects and derive a new infinite class of non-Abelian bosonizations.
In Section \ref{sec:string}
we embed our 4d CS setup into string theory and
interpret the thermodynamic limit of the lattice models 
as polarization of D-branes.
Finally, in Section \ref{sec:discussion}, we end with discussions and comments on future directions. We include many Appendices for reviews and detailed computations relevant to the results in the main text. 

\section{Main Ideas and Highlights of the Paper}
  \label{sec:idea}

While the paper is long and sometimes technical, the main idea of the paper is very simple. 

In the 4d CS theory, integrable field theories are obtained by the 4d CS theory on $\mathbb{C} \times C$ together with some 2d defects filling $\mathbb{C}$ and point-like along $C$ \cite{Costello:2019tri};
we obtain the standard integrable field theories when we reduce the theory along $C$, which curve plays the role of the spectral curve.

When we discretize the integrable field theory, we can first go back to the 4d-2d coupled system and then try to discretize the defect. This is simply to discretize the 2d defect into a set of 1d defects, so that we obtain a setup where the 4d CS theory couples to the 1d defects. 

It turns out that in many cases of interest the 1d defects in question are given by Wilson lines of the 4d CS theory (when we start with order defects as the 2d defects). We thus have a lattice of Wilson lines, which is known to reproduce the integrable lattice models \cite{Costello:2013zra,Costello:2017dso,Costello:2018gyb}. 

\begin{figure}[htbp]
    \centering    
    \includegraphics[width=0.95\linewidth]{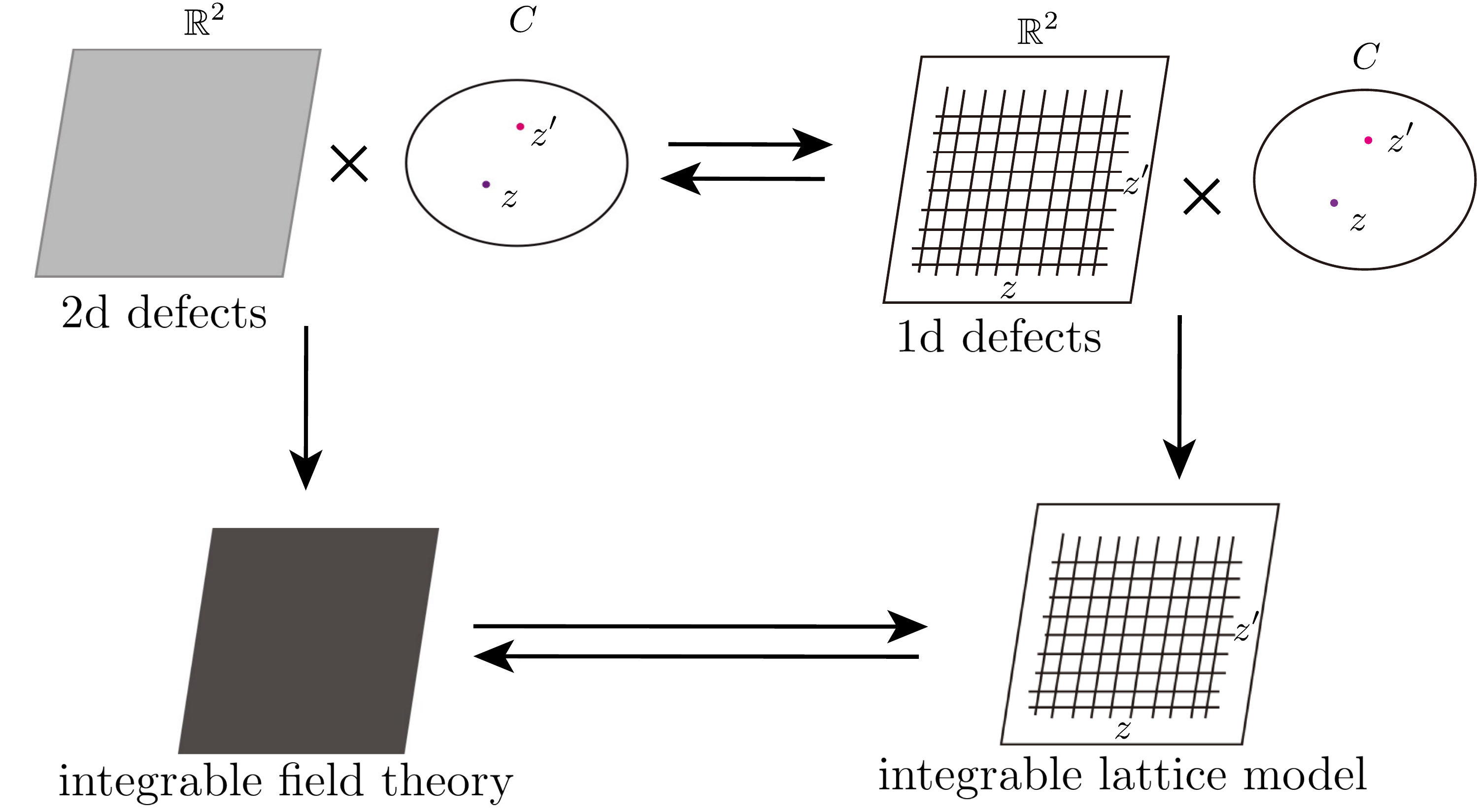}
    \caption{From from the perspectives of the 4d CS theory,
    the lattice discretization of the integrable field theory is understood as a discretization of 2d surface defects into 1d defects. When we reduce along the spectral curve $C$ we obtain the more traditional discretization of integrable field theories into integrable lattice models.}
    \label{thermlimitcy}
\end{figure}

We can keep track of this argument and discuss the relations between integrable structures in integrable lattice models and those in integrable field theories, as discussed in the following diagram and the associated figure (Figure \ref{thermlimitcy}).

\begin{figure}[htbp]
\centering
\begin{tikzpicture}
    \node[rectangle, draw, align=center](42) at (0,4) {4d CS \\+ 2d defects};
    \node[rectangle, draw, align=center](41) at (8,4) {4d CS\\ + 1d defects};
    \node[rectangle, draw, align=center](2) at (0, 0) {2d Integrable\\ Field Theory};
    \node[rectangle, draw, align=center](1) at (8,0) {1d Integrable\\ Lattice Model}; 
    \draw [->, >={Latex[round]}, shift=({5pt, 5pt})] ([yshift=-16mm]42)--([yshift=-16mm]2) node[midway, left] {Integration along $C$};
    \draw [->, >={Latex[round]}, shift=({5pt, 5pt})] (41)--(1) node[midway, right] {Path integral}; 
    \draw [->, >={Latex[round]}, shift=({5pt, 5pt})] ([yshift=-3mm] 1.west) -- ([yshift=-3mm]  2.east) node[midway, below] {Thermodynamic limit};
    \draw [->, >={Latex[round]}, shift=({5pt, 5pt})] ([yshift=3mm] 2.east) -- ([yshift=3mm]  1.west) node[midway, above] {Discretization};
    \draw [->, >={Latex[round]}, shift=({5pt, 5pt})] ([yshift=-3mm] 41.west) -- ([yshift=-3mm]  42.east) node[midway, below] {Thermodynamic limit of defects};
    \draw [->, >={Latex[round]}, shift=({5pt, 5pt})] ([yshift=3mm] 42.east) -- ([yshift=3mm]  41.west) node[midway, above] {Discretization of defects};
\end{tikzpicture}
\end{figure}

The highlights of our paper include the following:\footnote{We have included  a concise summary of the setups for this paper in Section \ref{sec:summary}.}
\begin{itemize}
\item Our discussion of integrable discretization from the 4d CS theory is very general and systematic, and encompasses infinite examples of models. This is in contrast with previous works, which often dealt with 
particular integrable models.  Examples of known models discussed in this paper include the Faddeev-Reshetikhin model, the Zakharov-Mikhailov model, the massless Thirring model, sigma models on flag manifolds, as well as the trigonometric and elliptic generalizations of these models. We emphasize, however, that our discussion is much more general than any of these specific models.

\item Our discretization procedure can be formulated 
directly in the 4d-2d coupled system, and can be discussed separately for each individual defect. In particular, most of the defects we consider are actually free theories, where the discretization is straightforward. We also discuss a very general class of defects described algebraically by vertex algebras, where we do not have known Lagrangian descriptions.

\item We find that many of the defects discussed in this paper, once discretized, are dual to Wilson lines. This means that there is a huge web of dualities between the defects. When we combine multiple defects inside the spectral curve, we obtain even larger web of dualities,
which contain known examples of dualities (such as the bosonization) as a special example. 

\item We study the anomaly inflow mechanism whereby gauge anomalies of chiral and anti-chiral surface defects are cancelled via quantum corrections to the meromorphic one-form of the 4d CS theory.

\item As a byproduct of the anomaly inflow analysis, we are led to the study of bosonizations of chiral/anti-chiral free-fermion defects and construct an infinite class of new bosonization statements between massless Thirring-type fermion models and coupled WZW models. The latter can be realized via disorder defects in the 4d CS theory, while the former can be discretized in the limit where the 4d CS coupling, $\hbar$, is small. 

\item We find an embedding of our setup into string theory, where the thermodynamic limit of the Wilson lines can be regarded as polarization of D-branes. We hope that this discussion will facilitate future research concerning connections to different aspects of integrability in string theory. 
\end{itemize}

\section{Review of 4d Chern-Simons Theory}
    \label{sec:4dCS_review}

In this section we begin with a brief summary of the 4d CS theory, highlighting the similarities and differences for the 
cases of integrable field theories and integrable lattice models. Readers who are familiar with the papers \cite{Costello:2017dso,Costello:2018gyb,Costello:2019tri} are encouraged to skip this section and come back here only when necessary.

\subsection{Integrable Field Theories from 4d Chern-Simons Theory}\label{sec:reviewint}

Let us begin by describing the construction of 2d integrable classical field theories from the 
4d CS theory, following the discussion in \cite{Costello:2019tri}.

We will consider a 4d spacetime of the product form 
$\cM\times C$, where both $\cM$ and $C$ are 2d manifolds with 
complex structures. Their local complex coordinates are denoted by $w$ and $z$, respectively.
For most of our discussions the manifold $\Sigma$ can be taken to be $\mathbb{C}$, except we sometimes compactify $\Sigma$ into a cylinder $S^1\times \mathbb{R}$ or a torus $\mathbb{T}^2$.

To define the 4d CS theory we need to fix a 
meromorphic one-form $\omega$ on $C$. In the local coordinate $z$, we have
\begin{align}
    \omega=\varphi(z)\,dz\,,
\end{align}
where $\varphi(z)$ is a meromorphic function on $C$.
In this paper, we primarily consider the following three possibilities, corresponding to the 
rational, trigonometric, elliptic classification of integrable models \cite{Belavin-Drinfeld,Costello:2017dso}:\footnote{When we include chiral/anti-chiral defects there is a one-loop correction to this formula for the holomorphic one-form $\omega$, as we will explain in Section \ref{sec:anomaly}.}
\begin{align}
\begin{cases}
    C=\mathbb{C} \;, & \omega = dz \:,\\
    C= \mathbb{C}^\times  \;,& \omega = \frac{dz}{z} \;,\\
    C=\mathbb{C}/(\mathbb{Z} + \tau \mathbb{Z}) \;, & \omega=dz \;.
\end{cases}
\end{align}
To simplify the presentation we will for the most part discuss the rational case $C=\mathbb{C}$, unless explicitly stated otherwise.
We emphasize, however, that our discussion is general and applies to all three cases above.

The action of the 4d CS theory is given by \cite{Costello:2013zra,Costello:2017dso}
\begin{align}
    S_{\textrm{CS}} = \frac{1}{2 \pi \hbar} \int_{\cM \times C } \omega \wedge \textrm{CS}(A) \,.\label{4dCS-action}
\end{align}
Here we have chosen a gauge group $G$, whose complexification we denote by $G_{\mathbb{C}}$ (we denote the associated Lie algebras as $\mathfrak{g}$ and $\mathfrak{g}_\mathbb{C}$, respectively) and $A\in\mathfrak{g}_{\mathbb{C}}$ is a partial complex connection of the form\footnote{Since $\omega$ is a holomorphic one-form, the 4d CS action (\ref{4dCS-action}) is invariant under the $\U(1)$ gauge transformation $A\mapsto A+\chi\,dz$\,.
By using this gauge symmetry, we can take a choice $A_{z}=0$\,. 
}
\begin{equation}
    A=A_{w}dw+A_{\wbar}d\wbar+A_{\zbar}d\zbar\,,
\end{equation}
and $\textrm{CS}(A)$ is the CS 3-form defined as
\begin{align}
    \textrm{CS}(A) :=  \Tr\left[A\wedge \left(dA+\frac{2}{3}A\wedge A\right)\right]\,.
\end{align}
The coupling constant $\hbar$ has dimensions of length and will play the role of the expansion parameter in the perturbation theory of the 4d theory. 

We obtain 2d integrable classical field theories by including surface
defects spreading along the curve $\cM$ \cite{Costello:2019tri}. The surface defects are broadly speaking classified into two different types: order defects and disorder defects. Order defects are themselves 2d field theories
which couple to the 4d CS theory; the $G$-global symmetries
of the former are gauged by the bulk 4d theory.
 Disorder defects by contrast specify singularities of the gauge fields of the 4d
CS theory, and are related to the zeros of the meromorphic one-form $\omega$.

In the following, we will concentrate mostly on the order defects.
We require the 2d defect field theory to (1)  be holomorphic or anti-holomorphic (chiral or anti-chiral)
and (2) have a $G$ global symmetry with an associated current $\cJ_{\mu}$;
otherwise the 2d field theory at a defect can be rather general,\footnote{We can try to include more general defects inside the formalism. For example, we expect to obtain
 a non-chiral order defect as a scaling limit of one chiral and one anti-chiral defect, where the two defects collider with each other.} 
and can be any chiral half of the conformal field theory, for example.\footnote{The theory is integrable in the sense that it admits a Lax connection, as we will explain shortly. This does not necessarily mean that the whole of the theory is integrable. As an extreme example, when the surface defect couples trivially (i.e., decouples) from the 4d CS theory, the decoupled sector from the defect can be non-integrable.}

The 2d defect couples to the bulk 4d CS theory via the current\footnote{We have not included a factor of $\hbar_{\rm 2d}$ in the coupling $\int A J$. This is different from the conventions in \cite{Costello:2019tri}.}: for ``chiral defects''
 \begin{align}
    S_{\textrm{4d--2d}} \supset \int_{\cM\times \{z^+\}} A_{\wbar} \cJ_{w}\,,
\label{chiral_defect}
\end{align}
and similarly ``anti-chiral defects''
 \begin{align}
    S_{\textrm{4d--2d}}
    \supset
    \int_{\cM\times \{z^-\}} A_{w} \bar{\mathcal{J}}_{\wbar}\,.
    \label{anti-chiral_defect}
\end{align}
 
The current $\mathcal{J}=\mathcal{J}^a t_a$ is a current for the global $G$-symmetry of the defect,
which is then gauged when the defect couples to the four-dimensional bulk.
The current satisfies the commutation relation 
\begin{align}\label{defect_JJ}
    [\mathcal{J}^a, \mathcal{J}^b] = f_{ab}{}^c \mathcal{J}^c  \;,
\end{align}
where $t^a$ are generators of $G$ satisfying $[t^a, t^b]=f^{ab}{}_c t^c$.\footnote{In this paper we choose the Lie algebra generators to be anti-Hermitian.}
 We will discuss many examples of surface defects in the rest of this paper. One simple example is a free chiral fermion $\psi$ in a representation $\rho$ in a Lie group $G$\,. The conserved current is
\begin{align} \label{defect_fermion}
    \mathcal{J}^a=\psi^* \rho(t^a) \psi =\psi_i^* \rho(t^a)^{i}{}_{j} \psi^j\,,
\end{align}
where $\rho(t^a)$ denotes the representation matrix of $t^a$ in $\rho$\,,
and $\psi, \psi^*$ satisfy the canonical anti-commutation relation
$\{\psi^i, \psi^*{}_j\}= \delta^i{}_j$.  
We can also consider a conserved current that includes the $N_F$ flavor degree of freedom by replacing $\psi$ with $\psi_I\,(I=1,\dots, N_F)$; this is a special situation where the representation $\rho$ is a direct sum of $N_F$ copies of a representation.

Since the defect theory is holomorphic or anti-holomorphic,
the symmetry algebra $\mathfrak{g}$  enhances to the loop algebra $\mathfrak{g}[[w, w^{-1}]]$, which is then centrally extended to
the current algebra (Kac-Moody algebra) for the Lie algebra $\mathfrak{g}$ in the quantum theory.
We obtain the commutation relation 
\begin{align}\label{defect_KM}
    [\mathcal{J}^a_m, \mathcal{J}^b_n] =  f^{ab}{}_c \mathcal{J}^c_{m+n} +  k m \delta^{ab} \delta_{m+n,0} \;,
\end{align}
where $k$ is the level and $\mathcal{J}^a_n$ (with $n\in \mathbb{Z}$) contain the original current as $\mathcal{J}^a=\mathcal{J}^a_0$.

\begin{figure}[!ht]
\begin{center}
  \begin{tikzpicture}[fill=white,draw=black]
    \pic{torus={2cm}{4mm}{70}} node[above=7mm,font=\tt,blue]{};
       \node[label=right:{$z_1$},shape=circle,fill=black, scale=0.3] at (-0.9,-0.5) {};
       \node[label=right:{$z_2$},shape=circle,fill=red, scale=0.3] at (-0.2,-0.7) {};
       \node[label=right:{$z_3$},shape=circle,fill=black, scale=0.3] at (0.6,-0.8) {};
       \node[label=right:{$z_4$},shape=circle,fill=red, scale=0.3] at (1.8,-0) {};
  \end{tikzpicture}
  \label{torus}
  \caption{Order surface defects located at various points on $C$. Black dots represent chiral defects while red dots represent anti-chiral defects. }\label{gentor}
\end{center}
\end{figure}
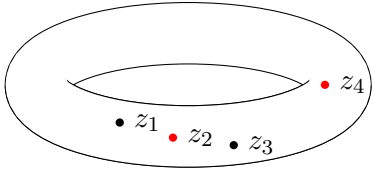

Let us consider a system of $n_+$ chiral surface defects at $z_1,\dots,z_{n_+}$ and $n_-$ anti-chiral defects at $z_1',\dots, z_{n_-}'$, such that the total number of defects is $n=n_++n_-$.
The action for the coupled 4d--2d system is then \cite{Costello:2019tri} (see Figure \ref{gentor})\begin{align} 
\begin{split}
S_{\textrm{4d--2d}}=
    \frac{1}{2\pi \hbar} \int_{\cM \times C} \omega\wedge \textrm{CS}(A) & + 
    \sum_{\alpha=1}^{n_+} \left(
    \int_{\cM \times \{z_\alpha^+\} }   \frac{1}{\hbar_{\rm 2d}} \mc{L}_\alpha\left(\phi_\alpha\right) 
    + A_{\wbar} \cJ_{w}
\right) \\
&+\sum_{\alphabar=1}^{n_-} \left(
    \int_{\cM \times \{z_\alpha^-\} }  \frac{1}{\hbar_{\rm 2d}} 
    \mc{L}_{\alphabar} 
    \left(  \phi_{\alphabar} \right)
    + A_{w} \bar{\cJ}_{\wbar}
    \right)\,.  
    \label{2d-4d}
\end{split}
\end{align}
 In these equations the overall constant $\hbar_{\rm 2d}$ is the ``2d Planck constant,'' which we choose to be dimensionless. 

\subsubsection*{2d Integrable Field Theories from 4d CS Theory}

The reduction of this coupled 4d--2d system to a classically integrable 2d field theory is derived by integrating out the four-dimensional gauge field along the complex manifold $C$ \cite{Costello:2019tri}. 

Let us briefly explain this reduction. 
To write down an explicit form of the gauge field propagator, we choose the holomorphic gauge,
\begin{equation}
    \partial_zA_{\zbar}=0\,.
\end{equation}
This is related to an analogue of the Lorentz gauge, $\partial_{\wbar} A_w + \partial_w A_{\wbar} + g^{z \zbar} \partial_{z} A_{\zbar} =0$, by rescaling the K\"ahler metric on $C$ such that the volume of $C$ becomes small. This gauge can be further simplified. For example, for $C=\mathbb{C}$, taking into account the Dirichlet boundary condition on the gauge fields at infinity, the gauge fixing condition further reduces to 
\begin{equation}
    A_{\zbar}=0\,,
\end{equation}
since global anti-holomorphic functions on a connected, compact Riemann surface are constants. 
The propagator in this gauge only involves gauge field components along $\cM$, and takes the form
\begin{equation}
\int_{\Sigma} d^2 w  \, \langle A_{a \wbar} (z)A_{bw}(z') \rangle =  r_{ab}(z,z')  \,.
\end{equation}
Here, we have made use of an orthonormal basis, $\{ t_a\}$, of the Lie algebra 
$\mf{g}$ with respect to the Killing form. The expression on the right-hand side is nothing but the classical $r$-matrix that solves the classical Yang-Baxter equation
\begin{align}\label{cYBE}
    [r_{12}(z_{12}),r_{13}(z_{13})+r_{23}(z_{23})]+[r_{13}(z_{13}),r_{23}(z_{23})]=0\,,
\end{align}
where $z_{ij}:=z_i-z_j$\,.

We can obtain an effective theory on $\Sigma$ either by integrating out the gauge field along the complex manifold $C$
in the path integral, or by solving the classical equations of motion along $C$ in the 4d CS theory \cite{Costello:2019tri}.
We then obtain 
a 2d field theory with the interaction term
\begin{equation} 
	\sum_{\alpha = 1}^{n_+} \sum_{\alphabar = 1}^{n_-}\hbar\, r_{ab}(z_\alpha- z'_{\alphabar}) \cJ_a^\alpha \br{\cJ}_b^{\alphabar} \,.
\end{equation}  
The total resulting 2d action is thus 
\begin{equation} 
S^{\rm eff}_{{\rm 2d}}=
\int_{\cM } d^2 w \left[ 
    \sum_{\alpha=1}^{n_+} \frac{1}{\hbar_{\rm 2d}}  \mc{L}_{\alpha}\left(\phi^{\alpha} \right) 
    +\sum_{\alphabar=1}^{n_-}\frac{1}{\hbar_{\rm 2d}}   \br{\mc{L}}_{\alphabar}\left(\phibar^{\alphabar}\right) 
    +\sum_{\alpha = 1}^{n_+} \sum_{\alphabar = 1}^{n_-} {\hbar}\, r_{ab}(z_\alpha- z'_{\alphabar}) \cJ_a^\alpha \br{\cJ}_b^{\alphabar}
    \right] \,.
    \label{acct}
 \end{equation}
The precise form of the classical $r$-matrix depends on the choices of $C$ and $\omega$, and leads to to three classes of integrable field theories of this type, i.e., rational, trigonometric, and elliptic. 
In the rational case (i.e. $C=\mathbb{C}$ and $\omega=d z$), for example, we have
\begin{equation}
    r_{ab}(z,z')=\frac{\delta_{ab}}{z-z'} \,.
    \label{rational_r}
\end{equation}
whereby theory \eqref{acct} is a current-current deformation of the decoupled defect theories.
We shall focus on classical $r$-matrices with rational spectral parameter dependence to illustrate key ideas of lattice discretization of 2d ultralocal integrable field theories in the next subsection.  

\subsubsection*{Lax Connections}

In the following
we find it convenient to use Lorentzian signature on $\cM$, such that the coordinates $w$ and $\bar{w}$ 
are equivalent to the lightcone coordinates $\sigma^+=\tau+\sigma$ and $\sigma^-=\tau-\sigma$, respectively, via a Wick rotation such that $ w=\tau-i \sigma$ and $\bar{w}=\tau+i \sigma$.
The associated metric $\eta_{\mu\nu}$ has off-diagonal components $\eta_{+-}=\eta_{-+}=-1/2$,
and the canonical volume form is $d^2 \sigma =  d\tau d\sigma  =-\frac{1}{2} d\sigma^+ d\sigma^-$. 

Classical integrability of integrable field theories follows from the existence of a Lax connection $\mathscr{L}(\tau, \sigma, z)$ satisfying a zero-curvature condition. 
This Lax connection gives rise to an infinite number of conserved nonlocal charges via the trace of the monodromy matrix:   
\begin{equation}\label{lax}
    W_{\cR}(z)=\textrm{Tr}_{\cR}\,\mathcal{P}\,\textrm{exp }\!\!\!\int_{\{\tau\}\times S^1} d\sigma \,\mathscr{L}(\tau, \sigma, z)\,,
\end{equation}
where we have compactified $\Sigma$
along the spatial ($\sigma$) direction into a cylinder $\mathbb{R} \times S^1$\,, and the symbol $\cP$ denotes the path-ordering operator along $\sigma$. This operator is associated with a representation $\cR$ of $\mathfrak{g}$ and is supported at a point in time, $\tau$, and located at a point $z$ in the spectral curve; the dependence on $\tau$ disappears due to the flatness of $\mathscr{L}$. 
From the perspective of the 4d CS theory, the Lax connection consists of gauge field components along the topological surface $\cM$, and the zero-curvature condition is precisely the equation of motion of the 4d CS theory for the gauge field component $A_{\bar{z}}$, and \eqref{lax} is precisely the Wilson line in the theory, supported along the spatial direction. 

We first focus on the simplest case: rational ($C=\mathbb{CP}^1$ and $\omega=dz$) and $n_+=n_-=1$\,.
In this case, we rename the chiral conserved currents $\cJ_1$ and $\bar{\cJ}_1$ on each defect with $\cJ_+$ and $\cJ_-$\,, respectively.
Then, the associated Lax connection is of the form  
\begin{align}
    &\mathscr{L} = \mathscr{L}_{+}(z) d\sigma^+ + \mathscr{L}_{-}(z) d \sigma^- \;,\\
    &\mathscr{L}_{\pm}(z)=\pm \frac{\hbar}{z-z_{\pm}} \mathcal{J}_{\pm}\,,\label{Lax-ultra}
\end{align}
where $z_+:=z_1$ and $z_-:=z_1'$. 
We also defined $\mathscr{L}_{\pm}=(\mathscr{L}_{\tau}\pm \mathscr{L}_{\sigma})/2$.
Since the associated 2d action (\ref{acct}) only depends on the difference of $z_+$ and $z_-$\,, the combination $z_+-z_-$ can be regarded as the coupling constant of the effective two-dimensional theory.
The Lax connection obeys the on-shell flatness condition
\ie
    \partial_{+} \mathscr{L}_{-}-\partial_{-} \mathscr{L}_{+}+\left[\mathscr{L}_{+}, \mathscr{L}_{-}\right]=0\,.\label{flatness}
\fe

For many models discussed in this paper, 
$\mathcal{J}_{\pm}$ satisfy the following classical Poisson brackets:
\ie \label{pp}
    \{\cJ_{\pm}^{a}(\sigma_1),\cJ_{\pm}^{b}(\sigma_2) \}
    &=f^{ab}{}_{c}\,\cJ_{\pm}^{c}(\sigma_2)\delta(\sigma_1-\sigma_2)\,,\\ \{\cJ_{+}^{a}(\sigma_1),\cJ_{-}^{b}(\sigma_2) \}
    &=0\,.
\fe
Here, we have a pair of Poisson commuting classical affine Kac-Moody algebras, each with no central extension (which would involve a derivative of the Dirac delta function). 

From \eqref{Lax-ultra} and \eqref{pp} we can derive the Poisson brackets between a pair of Lax connections to be 
\ie\label{PB-Lax}
    &\{\cL_{\sigma}(\sigma_1;z_1) \overset{ \otimes}{,}\cL_{\sigma}(\sigma_2;z_2)\} \\ &\qquad= \left[r(z_1,z_2), \cL_{\sigma}(\sigma_1;z_1)\otimes 1+1\otimes \cL_{\sigma}(\sigma_2;z_2)\right ]\delta(\sigma_1-\sigma_2)\,.
\fe
Here, we have defined the Poisson bracket using tensorial notation, that is,
\begin{align}
     \{X\overset{ \otimes}{,}Y \}:=  \{X\otimes 1,1\otimes Y\nonumber \}=
     \{X^a,Y^b\}\, t^a\otimes t^b\,.
\end{align}
The Poisson bracket \eqref{PB-Lax} does not contain derivatives of the delta function, and such theories are often called ``ultralocal'' integrable field theories.

More generally, for arbitrary values of $n_+$ and $n_-$, and the trigonometric ($C=\mathbb{C}^{\times}$ and $\omega = dz/z$) and elliptic ($C=\mathbb{C}/(\mathbb{Z} + \tau \mathbb{Z})$ and $\omega = dz$) cases, the components of the Lax connection 
$\mathscr{L}(z)=
\mathscr{L}^a(z) t_a$ take the form 
\ie 
\mathscr{L}^a(z)
    = \hbar\,\sum_{\alpha=1}^{n_{+}} \cJ^{\alpha}_b\left(w\right) r^{ab}\left(z, z_\alpha\right) d w
    - \hbar\, \sum_{\alphabar=1}^{n_{-}} \bar{\cJ}_b^{\alphabar}\left(w\right) r^{ab}\left(z, z_{\alphabar}\right) d \bar{w} \;,
\fe 
where $r(z,z')= r^{ab}(z,z') t_a \otimes t_b$ is either the trigonometric or elliptic solution of the classical Yang-Baxter equation. This form of the Lax connection can be computed either from tree-level Feynman diagram computations in the 4d CS theory, or by solving the four-dimensional gauge field along $C$ with source terms originating from the defects \cite{Costello:2019tri}.

The Poisson bracket is a starting point for the Hamiltonian quantization of the theory. The Hamiltonian formulation of the 4d CS theory can also be studied in terms of affine Gaudin models, see e.g.\  \cite{Vicedo:2017cge,Delduc:2018hty,Delduc:2019bcl,Vicedo:2019dej}.
Here, one usually considers a large class of integrable field theories that arise from \textit{disorder} defects in the 4d CS theory, defined with meromorphic one-form $\omega$ containing both poles and zeroes. The disorder defects correspond to singular behaviour of the gauge fields close to the zeroes of $\omega$. The Poisson brackets for such integrable field theories involve derivatives of a Dirac delta function, and are thus referred to as being ``non-ultralocal''. The latter property obstructs a straightforward discretization (and quantization) of these integrable field theories.

\subsection{Integrable Lattice Models from 4d Chern-Simons Theory}\label{sec:lattice-cs}

Let us next briefly recall the construction of integrable lattice models in the 4d CS theory, following the work of \cite{Costello:2017dso,Costello:2018gyb}. These are models with Boltzmann weights given by elements of R-matrices that satisfy the Yang-Baxter equation with spectral parameters.
The fundamental reason for the simple description of R-matrices in the 4d CS theory is the infrared freedom of the theory, coupled with its diffeomorphism invariance along $\Sigma$. In a typical gauge theory, gluon exchange between parallel Wilson lines cannot be ignored. However, since the 4d CS theory is infrared-free, one can scale up the metric of $\Sigma$ (that becomes relevant to the description of the theory after gauge-fixing), and the gluon exchange becomes irrelevant.

\begin{figure}[ht!]
 \centering
    \includegraphics[width=0.8\textwidth, trim = 0 70 0 20]{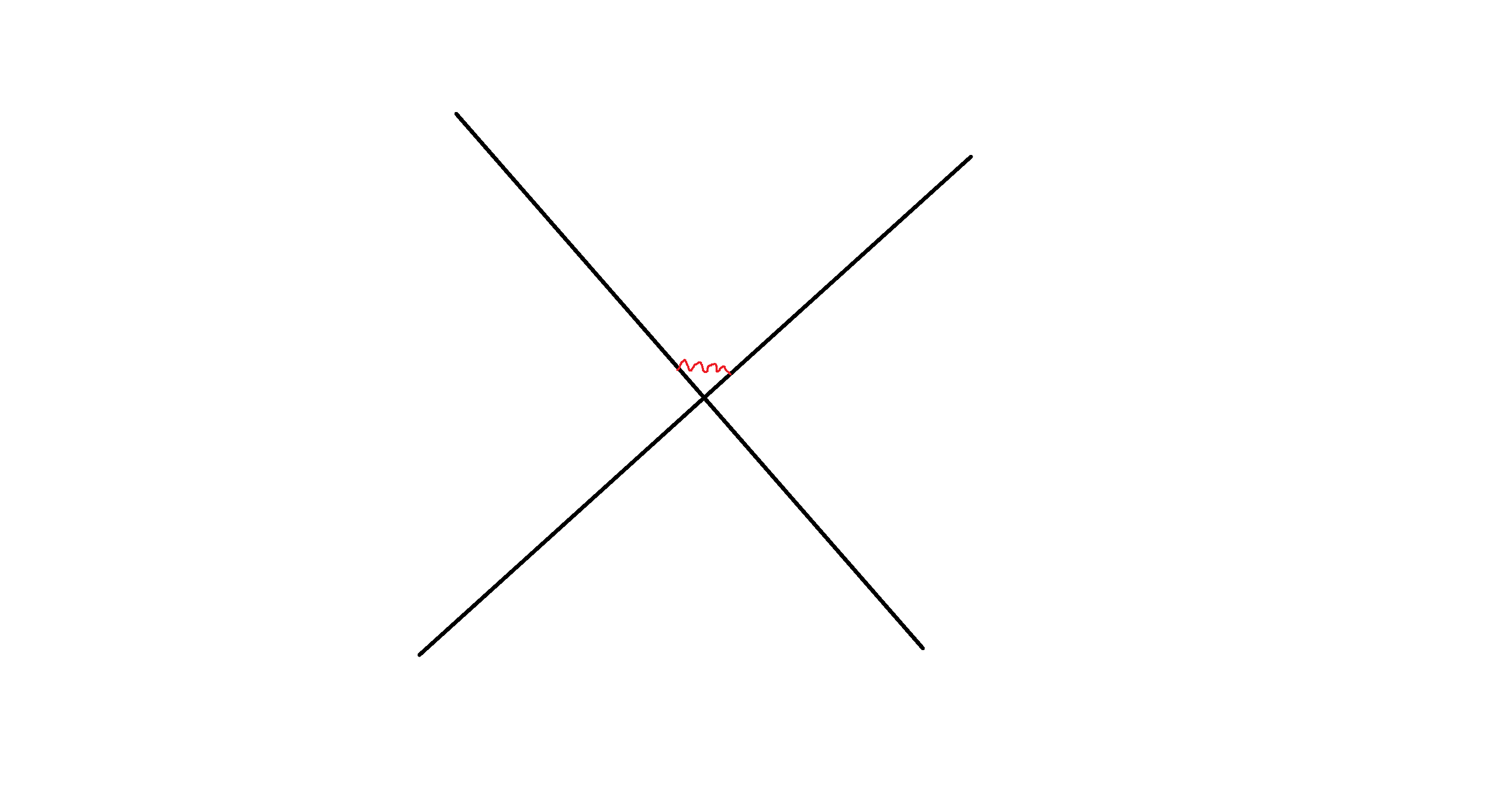}
    \caption{Tree-level gluon exchange between crossed Wilson lines}\label{inters}
 \end{figure}

The only non-trivial contributions arise from gluon exchange close to \textit{crossings} of Wilson lines, as depicted in Figure \ref{inters}. This is why the 4d CS theory is able to describe integrable lattice models: all interactions are confined to intersections of Wilson lines. These interactions can be computed to be Boltzmann weights (R-matrix elements) of the lattice models. In the case where $C=\mathbb{C}$, which corresponds to rational integrable lattice models, the correlation function of two Wilson lines in the configuration shown in Figure \ref{inters}, respectively located at $z_1$ and $z_2$ on $C$ and in representations $\rho_1$ and $\rho_2$, can be computed to be 
\ie \label{R_r}
    R_{12}(z_1,z_2)=I+\frac{\hbar\,c_{\rho_1, \rho_2}}{z_1-z_2}+\mathcal{O}\left(\hbar^2\right)\,,
\fe 
where $c_{\rho_1,\rho_2}$ is the quadratic Casimir element
evaluated in a tensor product representation $\rho_1\otimes \rho_2$:
\begin{align}
    c_{\rho_1,\rho_2}:=\sum^{\textrm{dim}\,\mathfrak{g}}_{a=1}\rho_1(t^a)\otimes \rho_2(t_a)\,.
\end{align}

\begin{figure}[ht!]
 \centering
    \includegraphics[width=0.8\textwidth, trim=0 50 0 0]{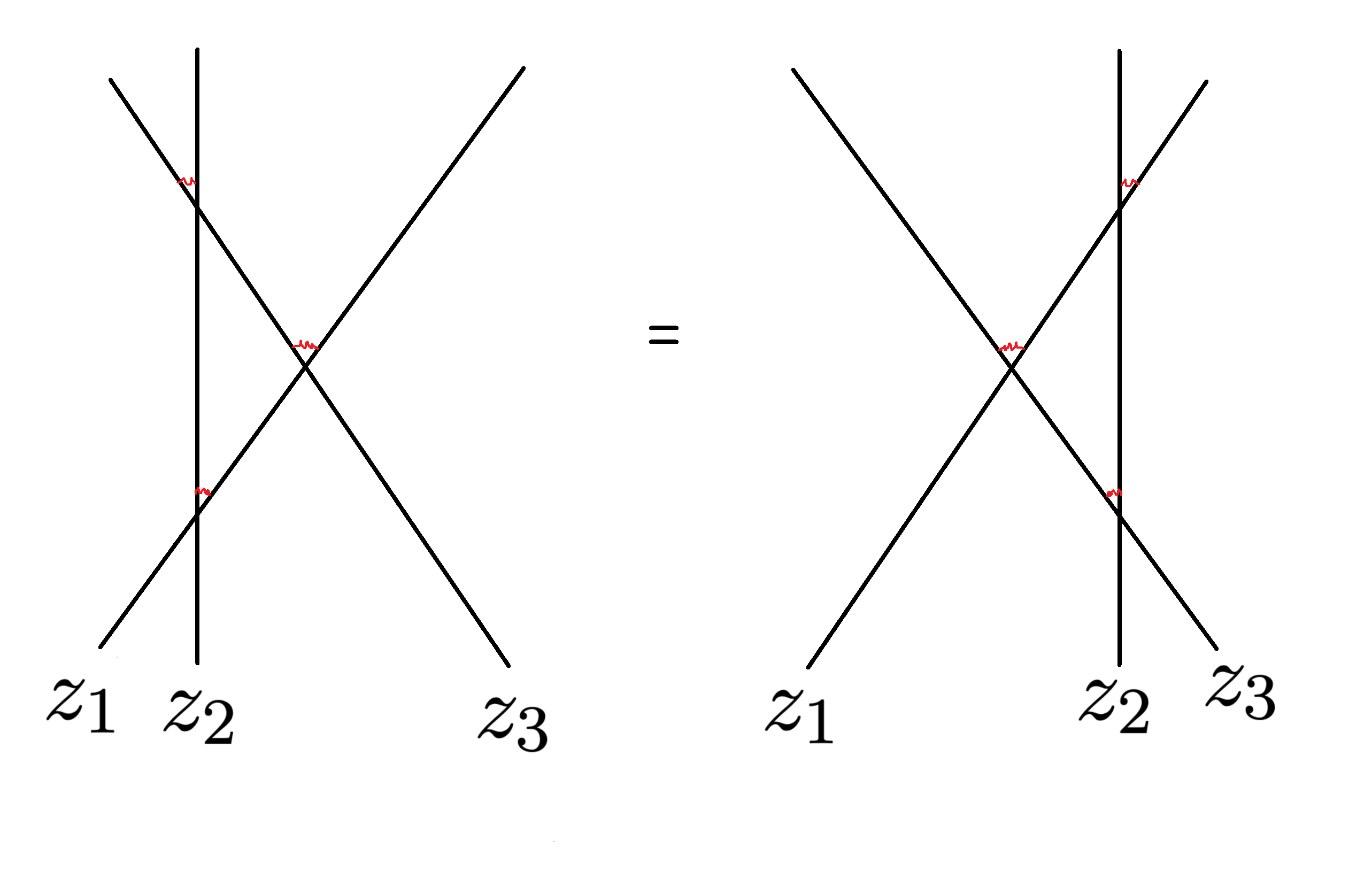}
    \caption{The Yang-Baxter equation}\label{YBE}
 \end{figure}

In fact, the topological invariance along $\Sigma$ ensures that the two diagrams in Figure \ref{YBE} are equivalent, which in turn implies the  Yang-Baxter equation,
\begin{equation}\label{ybe1}
    R_{12}(z_1,z_2)R_{13}(z_1,z_3)R_{23}(z_2,z_3)=R_{23}(z_2,z_3)R_{13}(z_1,z_3)R_{12}(z_1,z_2)\,,
\end{equation}
that ensures the integrability of these lattice models. This equation reduces to the classical Yang-Baxter equation \eqref{cYBE} when we expand the $R$-matrix as
\begin{align}
    \label{R_to_r}
    R(z) = \textrm{Id} + \hbar\, r(z) +\mathcal{O}(\hbar^2) \;,
\end{align}
where $r(z)$ is the classical $r$-matrix introduced previously.

We have avoided obtaining a more general equation where the regions between the Wilson lines would also be labeled by certain parameters (as in Section 11 of \cite{Costello:2017dso}). The reason for this is that one chooses a boundary condition at infinity, $A=0$, such that there is only one classical solution to expand around in perturbation theory, that does not have any continuous deformations. The regions between the Wilson lines would generally be labeled by a basis vector of the Hilbert space, $\mathcal{H}$, of the quantum states of the theory. But since the space of classical solutions modulo gauge transformations is a point, $\mathcal{H}$ is one-dimensional, and there are no labels in between Wilson lines.

Nevertheless, due to the infrared-freedom of the theory, away from crossings, each Wilson line is labeled by a state (i.e., a basis vector) in its associated representation. This is what leads us to configurations in the 4d CS theory 
involving nets of Wilson lines,
where all quantum interactions are localized to crossings of Wilson lines, and build up Boltzmann weights (that depend on the adjacent basis vectors) such that the theory describes an integrable lattice model.

The above discussion generalizes to the realization of trigonometric lattice models when $C=\mathbb{\C}^{\times}$, with Manin triple boundary conditions at the poles of the meromorphic one-form, $\omega=dz/z$, ensuring a unique classical solution with no continuous deformations \cite[Section 9]{Costello:2017dso}. In addition, the discussion also generalizes to the realization of elliptic lattice models when $C$ is an elliptic curve, with the restriction of $G=\PGL(N)$ where $N\geq 2$, as well as a distinguished choice of $G$-bundle, to ensure a unique classical solution with no continuous deformations \cite[Section 10]{Costello:2017dso}.  

\section{Lattice Discretizations of Integrable Field Theories}
     \label{sec:disc_FT}

The goal of this section is to illustrate how a certain type of 2d classical integrable field theory is discretized into a lattice model within the context of the 4d CS theory.

\subsection{Discretization of Order Surface Defects}\label{discretout}

In this section, we shall explain our approach to the discretization of integrable field theories from the perspective of the 4d CS theory.

In general, we may discretize chiral and anti-chiral surface operators of the form
\begin{align}\label{aabb}
\begin{split}
    &\frac{1}{\hbar_{\rm 2d}} \int_{\cM \times \{z_\alpha^{+}\} }\mathcal{L}_{\alpha}\left(\phi^{\alpha},\partial_-\phi^{\alpha}\right)|_++\cJ^{\alpha} A_{-} \quad\left(\alpha=1, \ldots n_{+}\right) \;,\\
    &\frac{1}{\hbar_{\rm 2d}} \int_{\cM \times \{z_{\alphabar}^{-}\} }\overline{\mathcal{L}}_{\alphabar}\left(\phibar^{\alphabar},\partial_+\phibar^{\alphabar}\right)|_-  +\bar{\cJ}^{\alphabar} A_{+} \quad\left(\alphabar=1, \ldots n_{-}\right) \;.
\end{split}
\end{align}
Here, the notation $|_+$ and $|_-$ denotes quantities that transform as components of one-forms defined along the $+$ and $-$ lightcone directions, respectively.
Let us try to discretize one of these surface operators, which has the chiral form given in the first line of \eqref{aabb}. 

For such a surface operator located at a point $z_+\in \mathbb{CP}^1$, and with dependence on a set of fields $\phi^{\alpha}$ and partial derivatives of these fields, but only with respect to $\sigma^-$, we shall discretize it along the $\sigma^+$ direction. We achieve this discretization by replacing the integral over $\sigma^+$ in the defect action 
 \ie
    &\frac{1}{\hbar_{\rm 2d}} \int_{\cM \times \{z_+\} }d\sigma^+ d\sigma^- \mc{L}\left(\phi^{\alpha}, \partial_-\phi^{\alpha}; A_{-} |_{z_+}\right)\,
    \label{cdefect}
\fe
by a Riemann sum over lattice points 
$\sigma^{+}=\sigma^{+}_n:=\Delta n$ ($n\in\mathbb{Z}$)  that are equidistant, with lattice spacing $\Delta$. The resulting action is then a sum over those of the 1d defects along the lightcone direction, which we denote at $\mathbb{R}_{-}$:
\begin{equation}
\begin{aligned}
     &\frac{1}{\hbar_{\rm 1d}}\sum_{i} \int_{ \{\sigma_i^+\} \times \mathbb{R}_{-}  \times \{z_+\} }d\sigma^- \mc{L}\left(\phi^{\alpha}_i  , \partial_-\phi^{\alpha}_i; A_{-,i} \right)\\
    & = \frac{1}{\hbar_{\rm 1d}}\sum_{i=1}^{N}\int_{\{\sigma_i^+\} \times \mathbb{R}_{-} \times \{z_+\} }d\sigma^-\bigg( \mc{L}\left(\phi^{\alpha}_i, \partial_- \phi^{\alpha}_i \right) +J^{\alpha}_i A_{-,a} |_{\sigma_i^+,z_+}\bigg) \;,
\end{aligned}
\end{equation}
where we defined 
\begin{align}
  \phi^{\alpha}_n := \phi^{\alpha} |_{\sigma^+=\sigma^{+}_n } \;,
  \quad A_{-,n} := A_{-} |_{\sigma^+=\sigma^{+}_n } \;,
  \quad J^{\alpha, a}_n :=\mathcal{J}^{\alpha, a} |_{\sigma^+=\sigma^{+}_n} \;,
\end{align}
and the one-dimensional Planck constant $\hbar_{\rm 1d}$ by
\begin{equation}
    \frac{1}{\hbar_{\rm 1d}}=\frac{\Delta}{\hbar_{\rm 2d}}\,.
\end{equation}

We now repeat the same procedure for an anti-chiral surface operator located at a point $z_-$ and with dependence on a set of fields denoted $\phibar$, i.e., 
\begin{equation}
\frac{1}{\hbar_{\rm 2d}} \int_{\cM \times \{z_-\} }d\sigma^+ d\sigma^- \mc{L}\left(\phibar^{\alphabar}, \partial_+ \phibar^{\alphabar}; A_{+} |_{z_-}\right),
\end{equation}
but discretize the $\sigma^-$ direction instead into $\sigma^{-}=\sigma^{-}_n:= n \Delta$. Thereby, we arrive at another infinite set of line defects oriented along the $\sigma^+$ direction, i.e., 
\begin{equation}
  \frac{1}{\hbar_{\rm 1d}}
  \sum_{j=1}^{N}\int_{\mathbb{R}_{+} \times \{\sigma_j^-\} \times  \{z_-\} }d\sigma^+
    \bigg( 
      \mc{L}\left(\phibar^{\alphabar}_j,   \partial_+\phibar^{\alphabar}_j\right) 
      +J\left(\phibar^{\alphabar}_j \right) A_{+}|_{\sigma^-_j, z_-} 
    \bigg) \;,
\end{equation}
where we defined $\phibar_l= \phibar_l|_{\sigma_j^-}$.

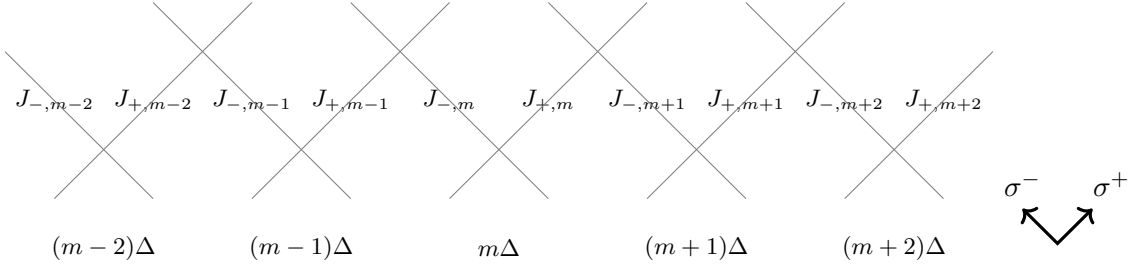
\begin{figure}
\begin{center}
\begin{tikzpicture}[scale=0.65]
\node at (-10,0) {\footnotesize $J_{-,m-2}$};
\node at (-8,0) {\footnotesize $J_{+,m-2}$};
\node at (-6,0) {\footnotesize $J_{-,m-1}$};
\node at (-4,0) {\footnotesize $J_{+,m-1}$};
\node at (-2,0) {\footnotesize $J_{-,m}$};
\node at (0,0) {\footnotesize $J_{+,m}$};
\node at (2,0) {\footnotesize $J_{-,m+1}$};
\node at (4,0) {\footnotesize $J_{+,m+1}$};
\node at (6,0) {\footnotesize $J_{-,m+2}$};
\node at (8,0) {\footnotesize $J_{+,m+2}$};
\draw[gray] (-10,-2) -- (-6,2);
\draw[gray] (-8,-2) -- (-11,1);
\draw[gray] (-6,-2) -- (-2,2);
\draw[gray] (-4,-2) -- (-8,2);
\draw[gray] (-2,-2) -- (2,2);
\draw[gray] (2,-2) -- (6,2);
\draw[gray] (6,-2) -- (9,1);
\draw[gray] (0,-2) -- (-4,2);
\draw[gray] (4,-2) -- (0,2);
\draw[gray] (8,-2) -- (4,2);
\node at (-9,-3) {\footnotesize $(m-2)\Delta$};
\node at (-5,-3) {\footnotesize $(m-1)\Delta$};
\node at (-1,-3) {\footnotesize $m\Delta$};
\node at (3,-3) {\footnotesize $(m+1)\Delta$};
\node at (7,-3) {\footnotesize $(m+2)\Delta$};
\begin{scope}[xshift=9.6cm,yshift=-1.5cm]
\draw[very thick,<->,rotate=45] (-0.5,-0.5) -- (-0.5,-1.5) -- (0.5,-1.5);
\node at (1.8,-0.2) {$\si^+$};
\node at (0,-0.2) {$\si^-$};
\end{scope}
\end{tikzpicture}
\caption{\footnotesize The lightcone lattice where the discrete modes of the currents are supported along the null segments.}
\label{f2-1} 
\end{center}
\end{figure}

The currents are now defined in terms of discrete modes, $J^{\alpha, a}_{\pm,m}$. To be precise, there is a pair of modes  $J^{\alpha, a}_{\pm,m}$ supported on the null links $\si^+=m\Delta$ and $\si^-=-m\Delta$ associated with each spatial lattice point $\si=m\Delta$. 
The discretized surface operators obtained in this way is depicted in Figure \ref{f2-1}. 
The discrete modes of the currents satisfy the Poisson brackets
\ie\label{discp}
\{J_{\pm,m}^a, J_{\pm,n}^b\}=\frac 1\Delta f^{ab}{}_c J^c_{\pm,n}\delta_{mn}\,,
\fe
where the Dirac delta distribution in \eqref{pp} has been replaced by $\delta_{mn}/\Delta$\,.

In general, the realization of lattice currents depends on the underlying integrable field theory.
We will discuss many examples of surface defects in Sections \ref{sec:disc_defect} and \ref{sec:VA}.

For example, if we introduce the lattice fermions $\psi_n=(\psi_n{}^i)\,, \psi_n^*=(\psi_n^*{}_j)$ in the Hilbert space in a representation $\rho$ of the 4d gauge group, $G$, satisfying the canonical anti-commutation relations
\begin{align}
    \{\psi_m{}^i, \psi^*_n{}_j \}=\frac{1}{\Delta} \delta_{mn} \delta^{i}_j\,,
\end{align}
the corresponding lattice current can be written as
\begin{align}
    J^a_n=\psi^*_n \rho(t^a) \psi_n =\psi^*_{n,i} \rho(t^a)^i{}_{j} \psi_n{}^j\,,
\end{align}
and satisfies the commutation relations of $\mathfrak{g}$\,.

Some readers might worry that our discretization procedure runs into problems associated
with the doubling of the chiral fermions \cite{Nielsen:1980rz,Nielsen:1981xu,Nielsen:1981hk,Hellerman:2021fla}.
Our setup, however, is not to discretize the full 2d spacetime into lattices, but 
rather to discretize only one of the lightcone directions while keeping the orthogonal direction intact;
the chiral (or anti-chiral) theory has the kinetic term which does not involve any $\sigma^{+}$ ($\sigma^{-}$) derivatives
and our discretization is only along the $\sigma^{+}$ ($\sigma^{-}$) direction.
This ensures the consistency of our procedure.

\subsection{Lightcone Lattice Discretization}

In the discussion above 
we have discretized the 2d defects into a set of 1d defects. Since the chiral/anti-chiral defects depend on different coordinates $\sigma^{+}$ and $\sigma^{-}$,
we obtain 1d defects both along $\sigma^{-}$ and $\sigma^{+}$ directions, located at
\begin{align}
    &\textrm{(chiral defect)}: \si^+=m\Delta\ , \qquad  m\in\mathbb Z\,, \\
    &\textrm{(anti-chiral defect)}:  \si^-=n\Delta\ ,\qquad  n\in\mathbb Z\, .
\end{align}

We have found from the 4d CS theory that lightcone lattices naturally arise from the discretization of chiral and anti-chiral surface operators.
In the literature such a discretization has a long history (see e.g., \cite{Faddeev:1985qu,Destri:1987ze,Destri:1987hc,Destri:1987ug,Faddeev:1996iy}). It is worth mentioning that lightcone discretization of integrable field theories plays a crucial role in the quantization of such theories in the approach of the quantum inverse scattering method of Faddeev and Reshetikhin \cite{Faddeev:1985qu}.
Our discussion, however, is not restricted to a particular class of models, is much more general than those in the earlier references, and is formulated inside the framework of the 4d CS theory.

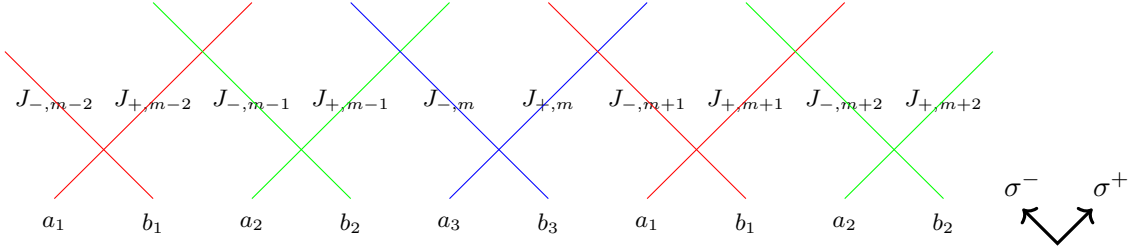
\begin{figure}
\begin{center}
\begin{tikzpicture}[scale=0.65]
\node at (-10,0) {\footnotesize $J_{-,m-2}$};
\node at (-8,0) {\footnotesize $J_{+,m-2}$};
\node at (-6,0) {\footnotesize $J_{-,m-1}$};
\node at (-4,0) {\footnotesize $J_{+,m-1}$};
\node at (-2,0) {\footnotesize $J_{-,m}$};
\node at (0,0) {\footnotesize $J_{+,m}$};
\node at (2,0) {\footnotesize $J_{-,m+1}$};
\node at (4,0) {\footnotesize $J_{+,m+1}$};
\node at (6,0) {\footnotesize $J_{-,m+2}$};
\node at (8,0) {\footnotesize $J_{+,m+2}$};
\draw[red] (-10,-2) -- (-6,2);
\draw[green] (-6,-2) -- (-2,2);
\draw[blue] (-2,-2) -- (2,2);
\draw[red] (2,-2) -- (6,2);
\draw[green] (6,-2) -- (9,1);
\draw[red] (-8,-2) -- (-11,1);
\draw[green] (-4,-2) -- (-8,2);
\draw[blue] (0,-2) -- (-4,2);
\draw[red] (4,-2) -- (0,2);
\draw[green] (8,-2) -- (4,2);
\node at (-10,-2.5) {\footnotesize $a_1$};
\node at (-6,-2.5) {\footnotesize $a_2$};
\node at (-2,-2.5) {\footnotesize $a_3$};
\node at (2,-2.5) {\footnotesize $a_1$};
\node at (6,-2.5) {\footnotesize $a_2$};
\node at (-8,-2.5) {\footnotesize $b_1$};
\node at (-4,-2.5) {\footnotesize $b_2$};
\node at (0,-2.5) {\footnotesize $b_3$};
\node at (4,-2.5) {\footnotesize $b_1$};
\node at (8,-2.5) {\footnotesize $b_2$};
\begin{scope}[xshift=9.6cm,yshift=-1.5cm]
\draw[very thick,<->,rotate=45] (-0.5,-0.5) -- (-0.5,-1.5) -- (0.5,-1.5);
\node at (1.8,-0.2) {$\si^+$};
\node at (0,-0.2) {$\si^-$};
\end{scope}
\end{tikzpicture}
\caption{\footnotesize A possible discretization with three chiral and three anti-chiral defects ($n_+=3, n_-=3$).}
\label{f2-2} 
\end{center}
\end{figure}

\begin{figure}
\begin{center}
\begin{tikzpicture}[scale=0.65]
\node at (-10,0) {\footnotesize $J_{-,m-2}$};
\node at (-8,0) {\footnotesize $J_{+,m-2}$};
\node at (-6,0) {\footnotesize $J_{-,m-1}$};
\node at (-4,0) {\footnotesize $J_{+,m-1}$};
\node at (-2,0) {\footnotesize $J_{-,m}$};
\node at (0,0) {\footnotesize $J_{+,m}$};
\node at (2,0) {\footnotesize $J_{-,m+1}$};
\node at (4,0) {\footnotesize $J_{+,m+1}$};
\node at (6,0) {\footnotesize $J_{-,m+2}$};
\node at (8,0) {\footnotesize $J_{+,m+2}$};
\draw[red] (-10,-2) -- (-6,2);
\draw[blue] (-6,-2) -- (-2,2);
\draw[green] (-2,-2) -- (2,2);
\draw[red] (2,-2) -- (6,2);
\draw[blue] (6,-2) -- (9,1);
\draw[red] (-8,-2) -- (-11,1);
\draw[blue] (-4,-2) -- (-8,2);
\draw[green] (0,-2) -- (-4,2);
\draw[red] (4,-2) -- (0,2);
\draw[blue] (8,-2) -- (4,2);
\node at (-10,-2.5) {\footnotesize $a_1$};
\node at (-6,-2.5) {\footnotesize $a_3$};
\node at (-2,-2.5) {\footnotesize $a_2$};
\node at (2,-2.5) {\footnotesize $a_1$};
\node at (6,-2.5) {\footnotesize $a_3$};
\node at (-8,-2.5) {\footnotesize $b_1$};
\node at (-4,-2.5) {\footnotesize $b_3$};
\node at (0,-2.5) {\footnotesize $b_2$};
\node at (4,-2.5) {\footnotesize $b_1$};
\node at (8,-2.5) {\footnotesize $b_3$};
\begin{scope}[xshift=9.6cm,yshift=-1.5cm]
\draw[very thick,<->,rotate=45] (-0.5,-0.5) -- (-0.5,-1.5) -- (0.5,-1.5);
\node at (1.8,-0.2) {$\si^+$};
\node at (0,-0.2) {$\si^-$};
\end{scope}
\end{tikzpicture}
\caption{\footnotesize  An alternate discretization with three chiral and three anti-chiral defects ($n_+=3, n_-=3$).}
\label{f2-3} 
\end{center}
\end{figure}
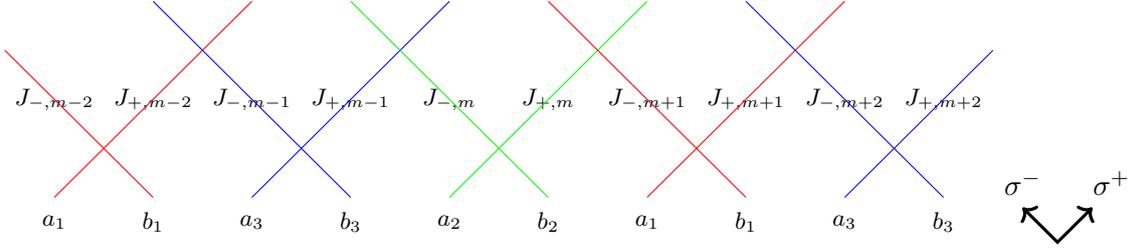

Care is needed when we have multiple chiral or anti-chiral defects (i.e.\ $n_+>1$ or $n_->1$). In this case, 
all the chiral (anti-chiral) defects are located on the same lightcone ray in the two-dimensional plane,
which looks singular as a two-dimensional picture. One should keep in mind, however, that the surface defects are separated along the spectral curve $C$, and there are no actual overlaps or divergences. Moreover, the 4d CS theory is topological along the
direction $\Sigma$, and hence we can freely move around the 1d defects without affecting any physical answers, to obtain a statistical lattice as in Figure \ref{f2-1}.
Note that this means in addition that we can permute the chiral (or anti-chiral) defects without affecting any physics (see Figures \ref{f2-2} and \ref{f2-3} for examples). If one wishes to obtain a statistical lattice, we can for example choose 
\begin{align}
    &\textrm{(chiral defect)}: \si^+_{\alpha}=\left( m + \frac{\alpha}{n_+} \right) \Delta  \ , \qquad  m\in\mathbb Z\,, \\
    &\textrm{(anti-chiral defect)}:  \si^-_{\alphabar} =\left( n + \frac{\alphabar}{n_-} \right) \Delta \ ,\qquad  n\in\mathbb Z\, .
\end{align}

\begin{figure}
\begin{center}
\begin{tikzpicture}[scale=0.65]
\node at (-10,0) {\footnotesize $J_{-,m-2}$};
\node at (-8,0) {\footnotesize $J_{+,m-2}$};
\node at (-6,0) {\footnotesize $J_{-,m-1}$};
\node at (-4,0) {\footnotesize $J_{+,m-1}$};
\node at (-2,0) {\footnotesize $J_{-,m}$};
\node at (0,0) {\footnotesize $J_{+,m}$};
\node at (2,0) {\footnotesize $J_{-,m+1}$};
\node at (4,0) {\footnotesize $J_{+,m+1}$};
\node at (6,0) {\footnotesize $J_{-,m+2}$};
\node at (8,0) {\footnotesize $J_{+,m+2}$};
\draw[red] (-10,-2) -- (-6,2);
\draw[green] (-6,-2) -- (-2,2);
\draw[red] (-2,-2) -- (2,2);
\draw[green] (2,-2) -- (6,2);
\draw[red] (6,-2) -- (9,1);
\draw[green] (-8,-2) -- (-11,1);
\draw[red] (-4,-2) -- (-8,2);
\draw[green] (0,-2) -- (-4,2);
\draw[red] (4,-2) -- (0,2);
\draw[green] (8,-2) -- (4,2);
\node at (-10,-2.5) {\footnotesize $a_1$};
\node at (-6,-2.5) {\footnotesize $a_2$};
\node at (-2,-2.5) {\footnotesize $a_1$};
\node at (2,-2.5) {\footnotesize $a_2$};
\node at (6,-2.5) {\footnotesize $a_1$};
\node at (-8,-2.5) {\footnotesize $b_1$};
\node at (-4,-2.5) {\footnotesize $b_2$};
\node at (0,-2.5) {\footnotesize $b_1$};
\node at (4,-2.5) {\footnotesize $b_2$};
\node at (8,-2.5) {\footnotesize $b_1$};
\begin{scope}[xshift=9.6cm,yshift=-1.5cm]
\draw[very thick,<->,rotate=45] (-0.5,-0.5) -- (-0.5,-1.5) -- (0.5,-1.5);
\node at (1.8,-0.2) {$\si^+$};
\node at (0,-0.2) {$\si^-$};
\end{scope}
\end{tikzpicture}
\caption{\footnotesize A discretization with two chiral and two anti-chiral defects ($n_+=2$, $n_-=2$).}
\label{4} 
\end{center}
\end{figure}
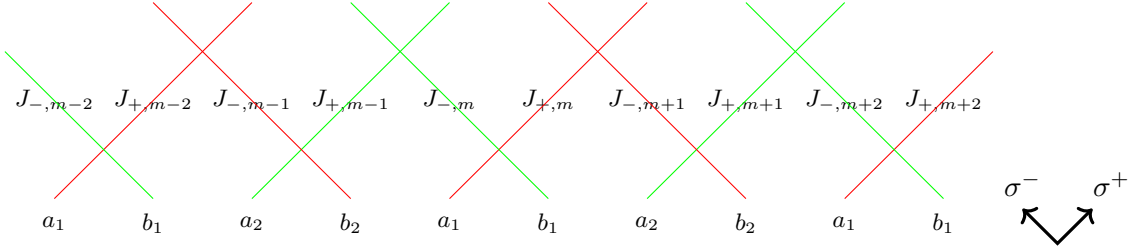

\subsection{Quantization into Spin Chain} 

Having discussed the discretization of the theory,
we can now discuss the quantization of the resulting discretized theory (whose Lagrangian we already know).
The complete path integral now takes the form\footnote{We are in the Lorentzian signature, where the weight for the path integral is defined by $\exp(iS)$ for an action $S$.}
\begin{equation}
\begin{aligned}
    &\int \mathcal{D} A \exp\bigg(\frac{i}{2\pi \hbar} \int_{\cM \times C} \d z\wedge  \textrm{CS}(A)\bigg) \\ 
    \times 
    &\prod_{\alpha=1}^{n_+}\prod_{i}^N \int \mathcal{D}\phi^{\alpha}_i
    \exp \bigg( \frac{i}{\hbar_{\rm 1d}}\int_{ \{\sigma_i^+\} \times \mathbb{R}  \times \{z_+\} }d\sigma^- \mc{L}\left(\phi^{\alpha}_i , \partial_-\phi^{\alpha}_i; A_{-} |_{\sigma_i^+,z_+}\right) \bigg)\\
     \times 
     &\prod_{\alphabar=1}^{n_-}\prod_{j}^N  \int \mathcal{D}\phibar^{\alphabar}_{j}
     \exp \bigg( \frac{i}{\hbar_{\rm 1d}}\int_{ \mathbb{R} \times \{\sigma_j^-\} \times \{z_-\} }d\sigma^+ \mc{L}\left(\phibar^{\alphabar}_j, \partial_+\phibar^{\alphabar}_j ;A_+|_{\sigma_j^-,z_-}\right)\bigg) \;.
    \end{aligned}
\end{equation}

To quantize the discretized integrable field theory, we ought to replace the Poisson brackets in \eqref{discp} by commutators, i.e., by making the replacement $\{\ ,\ \}\rightarrow [\ ,\ ]$\,. This gives us the canonical commutator
\ie\relax
[J_{\pm,m}^a,J_{\pm,n}^b]=\frac{1}{\Delta} f^{ab}{}_c J^c_{\pm,n}\delta_{mn}\,,
\fe
and hence each of the modes 
$J_{\pm,m}^a$ is in the adjoint representation of $\mathfrak{g}$.

\subsubsection*{Duality with Wilson Lines and Spin Chains}

\begin{figure}[htbp]
\centering
\begin{tikzpicture}
    \node[rectangle, draw, align=center](CA) at (0,4) {Coadjoint Orbit \\ Defect\\ (Section \ref{sec:coadjoint})};
    \node[rectangle, draw, align=center](WL) at (0, 0){Wilson Line};
    \node[rectangle, draw, align=center](FF) at (-5, -3) {Free Fermion\\ Defect\\ (Section \ref{sec:free_fermion})};
    \node[rectangle, draw, align=center](BG) at (5,-3){Free/Curved \\ $\beta\gamma$ Defect\\ (Sections \ref{sec:fbg-sec}, \ref{sec:curved_betagamma}
    and \ref{sec:curved-bg})}; 
    \draw [<->]  (CA)--(WL) node[midway] {Section \ref{sec:WL-CA}};
    \draw [<->]  (FF)--(WL) node[midway, right] {Section \ref{sec:WL-FF}};
    \draw [<->]  (BG)--(WL) node[midway, left] {Section \ref{sec:WL-BG}};
    \draw [<->]  (CA)--(BG) node[midway, right] {Section \ref{sec:coadjointbeta}};
    \draw [<->]  (FF)--(BG) node[midway, below] {Section \ref{sec:fermion_betagamma}};
    \draw [<->]  (CA)--(FF) node[midway, left]{};
    \end{tikzpicture}
    \caption{Many of the 1d defects can be turned into Wilson lines, as we will discuss in later sections of this paper. We will also discuss the relations between the defects directly.}
    \label{1d_as_WL}
\end{figure}
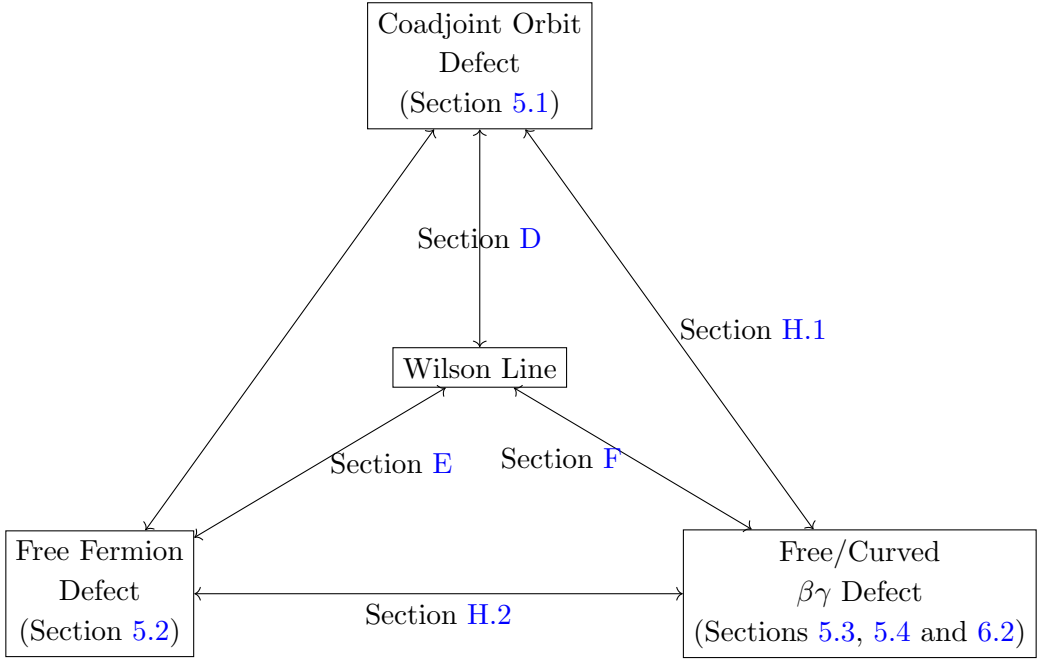

We thus arrive at a 4d CS theory path integral with insertions of line operators described by path integrals of 1d theories. As we shall see in concrete examples in Section \ref{sec:disc_defect} and Section \ref{sec:VA}, these 1d systems can often be identified with gauge-invariant Wilson lines using a convenient quantization scheme, once we adopt the periodic boundary conditions on the lattice.
In such cases, we find the path integral to be of the form
\begin{equation}
\begin{aligned}
    &\int \mathcal{D}A \exp\bigg(\frac{i}{2\pi \hbar} \int_{ \mathbb{C} \times C} \d z\wedge  \textrm{CS}(A)\bigg) \\ \times 
    &\prod_{\alpha=1}^{n_+}\prod_{i=1}^{N} \textrm{Tr}_{\rho_{+, i}}\mathcal{P}\exp \bigg( \frac{1}{\hbar_{\rm 1d}}\int_{\{\sigma_i^+\} \times  S^1_{-} \times  \{z_+\} }d\sigma^-  A_{-} |_{\sigma_i^+,z_+} \bigg)\\
     \times &\prod_{\alphabar=1}^{n_-}
    \prod_{j=1}^{N}\textrm{Tr}_{\rho_{-,j}}\mathcal{P}\exp \bigg( \frac{1}{\hbar_{\rm 1d}}\int_{S^1_{+} \times \{\sigma_j^-\} \times \{ z_-\} }d\sigma^+  A_+|_{\sigma_j^-,z_-}\bigg) \;.
    \end{aligned}
\end{equation}
Here, the traces are taken in representations $\rho_{+,i}$ and $\rho_{-,j}$. In other words,
the discretized defects can be turned into Wilson lines, as shown in Figure \ref{1d_as_WL}.

Since the 1d defects are identified with Wilson lines,
the resulting setup is an integrable lattice model made up of Wilson lines, which is exactly the setup for integrable lattice models reviewed in Section \ref{sec:lattice-cs}.
To simplify the presentation let us specialize to the case $n_{+}=n_{-}=1$.
The Hilbert space $\mathcal{H}$ of the discretized integrable field theory is defined in terms of tensor products of modules for representations of $\mathfrak{g}$:
\begin{equation}\label{hilb}
    \mathcal{H}=\bigotimes_{i,j} (V_{+,i} \otimes V_{-,j})\,,
\end{equation}
where we denoted representation space for $\rho_{+,i}$ ($\rho_{-,j}$) by $V_{+,i}$ ($V_{-,j}$).
As in Figure \ref{4} 
the chiral and anti-chiral defects are placed at even and odd sites of a spin chain (that arises from the lattice at a fixed point in time), respectively, which leads to the
natural labeling 
\begin{equation}\label{hilb2}
    \mathcal{H}=\bigotimes_{n} V_n\,,
\end{equation}
with $V_{2m-1}:=V_{+,i}$ and $V_{2m}:=V_{-,j}$
and correspondingly $\rho_{2m-1}:=\rho_{+,i}$ and $\rho_{2m}:=\rho_{-,j}$.

Note that the label $i$ in \eqref{hilb} runs over all the integers and $\mathcal{H}$ is infinite-dimensional.
Instead, we often impose the periodic boundary condition along the spatial direction,
so that we have $V_{n+2N} = V_n$ for an integer $N$, so that we obtain a spin chain with 
the periodic boundary condition.

\subsubsection*{R-matrix versus Lax Connection}

Having defined the spin chain, we can now discuss the 
relations of the integrability of the 
field theories and that of the discretized spin chains.

As we have already discussed around \eqref{lax}, the integrability of quantum field theory 
originates from the existence of the Wilson line along the spatial direction.
Note that this Wilson line is included in addition to the Wilson lines discussed above along the lightcone direction.
Assuming that this auxiliary Wilson line is in the representation $\rho_A$ 
with associated representation space $V_A$ (``$A$'' for ``auxiliary''),
we now have an enlarged Hilbert space 
$\mathcal{H}\otimes V_A$.

\begin{figure}
\begin{center}
\begin{tikzpicture}[scale=0.63]
\draw[gray,->] (-10,-2) -- (-6,2);
\draw[gray,->] (-8,-2) -- (-11,1);
\draw[gray,->] (-6,-2) -- (-2,2);
\draw[gray,->] (-4,-2) -- (-8,2);
\draw[gray,->] (-2,-2) -- (2,2);
\draw[gray,->] (2,-2) -- (6,2);
\draw[gray,->] (6,-2) -- (9,1);
\draw[gray,->] (0,-2) -- (-4,2);
\draw[gray,->] (4,-2) -- (0,2);
\draw[gray,->] (8,-2) -- (4,2);
\draw[very thick, black, ->] (-11.5,0) -- (10,0)  ;
\node at (10,0.4) {\footnotesize $z$};
\node at (-10,-2.5) {\footnotesize $-\nu$};
\node at (-6,-2.5) {\footnotesize $-\nu$};
\node at (-2,-2.5) {\footnotesize $-\nu$};
\node at (2,-2.5) {\footnotesize $-\nu$};
\node at (6,-2.5) {\footnotesize $-\nu$};
\node at (-8,-2.5) {\footnotesize $\nu$};
\node at (-4,-2.5) {\footnotesize $\nu$};
\node at (0,-2.5) {\footnotesize $\nu$};
\node at (4,-2.5) {\footnotesize $\nu$};
\node at (8,-2.5) {\footnotesize $\nu$};
\begin{scope}[xshift=9.6cm,yshift=-1.5cm]
\draw[very thick,<->,rotate=45] (-0.5,-0.5) -- (-0.5,-1.5) -- (0.5,-1.5);
\node at (1.8,-0.2) {$\si^+$};
\node at (0,-0.2) {$\si^-$};
\end{scope}
\end{tikzpicture}
\caption{\footnotesize Wilson line configuration giving rise to inhomogeneous spin chain with alternating inhomogeneities.}
\label{1-1} 
\end{center}
\end{figure}
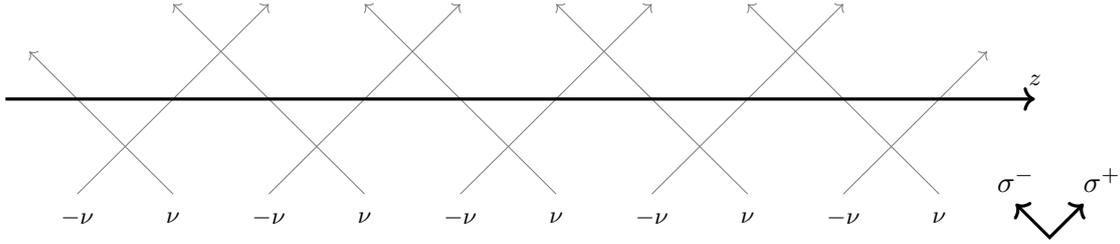

This additional Wilson line $W_A$ gives rise to the configuration shown in Figure \ref{1-1},
where $W_A$ crosses with the other Wilson lines along the lightcone directions.
Since the interactions localized at the crossing of the two Wilson lines generate the R-matrix (as explained in Section \ref{sec:lattice-cs}),
we obtain the product of R-matrices along $W_A$: 
\begin{equation} \label{T_A}
    T(z) = \cdots R_{\rho_A, \rho_-}(z-z_-)R_{\rho_A, \rho_+}(z-z_+)R_{\rho_A, \rho_-}(z-z_-)R_{\rho_A, \rho_+}(z-z_+)\cdots\,.
\end{equation}
(There are corrections to this formula from the framing anomaly, which however are not important for the understanding of what follows;
see Appendix \ref{sec:framing_anomaly}.)
This product is known as the total monodromy matrix, and its matrix elements are operators that act on the Hilbert space, $\mathcal{H}$ defined in \eqref{hilb}, and the matrix itself acts on the auxiliary space $V_A$.

The total monodromy matrix is one of the central ingredients for the quantum inverse scattering method for spin chains.
Let us consider two parallel Wilson lines $W_A$ and $W_{A'}$ along the spatial directions, and let us compactify the spatial direction
so that we are taking the trace of the monodromy matrix along the auxiliary Hilbert space $V_A$.
We can use the topological invariance along $\cM$ to exchange the two Wilson lines, realizing the Yang-Baxter equation along the way. This implies that 
\ie\relax
    [\Tr_{A} T(z),\Tr_{A'} T(z')]=0 \;.
\label{iuu2}
\fe
The conserved charges include the lightcone components of energy and momentum. These are unitary operators that can be expressed as 
\ie
    U_+=\Tr_AT(\nu)\,,\qquad U_-^\dagger=\Tr_AT(-\nu)\,.\label{lc-tmat}
\fe
These operators generate discrete shifts along the lightcone directions, i.e., $U_+$ generates shifts of the form 
$\si^+\to \si^++\Delta$ and $U_-^{\dagger}$ generates shifts of the form $\si^-\to \si^--\Delta$.

Now let us discuss how the monodromy matrix for the spin chain is related to the monodromy operator 
before discretization. 
It turns out that we can make identifications
\ie
    &\mathcal{P}\, {\text{exp}}\Big[-\int_{m\Delta}^{(m-1)\Delta}d\si^-\,\Lax_-(z)\Big]\sim R_{\rho_A, \rho_{2m-1}}(z,\nu) \,,\\
    &\mathcal{P}\,{\text{exp}}\Big[-\int_{m\Delta}^{(m+1)\Delta}d\si^+\,\Lax_+(z)\Big] \sim R_{\rho_A, \rho_{2m}}(z,-\nu) \,,
    \label{bxx2}
\fe
 where $R_{\rho_1, \rho_2}(z_1,z_2)$ is the R-matrix in the tensor product $\rho_1\otimes \rho_2$ satisfying the Yang-Baxter equation \eqref{ybe1}. 
 By taking a product over these matrices, we obtain the relation between the monodromy operators:
\ie
    &\mathcal{P}\, 
    {\text{exp}}\Big[-\int_{C}\,\Lax(z)\Big]\sim \prod_m  R_{\rho_A, \rho_{2m-1}}(z,\nu) R_{\rho_A, \rho_{2m}}(z,-\nu) =T(z, \nu)\,.
\fe
where the contour $C$ is along the $\sigma$-direction.
Note that the monodromy is invariant under smooth deformations of the contour, since the connection $\mathcal{L}$ is flat. When we compactify the $\sigma$-direction, $C$ wraps along the winding cycle along $\sigma$.

 The identification \eqref{bxx2} is very natural in the 4d CS theory, since the monodromy operators
before and after discretization are the same operator, i.e, the Wilson line.

This identification  suggests that one can quantize the Lax connection \eqref{lax}
into a well-defined operator in the quantized Hilbert space of the 2d integrable field theory.
This will imply the quantum integrability of the integrable field theory. While we leave a complete discussion of 
such a quantization for future work, we can already check the relation 
 \eqref{bxx2} in the semiclassical limit.  For this purpose, let us expand the left-hand side of \eqref{bxx2}. One then obtains (recall \eqref{Lax-ultra} for the rational case, which we discuss here)
\ie
    &\mathcal{P}\,{\text{exp}}\Big[-\int_{m\Delta}^{(m-1)\Delta}d\si^-\,\Lax_-(z)\Big]= 1 +  \Delta\hbar \, \frac{1}{z-\nu} J^a_{-,m} \rho_{A}(t_a)+ \ldots \,,\\
    &\mathcal{P}\, {\text{exp}}\Big[-\int_{m\Delta}^{(m+1)\Delta}d\si^+\,\Lax_+(z)\Big] =  1 +   \Delta\hbar \, \frac{1}{z+\nu}  J^a_{+,m}\rho_{A}(t_a)+ \ldots\,.
    \label{bxx}
\fe 
Now, the discretized current can be regarded as an operator acting on the auxiliary Hilbert space as 
\ie\label{currentmat}
    J_{-,m}&=J_{-,m}^a \rho_A(t_a)=\frac{1}{\Delta}  \rho_{-,m}(t^a)\otimes \rho_A (t_a)\,,\\
    J_{+,m}&= J_{+,m}^a \rho_A(t_a)=\frac{1}{\Delta} \rho_{+,m}(t^a)\otimes \rho_A(t_a)\,.
\fe
By using this relation for the current,
we obtain
\ie
    &\mathcal{P}\, {\text{exp}}\Big[-\int_{m\Delta}^{(m-1)\Delta}d\si^-\,\Lax_-(z)\Big]= 1 +  \hbar \,r_{\rho_A, \rho_{-,m}}(z,\nu)  +\ldots \,,\\
    &\mathcal{P}\, {\text{exp}}\Big[-\int_{m\Delta}^{(m+1)\Delta}d\si^+\,\Lax_+(z)\Big] =  1 +   \hbar\, r_{\rho_A, \rho_{+,m}}(z,-\nu) +\ldots\,,
    \label{bxx3}
\fe 
which coincides with the expansion of the right-hand side of \eqref{bxx2} (recall \eqref{R_r} and \eqref{R_to_r}).

\subsection{Going Backwards: Thermodynamic Limit of Lattice Models} \label{sec:backwards}

In the previous subsections we discretized the field theory and then quantized the resulting lattice model.
Let us now discuss the opposite direction,
namely to discuss the thermodynamic limit of the lattice model, where we will be led back to the realm of 
integrable field theories.

We can go back to the field theory\footnote{To go back to the original field theory before discretization, one needs to ensure $\xi \gg \Delta$, where $\xi$ is the typical correlation length of the theory. This is ensured when the system is near critically, while more care is needed for general massive theories.} by taking the thermodynamic limit
\begin{align} \label{limit_1}
    \Delta \to 0\,, \quad \hbar_{\rm 1d}: \,\, \textrm{fixed} \,.
\end{align}
This limit sends the 2d Planck constant as $\hbar_{\rm 2d} = \Delta \hbar_{\rm 1d}\to 0$, and hence is the semiclassical limit of the two-dimensional field theory.\footnote{When we are interested in 
the quantization of two-dimensional field theories, 
we can also consider a double-scaling limit
\begin{align}
    \Delta \to 0, \quad \hbar_{\rm 1d}\to \infty, \quad
    \hbar_{\rm 2d} : \,\, \textrm{finite} \,,
\end{align}
so that the 2d Planck constant is kept finite.}
In the following we discuss the limit \eqref{limit_1} and explain the relation between the integrable structures before and after the limit.

\subsubsection*{Discretized Poisson Brackets from RLL Relation}

In the quantum spin chain, the integrability of the spin chain is described by the so-called 
RLL relation, where $`L'$ in this context denotes the Lax connection of the one-dimensional spin chain:
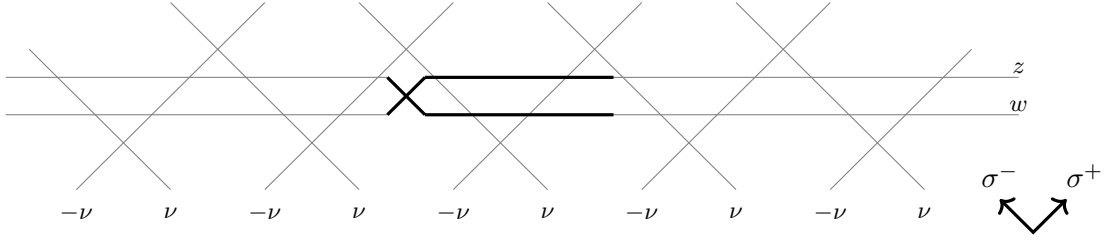
\begin{figure}
\begin{center}
\begin{tikzpicture}[scale=0.62]
\draw[gray] (-10,-2) -- (-6,2);
\draw[gray] (-8,-2) -- (-11,1);
\draw[gray] (-6,-2) -- (-2,2);
\draw[gray] (-4,-2) -- (-8,2);
\draw[gray] (-2,-2) -- (2,2);
\draw[gray] (2,-2) -- (6,2);
\draw[gray] (6,-2) -- (9,1);
\draw[gray] (0,-2) -- (-4,2);
\draw[gray] (4,-2) -- (0,2);
\draw[gray] (8,-2) -- (4,2);
\draw[gray] (-11.5,0.4) -- (-3.4,0.4)  ;
\draw[gray] (-11.5,-0.4) -- (-3.4,-0.4)  ;
\draw[gray] (-2.6,0.4) -- (10,0.4)  ;
\draw[gray] (-2.6,-0.4) -- (10,-0.4)  ;
\draw[very thick, black] (-3.4,-0.4) -- (-2.6,0.4)  ;
\draw[very thick, black] (-3.4,0.4) -- (-2.6,-0.4)  ;
\draw[very thick, black] (-2.6,-0.4) -- (1.4,-0.4)  ;
\draw[very thick, black] (-2.6,0.4) -- (1.4,0.4)  ;
\node at (10,0.6) {\footnotesize $z$};
\node at (10,-0.2) {\footnotesize $w$};
\node at (-10,-2.5) {\footnotesize $-\nu$};
\node at (-6,-2.5) {\footnotesize $-\nu$};
\node at (-2,-2.5) {\footnotesize $-\nu$};
\node at (2,-2.5) {\footnotesize $-\nu$};
\node at (6,-2.5) {\footnotesize $-\nu$};
\node at (-8,-2.5) {\footnotesize $\nu$};
\node at (-4,-2.5) {\footnotesize $\nu$};
\node at (0,-2.5) {\footnotesize $\nu$};
\node at (4,-2.5) {\footnotesize $\nu$};
\node at (8,-2.5) {\footnotesize $\nu$};
\begin{scope}[xshift=9.6cm,yshift=-1.5cm]
\draw[very thick,<->,rotate=45] (-0.5,-0.5) -- (-0.5,-1.5) -- (0.5,-1.5);
\node at (1.8,-0.2) {$\si^+$};
\node at (0,-0.2) {$\si^-$};
\end{scope}
\end{tikzpicture}
\caption{\footnotesize The configuration that realizes the LHS of \eqref{fcr}.}
\label{2} 
\end{center}
\end{figure}

\begin{equation}
    R_{12}(z-w)L_{n,1}(z)L_{n,2}(w)=L_{n,2}(w)L_{n,1}(z)R_{12}(z-w)\ ,
\label{fcr}
\end{equation}

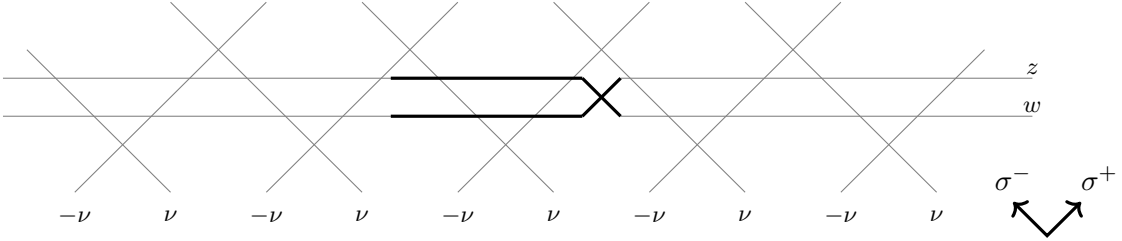
\begin{figure}
\begin{center}
\begin{tikzpicture}[scale=0.63]
\draw[gray] (-10,-2) -- (-6,2);
\draw[gray] (-8,-2) -- (-11,1);
\draw[gray] (-6,-2) -- (-2,2);
\draw[gray] (-4,-2) -- (-8,2);
\draw[gray] (-2,-2) -- (2,2);
\draw[gray] (2,-2) -- (6,2);
\draw[gray] (6,-2) -- (9,1);
\draw[gray] (0,-2) -- (-4,2);
\draw[gray] (4,-2) -- (0,2);
\draw[gray] (8,-2) -- (4,2);
\draw[gray] (-11.5,0.4) -- (0.6,0.4)  ;
\draw[gray] (-11.5,-0.4) -- (0.6,-0.4)  ;
\draw[gray] (1.4,0.4) -- (10,0.4)  ;
\draw[gray] (1.4,-0.4) -- (10,-0.4)  ;
\draw[very thick, black] (1.4,-0.4) -- (0.6,0.4)  ;
\draw[very thick, black] (1.4,0.4) -- (0.6,-0.4)  ;
\draw[very thick, black] (0.6,-0.4) -- (-3.4,-0.4)  ;
\draw[very thick, black] (0.6,0.4) -- (-3.4,0.4)  ;
\node at (10,0.6) {\footnotesize $z$};
\node at (10,-0.2) {\footnotesize $w$};
\node at (-10,-2.5) {\footnotesize $-\nu$};
\node at (-6,-2.5) {\footnotesize $-\nu$};
\node at (-2,-2.5) {\footnotesize $-\nu$};
\node at (2,-2.5) {\footnotesize $-\nu$};
\node at (6,-2.5) {\footnotesize $-\nu$};
\node at (-8,-2.5) {\footnotesize $\nu$};
\node at (-4,-2.5) {\footnotesize $\nu$};
\node at (0,-2.5) {\footnotesize $\nu$};
\node at (4,-2.5) {\footnotesize $\nu$};
\node at (8,-2.5) {\footnotesize $\nu$};
\begin{scope}[xshift=9.6cm,yshift=-1.5cm]
\draw[very thick,<->,rotate=45] (-0.5,-0.5) -- (-0.5,-1.5) -- (0.5,-1.5);
\node at (1.8,-0.2) {$\si^+$};
\node at (0,-0.2) {$\si^-$};
\end{scope}
\end{tikzpicture}
\caption{\footnotesize The configuration that realizes the RHS of \eqref{fcr}.}
\end{center}
\end{figure}
This relation, known as the RLL relation, is a version of the Yang-Baxter equation,
and is part of the defining relations of the infinite-dimensional algebra (such as the Yangian for the rational case $C=\mathbb{C}$) governing integrable models (see \cite{Costello:2018gyb} for discussion in the context of the 4d CS theory). For later purposes, we rewrite the RLL relation as
\begin{align}
\begin{split}
    [L_{n,1}(z),L_{n,2}(w)]&=\big(1-R_{12}(z-w)\big)L_{n,1}(z)L_{n,2}(w)\\ & \qquad -L_{n,2}(w)L_{n,1}(z)\big(1-R_{12}(z-w)\big)\ .
\end{split}
   \label{fcr2}
\end{align}

We can now discuss the semiclassical limit of the RLL relation.
First, the semiclassical limit of the R-matrix $R$ is given by the classical r-matrix $r$
\begin{equation}\label{limit1}
    R(z)\longrightarrow1+i\hbar \, r(z)+\cdots \,.
\end{equation}
Similarly, the L-operator is also associated with the crossings of Wilson lines, and
we obtain
\begin{equation}\label{limit2}
    L_{n,A}(z)\longrightarrow 1+\frac{i\hbar\,c_{2n-1,A}}{z+\nu}+\frac{i\hbar\,c_{2n,A}}{z-\nu}+\cdots\,,
\end{equation}
where $c_{m,A}=\rho_m(t^a)\otimes \rho_A(t^a)$.
Moreover, we can straightforwardly generalize the relation \eqref{bxx2} for the $R$-matrix into 
that for the $L$-operator:
\begin{equation}
    L_{n,A}(z) \sim \mathcal{P}\,{\text{exp}}\Big[-\int_{n\Delta }^{(n+1)\Delta }dx\,\Lax(x;z)\Big]=1-\Delta\Lax_n(z)+\cdots\,,
\end{equation}
so that  we have
\begin{equation}\label{limit3}
    [L_{n,1}(z) \overset{ \otimes}{,} L_{n,2}(w)]\longrightarrow -i\hbar^2\Delta^2\{\Lax_{n,1}(z),\Lax_{n,2}(w)\}\ ,
\end{equation}
in the semiclassical limit.
By substituting the semiclassical expressions \eqref{limit1}, \eqref{limit2}, \eqref{limit3} into
\eqref{fcr2}, we obtain
\begin{equation}
    \{\Lax_{n,1}(z) , \Lax_{n,2}(w)\}= \frac1\Delta[r_{12}(z-w),\Lax_{n,1}(z)+\Lax_{n,2}(w)]\ ,
\end{equation}
or equivalently
\begin{equation}
    \{\Lax_{n}(z) \overset{ \otimes}{,}\Lax_{n}(w)\}= \frac1\Delta[r_{12}(z-w),\Lax_{n}(z) \otimes 1 + 1\otimes \Lax_{n,2}(w)]\ .
\end{equation}
This is nothing but the discrete version of the Poisson bracket algebra of the integrable field theories we considered previously in \eqref{PB-Lax}.

\section{Discretizations of Surface Defects}
  \label{sec:disc_defect}

We now turn to concrete examples of 2d defects. 
We will in turn discuss the coadjoint orbit defect (Section \ref{sec:coadjoint}),
free fermion defect (Section \ref{sec:free_fermion}), 
free $\beta\gamma$ system defect (Section \ref{sec:fbg-sec})
and finally curved $\beta\gamma$ system defect (Section \ref{sec:curved_betagamma}).

In order to simplify the presentation, we will in this section 
mostly discuss the case of one chiral ($n_{+}=1$) and
one anti-chiral ($n_{-}=1$) defect,
and drop the corresponding indices $\alpha$ ($\alphabar$) labeling different 
chiral (anti-chiral) defects. The generalization to 
multiple chiral and anti-chiral defects is straightforward, and we have already described the 
general results in Section \ref{sec:disc_FT}. In addition, although we shall often use integrable field theories of rational type to elucidate our results, our techniques also apply to trigonometric and elliptic integrable field theories, as explained in Appendix \ref{sec:trig_ell}.

\subsection{Coadjoint Orbit Defects}\label{sec:coadjoint}

In this subsection, we shall investigate the discretization of coadjoint orbit defects, which are the defects required for the realization of the Faddeev-Reshetikhin model (which is reviewed in Appendix \ref{sec:frapp}) and its generalizations.

\subsubsection{Coadjoint Orbit Defect and Faddeev-Reshetikhin Model}

The actions of the chiral and anti-chiral coadjoint orbit surface defects respectively take the form
\ie
    S^{\textrm{2d-coadj}}_{+} = \int_{\Sigma \times \{z_{+}\} }\Tr\left(\Lambda \cG_{(+)}^{-1}D_-\cG_{(+)}\right)\,d\sigma^+ d\sigma^-,
\fe
and 
\ie
    S^{\textrm{2d-coadj}}_{-} = \int_{\Sigma \times \{z_{-}\} }\Tr\left(\Lambda \cG_{(-)}^{-1}D_{+}\cG_{(-)}\right)\,d\sigma^+ d\sigma^-,
\fe
where $z_+$ and $z_-$ indicate the locations of the defects on $C$, $D_{\pm}=\partial_{\pm} + A_{\pm}$ is a covariant derivative along the lightcone directions, $\mathcal{G}_{(\pm)}$ are $G$-valued fields, and $\Lambda \in \mathfrak{h}$, where $\mathfrak{h}$ is the Cartan subalgebra of the Lie algebra $\mathfrak{g}$ that generates $G$.   
We can check the gauge invariance of this action under local transformations of the form\footnote{Note that gauge invariance itself does not require that $\Lambda\in \mathfrak{h}$, instead it can be any element of $\mathfrak{g}$.} 
\ie 
    \mathcal{G}_{(\pm)} \mapsto U \mathcal{G}_{(\pm)}\,,\qquad A \mapsto U A U^{-1}-d U U^{-1} \, .
\fe
The Lagrangian of these surface defects are similar in the form to the Lagrangian used to describe Wilson loops as a path integral over degrees of freedom describing the orbit $\cG_{(\pm)}\Lambda\cG_{(\pm)}^{-1}$, which justifies the terminology we use for these surface defects. 
In the case where $G$ is compact, this description of Wilson loops is natural from the perspective of the one-to-one correspondence between 
irreducible finite dimensional representations of $G$ and integral coadjoint orbits  \cite{kirillov2004lectures}. To be precise, an irreducible representation of $G$ is determined uniquely by a highest weight $\lambda \in \mathfrak{h}^*$, and associated to each such $\lambda$ is an orbit of the coadjoint action in the space $\mathfrak{g}^*$.
For these loop operators, one may also consider a separate class of gauge transformations of the form 
\ie
    \mathcal{G}_{(\pm)} \mapsto \mathcal{G}_{(\pm)} \mathcal{T}^{-1},
\fe 
where $\mathcal{T}\in H_\Lambda$, the subgroup of $G$ that preserves $\Lambda$ under the adjoint action, i.e., 
\ie 
    H_\Lambda=\left\{h \in G ; h \Lambda h^{-1}=\Lambda\right\}.
\fe
Invariance of the action under a large gauge transformation in this subgroup may be enforced by requiring that $\Lambda$ can be identified with an integral weight \cite{Alekseev:1988vx,Alekseev:2015hda}.

\subsubsection{Lattice Discretization}

We shall now discretize these surface operators using the general procedure outlined in the previous section.
For the first surface operator located at $z_+$, we shall discretize it along the $\sigma^+$ direction, by setting
\ie
    S^{\textrm{2d-coadj}}_{+}& = \frac{1}{\hbar_{\rm 2d}}\int_{\Sigma\times \{z_+\} }\Tr\left(\Lambda \cG_{(+)}^{-1}D_{-} \cG_{(+)}\right)\,d\sigma^+ d\sigma^-\\
    & \quad 
    \to S^{\textrm{1d-coadj}}_{+} = \frac{1}{\hbar_{\rm 1d}}   \sum_i \int_{\{\sigma^+_i\} \times \mathbb{R}_{-} \times \{z_+\} }\Tr\left(\Lambda \cG_{i(+)}^{-1}D_{-} \cG_{i(+)}\right)\, d\sigma^-\,,
\fe
where we have used $\cG_{i(+)} $ and $\cG^{-1}_{i(+)} $ to respectively 
denote $\cG_{(+)}|_{\sigma_i^+}$ and $\cG^{-1}_{(+)}|_{\sigma_i^+} $.
We repeat the same procedure for another surface defect located at $z_-$\,, whereby we arrive at another infinite set of line defects
\ie 
    S^{\textrm{1d-coadj}}_{-} = \frac{1}{\hbar_{\rm 1d}} \sum_{j} \int_{\mathbb{R}_{+} \times \{\sigma^-_j\} \times \{ z_-\} }\Tr\left(\Lambda \cG_{j(-)}^{-1}D_{+} \cG_{j(-)}\right)\, d\sigma^+\,,
\fe
where  $\cG_{j(-)} $ and $\cG^{-1}_{j(-)} $ respectively
denotes $\cG_{(-)}|_{\sigma_j^-}$ and $\cG^{-1}_{(-)}|_{\sigma_j^-} $.

Given that the defect actions are exponentiated in the path integral, the partition function of the lattice model takes the form 
\begin{equation}\label{infmesh}
\begin{aligned}
    & Z^{\textrm{1d-coadj}}=\int \mathcal{D}A\, 
    \prod_{i=1}^{N}   \mathcal{D}\cG_{i(+)}    \prod_{j=1}^{N}   \mathcal{D}\cG_{j(-)}
    \exp\bigg(\frac{i}{2\pi \hbar} \int_{\Sigma \times C} \d z\,\wedge \textrm{CS}(A)\bigg) \\
    & \quad \times\exp\left(i\int_{ S^1_{-}\times \{z_+\} }\Tr\left(\Lambda \cG_{i(+)}^{-1}D_{-} \cG_{i(+)}\right)\, d\sigma^- 
   +i\int_{S^1_{+}\times  \{ z_-\} }\Tr\left(\Lambda \cG_{j(-)}^{-1}D_{+} \cG_{j(-)}\right)\, d\sigma^+ \right)\,,   
    \end{aligned}
\end{equation}
where we have adopted the periodic boundary conditions to obtain $S^1$ for both light cone directions.

We can re-express the line defects that appear in \eqref{infmesh}
 as Wilson line operators, as long as $\Lambda$ can be identified with the highest weight of an irreducible representation of $G$\,.\footnote{If one is interested in the compact form of $G_{\mathbb{C}}$ a reality condition such as
\ie\label{reality}
    \overline{\cG_{(\pm)}}&=\cG_{(\pm)}\,,\qquad 
    \overline{A|_{\mathbb{C}}}=A|_{\mathbb{C}}\,,
\fe 
can be imposed,
where $A|_{\mathbb{C}}$ denotes the components of the gauge field along $\mathbb{C}$. As a result, the surface defect fields and gauge fields along $\mathbb{C}$ are valued in the real subgroup $G$ and real subalgebra $\mathfrak{g}$ of $G_{\mathbb{C}}$ and $\mathfrak{g}_{\mathbb{C}}$, respectively. } This can be achieved via geometric quantization, or equivalently, via coherent state quantization. These results are reviewed in Appendix \ref{sec:WL-CA}.

Thus, the partition function of the coupled 4d--2d system of the 4d CS theory and the coadjoint orbit surface operators given in \eqref{action-defect} can be expressed as   
\begin{equation}\label{lattice}
\begin{aligned}
   Z^{\textrm{1d-coadj}}= &\int \mathcal{D}A \exp\bigg(\frac{i}{2\pi \hbar} \int_{\cM \times C} \d z\,\wedge \textrm{CS}(A)\bigg) \\ \times &\prod_{i=1}^{N} \textrm{Tr}_{\rho_-}\mathcal{P}\exp \bigg( i\int_{ S^1  }d\sigma^- \cA_{-,i}^{(z_+)} \bigg)\prod_{j=1}^{N}\textrm{Tr}_{\rho_+}\mathcal{P}\exp \bigg( i\int_{S^1  }d\sigma^+ \cA_{+,j}^{(z_-)}\bigg)\,,
    \end{aligned}
\end{equation}
where $\cA_{-,i}^{(z_+)}=A_-(\sigma,z)\lvert_{\sigma^+=\sigma_i^+,z=z_{+}}$ and $\cA_{+,i}^{(z_-)}=A_+(\sigma,z)\lvert_{\sigma^-=\sigma_i^-,z=z_{-}}$\,.
A discretization of the Faddeev-Reshetikhin model can thus be understood in terms of discretizations of coadjoint orbit surface defects to Wilson loops followed by a regularization
where the number of Wilson lines in both lightcone directions, $N$, is taken to be finite. The representations of these Wilson lines are determined by the highest weight that corresponds to $\Lambda$.

\subsubsection*{Connections to Other Works}

The Faddeev-Reshetikhin model can be derived from the 4d CS theory 
by coupling the 4d CS theory to two coadjoint orbit defects, each supporting fields valued in the complexified Lie group $G_{\mathbb{C}}$, denoted $\cG_{(+)}$ and $\cG_{(-)}$ \cite{Fukushima:2020tqv}. 
The 4d--2d action is given by
\ie
    S[A,\cG_{(\pm)}]&=\frac{1}{2\pi \hbar} \int_{\cM\times C} \d z\wedge  \textrm{CS}(A)+\frac{1}{\hbar_{\rm 2d}}\int_{\cM\times \{z_+\} }\Tr\left(\Lambda \cG_{(+)}^{-1}D_{-} \cG_{(+)}\right)\,d\sigma^+\wedge d\sigma^-\\
        &\quad+\frac{1}{\hbar_{\rm 2d}}\int_{\cM\times \{z_-\} }
    \Tr\left(\Lambda\,\cdot\cG_{(-)}^{-1}D_{+} \cG_{(-)}\right)\,d\sigma^+\wedge d\sigma^-\,,
\label{action-defect}
\fe
where $\Lambda$ is an element of the real Lie subalgebra of the complex Lie algebra $\mathfrak{g}_{\mathbb{C}}$\,. 
In the case $G_{\mathbb{C}}=\SL(2,\mathbb{C})$\,, we take $\Lambda=-i\sigma^3/2$, where $\sigma^3=\left(\begin{array}{cc}
1 & 0\\
0 & -1
\end{array}\right)$ is the Pauli matrix.

Caudrelier, Stoppato and Vicedo \cite{Caudrelier:2020xtn} have shown that the 2d Zakharov-Mikhailov action, an ultralocal integrable field theory that includes the Faddeev-Reshetikhin model as a special case, can also be derived from the 4d CS theory coupled to order surface operators:
\begin{align} \label{defect 4d action}
\frac{1}{2\pi \hbar} \int_{\R^2 \times C} \d z\wedge \textrm{CS}(A) + S_{\rm defect}\big( A, \{ \phi_m \}_{m=1}^{N_1}, \{ \chi_n \}_{n=1}^{N_2} \big)\,,
\end{align}
where 
\begin{align} \label{defect 2d action}
    S_{\rm defect}\big( A, \{ \phi_m \}_{m=1}^{N_1}, \{ \chi_n \}_{n=1}^{N_2} \big) \coloneqq 
    &- \frac{1}{\hbar_{\rm 2d}}\sum_{m=1}^{N_1} \int_{\R^2 \times \{ a_m \}} \Tr \big( \phi_m^{-1} D_+ \phi_m U^{(0)}_m \big) d \si^+ d\si^-\notag\\
    &\; - \frac{1}{\hbar_{\rm 2d}}\sum_{n=1}^{N_2} \int_{\R^2 \times \{ b_n \}} \Tr \big( \chi_n^{-1} D_- \chi_n V^{(0)}_n \big) d\sigma^+ d\sigma^-\,,
\end{align}
where $\phi_m$ and $\chi_n$ are fields valued in $G_{\mathbb{C}}$, while $U^{(0)}_m$ and $V^{(0)}_n$ are constant non-dynamical elements valued in the associated Lie algebra, $\mathfrak{g}_{\mathbb{C}}$.
Note that in \cite{Caudrelier:2020xtn}, a boundary term was added to the 4d CS action to regularize the action, to ensure a locally integrable action at infinity on $C=\mathbb{CP}^1$. Here, we instead impose the boundary condition that the gauge field goes to zero at infinity, whereby such a boundary term is unnecessary. 

The 4d-2d coupled system considered by \cite{Caudrelier:2020xtn} is similar in form to the system with action \eqref{action-defect} considered in \cite{Fukushima:2020tqv}. The main difference is that $U_m^{(0)}$ and $V_n^{(0)}$ are independent variables, and also that there are multiple copies of the surface operators. One can generalize the previous arguments, e.g., when $N_1=N_2=1$, where the discretization would lead to two sets of Wilson lines in \textit{different} representations, with $U_m^{(0)}$ and $V_n^{(0)}$ as highest weights. For general values of $N_1$ and $N_2$ the discretization can be identified with a more general null lattice where there is a repeating pattern for every $N_1$ and $N_2$ Wilson line for each lightcone coordinate, respectively. An example of this for $N_1=2$ and $N_2=2$ is shown in Figure \ref{4}.

\subsection{Fermionic Defects}\label{sec:free_fermion}

Let us next turn to the discretization of defects described by chiral/anti-chiral free-fermion surface defects. Such defects can realize 2d massless Thirring model and 2d chiral Gross-Neveu models \cite{Costello:2019tri}.

\subsubsection{Fermionic Defects and Massless Thirring Model}\label{Th-4dCS}

Before discussing a lattice discretization, we first give a brief derivation of the massless Thirring and chiral Gross-Neveu models from the 4d CS theory with two order defects.  

Let us suppose the 4d CS theory with gauge group $G_{\mathbb{C}}$ coupled to the 2d free massless fermions whose the actions
are given by (see Appendix \ref{convention} for conventions of fermions):
\begin{align}
    S_{+}^f[A_-,\psi_L]&=\frac{1}{\hbar_{\rm 2d}}\int_{\cM\times \{z_{+}\}} d^2\sigma\, \psi_{L}^* i(\partial_-+ A_-)\psi_{L}
      \,,\label{fermaction1}\\
    S_{-}^f[A_+,\psi_R]&=\frac{1}{\hbar_{\rm 2d}}\int_{\cM\times \{z_{-}\}}d^2\sigma\,\psi_{R}^* i(\partial_{+}+A_+)\psi_{R}\,,\label{fermaction2}
\end{align}
where $\psi_L, \psi_{R}$ are left and right Majorana-Weyl fermions defined on $\cM\times \{z_{\pm}\}$ living in a representation $\rho$ of the gauge group $G_{\mathbb{C}}$.

The 2d effective theory is the massless Thirring/Gross-Neveu model whose action is given by
\begin{align}
    S_{\text{Th}}[\psi]=\frac{1}{\hbar_{\rm 2d}}\int_{\cM}d^2\sigma\left(\psi^{*}_{L}i\partial_-\psi_{L}
    +\psi^{*}_{R}i \partial_+\psi_{R}+\frac{1}{4\hbar_{\rm 2d}}\frac{i \hbar}{z_{+}-z_{-}}(\psi^{*}_{L}\rho(t_a)\psi_{L})(\psi^{*}_{R} \rho(t^a)\psi_{R})\right)\,.\label{Th-ac}
\end{align}
In terms of a Dirac spinor $\Psi$, the action (\ref{Th-ac}) can be rewritten as
\begin{align}
    S_{\text{Th}}[\Psi]=\frac{1}{\hbar_{\rm 2d}}\int_{\cM}d^2\sigma\left(\bar{\Psi}i\gamma^{\mu}\partial_{\mu}\Psi
    -\frac{1}{4\hbar_{\rm 2d}}\frac{i\hbar}{z_{+}-z_{-}}(\bar{\Psi}\gamma^{\mu}\rho(t_a)\Psi ) (\bar{\Psi}\gamma_{\mu}\rho(t^a)\Psi) \right)\,,\label{PTh-ac}
\end{align}
where $\bar{\Psi}=\Psi^{\dagger}\gamma^0=(\psi_R^*,\psi_L^*)/\sqrt{2}$\,.
In the derivation of this action, an appropriate reality condition for $G_{\mathbb{C}}$ is imposed.
It is noted that the coupling constant of the massless Thirring/Gross-Neveu models can be identified with the inverse of the distance $z_{+}-z_{-}$ between two order defects, and it must be pure imaginary to obtain a real coupling constant.
The associated Lax pair is
\begin{align}
    \cL=\frac{\hbar}{4\hbar_{\rm 2d}}\frac{\psi^{*}_{L}\rho(t^a)\psi_L\,\rho(t_a)}{z-z_{+}}d\sigma^+-\frac{\hbar}{4\hbar_{\rm 2d}}\frac{\psi^{*}_{R}\rho(t^a)\psi_R\,\rho(t_a)}{z-z_{-}}d\sigma^-\,.\label{Th-Lax}
\end{align}
This chiral fermionic model is known to be ultralocal at the classical level, and as in the Faddeev-Reshetikhin model model, we can apply the lightcone lattice discretization to the chiral fermion models \cite{Destri:1987ze,Destri:1987hc,Destri:1987ug} (For a review on this subject, see \cite{deVega:1989wym}).

The above construction of chiral fermion models in the 4d CS theory can be extended to the multi-flavor case. The relevant chiral surface defect actions with $K$ fermions $\psi_{L,I}\,, \psi_{R,I}$ $(I=1,\dots,K)$ are given by 
\begin{align}
    S_{+}^f[A_-,\psi_L]&=\frac{i}{\hbar_{\rm 2d}}\sum_{I=1}^{K}\int_{\cM\times \{z_{+}\}} \!\!d^2\sigma\, \psi^{*}_{L,I} (\partial_-+A_-)\psi_{L,I}\,,
    \label{Nfermaction1}\\
    S_{-}^f[A_+,\psi_R]&=\frac{i}{\hbar_{\rm 2d}}\sum_{I=1}^{K}\int_{\cM\times \{z_{-}\}} \!\!d^2\sigma\, \psi^{*}_{R,I} (\partial_++A_+)\psi_{R,I} \,.
    \label{Nfermaction2} 
\end{align}
The corresponding effective 2d integrable field theory
has the action
\begin{align}
    S_{\text{Th}}[\psi]&=\frac{1}{\hbar_{\rm 2d}}\int_{\cM}d^2\sigma\left(\sum_{I=1}^{K}(\psi^{*}_{L,I}i\partial_-\psi_{L,I}
    +\psi^{*}_{R,I}i \partial_+\psi_{R,I})-\frac{\hbar_{\rm 2d}}{4}\frac{i \hbar }{z_{+}-z_{-}}\cJ^a_{+}\cJ_{a-}\right)\,,\label{NTh-ac}
\end{align}
and the Lax pair 
\begin{align}
    \cL&=\frac{\hbar}{4}\frac{\cJ^a_{+}\,\rho(t_a)}{z-z_{+}}d\sigma^+-\frac{\hbar}{4}\frac{\cJ^a_{-}\,\rho(t_a)}{z-z_{-}}d\sigma^-\,,\label{NTh-Lax} \\
\cJ^a_{+}&=\frac{1}{\hbar_{\rm 2d}}\sum_{I=1}^{K}\psi^{*}_{L,I}\rho(t^a)\psi_{L,I}\,,\qquad \cJ^a_{-}=\frac{1}{\hbar_{\rm 2d}}\sum_{I=1}^{K}\psi^{*}_{R,I}\rho(t^a)\psi_{R,I}\,.
\end{align}
In Section \ref{sec:duality-boson}, we will discuss a realization of the nonabelian bosonization of the multi-flavor massless Thirring model in the 4d CS theory.

\subsubsection{Lattice Discretization}

In \cite{Destri:1987hc,Destri:1987ug}, 
a lightcone lattice discretization of  the 2d integrable field theory (\ref{Th-ac}) has been considered.
 We shall now derive this discretization of the massless Thirring model from the viewpoint of the 4d CS theory.

The path integral for the discretized theory reads
\begin{align}
    Z_{f}^{\text{Th}}(z_{L,R})&= \int \cD A \cD\Psi_L\cD\Psi_{R}\exp\left(iS_{f}[A,\Psi_L,\Psi_R]\right)\no\\
    &=\int \cD A\,\prod_{n}Z_{L,n}[\cA_{-,n}^{L}]
    \prod_{m}Z_{R,m}[\cA_{+,m}^{R}] \exp\left(iS_{\rm CS}[A]\right) \,,\label{Th-part}
\end{align}
where $Z_{L,n}[\cA_{-,n}]$ and $Z_{R,m}[\cA_{+,m}]$ are
\begin{align}
    Z_{L,n}[\cA_{-,n}]&=\int \cD\Psi_{n,L}\exp\left(-\int_{S^1\times \{z_{+}\}}d\sigma^-\,\Psi^{\dagger}_{L,n} (\partial_-+\cA_{-,n})\Psi_{L,n}\right)\,,\\
    Z_{R,m}[\cA_{+,m}]&=\int \cD\Psi_{m,R}\exp\left(-\int_{S^1\times \{z_{-}\}}d\sigma^+\,\Psi^{\dagger}_{R,m} (\partial_++\cA_{+,m})\Psi_{R,m}\right)\,.
\end{align}
Here we have chosen anti-periodic boundary conditions in both lightcone directions.

By using the results in Section \ref{f-Wilson}, the partition functions $Z_{L,n}[\cA_{-,n}]$ and $Z_{R,m}[\cA_{+,m}]$ can be written as Wilson lines
\begin{align}
     Z_{L,n}[\cA_{-,n}]&=W_{L,n}[\cA_{-,n}]=\Tr_{\cH_{n,f}}\cP\exp\left(-\oint_{S^1} d\sigma^-\,\cA_{-,n}^{a}\,\hat{\rho}_f(t_a)\right)\,,\\
   Z_{R,m}[\cA_{+,m}]&=W_{R,m}[\cA_{+,m}]=\Tr_{\cH_{m,f}}\cP\exp\left(-\oint_{S^1} d\sigma^+\,\cA_{+,m}^{a}\,\hat{\rho}_f(t_a)\right)\,.
\end{align}
Here, $ \hat{\rho}_f(t_a)$ is the Schwinger representation of $t_a \in \mathfrak{u}(N)$ in terms of the fermionic creation and annihilation operators $\hat{c}^i\,, \hat{c}_i^{\dagger}\,(i=1,2,\dots, N)$:
\begin{align}
    \hat{\rho}_f(t_a)=\hat{c}_i^{\dagger}\rho(t_a)^{i}{}_{j}\hat{c}^j\,,
\end{align}
where $\rho(t_a)^{i}{}_{j}$ is the fundamental representation of $t_a\in  \mathfrak{u}(N)$\,, and the operators $\hat{c}^i\,, \hat{c}_i^{\dagger}$ satisfy the anticommutation relations
\begin{align}
    \{\hat{c}^j, \hat{c}^{\dagger}_{k}\}=\delta^j_k\,,\qquad 
    \{\hat{c}^j, \hat{c}^k\}=0\,,\qquad 
    \{\hat{c}^{\dagger}_{j}, \hat{c}^{\dagger}_{k}\}=0\,.
\end{align}
The trace $\Tr_{n, \cH_{f}}$ is taken over the Fock space $\cH_{n,f}$ that is generated by acting the creation operators $\hat{c}_i^{\dagger}$ on the Fock vacuum $\lvert 0\rangle_{f}$ satisfying $\hat{c}^i\lvert 0\rangle_{f}=0$ for all $i$.

Then, the lattice discretization of the partition function (\ref{Th-part}) is  given as an expectation value of the product of Wilson lines 
\begin{align}
    Z_{f}^{\text{2d-Th}}(z_{\pm})
    \to Z_{f}^{\text{1d-Th}}(\Delta,z_{\pm})&=\int \mathcal{D}A\,\prod_{n} W_{L,n}[\cA_-]\prod_{m} W_{R,m}[\cA_+]\,\exp\left(i\,S_{\rm CS}[A]\right)\no\\
    &=\left\langle\prod_{n} W_{L,n}[\cA_-]\prod_{m} W_{R,m}[\cA_+] \right\rangle\,.\label{ZFlat-1}
\end{align}
Thus, the lattice discretization of the massless Thirring model can also be understood in terms of regularization by Wilson lines in the light cone direction. When we compactify $\Sigma$ into a torus by periodic boundary conditions, the expectation value (\ref{ZFlat-1}) is equivalent to the partition function of the rational vertex model on a torus \cite{Costello:2013zra}, that is, each Wilson line intersection is a rational $R$-matrix element \cite{Costello:2019tri}:
\begin{align}
    Z_{f}^{\text{1d-Th}}(\Delta,z_{\pm })=Z^{\text{r-vertex}}(\Delta,z_{\pm})\,.
\end{align}
Our lattice discretization by Wilson lines in the lightcone direction corresponds to the lightcone discretization of the work of Destri and de Vega \cite{Destri:1987hc,Destri:1987ug} (see also \cite{Destri:1987ze}).

We can go backwards and discuss the continuum limit of $Z_{f}^{\text{lat}}(\Delta,z_{\pm})$ as in Section \ref{sec:backwards}.
As discussed in \cite{Destri:1987ug}, massless Thirring model is obtained by taking the bare scaling limit of a gapless lattice model, where we keep the spectral parameter $z$ fixed as we take $\Delta \to 0$.
For the partition function $Z_{f}^{\text{lat}}(\Delta,z_{\pm})$\,, this scaling limit should be translated to
\begin{align}
Z_{f}^{\text{2d-Th}}(z_{\pm})=  \lim_{\Delta \to 0} Z_{f}^{\text{1d-Th}}(\Delta,z_{\pm})\,,
\end{align}
where $z_{+}$ and $z_{-}$ do not depend on the lattice spacing $\Delta$\,.
We can consider different scaling limits
where we also scale the spectral parameter $z$ simultaneously with $\Delta\to 0$,
and we expect that such a limit gives a massive QFT. It would be an interesting problem to understand the scaling limits in more detail in our framework.

\subsubsection{Extension to Arbitrary Irreducible Representations}\label{genrep}

The Wilson loop, which appeared in the lattice discretization of the previous subsection, is an operator in a reducible representation of $\SU(N)$. 
Indeed, a given state $\lvert \Psi\rangle$ has a finite expansion on the fermionc creation operators $ \hat{c}_j^{\dagger}$ of the form 
\begin{align}
    \lvert\Psi\rangle=\Psi_0 \lvert 0\rangle_f + \Psi_j\hat{c}^{\dagger}_j\lvert 0\rangle_f +\cdots +\frac{1}{N!}\Psi_{j_1\dots j_{N}}\hat{c}_{j_1}^{\dagger}\dots \hat{c}^{\dagger}_{j_{N}}\lvert 0\rangle_f \,.\label{f-SU(N)state}
\end{align}
Each coefficient $\Psi_{j_1\dots j_{k}}$ transforms as an antisymmetric tensor product of the fundamental representation of $\SU(N)$ i.e. the $k$-th antisymmetric representation of $\SU(N)$. 
Thus, the Wilson loop is found to be in a finite reducible representation of $\SU(N)$ corresponding to the state (\ref{f-SU(N)state}).

The Wilson loop operator can in general be defined for any irreducible representation of a given Lie group.
In the following, we will discuss an extension to the Wilson loop operator in finite-dimensional irreducible representations of $\SU(N)$ and give the corresponding order defect actions.

\subsubsection*{$k$-th Anti-Symmetric Representation}

For simplicity, we first consider the $k$-th antisymmetric representation of $\SU(N)$ by following the procedure of \cite{Bastianelli:2013pta} (see also \cite{Tong:2014yla,Gomis:2006sb}).
From the above expansion (\ref{f-SU(N)state}), this can be done by projecting the Fock space $\cH_f$ onto the restricted Hilbert space $\cH_{f}^{(k)}$ that contains only $k$ fermion excitation states, i.e.,
\begin{align}
    \hat{n}_f\lvert \Psi \rangle =k \lvert \Psi \rangle\,,\qquad \lvert \Psi \rangle\in \cH_{f}^{(k)}\,,\qquad \hat{n}_f:=\sum_{i=1}^{N}\hat{c}^{\dagger}_i\hat{c}^i\,.\label{excite-cont}
\end{align}
Hence, the Wilson loop operator for the $k$-th antisymmetric representation of $\SU(N)$ is defined by
\begin{align}
    W_f^{(k)}[A]=\Tr_{\cH_{f}^{(k)}}\left(\cP\exp\left(-\oint dt A^a_t\hat{\rho}_f(t_a)\right)\right)\,,\label{WL-f-ka}
\end{align}
where the trace $\Tr_{\cH_{f}^{(k)}}$ is taken over $\cH_{f}^{(k)}$\,.
To obtain the path integral representation of this operator, we divide the path of the Wilson loop into $N$ segments with a small length $\varepsilon$ as usual, and in particular, it is convenient to insert the projection operators $ \hat{P}_{(k)}: \cH_f\to \cH_f^{(k)}$ in each segment,
\begin{align}
    W_f^{(k)}[A]&=\Tr_{\cH_{f}}\left(\hat{P}_{(k)}\cP\exp\left(-\oint dt A_t^a\hat{\rho}_f(t_a)\right)\right)\no\\
    &:=\lim_{\epsilon\to 0}\Tr_{\cH_{f}}\left(\prod_{j=1}^{N}\hat{P}_{(k)}(t_j)\exp\left( -\epsilon\,A^a(t_j)\hat{\rho}_f(t_a)\right)\right)\,,
\end{align}
where we use the integral representation of the projection operators 
\begin{align}
   \hat{P}_{(k)}(t_j)&=\int \frac{d\tilde{A}^f(t_j)}{2\pi}\,\exp\left(i \epsilon\,\tilde{A}^f(t_j)(\hat{n}_{f,j}-k)\right)
    \,.
\end{align}
By using the coherent states for fermionic operators, we can rewrite the Wilson loop operator 
as the partition function of the 1d free chiral fermion system (see Appendix \ref{f-Wilson} for details) 
\begin{align}
    W_f^{(k)}[A]&=\int \cD \psi \cD \psi^{*} \cD \tilde{A}^f\,\exp\left(i\,S_f^{(k)}\right)\exp\left(-ik_{\text{eff}}\int dt\,\tilde{A}_t^f\right)\,,\label{Z-f-pat}\\
    S_f^{(k)}&=\int dt\,\psi^{*}_ii(\delta^{ij}\partial_t+A_t^a\rho(t_a)^i{}_j-i\delta^{ij}\tilde{A}_t^f)\psi^j \,,
\end{align}
where $\psi^j$ is a Grassmann field in the fundamental representation of $\SU(N)$, and $\psi_j^*$ is the complex conjugate of $\psi^j$.
The path integral formulation employs the Weyl ordering at the operator level (see e.g., \cite{Sato:1976hy}), so the coefficient $k$ of $\tilde{A}_t^f$ suffers the following shift: 
\begin{align}
  k \to  k_{\text{eff}}=k-\frac{N}{2}\,.
\end{align}
Note that the path integral (\ref{Z-f-pat}) is invariant under the $\U(1)$ gauge transformation
\begin{align}
\psi \to e^{i\varphi} \psi\,,\qquad \psi^{*} \to e^{-i\varphi} \psi^{*}\,,\qquad \tilde{A}^f_t\to \tilde{A}^f_t +\partial_t \varphi\,.
\end{align}
Hence, we regard the Lagrange multiplier $\tilde{A}^f_t$ as a $\U(1)$ gauge field, and  $k_{\text{eff}}$ appears as the coefficient of a 1d Chern-Simons term $\int dt\,\tilde{A}^f_t$\,. This observation reflects the fact that the constraint (\ref{excite-cont}) is a first-class constraint.
One might argue that $k_{\text{eff}}$ should be quantized and that $N$ cannot be a natural number. However, as we will see in Section \ref{sec:triality}, there is an anomalous contribution from the path integral with respect to $\psi$, $\psi^*$, which cancels out with $-N/2$ in $k_{\text{eff}}$ and hence $N$ can be any natural number.

\subsubsection*{General Irreducible Representation}

We can extend the above discussion to the case of arbitrary irreducible representation by using the results in \cite{Gomis:2006sb,Corradini:2016czo}.
Let us first define the Wilson loop operator for an irreducible representation $\rho$ of $\SU(N)$ which is characterized by the following Young tableau with $k_I (<N)$ boxes in the $I$-th column $(1 \leq  I \leq  K)$:
\begin{center}
\begin{ytableau}
       \none &  &  & & &  & & & \none \\
  \none  &  &  & & & \none & \none & \none & \none \\
  \none  &  &  & \none & \none & \none & \none & \none &  \none \\
  \none  &  & \none & \none & \none & \none & \none & \none &  \none \\
  \none  &  & \none & \none & \none & \none & \none & \none & \none \\
  \none & \none[k_1] & \none[k_2] & \none[k_3] & \none[$\dots$] & \none  & \none[$\dots$] & \none[k_K]  
  \label{youngt}
\end{ytableau}
\end{center}
To this end, we need to construct a Hilbert space that realizes the irreducible representation $\rho$.
As discussed in \cite{Gomis:2006sb,Corradini:2016czo}, this can be done by first introducing the fermionic creation and annihilation operators $\hat{c}_{i}^{I}$ and $\hat{c}^{\dagger\,i}_{I}\,(i=1,\dots,N)$ with the flavor indices $I\,, J=1,\dots, K$\,, which is equal to the number of column of the Young tableau.
The operators satisfy the anticommutation relations
\begin{align}\label{commrel}
    \{\hat{c}_i^{I}, \hat{c}^{\dagger\,j}_{J}\}=\delta_i^j\delta_{J}^{I}\,,\qquad 
    \{\hat{c}_i^{I}, \hat{c}_j^{J}\}=0\,,\qquad 
    \{\hat{c}^{\dagger\,i}_{I}, \hat{c}^{\dagger\,j}_{J}\}=0\,,
\end{align} 
and enables us to construct the Schwinger representation $\hat{\rho}_f(t_a)$ of $t_a \in \SU(N)$ defined by 
\begin{align}
    \hat{\rho}_f(t_a)=\sum_{I=1}^{K} \hat{c}^{I\dagger}_i\rho(t_a)^{i}{}_{j}\hat{c}^j_{I}\,.\label{f-sch-rep}
\end{align}
Let $\cH_f$ be the Fock space generated by $\hat{c}_{i}^{I}\,,\hat{c}^{\dagger\,i}_{I}$ with the Fock vacuum $\lvert 0\rangle_{f}$ which is defined by $ \hat{c}_{i}^{I}\lvert 0\rangle_{f}=0$\,, and the representation realized on this space is reducible as before.
Let $\hat{P}_{\rho}$ be the projection operator that projects onto the subspace of $\cH_f$ realizing the irreducible representation $\rho$.
Then, by using the operator $\hat{P}_{\rho}$, we can define the Wilson loop operator in the same way as for the $k$-th antisymmetric representation:
\begin{align}
     W_{\rho}[A]=\Tr_{\cH_f}\left(\hat{P}_{\rho}\cP \exp\left(-\oint A^a \hat{\rho}_f(t_a)\right) \right)\,.\label{WL-f-gen}
\end{align}
The projection operator $\hat{P}_{\rho}$ can be written down by following the discussion in \cite{Corradini:2016czo} and introducing auxiliary fields:
a state $\lvert \Psi \rangle \in \cH_f$ in the irreducible representation $\rho$ can be characterized by the constraints \cite{Corradini:2016czo}
\begin{align}
    \hat{L}^I{}_{I}\lvert \Psi \rangle=k_{I}\lvert \Psi \rangle\,,  \qquad   \hat{L}^{J}{}_{K}\lvert\Psi\rangle=0\,,\qquad J>K\,,\label{irep-const}
\end{align}
where $\hat{L}^{I}{}_{J}$ are generators of $\mathfrak{u}(K)$ defined by
\begin{align}
    \hat{L}^{I}{}_{J}=\sum_{i=1}^{N}\hat{c}^{i\dagger}_{J}\hat{c}_i^{I}\,,\label{Lf-def}
\end{align}
and satisfy the commutation relations 
\begin{align}
    [\hat{L}^{I}{}_{J},\hat{L}^{K}{}_{L}]=\delta^{I}_{L} \hat{L}^{K}{}_{J}-\delta^{K}_{J} \hat{L}^{I}{}_{L}\,.
\end{align}
The first constraint of (\ref{irep-const}) specifies the length of each column in the Young tableau $\rho$ as in the $k$-th antisymmetric representation case, and the second realizes symmetrization between different columns (For the details, see \cite{Corradini:2016czo} and Appendix \ref{f-Wilson-irep}).
Note that the generators $\hat{L}^I{}_{I}$ and $\hat{L}^{J}{}_{K}\,(J>K)$ form the commutation relations of the Borel subalgebra $\mathfrak{b}$ of $\mathfrak{sl}_K$,
\begin{align}
\begin{split}
 [\hat{L}^{I}{}_{I},\hat{L}^{J}{}_{J}]&=0\,,\\
    [\hat{L}^{I}{}_{J},\hat{L}^{K}{}_{K}]&=- \delta_{JK}\hat{L}^{I}{}_{K}\,,\qquad I>K\,, I>J\,,\\
    [\hat{L}^{I}{}_{J},\hat{L}^{K}{}_{K}]&=\delta_{IK}\hat{L}^{K}{}_{J}\,,\qquad K>J\,, I>J\,.
\end{split}
\label{borel-comm}
\end{align}
This indicates that the constraints (\ref{irep-const}) are first-class.
Hence, it enable us to introduce the axial gauge field $\tilde{A}^f \in \mathfrak{b}$, whose the non-zero components are $(\tilde{A}^f)_{IJ}$ with $I\geq J$, to define the projection operator $\hat{P}_{\rho}$ as
\begin{align}
    \hat{P}_{\rho}
    &=\int_{G_{\mathbb{C}}/B} d\tilde{A}^f
    \exp\biggl[i\sum_{I \geq J} (\tilde{A}^f)_{IJ}\left(\left(\hat{L}^{I}{}_{J}\right)_{W}-\frac{1}{2}\delta_{IJ}k_{\text{eff},I}\right)\biggr]\,,\label{proj-f}
\end{align}
where the integral is over the 
flag variety ${G}_{\mathbb{C}}/{B} = {G}/{T}$ (${B}$ is the Borel subgroup whose associated Lie algebra is $\mathfrak{b}$, and $\textrm{T}=\U(1)^K$ is the Cartan subgroup). 
The bracket $(\,\cdot\,)_{W}$ stands for the Weyl ordering, and $k_{\text{eff},I}$ is the integer $k$ shifted by $-N/2$ due to Weyl ordering,
\begin{align}
    k_{\text{eff},I}=k_I-\frac{N}{2}\,.
\end{align}

We can rewrite (\ref{WL-f-gen}) into the path integral representation by employing the coherent state method as in the case of the antisymmetric representation.
The resulting expression is \cite{Corradini:2016czo}
\begin{align}
    W_{\rho}^{f}[A]&=\int \cD\psi\cD \psi^{*}\cD \tilde{A}^f\,\exp \left(i S_f\right)\exp\left(-i\int dt\,\sum_{I=1}^{K}k_{\text{eff},I}\,(\tilde{A}^f_{t})_{II}\right)\,, \label{f-part0}\\
S^f&=\int_{t_0}^{t_f} dt\,\sum_{I,J=1}^{K}\sum_{i,j=1}^{N}\psi^{I*}_i(i\delta^{IJ} \delta^{i}_{j}\partial_t+\delta^{IJ}A_t^a\rho(t_a)^{i}{}_{j}+\delta^{ij}(\tilde{A}_t^f)_{IJ})\psi_{J}^j\,.\label{1d-f-action-g00}
\end{align}
The measure $\cD \tilde{A}^f$ for the flavor gauge group is given by
\begin{align}
     \cD \tilde{A}^f&=
     \prod_{I>J}\cD \tilde{A}^f_{IJ}\prod_{I=1}^{K}\cD \tilde{A}^f_{II}\,.
\end{align}

\subsection{\texorpdfstring{Free $\beta\gamma$ Defects}{Free beta-gamma Defects}}\label{sec:fbg-sec} 

Let us next consider defects associated with free $\beta\gamma$ systems.
The chiral free $\beta\gamma$ system has the explicit form 
\ie 
    S^{\textrm{2d-$\beta\gamma$}}_{+}
    =\int_{\mathbb{C} \times \{z_+\}} \beta^i \bar{\partial}_A \gamma_i=\int_{\mathbb{C} \times \{z_+\}} dw d\bar{w}\left(\beta^i {\partial}_{\bar{w}} \gamma_i +A_{\bar{w}}^a\beta^i \rho(t_a)_i{}^{j} \gamma_j\right)\,,\label{cfbg}
\fe
while the anti-chiral free $\beta\gamma$ system has the explicit form 
\ie 
    S^{\textrm{2d-$\beta\gamma$}}_{-}
    =\int_{\mathbb{C} \times \{z_-\}} \overline{\beta}^i \partial_A \overline{\gamma}_i
    =\int_{\mathbb{C} \times \{z_-\}} dw d\bar{w}\left(\overline{\beta}^i {\partial}_{{w}} \overline{\gamma}_i + A_{w}^a\overline{\beta}^i \rho(t_a)_i{}^{j} \overline{\gamma}_j\right)\,,
    \label{acfbg}
\fe
where $\rho$ is a representation of the gauge group $G$.

As shown in \cite{Costello:2019tri}, when we consider the 4d CS theory with gauge group $G_{\mathbb{C}}=\GL(N,\mathbb{C})$ 
coupled with chiral and anti-chiral free $\beta\gamma$ defects to 
the fundamental representation of the $\mathfrak{gl}(N,\mathbb{C})$:
\ie
    S^{4 \mathrm{d}\text{-}2 \mathrm{d}}
    =\frac{1}{2 \pi\hbar} \int_{\mathbb{C} \times \mathbb{C}} \mathrm{d} z\wedge  \mathrm{CS}(A)+\frac{1}{\hbar_{\rm 2d}}\int_{\mathbb{C} \times \{z_+\}} \beta^i \bar{\partial}_A \gamma_i+\frac{1}{\hbar_{\rm 2d}}\int_{\mathbb{C} \times \{z_-\}} \bar{\beta}_i \bar{\partial}_A \bar{\gamma}^i\,.
\fe 
The associated 2d sigma model is derived by
integrating out the gauge field $A$, which leads to the classical action 
\ie
    S^{\textrm{2d-\textrm{eff}}}
    =\frac{1}{\hbar_{\rm 2d}}\int_{\mathbb{C}}\left( \beta^i \bar{\partial} \gamma_i+ \bar{\beta}_i \partial \bar{\gamma}^i+\frac{i}{2}\frac{\hbar}{z_+-z_-}  \beta^i \gamma_j \bar{\beta}_i \bar{\gamma}^j\right) dw d\bar{w}\,, \label{2d-fbg}
\fe
where we used the identity $(\rho(t_a) \otimes \rho(t^a))_{ij,kl}=\delta_{ik}\delta_{jl}$. As described in \cite{Costello:2019tri}, integrating out the auxiliary fields $\beta^i$ or $\bar{\beta}_i$ results in a sigma model with target space $\mathbb{C}^N \backslash\{0\} \cong S^{2 N-1} \times \mathbb{R}$. In other words, we obtain a sigma model on $S^{2N-1}$ coupled to a free boson. 

\subsubsection*{Discretization and Relation with Wilson Loop}

For the chiral free $\beta\gamma$ defect, the familiar discretization procedure leads us to a 1d quantum mechanical system with action of the form 
\ie 
    S^{\textrm{1d-$\beta\gamma$}}_{+}
    =\int_{\mathbb{R}_{-} \times \{z_+\}} \beta^i \bar{\partial}_A \gamma_i=\int_{\mathbb{R}_{-} \times \{z_+\}} d\sigma^-\,\left(\beta^i {\partial}_{-} \gamma_i + A_{-}^a\beta^i \rho(t_a)_i^{\textrm{ }j} \gamma_j\right)\,.\label{1dbg-action}
\fe
Likewise, the anti-chiral free $\beta \gamma$ system can be discretized to 1d quantum mechanical systems of the form  
\ie 
    S^{\textrm{1d-$\beta\gamma$}}_{-} 
    =\int_{\mathbb{R}_{+} \times \{z_-\}} \bar{\beta}^i {\partial}_A \bar{\gamma}_i=\int_{\mathbb{R}_{+} \times \{z_-\}} d\sigma^+\,\left(\bar{\beta}^i {\partial}_{+} \bar{\gamma}_i + A_{+}^a\bar{\beta}^i \rho(t_a)_i^{\textrm{ }j} \bar{\gamma}_j\right)\,.
    \label{1dbg-action-minus}
\fe

Now, the fields $\beta$ and $\gamma$ are complex-valued, and just like for the bulk 4d CS fields, require the specification of an integration cycle for the path integral when working beyond perturbation theory. For convenience of analysis, we shall choose the integration cycle such that $\beta^i$ and $\gamma_i$ are both real.

As in other examples, the partition function of the 1d free $\beta\gamma$ system (\ref{1dbg-action})
\begin{align}
    Z^{\textrm{1d-$\beta\gamma$}}
    =\int \mathcal{D}\beta \mathcal{D}\gamma\,\exp\left(i\int_{\mathbb{R} } dt\,\left(\beta^i {\partial}_{t} \gamma_i + A_{t}^a\beta^i \rho(t_a)_i^{\textrm{ }j} \gamma_j\right)\right)\label{parti-bg}
\end{align}
can be rewritten as the trace of a Wilson loop
\begin{align}
    W_{\rho}
    =\Tr_{\cH_{\beta\gamma}}\left(\cP\exp\left(i\int_{S^1} dt\,A^a\hat{\rho}(t_a)\right)\right)\,,\label{W-bg}
\end{align}
where $\hat{\rho}(t_a)$ is the Schwinger representation of $t_a\in \mathfrak{gl}(N,\mathbb{C})$ given by
\begin{align}
   \hat{\rho}(t_a)
   =\hgamma_i \rho(t_a)^i{}_{j} \hbeta^j\,,
\end{align}
and where we adopt periodic boundary conditions to obtain $S^1$ for both lightcone directions.
Here, the trace of (\ref{W-bg}) is taken over the Fock space $\cH_{\beta\gamma}$ which is generated by quantum operators $\{\hat{\gamma}_i\,, \hat{\beta}^i\}\,(i=1\,,\dots\,,N)$ satisfying the canonical commutation relations, also known as the Heisenberg-Weyl algebra
\begin{equation}
    [\hgamma_i,\hbeta^j]=i\delta_i^{j}\,,
    \qquad 
    [\hgamma_i,\hgamma_j]=0 \,, 
    \qquad 
    [\hbeta^i,\hbeta^j]=0\,.
    \label{cano-comm}
\end{equation}
In Appendix \ref{beta-gamma-Wilson}, we  show the equivalence between (\ref{parti-bg}) and (\ref{W-bg}) by employing coherent states for the Heisenberg-Weyl algebra, and then (\ref{parti-bg}) corresponds to quantum mechanics on the coadjoint orbit of the Heisenberg-Weyl group with the highest weight which corresponds to the Fock vacuum $|0\rangle_b$ of $\cH_{\beta\gamma}$\,.
As a result, the 2d integrable field theory (\ref{2d-fbg}) can also be lattice discretized by considering a discretization of the free $\beta\gamma$ systems (\ref{cfbg}), (\ref{acfbg}). 

Note that the representation we get is an infinite direct sum of symmetric representations. If one wants to construct a Wilson loop operator associated with an irreducible representation, one needs to project it to a subspace of the Fock space $\cH_{\rho\gamma}$ that realizes the irreducible representation, as in the case of fermions.

We may pick an alternate contour where the $\beta$ and $\gamma$ fields are not real and independent, but rather complex conjugates of each other, i.e., $\beta_i=z_i^*$ and $\gamma^i=z^i$. The quantization of a line defect with these degrees of freedom can be achieved using Schwinger bosons, as described in Appendix \ref{beta-gamma-Wilson}. The associated integrable field theory, for general gauge group $G$, can be described as generalizations of the bosonic version of the massless Thirring model. 

\subsubsection*{Extension to Arbitrary Irreducible Representations}

Analogous to the case of discretized fermion actions discussed in Section \ref{genrep}, we can construct Wilson lines in arbitrary irreducible representations of $\GL(N,\mathbb{C})$ using one-dimensional actions with bosonic degrees of freedom with appropriate flavor symmetry. 
Let us suppose the irreducible representation $\rho$ specified by the Young tableau with with $l_{I}$ boxes in the $I$-th row ($1 \leq I  \leq L$):
\begin{center}
\begin{ytableau}
   \none &  &  & & &  & & & \none[l_1] \\
  \none  &  &  & & & \none & \none & \none & \none[l_2] \\
  \none  &  &  & \none & \none & \none & \none & \none &  \none[$\vdots$] \\
  \none  &  & \none & \none & \none & \none & \none & \none &  \none[$\vdots$] \\
  \none  &  & \none & \none & \none & \none & \none & \none & \none[l_L] 
  \label{young-bf-1}
\end{ytableau}
\end{center}
The Hilbert space supporting the representation $\rho$ can be constructed in the same way as in the fermionic case by introducing the creation and annihilation operators $\hat{\beta}^i_{I}\,, \hat{\gamma}_i^{I}\,(i=1,\dots, N; I=1,\dots, L)$ of $\beta\,,\gamma$ fields with $L$ flavor, which are in the anti-fundamental and fundamental representations of $\GL(N,\mathbb{C})$\,, respectively.
Indeed, a state $\lvert \Psi \rangle \in \cH_{\beta\gamma}$ in the representation $\rho$ by imposing the first-class constraint \cite{Corradini:2016czo} (see Appendix \ref{f-Wilson-irep} for similar discussion for  fermions):
\begin{align}
    \hat{L}^{I}{}_{I}\lvert \Psi \rangle=l_{I}\lvert \Psi \rangle\,,  \qquad   \hat{L}^{I}{}_{J}\lvert\Psi\rangle=0\,,\qquad I>J\,,\qquad \hat{L}^{I}{}_{J}=\sum_{i=1}^{N}\hat{\beta}^{i}_{J}\hat{\gamma}_i^{I}\,.\label{irep-const-b}
\end{align}
Once the projection operator corresponding to the first-class constraint (\ref{irep-const-b}) is constructed, 
we can obtain the one-dimensional path integral for the Wilson loop operator which takes the form 
\begin{align}
   W_{\rho}[A]
   &=\int \cD{\bf z} \cD {\bf z}^*\cD \tilde{A}^b\,\exp \left(iS^b\right)\exp\left(-i\int_{S^1} dt\,\sum_{I=1}^{L}l_{\text{eff}, I}\,(\tilde{A}_{t}^b)_{II}\right)\,,\label{b-part-ge}\\
   S^b
   &=\int_{t_0}^{t_f}dt\sum_{\alpha,\beta=1}^{L}\sum_{i,j=1}^{N}z_i[I]^*\left(\delta^{IJ}\delta^{i}_{j}\partial_t +\delta^{IJ}A^a_t  \rho(t_a)^{i}{}_{j}+\delta^{ij}(\tilde{A}_t^b)_{IJ}\right)z^j[J]\,,\label{constraints}
\end{align}
where the auxiliary gauge field $(\tilde{A}_t^b)_{IJ}$ takes values in the Borel subalgebra $\mathfrak{b}_L$ of $\mathfrak{sl}_L$ and the non-zero components are $(\tilde{A}_t^b)_{IJ}$ with $I \geq J$\,.
Here, the flavor indices are $I\,,J\,,\ldots$, which in the present context counts the number of rows of the Young tableau of interest (in the case of fermions, the flavor index counted the number of columns).  In addition, ${l}_{\text{eff},I}$ measures the size of each row
\begin{align}
    l_{\text{eff},I}=l_{I}+\frac{N}{2}\,.
\end{align}
Note the shift $N/2$ is a quantum effect that comes from Weyl ordering.
The constraint that arises from the second term of \eqref{constraints} ensures the antisymmetrization between different rows. 

\subsection{\texorpdfstring{Curved $\beta\gamma$ Defects}{Curved beta-gamma Defects}}
    \label{sec:curved_betagamma}

In this subsection, we shall be concerned with the coupling of the 4d CS theory to curved $\beta\gamma$ defects, and the discretization of these defects. 

These defects are associated with a K\"ahler manifold, $X$, which enjoys a holomorphic $G$-action. The defect is defined in terms of fields $\gamma: \mathbb{C} \rightarrow X$ and $\beta \in \Omega^{1,0}\left(\mathbb{C}, \gamma^* T^* X\right.$) (here, $T^* X$ is the holomorphic cotangent bundle of $X$).  To define the coupling to 4d CS gauge fields, we employ the map $\rho: \mathfrak{g} \rightarrow \operatorname{Vect}(X)$ which is the Lie algebra homomorphism from $\mathfrak{g}$ to the Lie algebra of holomorphic vector fields on $X$.
These holomorphic vector fields generate the infinitesimal $G$-action on $X$. Picking local holomorphic coordinates $u_1, \ldots, u_n$ on $X$, and a basis $t_a$ of $\mathfrak{g}$, the holomorphic vector fields can be expressed as
\ie
    \rho\left(t_a\right)=\sum \rho_{a, i}(u) \partial_{u_i} \;,
\fe
where $\rho_{a, i}(u)$ are holomorphic functions of $u_j$. The explicit, local form of the surface defect action is 
\ie\label{chiralbg}
    S^{\textrm{2d-c$\beta\gamma$}}_{+} 
    =\int_{\cM\times \{z_+\}} \beta^i {D}_+\,
    \gamma_i\,d^2\si 
    := \int_{\cM\times \{z_+\}} \left(\beta^i \partial_{+} \gamma_i+A_{a, +} \beta^i \rho_{a, i}(\gamma) \right)d^2\si \,.
\fe
To realize integrable field theories, \cite{Costello:2019tri} also introduced a complex-conjugate $\bar{\beta}\bar{\gamma}$ system located at another point on $C$, where $\bar{\gamma}$ is a map to $\bar{X}$ (defined to be $X$ with the opposite complex structure) and $\bar{\beta}\in \Omega^{0,1}\left(\mathbb{C}, \bar{\gamma}^* T^* \bar{X}\right)$. The explicit surface defect action is
\ie
    S^{\textrm{2d-c$\beta\gamma$}}_{-} =
    \int_{\cM\times \{z_-\}} \bar{\beta}^i D_-\,
    \bar{\gamma}_i\,d^2\si  := \int_{\cM\times \{z_-\}} \left(\bar{\beta}^i \partial_- \bar{\gamma}_i +A_{a, -} \bar{\beta}^i \bar{\rho}_{a, i}(\bar{\gamma})\right)\,d^2\si \,.
\fe

The full 4d--2d coupled system, with holomorphic and anti-holomorphic surface defects at points $z_+$ and $z_1$ on $C$, respectively, is 
\begin{equation}
    S^{\textrm{4d--2d c$\beta\gamma$}}
    =\frac{1}{2 \pi \hbar} \int_{\cM \times \mathbb{C}} \mathrm{d} z\wedge \mathrm{CS}(A)+\int_{{\cM} \times \{z_+\}} \beta^i D_+ \gamma_i\,d^2\si+\int_{\cM\times \{z_-\}} \bar{\beta}_i D_- \bar{\gamma}^i\,d^2\si\,.
\end{equation}
Integrating out the gauge field gives us
\ie
    S^{\textrm{2d-c$\beta\gamma$}}
    =\int_{\cM} \left(\beta^i \partial_{+} \gamma_i+\bar{\beta}^i \partial_- \bar{\gamma}_i -\frac{i}{4}\frac{\hbar}{z_+-z_-} \sum \beta^i \bar{\beta}^j \rho_{a, i}(\gamma) \bar{\rho}_{a, j}(\bar{\gamma})\right) d^2\si \,,
\fe
and further integrating out $\beta$ and $\overline{\beta}$ gives a sigma model with K\"ahler target space, $X$. 

\subsubsection*{Geometric Quantization and Wilson Lines}

Let us apply the discretization procedure to the holomorphic defect, for which we must first analytically continue $\overline{w}$ to the lightcone coordinate $\sigma^-$. The quantum mechanical systems we obtain via discretization have actions of the form
\ie 
    \int_{S^1} \beta^i D_- \gamma_i \mathrm{~d} \sigma^{-}\,,
\fe
after adopting periodic boundary conditions for the lightcone direction. 

We immediately observe the familiar action ($L \sim p \dot{q}$) of a quantum mechanical system, with Hamiltonian equal to $A^a_{-} \beta^i \rho_{a,i}(\gamma)$\,. Our aim is thus to quantize the action of each line operator, such that the composite field $\beta^i \rho_{a,i}(\gamma)$ acts on the resulting Hilbert space in some representation of $\mathfrak{g}$\,.

Let us discuss the specific example of $X=\mathbb{CP}^1$ for simplicity, so that there is only one $\beta\gamma$ pair.  
Following the conventions of \cite{Costello:2019tri}, the holomorphic vector fields $\rho_a(\gamma)$ are described in a local complex coordinate $\gamma$ as $\rho_1=i \gamma \partial_{\gamma}\,, \rho_2=\frac{1}{2}\left(-\partial_{\gamma}-{\gamma}^2 \partial_{\gamma}\right)\,, \rho_3=\frac{1}{2}\left(i \partial_{\gamma}-i {\gamma}^2 \partial_{\gamma}\right)$\,. 
These satisfy $    [\rho_i, \rho_j]=\epsilon_{ijk}\rho_{k}\,,$
where $i\,,j=1\,,2\,,3$ and the antisymmetric tensor $\epsilon_{ijk}$ is normalized as $\epsilon_{123}=1$\,.
We can rewrite the Hamiltonian as 
\ie 
    A_-^H\beta \rho_H + A_-^F\beta \rho_F + A_-^E\beta \rho_E\,,
\fe
where 
$A^H=-iA^1\,, A^F=\frac{1}{2}(iA^3-A^2)\,, A^E=\frac{1}{2}(iA^3+A^2)\,,$
and
\ie 
    \rho^H&=-\gamma\,,\qquad
    \rho^F= 1\,, \qquad
    \rho^E= -\gamma^2\,.
\fe
Let us also define $\sigma_E=\beta \rho^E$, $\sigma_F=\beta \rho^F$ and $\sigma_H=\beta \rho^H$.

To quantize the theory in a coordinate chart of $\mathbb{CP}^1$,
we can impose the canonical commutation relations 
 \begin{equation}
     [\hat{\gamma},\hat{\beta}]=1\,,\qquad [\hat{\gamma},\hat{\gamma}]=0\,,\qquad [\hat{\beta},\hat{\beta}]=0 \,,
 \end{equation}
so that $\hat{\beta}$ can be regarded as a 
 differential operator $-\partial/\partial \gamma$. 
 In this representation (oscillator or Schwinger representation), the fundamental representation of $\SL(2,\mathbb{C})$ is given by
     \ie \label{deffdiff-1}
      \sigma_E&=\gamma^2\frac{\partial}{\partial \gamma}+2\gamma a\,,\\
     \sigma_F&=-\frac{\partial}{\partial \gamma}\,,\\
     \sigma_H&=\gamma\frac{\partial}{\partial \gamma}+ a\,,
 \fe 
where $E, F, H$ are the Chevalley generators of $\SL(2, \mathbb{C})$ satisfying the commutation relations
$[E, F] = 2 H\,, [H, E] =  E\,, [H, F] = - F$
and $a$ is in general complex, and accounts for the ambiguity in operator ordering.\footnote{This representation is also known as the Dyson-Maleev representation.} 
For $a\ne 0$ we say that we have a set of \textit{twisted} differential operators.
Note that the non-zero value of $a$ is needed to obtain a non-trivial Casimir, i.e.,  
\ie \label{casi1}
    \sigma_H\sigma_H+\frac{\sigma_E\sigma_F}{2}+\frac{\sigma_F\sigma_E}{2}=a(a+1)\,.
\fe

 Consider $\mathbb{CP}^1$ with charts parametrized by $\gamma$ and $\gamma'$, where $\gamma=1/\gamma'$. 
 For twisted differential operators defined on $\mathbb{CP}^1$ to be globally well-defined  (see, e.g., page 273 of \cite{hotta2007d}), the transition function 
 \ie 
     \partial_{\gamma'}=-\gamma^2 \partial_{\gamma}
 \fe
 ought to be deformed to 
 \ie
     \partial_{\gamma'}=-\gamma^2  \partial_{\gamma }-2a \gamma\,,
 \fe
whereby 
\ie 
    \label{CP1_DO}
    \sigma_E&=\gamma^2\frac{\partial}{\partial \gamma}+2\gamma a && \rightarrow -\frac{\partial}{\partial \gamma'}\,,\\
    \sigma_F&=-\frac{\partial}{\partial \gamma} &&\rightarrow \gamma'^2\frac{\partial}{\partial \gamma'}+2\gamma a\,, \\
    \sigma_H&=\gamma\frac{\partial}{\partial \gamma}+ a && \rightarrow -\gamma'\frac{\partial}{\partial \gamma'}- a\,.
 \fe 
 These transformation rules define a sheaf of twisted differential operators.
 Note that the gluing equations in \eqref{CP1_DO}
 corresponds to an outer automorphism $(E, F, H)\to (F, E, -H)$ of $\mathfrak{sl}_2$.

  Algebras of twisted differential operators defined on $\mathbb{CP}^1$, and more generally, flag varieties, were analyzed by Bernstein and Beilinson \cite{bernstein1981localisation, beilinson1993proof} precisely for the purpose of studying the representation theory of simple complex Lie algebras. Their result can be stated as follows \cite{kashiwara1988representation}: for a flag variety $X$ of a reductive group $G$, and a choice of weight $\lambda$, let $\chi_\lambda$ be the corresponding character of the center of the universal enveloping algebra $\U(\mathfrak{g})$. Defining the associated Verma module $\U_{\lambda}(\mathfrak{g})=\U(\mathfrak{g})/\U(\mathfrak{g})\textrm{Ker }\chi_\lambda$, a twisted ring of differential operators $D_{\lambda}$ can be defined on $X$ such that 
\ie
    \Gamma\left(X ; D_\lambda\right)=\U_\lambda(\mathfrak{g})\,. 
\fe
This is the algebraic counterpart of the statement that the discretized curved $\beta\gamma$ system can be described by a Wilson line.

\section{Discretizations of Vertex Algebra Defects}
    \label{sec:VA}

\subsection{Lattice Discretization and Zhu's Algebra}\label{sec:lattice-Zhu}

\subsubsection*{General Vertex Algebra}

In Section \ref{sec:disc_defect} the defect theories are free theories with explicit Lagrangians, and we have first discussed the discretization at the Lagrangian level.
It is clear from the discussion of \ref{sec:reviewint} that this is not a necessity: we can couple the 4d CS theory to more general chiral or anti-chiral defects, each with a global symmetry $G$, and we will still obtain an integrable model.

As examples of more general defects, we discuss chiral (anti-chiral) defects described algebraically by a Vertex Algebra (VA). This is an algebraic framework for describing the chiral half of the CFT (see Appendix \ref{vz} for review), and in this setup we do not necessarily have a Lagrangian description.\footnote{A vertex algebra becomes the Vertex Operator Algebra (VOA) when we include the stress-energy tensor. The axioms of the vertex algebra, however, are sufficient for the purposes of this section.}

For the coupling to the 4d CS theory, the defect needs to have a global $G$-symmetry. The algebraic counterpart of this statement is that the VA contains the current algebra of $G$ as a subalgebra, and in the following we will assume this condition throughout. In fact, as we shall show explicitly, all the chiral/anti-chiral defects that we have studied thus far support affine Kac-Moody algebras. 

When we discretize the integrable field theory,
the question is then how to discretize the 
VA defect. From the discussion of discretization in Section \ref{sec:disc_FT}, we find that this amounts to 
the dimensional reduction of the theory from 2d to 1d.\footnote{This can be regarded as a field theory version of T-duality \cite{Taylor:1996ik,Yamazaki:2019prm}.}

We can be more explicit in this procedure. Suppose that we have a operator $\mathcal{O}$ of a VA defined on $\mathbb{C}$, with complex coordinate $w$, with the mode expansion 
\ie
    \mathcal{O}(w)=\sum_{n=0}^{\infty} \mathcal{O}_n w^{-\operatorname{deg}(\mathcal{O})+n}\,,
\fe
where we denoted the conformal weight of the operator $\mathcal{O}$ as $\operatorname{deg}(\mathcal{O})$. When we discuss dimensional reduction we first need to apply a conformal transformation to a cylinder and extract the zero mode, which is given by
\ie 
    o(\mathcal{O}) :=  \mathcal{O}_{\operatorname{deg}(\mathcal{O})-1}\,.
\fe
We thus conclude that the dimensionally-reduced algebra should be spanned by the zero modes $o(\mathcal{O})$.

Interestingly, the discussion above regarding the dimensional reduction of VA is known in the mathematical literature, where the dimensionally-reduced algebra is known as Zhu's algebra \cite{MR1317233} associated with the VA (see Appendix \ref{vz} and e.g., \cite{Brungs:1998ij} and \cite[Section 4.1]{Dedushenko:2019mzv}).

The product $\circ$ of the VA induces
an associative $\star$-product to the Zhu's algebra as (see Appendix \ref{vz})
\ie 
    o\left(\mathcal{O}_1\right) \star o\left(\mathcal{O}_2\right)
    :=o\left(\mathcal{O}_1 \circ \mathcal{O}_2\right)\,.
\fe

\subsubsection*{Current Algebra}
We can discuss the dimensional reduction of the current algebra, which we assumed to be a subalgebra of the defect VA.
The basic OPE of an affine VOA $V_{k}(\mathfrak{g})$ (at a non-critical level) is given by (see Appendix \ref{vz})
\ie\label{jjzhu}
    J_{a}(w) J_{b}(0) \sim \frac{k \delta_{ab}}{w^{2}}+\frac{f_{a b}{}^{c} J_{c}(0)}{w} + (J_aJ_b)\,,
\fe 
where $\delta_{ab}$ is the Killing form on $\mathfrak{g}$. 
Denoting $\left[J_{a}\right] \in \mathrm{Zhu}_{s}(V)$ as $j_{a}$ and using \eqref{star}, we can compute products in the Zhu algebra, 
\ie 
    \label{zhujj1}
    j_{a} \star j_{b}
    &=\textrm{Res}_w\bigg(\frac{(\hbar_{\star} w+1)}{w} \bigg(\frac{ f_{a b}{}^{c} J_{c}(0)}{w} + (J_aJ_b)\bigg)\bigg)\\
    &=\left(j_{a} j_{b}\right)+ \hbar_{\star} f_{ab}{ }^{c} j_{c}\,,
\fe
where $\left(j_{a} j_{b}\right)=\left[\left(J_{a} J_{b}\right)\right]$ denotes conformal normal ordering. 
Moreover, one can derive the commutation relation of the Lie algebra $\mathfrak{g}$. 
This follows since there are two expressions for the star product 
\ie
    a * b 
    & =\operatorname{Res}_w\left(Y(a, w) \frac{(\hbar_{\star}w+1)^{\operatorname{deg}(a)}}{w} b\right)  
    =\operatorname{Res}_w\left(Y(b, w) \frac{(\hbar_{\star}w+1)^{\operatorname{deg}(b)-1}}{w} a\right),
\fe
and using these, it can be shown that \cite{zhu1990vertex}
\ie 
    a * b-b * a 
    \equiv &  \operatorname{Res}_w\left(Y(a, w) \frac{(\hbar_{\star}w+1)^{\operatorname{deg}(a)}}{w} b\right) -\operatorname{Res}_w\left(Y(a, w) \frac{(\hbar_{\star}w+1)^{\operatorname{deg}(a)-1}}{w} b\right) \\
    =&  \operatorname{Res}_w\left(Y(a, w)(\hbar_{\star}w+1)^{\operatorname{deg}(a)-1}  b\right) ,
\fe
whereby we find that \eqref{jjzhu} implies 
\begin{equation}
    j_{a} \star j_{b}-j_{b} \star j_{a}=  f_{ab}{}^{c} j_{c}\,.
\end{equation}

In fact, we can recover the whole of the universal enveloping algebra $\U(\mathfrak{g})$:
we can define a  basis of $\U(\mathfrak{g})$ using the $\star$-product to define higher order products of $j_a$ (with fixed ordering), and using the associativity of the $\star$-product to define commutators between such higher-order products. The latter form a linearly independent basis of $\U(\mathfrak{g})$, according to the Poincar\"e-Birkhoff-Witt (PBW) theorem. 
We therefore conclude that  Zhu's algebra for the current algebra is the universal enveloping algebra $\U(\mathfrak{g})$.  

This analysis suggests that an order chiral/anti-chiral surface operator that supports an affine VOA ought to be discretizable to a Wilson line in a representation of $\U(\mathfrak{g})$.

We can summarize the discussion of this subsection in the following diagram:

\begin{figure}[htbp]
\centering
\begin{tikzpicture}
    \node[rectangle, draw, align=center](42) at (0,2) {(2d) vertex algebra};
    \node[rectangle, draw, align=center](41) at (6,2){(2d) current algebra};
    \node[rectangle, draw, align=center](2) at (0, 0) {(1d) Zhu's algebra};
    \node[rectangle, draw, align=center](1) at (6,0){(1d) Lie algebra $\mathfrak{g}$}; 
    \node[ align=center] at (3,2){$\largesupset$};
    \node[ align=center] at (3,0){$\largesupset$};
    \draw [->]  (42)--(2) node[midway, left] {dimensional reduction};
    \draw [->]   (41)--(1) node[midway, right] {dimensional reduction};
\end{tikzpicture}
\end{figure}

\subsubsection*{Free $\beta\gamma$ Vertex Algebra}

As a simple example, let us revisit the example of the free $\beta\gamma$ system in Section \ref{sec:fbg-sec} in the algebraic framework.

The free $\beta\gamma$ VA is described by 
the standard OPE between $\beta^i(w)$ and $\gamma_i(w')$:
 \ie 
    \label{stbg}
    \beta^i(w)\gamma_j(w') \sim - \frac{\delta^i_j}{w-w'} \,.
 \fe
For free $\beta \gamma$ systems, the currents that appear in the surface operator action have the Wakimoto free field realization  $\cJ^a=\beta_i \rho(t^a)^{i}{}_{j}\gamma^j$. This gives rise to a current algebra, and in particular to the familiar current algebra that arises from symplectic bosons when $\beta$ and $\gamma$ are complex conjugates of each other. Let us recall this construction.  

Given the surface operator of the form 
\ie
    \frac{1}{2\pi}\int_{\cM} d^2w (\beta_i \partial_{\wbar} \gamma^i +A_{\wbar}^a\beta_i \rho(t_a)^{i}{}_{j}\gamma^j)\,,
\fe 
 the free-field OPE implies the following OPE for the normal-ordered currents $\cJ_a=(\beta_i \rho(t_a)^{i}{}_{j}\gamma^j )$ of the $G$-symmetry: 
\ie
    \cJ_{a}(w) \cJ_{b}(w') 
    &\sim -\frac{\rho(t_a)_{ij}\rho(t_b)^{ji}}{(w-w')^2} 
        +\frac{\rho(t_a)_{ij}\gamma^j(w)\beta^{k}(w')\rho(t_b)_{ki}}{(w-w')} 
        -\frac{\beta^{k}(w)\rho(t_a)_{ki} \rho(t_b)^i_{j} \psi^{j}(w')}{(w-w')} \\
    &\sim -\frac{\rho(t_a)_{ij}\rho(t_b)^{ji}}{(w-w')^2}  + \frac{f_{ab}{}^c\cJ_c(w')}{w-w'}\,,\\
\fe 
where the parentheses indicate conformal normal-ordering.
Here we find a second-order pole in the OPE arising due to double contractions. 
The corresponding commutator can be computed to be an affine Kac-Moody algebra,
\begin{align}
    [\cJ_{an},\cJ_{bm}]=f_{ab}{}^c \cJ^c_{n+m}-c n\delta_{ab}\delta_{n+m,0}\,,
\end{align}
where the coefficient $c$ is determined by
\ie
    \rho(t_a)_{ij}\rho(t_b)^{ji}=c\delta_{ab}\,.
\fe 
and $c$ depends on the choice of gauge group and choice of representation, $\rho$. For example, if $\gamma$ is in the fundamental representation of $\SU(N)$, then $c=1$.

Crucially, since the anomaly term is a double pole, it does not contribute to the Lie algebra commutator one obtains from the zero modes of the affine Kac-Moody algebra.
However, as we shall see in Section \ref{akccalc}, the value of $c$ plays an important role in the gauge anomaly of chiral surface operators that support affine Kac-Moody algebras, and thereby in the quantum correction to the meromorphic one-form $\omega$ of the 4d CS theory.

\subsubsection*{Free Fermion Vertex Algebra}

Given the surface operator of the form 
\ie
    \frac{1}{2\pi}\int_{\cM} d^2w  (\psi^*_i \partial_{{\wbar}} \psi^i +A_{\wbar}^a\psi^*_i \rho(t_a)^{i}{}_{j}\psi^j)\,,
\fe 
we have the standard free-field OPE
\ie
    \psi^*_i(w) \psi^j(w') \sim \frac{\delta_{i}^{j}}{(w-w')}\,,
\fe 
which implies the following OPE for the normal-ordered currents $\cJ_a=\psi^*_i \rho(t_a)^{i}{}_{j}\psi^j$ of the $G$-symmetry:
\ie
    \cJ_{a}(w) \cJ_{b}(w') 
     &\sim \frac{\rho(t_a)_{ij}\rho(t_b)^{ji}}{(w-w')^2} +\frac{\rho(t_a)_{ij}\psi^j(w)\psi^{*}_{k}(w')\rho(t_b)^{k}{}_{i}}{(w-w')} -\frac{\psi^{*}_{k}(w)\rho(t_a)^{k}{}_{i} \rho(t_b)^i{}_{j} \psi^{j}(w')}{(w-w')} \\
    &\sim \frac{\rho(t_a)_{ij}\rho(t_b)^{ji}}{(w-w')^2}  + \frac{f_{ab}{}^c\cJ_c(w')}{w-w'}\,.
\fe 
Here, as in the free $\beta\gamma$ VA, we find a second-order pole in the OPE arising due to double contractions, but with an opposite sign. This is the Wakimoto free-field realization of an affine Kac-Moody algebra in terms of fermions.
The corresponding commutator can be computed to be 
\begin{align}
    [\cJ_{an},\cJ_{bm}]=f_{ab}{}^c \cJ_{c\,n+m}+cn\delta_{ab}\delta_{n+m,0}\,,
\end{align}
where $c$ is defined by
\ie
    \rho(t_a)_{ij}\rho(t_b)^{ji}=c\,\delta_{ab}\,.
\fe 
Once again, since the anomaly term is a double pole, it will not contribute to the Lie algebra commutator one obtains from the zero modes of the affine Kac-Moody algebra.

\subsubsection*{Coadjoint Orbit Defect}

In order to describe the coadjoint orbit defect 
\ie 
    \int_{\Sigma } \Tr\left(\Lambda\,\cdot\cG^{-1}D_{\wbar} \cG\right)\,d^2w
\fe 
in terms of a VA, we need to take into account the fact that the degrees of freedom of the defect are valued in the flag variety $G_{\mathbb{C}}/B$. In fact, we need a sheaf of VAs defined over each coordinate chart of $G_{\mathbb{C}}/B$. To facilitate this analysis, we can show that the coadjoint orbit defect action for $G_{\mathbb{C}}/B$ is equivalent to the curved $\beta\gamma$ defect  action with holomorphic vector fields taking the form of twisted differential operators; this is explained in Appendix \ref{betagammacoadjoint}.
As such, the analysis of VAs associated with curved $\beta\gamma$ defects, performed in the next section, is expected to encompass that of coadjoint orbit defects, as it involves the appropriate chiralization of sheaves of twisted differential operators.

\subsection{Sheaves of Vertex Algebras and Curved \texorpdfstring{$\beta\gamma$}{bg} Defects}\label{sec:curved-bg}

Let us now turn to the example of the curved $\beta\gamma$
system. To describe this example, we need to generalize our discussion of a VA to ``a sheaf of VA''. To avoid being pedantic, let us explain this concept in the specific example of the curved $\beta\gamma$
system.

The target space of the curved $\beta\gamma$-system is a curved manifold $X$, which can be covered by a set of local coordinate charts, each of which can be regarded as a copy of the flat space. The VA counterpart of this statement is that the free $\beta\gamma$ systems associated to each coordinate chart will be ``glued together'' appropriately, so that we will find an algebra defined globally on $X$. This is what is meant by the statement that we have a sheaf of VA.
As we shall recall in the following subsections, this system is naturally described in terms of a sheaf of (twisted) chiral differential operators; when discretized
we obtain a sheaf of (twisted) differential operators.
 
 \subsubsection{Chiral Differential Operator as Vertex Algebra}\label{bg-anomaly}

Sheaves of $\beta\gamma$ systems can be formulated mathematically in the language of chiral differential operators. Chiral differential operators were first studied by Malikov, Schechtman, and Vaintrob \cite{Malikov:1998dw}, with the aim of defining sheaves of VAs on curved manifolds. They provide a mathematically rigorous definition of the chiral part of a conformal field theory with curved target space. In what follows, we shall quantize the curved $\beta \gamma$ defects by coupling a generalization of chiral differential operators, namely \textit{twisted} chiral differential operators, to the 4d CS theory. This embedding of twisted chiral differential operators into the 4d CS theory is novel, and the twist in question is with respect to the 4d CS gauge field.\footnote{A different embedding of twisted chiral differential operators in the 4d CS theory was recently considered in \cite{Khan:2022vrx}. In that setup, the underlying $\beta \gamma$-system is a holomorphic surface defect along the holomorphic surface  $C$, while we consider such systems along $\Sigma$.  } 

In what follows, we shall focus on the chiral defect defined in \eqref{chiralbg} for $\Sigma=\mathbb{C}$. For ease of analysis, we shall employ the gauge $A_{\bar{w}}=0$ at the location of the defect, $z_+$. The equations of motion then imply that 
\ie 
    \partial_{\wbar}\gamma_i=0\,,\\
    \partial_{\wbar}\beta^i=0\,.
\fe
 In addition, in holomorphic gauge $A_{\zbar}=0$, the 4d CS equation of motion implies that 
 \ie
     \partial_{\wbar}A_w=0\,.
 \fe
 Thus, we may perform the Laurent expansions 
 \begin{align}
     \gamma_i(w)&=\sum_{m\in \mathbb{Z}}\frac{\gamma_{im}}{w^m}\,,
     \\
      \beta^i(w)&=\sum_{m\in \mathbb{Z}}\frac{\beta^i_{m}}{w^{m+1}}\,,
      \\
      \label{gauge3}
      A_w(w)&=\sum_{m\in \mathbb{Z}}\frac{A_{m}}{w^{m+1}}\,,
\end{align}
 which all converge in an annulus around $w=0$. In addition, the standard OPE between $\beta^i(w)$ and $\gamma_i$ holds, i.e.,
\ie 
    \beta^i(w)\gamma_j(w') \sim - \frac{\delta^i_j}{w-w'} \,,
\fe
while the OPE of $A_w$ with $\beta^i$ and $\gamma_i$ is non-singular.  
We shall further specify the gauge for the Cartan component of the gauge field \eqref{gauge3}, denoted $A_w^h$, by gauging away the non-singular terms in its Laurent expansion by a $w$-dependent gauge transformation. Such a gauge transformation would not affect the gauge that we have previously chosen locally for $A_{\wbar}$ and $A_{\zbar}$. 
 The annihilation operators, i.e., $A^h_{m}$ for $m>0$, then annihilate any state in the Hilbert space, and are thus effectively equal to zero as well since we are interested in nontrivial Hilbert spaces. We are thus left with 
 \ie \label{regs}
     A^h_w(w)=\frac{A^h_0}{w}
 \fe
 for the Cartan component of $A_w^h(w)$.
 
 Now, the OPE of conserved currents that arise from the $G$-symmetry of a 2d curved $\beta\gamma$ system with target space $X$ is known to have an anomalous term \cite{Malikov:1998dw,Witten:2005px}:
\ie 
    J_V(w)J_W(w')
    &\sim -\frac{\partial_jV^i(w)\partial_iW^j(w')}{(w-w')^2}-\frac{(V^i\partial_iW^j-W^i\partial_iV^j)\beta_j}{ w-w'}\\
    &\sim -\frac{\partial_jV^i\partial_iW^j(w)}{(w-w')^2}-\frac{(V^i\partial_iW^j-W^i\partial_iV^j)\beta_j}{ w-w'}
-\underbrace{\frac{(\partial_k\partial_jV^i)(\partial_iW^j\partial\gamma^k)}{
w-w'}}_{\textrm{anomalous term}}\,,\label{JJ-OPE}
\fe
where $J_V=-V^i\beta_i$, and where $V$ and $W$ are holomorphic Killing vector fields on $X$ that generate the $G$ symmetry.
The first term does not contribute to the Lie algebra commutator as it contains a double pole. 
As shown in Appendix \ref{opeal}, the algebra of zero modes is 
\ie\relax 
    [J_{V0},J_{W0}]= J_{[V,W]0}+\tilde{J}_{0}\,.\label{zeroal}
\fe 
However, this is not in the form we want, as the second term is the anomaly that spoils the closure of the Lie algebra. It is related to the obstruction to gluing $\beta\gamma$ systems contained in open sets of the target space, as reviewed in Appendix \ref{cdoapp}, and can be understood as a Lie algebra extension 
by $\tilde{J}_0$.

We will need to define a consistent global sheaf of chiral differential operators on the target space.
 We review how this is accomplished for K{\"a}hler manifolds, denoted $X$, in Appendix \ref{cdoapp}. The crucial point is such a definition is only possible when the first Pontryagin class of $X$ vanishes, i.e., $p_1(X)=0$, and this can be interpreted as the vanishing of an anomaly of the $\beta\gamma$ system \cite{Witten:2005px}.\footnote{There is another anomaly vanishing condition, which requires $c_1(\Sigma)c_1(X)=0$. This condition shall not play a major role in our analysis, since we shall only consider $\Sigma$ such that $c_1(\Sigma)=0$. } 

Having defined a global sheaf of chiral differential operators, we can go on to study its global sections. Let us focus on the example of $X=\mathbb{CP}^1$, with charts parametrized by $\gamma$ and $\gamma'$, where $\gamma=1/\gamma'$.  As explained in \cite{Witten:2005px}, the global sections in this case take the following form\footnote{The sign flip in the sign of $J_3$ follows from a rearrangement lemma, which states
\ie 
    \label{rearrange}
    ((AB)(CD)) =& (A(C(DB))) + (A(C([B,D]))) + (A(([B,C])D)) + (A([CD,B])) + (([CD,A])B) \\&+ ([(AB), (CD)]) \,.
\fe
We shall use the standard identity
\ie 
    ([A, B])=\sum_{n>0} \frac{(-1)^{n+1}}{n !} \partial^n\{A B\}_n(w) \,,
\fe
where OPEs are defined as
\ie 
    {A(z) B}(w) \equiv \sum_{n=1}^N \frac{\{A B\}_n(w)}{(z-w)^n} \,.
\fe 
Substituting $A=\frac{1}{\gamma}$, $B=\gamma$, $C=\beta$ and $D=\gamma$ into \eqref{rearrange}, we find
\ie 
    \left(\frac{1}{\gamma}\gamma\right)(\beta \gamma)
    =\left(\frac{1}{\gamma}(\beta (\gamma \gamma ))\right)+0 + 0 -\frac{\partial \gamma}{\gamma } -\frac{\partial \gamma}{\gamma }  +0 \,,
\fe
where the second term vanishes since $\gamma$ has a regular OPE with itself and hence $([\gamma,\gamma])=0 $, the third term vanishes since 
the $\gamma-\beta$ OPE has no field-dependent singular terms and hence $([\gamma,\beta])=0 $, and the last term vanishes since the OPE of $\beta\gamma$ with 1 is regular. Thus, we find that 
\ie 
    \label{finrearr}
    (\gamma \beta)
     = \left(\frac{1}{\gamma}(\beta (\gamma \gamma ))\right)-2\frac{\partial \gamma}{\gamma }  \,.
\fe 
This implies the gluing law of $J_3$. 
}
\ie \label{globse}
    J_{+}(w)&=-(\beta(\gamma \gamma)  )+2 \partial \gamma &&= \beta' \,,\\
    J_{-}(w)&=\beta &&= -(\beta'(\gamma'\gamma') )+2 \partial \gamma' \,,\\
    J_{3}(w)&=-(\gamma\beta)&&=(\gamma'\beta')\,,
\fe
where the parentheses indicate normal-ordering. 
One can show that these are globally well-defined via the CFT automorphisms 
\ie 
    \gamma'=\frac{1}{\gamma}\, \, , \quad
    \beta'= -(\beta(\gamma \gamma)  )+2 \partial \gamma 
\fe
that are used to glue fields across open sets (the term proportional to $\partial \gamma$ is necessary in the gluing law to ensure that the $\beta (w) \beta (w')$ OPE is preserved in both charts).
Their OPEs can be shown to be
\ie
    J_{+}(w) J_{-}\left(w^{\prime}\right) &\sim \frac{2 J_{3}}{w-w^{\prime}}-\frac{2}{\left(w-w^{\prime}\right)^{2}}\,,\\
    J_{3}(w) J_{3}\left(w^{\prime}\right) &\sim-\frac{1}{\left(w-w^{\prime}\right)^{2}}\,, \\
    J_{3}(w) J_{+}\left(w^{\prime}\right) &\sim \frac{J_{+}\left(w^{\prime}\right)}{w-w^{\prime}}\,, \\
    J_{3}(w) J_{-}\left(w^{\prime}\right) &\sim-\frac{J_{-}\left(w^{\prime}\right)}{w-w^{\prime}}\,,\label{critakm}
\fe
which implies that they satisfy an $\SL(2)$ current algebra at critical level. The zero mode algebra of this current algebra is indeed an ordinary Lie algebra. 

However, one can show that the rescaled Sugawara energy-momentum tensor in this case is zero, i.e., 
\begin{equation}
    S=\bigg((J_3J_3)+\frac{1}{2}(J_+J_-)+\frac{1}{2}(J_-J_+)\bigg)=0\,.
\end{equation}
Hence, we do not expect to obtain nontrivial representations of $\mathfrak{g}$ from the zero-mode algebra of the current algebra. This suggests that one should consider further quantum corrections that can modify the form of the global sections \eqref{globse}. Indeed, given that we are dealing with a gauged $\beta\gamma$-system, it it more natural to consider sheafs of \textit{twisted} CDOs, first studied by Arakawa et al.\ in \cite{arakawa2011algebras}. The global sections in this case obtain further quantum corrections from the presence of gauge fields. As we shall observe, sheaves of twisted CDOs are the natural chiralization of the sheaves of twisted differential operators, and are thus useful to describe VAs associated with curved $\beta\gamma$-defects. 

For twisted CDOs, the $\beta\gamma$ OPE remains the same, but the CFT automorphisms used to glue fields across open sets take the form 
\ie
    \label{bgtx}
    \tilde{\gamma} &=\frac{1}{\gamma}\,, \\
    \tilde{A}^h_w &=A^h_w\,, \\
    \tilde{\beta} &=-(\beta (\gamma \gamma) )+2 \partial \gamma-(\gamma A^h_w)\,.
\fe
In the present context, we have identified the twist operator with $A_w^h$, the Cartan component of the $w$-component of the 4d CS gauge field.
There are still currents that generate a critical level affine Kac-Moody algebra, but they take the form 
\ie 
    J_{+}(w)&=-(\beta(\gamma \gamma)  )+2 \partial \gamma -({\gamma} {A}^h_w ) &&= \tilde{\beta}\,,\\
    J_{-}(w)&= \beta &&= -(\tilde{\beta}(\tilde{\gamma} \tilde{\gamma})  )+2 \partial \tilde{\gamma} -(\tilde{\gamma} \tilde{A}^h_w) \,,\\
    J_{3}(w)&=-(\gamma\beta) -\frac{1}{2}{A}^h_w
    &&=(\tilde{\gamma} \tilde{\beta}) +\frac{1}{2}{\tilde{A}}^h_w
\,.
\fe
The rescaled energy-momentum tensor in this case is nonzero, i.e., 
\begin{equation}
    S(w)=\bigg((J_3J_3)+\frac{1}{2}(J_+J_-)+\frac{1}{2}(J_-J_+)\bigg)=\frac{1}{4} (A^h_w(w))^2-\frac{1}{2} \partial_w A^h_w(w)\,.
    \label{rEM}
\end{equation}

From the transformation law for $\beta$ in \eqref{bgtx}, we further observe, that $A_w$ can now be interpreted as the deformation parameter for affine deformations of $T^*\mathbb{CP}^1$, which are encapsulated in $H^1\left(\mathbb{CP}^1, \mathcal{O}(-2)\right)$. By \v{C}ech-Dolbeault isomorphism, this is equivalent to $H^{1,1}(\mathbb{CP}^1)$, implying an analogue of the Bohr-Sommerfeld quantization condition where periods of the K\"ahler form are not quantized in terms of integers, but may be valued in any complex number.

\subsubsection{Twisted Differential Operators as Zhu's Algebra}\label{sec:DO_Zhu}

Let  us now consider the discretization of the $\beta\gamma$ system.  On a cylinder with holomorphic coordinate 
$\tilde{w}=t+i\theta$, we obtain
\ie 
    \int_{\mathbb{R} \times S^1 } \beta^i \bar{D}_{\bar{\tilde{w}}}\,\gamma_i\,\mathrm{d}\tilde{w}\mathrm{d}\bar{\tilde{w}} := 
    \int_{\mathbb{R} \times S^1 } \left(\beta^i \partial_{\bar{\tilde{w}}} \gamma_i+A_{a, \bar{\tilde{w}}} \beta^i \rho_{a, i}(\gamma) \right)\mathrm{d}\tilde{w}\mathrm{d} \bar{\tilde{w}}\,,
\fe
and
\ie 
    \partial_{\bar{\tilde{w}}}=\frac{1}{2}(\partial_t-i\partial_{\theta})\,.
\fe 
Dimensionally reducing along the circular direction then gives the 1d action
\ie 
    \frac{1}{2}\int_{\mathbb{R}  } \beta^{i} \partial^{A}_t \gamma_{i}\,,
\fe
if we identify $A_{\bar{\tilde{w}}}$ with $A_t$. As we have already described in Section \ref{sec:lattice-Zhu},
the algebraic counterpart of the 1d theory ought to be  given by the sheaf of Zhu's algebra associated with the
sheaf of twisted chiral differential operators. 

Now, a result of \cite[Theorem 4.7]{arakawa2011algebras} states that the Zhu algebra of a sheaf of twisted chiral differential operators is equivalent to a sheaf of twisted differential operators.  
A technical point in the proof of \cite{arakawa2011algebras} is the requirement that only regular singularities are permitted in the Laurent expansion of $A^h_w(w)$, which plays the role of the central character. This is necessary (from \cite[Theorem 5.2]{arakawa2011algebras}) because a module over a sheaf of twisted chiral differential operators with a central character $\chi(z)$ can be identified with a module over a sheaf of twisted differential operators, defined with respect to a weight given by $\chi_0$, when $\chi(z)$ has a regular singularity, but is zero otherwise. Indeed, from \eqref{regs}, the form of $A^h_w$ satisfies this requirement in the gauge that we are in. The advantage of arriving at twisted differential operators by starting from twisted chiral differential operators is that we know which $\beta \gamma$ surface defects have vanishing anomalies and can be discretized consistently.

Let us try to understand the relationship between twisted chiral differential operators and twisted differential operators in greater detail, for the example of $X=\mathbb{CP}^1$.
An immediate observation that can be made is that if we set $A^h_w(w)=2a/w$ for a nonzero complex number, $a$, and perform the Laurent expansion 
\begin{equation}\label{aews}
    S(w)=\sum_{n=0}^{\infty}\frac{S_n}{w^{n+2}}
\end{equation}
of the rescaled energy-momentum tensor \eqref{rEM}, we arrive at $S_0=a(a+1)$, which is precisely the expression for the Casimir found in \eqref{casi1}.

In addition, since $A^h_w$ has regular OPEs with $\beta$, $\gamma$ and itself, $S(w)$ has a regular OPE with itself as well as the currents that generate the critical level affine Kac-Moody algebra. 
Moreover, the modes obtained from the Laurent expansion \eqref{aews}
also commute with the Laurent modes of the currents
that appear in the expansion 
\begin{equation}
    J^a(w)=\sum_{n=-\infty}^{\infty}\frac{J^a_n}{w^{n+1}}\,,
\end{equation}
i.e., 
\ie \relax
[S_n,J^a_m]=0\,,\qquad  [S_n,S_m]=0\,.
\fe
Note that the standard relation
\begin{equation}
    S_n=\frac{1}{2}\sum_a\sum_m:J^a_mJ^a_{n-m}:\,.
\end{equation}
among the Laurent modes also holds. In particular, the zero mode $S_0$ is
\ie 
S_0=\frac{1}{2}\sum_a\sum_m:J^a_mJ^a_{-m}:\,.
\fe
Defining a highest weight state $|\eta \rangle$ via
\begin{equation}
    J_0^3|\eta \rangle=\eta |\eta \rangle \,, 
\end{equation}
and 
\begin{equation}
    J^a_n|\eta\rangle =0\,, \quad \quad n>0\,.
\end{equation}
We note that $|\eta \rangle$ is also an eigenstate for $S_0$ since $[S_0,J_0^3]=0$.
Moreover, defining a highest weight module in the standard manner, i.e., 
\begin{equation}
    J^-_0J^-_0\ldots J^-_0|\eta\rangle\,,
\end{equation}
we find that $S_0$ indeed behaves as a Casimir operator since $[S_0,J_0^-]=0$, and since 
$S_0$ acts effectively as $\frac{1}{2}\sum_a:J^a_0J^a_0:$ on $|\eta \rangle$ due to normal ordering.
Thus, we find that we can define nontrivial  modules of conformal dimension zero that take the form of highest weight modules for the Lie algebra $\mathfrak{g}$.

We have so far focused on twisted chiral differential operators on $\mathbb{CP}^1$. Generalization of the above techniques to $X=\mathbb{CP}^n$ for $n>1$ is not possible, since the curved $\beta\gamma$ system is anomalous for $X=\mathbb{CP}^{n>1}$ which has a non-vanishing $p_1(X)$.
This is consistent with the result that sigma models on all $\mathbb{CP}^n$ for $n>1$ are known to non-integrable at the quantum level \cite{Abdalla:1982yd}.

\subsubsection*{Generalization to the Case of Flag Manifolds}

The above construction can straightforwardly be generalized to the case of flag manifolds of the form ${G}_{\mathbb{C}}/{B}$.\footnote{ Flag manifold sigma models have previously been related to integrable spin chains in \cite{Affleck:2021vzo,Bykov:2012am}.} Here, we can define global sections that generate the relevant affine Lie algebra, $\hat{\mathfrak{g}}_{-\sh^{\vee}}$ at critical level, since $p_1({G}_{\mathbb{C}}/{B})=0$. We may then obtain representations of $\mathfrak{g}$ as representations of the Zhu algebra of the globally-defined sheaf of twisted CDOs on ${G}_{\mathbb{C}}/{B}$, described in terms of twisted differential operators. It is natural to twist the ordinary sheaf of CDOs in this case with respect to the Cartan elements of the 4d CS field $A_w$. This agrees with an analogue of the Bohr-Sommerfeld quantization condition corresponding to $H^{1,1}({G}_{\mathbb{C}}/{B})=\mathfrak{h}$ (compare with discussion below \eqref{rEM}).

We can also discuss  twisted differential operators on even more general manifolds, such as partial flag varieties ${G}_{\mathbb{C}}/{P}$, where ${P}$ is a parabolic subgroup of ${G}_{\mathbb{C}}$. The associated representations of ${G}_{\mathbb{C}}$ are in general parabolic Verma modules. However, sigma models on some examples of such manifolds are known to non-integrable at the quantum level (for example, $\mathbb{CP}^2$ \cite{Abdalla:1980jt,Abdalla:1982yd}).
Hence, the thermodynamic limit of line operators defined as 1d gauged $\beta \gamma$ systems with such target spaces ought to give rise to anomalous surface operators, even if the line operators themselves are not anomalous (for example, recall from \cite{Costello:2017dso} that a Wilson line in any representation of $\SL(N,\mathbb{C})$ is quantizable). 

\section{Anomaly Inflow for Chiral Anomaly}
  \label{sec:anomaly}

For an order surface operator supporting an affine Kac-Moody algebra, we expect a gauge anomaly proportional to the level of this algebra, localized at the location of the defect.

We need to ensure that such anomalies are cancelled by the coupling to the 4d CS theory.
The mechanism that ensures this takes the form of 4d--2d anomaly-inflow, and as we shall see requires a quantum correction to the meromorphic one-form that enters the definition of the 4d CS theory. 
Hence, if we wish to discretize, e.g., the quantum massless Thirring model, we have to ensure that the
chiral anomaly is cancelled in this manner. 

We emphasize at this point that for the case of chiral fermions, the phenomenon of fermion doubling does not occur. This is because the discretization procedure does not involve the discretization of derivatives into difference operators, and no modification of dispersion relations occurs. Thus, the aim of anomaly cancellation is not to prevent doubling of degrees of freedom, but rather to ensure that we have a consistent quantum system, at least in perturbation theory. 

In this section, we shall first explain how gauge anomalies arise for any surface operator with degrees of freedom that to give rise to an affine Kac-Moody algebra. We then show how they are cancelled by coupling with the 4d CS theory.\footnote{MY would like to thank Kevin Costello for a related discussion circa 2018 \cite{CY_unpublished}. See also \cite{Gaiotto:2020fdr,Gaiotto:2020dhf}.}

\subsection{Anomaly for Chiral and Anti-Chiral Surface Defects}\label{akccalc}

Let us first discuss the anomalies localized at the surface defect. In Appendix \ref{chiralanomaly} we discuss this anomaly for the textbook example of chiral fermions.
Here we instead discuss more general defects with Kac-Moody symmetries.

Consider a 2d chiral defect with action 
\ie 
S_{\textrm{defect}}[X,A] = S[X] +  \int_{\Sigma}d^2w  A_{\wbar}^a\,\cJ_a\,.
\fe
Here we denote the field content of the surface defect collectively as $X$, and $ \int_{\Sigma} d^2w\,A_{\wbar}^a\,\cJ_a$ denotes the coupling of the 2d defect to the 4d bulk.
In general, given a such chiral surface defect, we can detect a gauge anomaly by performing an infinitesimal gauge transformation of the partition function $Z[A_{\wbar}]$\, as a functional of $A_{\wbar}$, in order to obtain the gauge transformation of the effective action
$W_+ [A_{\wbar}] = -i \textrm{ log }Z[A_{\wbar}]$. 
We obtain
\begin{align}
    \delta Z[A_{\wbar}]  
    =&\delta \left(\int \mathcal{D}X\,e^{iS[X]+i\int_{\Sigma} d^2w A_{\wbar}^a\cJ_a}\right)\no\\
    =&-i\int \cD X\,\bigg(  \int_{\Sigma}  d^2w' D_{\bar{w}'}\epsilon^a(w',\bar{w}') \cJ_a(w')\bigg)e^{iS[X]+i\int_{\Sigma} d^2w A_{a\wbar}\cJ^a}\,\\
    =&-i\int \cD X\,\bigg(\int_{\Sigma}  d^2w' \partial_{\bar{w}'}\epsilon^a(w',\bar{w}') \cJ_a(w')\bigg)e^{iS[X]+i\int_{\Sigma} d^2w A_{a\wbar}\cJ^a} 
\nonumber
\\ &-i \int \cD X\,\bigg(\int_{\Sigma}  d^2w'f^a{}_{bc}A^b_{\wbar}\epsilon^c(w',\bar{w}') \cJ_a(w')\bigg)e^{iS[X]+i \int_{\Sigma} d^2w A_{a\wbar}\cJ^a}\,. \label{tocancel}
\end{align}
Let us study the first term in \eqref{tocancel}. Taylor-expanding the exponential of the gauge field coupling in this term, contracting pairs of currents, and taking the VEV of a single current to be zero 
(since it is a composite field to which we apply the standard conformal normal-ordering prescription)
gives 
\ie \label{befjjj}
    &-\int \mathcal{D}X\bigg(i\int_{\Sigma} d^2w'\, i\int_{\Sigma} d^2w \,\partial_{\bar{w}'}\epsilon^a(w',\bar{w}') \\
    &\qquad \qquad \bigg(-   \frac{f_{ab}{}^c \cJ_c(w)}{2\pi  (w'-w)}+\frac{k_+ \delta_{ab}}{(2\pi )^2 (w'-w)^2} + \textrm{(regular)}\bigg)A^b_{
    \wbar}(w,\bar{w})\bigg)e^{iS_{\rm defect}[X,A]}\,,f
\fe 
where we have used the OPE\footnote{The unconventional factors of $-\frac{1}{2\pi }$ appear in this OPE since the defects we study have actions with overall normalization equal to $1$. For example, in the free fermion case, the underlying OPE between fermionic fields would be 
\ie 
\psi(w')\psi(w)=-\frac{1}{2\pi  }\frac{1}{w'-w}.
\fe
}
\ie \label{jjjope}
\cJ_a(w')\cJ_b(w) \sim -\frac{1}{2\pi }\frac{f_{ab}{}^c \cJ_c(w)}{w'-w}+\frac{1}{(2\pi )^2}\frac{k_+ \delta_{ab}}{(w'-w)^2}  \,.
\fe 
Integration by parts in \eqref{befjjj} picks up both the first order and second order pole in the OPE \eqref{jjjope}, with the first order pole contributing  the expression
\ie 
-i\int \mathcal{D}X \int_{\Sigma} d^2w \epsilon_a(w,\bar{w})  f^{ab}{}_c \cJ^c(w)A_{b\wbar}(w,\bar{w})e^{iS_{\textrm{defect}}[X,A]}\,,
\fe 
which cancels the second term in \eqref{tocancel}.

Using the identity 
\begin{equation}
   \frac{1}{2\pi i} \partial_{\wbar} \frac{1}{w^2}=\,\partial_w \delta^2(w)\,,
\end{equation}
we find that the remaining expression is
\ie 
k_+\int \mathcal{D}X\bigg(\int_{\Sigma} d^2w'\int_{\Sigma} \frac{d^2w}{2\pi i}\,\epsilon_a(w',\bar{w}') 
A^a_{
\wbar}(w,\bar{w})\bigg)\partial_{w'}\delta^2(w'-w)e^{iS_{\rm defect}[X,A]}\,.
\fe 
Integrating by parts again moves the derivative onto $\epsilon_a(w')$, upon which we can integrate over the delta function, to obtain
\ie 
 -k_+ \int \mathcal{D}X\bigg(\int_{\Sigma} \frac{d^2w}{2\pi i} \,\partial_{w}\epsilon_a(w,\bar{w}) 
A^a_{
\wbar}(w,\bar{w})\bigg)e^{iS_{\rm defect}[X,A]}\,.
\fe 
Hence,\footnote{The relationship of the 2d chiral anomaly to the correlation function of currents is analogous to the familiar relationship of the 4d chiral anomaly to the 1-loop triangle Feynman diagram.  } we deduce that the gauge variation of the effective action is given by 
\ie \label{c_var}
\delta W_{+} =
 \frac{k_{+}}{2\pi } \int_{\Sigma} d^2w \,\partial_{w}\epsilon_a(w,\bar{w}) 
A^a_{
\wbar}(w,\bar{w})\,.
\fe 
Similarly, the gauge variation of the effective action with an anti-chiral defect, with Kac-Moody symmetry with level $k_{-}$, 
is given by
\ie \label{a_var}
 \delta W_{-} =\frac{k_{-}}{2\pi } \int_{\Sigma} d^2w \,\partial_{\bar{w}}\epsilon_a(w,\bar{w}) 
A^a_{w}(w,\bar{w})\,.
\fe 

In addition, we ought to impose the Wess-Zumino consistency condition \cite{Wess:1971yu} to get the correct form of the anomaly for either defect. This corresponds to the requirement that, when employing the BRST gauge-fixing scheme, the BRST operator $Q_{\rm BRST}$ is nilpotent, i.e.,  $Q_{\rm BRST}^2=0$, where the BRST  transformations are
\ie \label{brst}
Q_{\rm BRST}A^a_{\mu} &= - D_{\mu}c^a\,,\\
Q_{\rm BRST}c^a & = \frac{1}{2}f^{abc}c_bc_c\,,
\fe 
for $\mu=w,\wbar$. This can be achieved by adding a local counter-term to the effective action that is proportional to $\int_{\Sigma} d^2w \textrm{Tr}(A_{\mu} A^{\mu})$, and whose BRST variation is proportional to
\ie 
\int_{\Sigma}d^2w\, \mathrm{Tr}(\partial_{\mu}c A^{\mu})\,.
\fe
Choosing the coefficient of the counter-term such that this variation becomes 
\ie 
-\frac{k_{+}}{4\pi }\int_{\Sigma}d^2w\textrm{Tr}(\partial_{w}c A_{\wbar}+ \partial_{\wbar}c A_{w})\,,
\fe
the anomaly which takes the form, e.g.,  $\frac{k_{+}}{2\pi } \int_{\Sigma} d^2w \,\textrm{Tr}(\partial_{w}c(w,\bar{w}) 
A_{
\wbar}(w,\bar{w}))$ is then modified to 
\ie 
-\frac{k_{+}}{4\pi }\int_{\Sigma}d^2w\textrm{Tr}(-\partial_{w}c A_{\wbar}+ \partial_{\wbar}c A_{w}) = \frac{k_{+}}{4\pi } \int_{\Sigma} \textrm{Tr}(dc \wedge A)\,.
\fe
This can be shown to be invariant under the transformation \eqref{brst}.
This follows since 
\ie 
&Q_{\rm BRST}\int_{\Sigma} (\epsilon^{\mu \nu} \partial_{\mu }c_aA^a_{\nu} )d^2w\\
=& \int_{\Sigma} (\epsilon^{\mu \nu} \partial_{\mu}c_a D_{\nu}c^a )d^2w + \int_{\Sigma} \epsilon^{\mu \nu} \partial_{\mu}\big( \frac{1}{2}f^{abc}c_b c_c \big)A_{a\nu} d^2w  \\
=& \int_{\Sigma} (\epsilon^{\mu \nu} \partial_{\mu}c_a \partial_{\nu}c^a )d^2w + \int_{\Sigma} (\epsilon^{\mu \nu} \partial_{\mu}c_a f^{abc}A_{\nu b}c_c )d^2w + \int_{\Sigma} \epsilon^{\mu \nu}  \frac{1}{2}f^{abc}\partial_{\mu}c_b c_c A_{a\nu} d^2w  \\ & +\int_{\Sigma} \epsilon^{\mu \nu}  \frac{1}{2}f^{abc}c_b \partial_{\mu} c_c A_{a\nu} d^2w\,,
\fe
where the first term vanishes due to integration by parts, while the remaining terms cancel upon using the Jacobi identity. 
Thus, the effective action which includes the counter-term is invariant under $Q_{\rm BRST}^2$, and the Wess-Zumino consistency condition is satisfied.

The $Q_{\rm BRST}$-invariant gauge anomalies for the chiral and anti-chiral defects are thus
\ie \label{c_var2}
    \delta W_{+} = \frac{k_{+}}{4\pi } \int_{\Sigma}  \,\textrm{Tr} (d\epsilon  \wedge A )\,.
\fe 
Similarly, upon adding the aforementioned local counter-term with appropriate coefficient, the gauge variation of the effective action with an anti-chiral defect, with Kac-Moody symmetry with level $k_{-}$, 
is given by
\ie \label{a_var2}
    \delta W_{-} =-\frac{k_{-}}{4\pi } \int_{\Sigma} \,\textrm{Tr} (d\epsilon  \wedge A )\,.
\fe 

\subsection{Gauge Anomaly Cancellation in 4d CS Theory}

We will now explain how the 2d gauge anomalies derived in the previous subsection can be cancelled by an anomaly inflow mechanism from the bulk 4d CS theory.

Under the infinitesimal gauge transformation $\delta A = - d \epsilon - [A, \epsilon]$, the 4d CS action  transforms as 
\begin{align}
    \delta S_{\rm CS}[A]=\frac{1}{2\pi \hbar}\int_{\cM\times \mathbb{CP}^1} d\omega\wedge \Tr\left(d\epsilon\wedge A\right)\,, 
\label{gauge-variation-2}
\end{align}
The transformation implies that the chiral anomalies can be canceled out by considering the shift in $\omega$\,,
\begin{align}
    \omega \mapsto \omega_{\text{eff}}=\omega-\frac{\hbar}{4\pi i}\left(\sum_{\alpha=1}^{n_{+}}\frac{k_{+,\alpha}}{z-z_{+,\alpha}}-\sum_{\beta=1}^{n_{-}}\frac{k_{-,\beta}}{z-z_{-,\beta}}\right)dz\,.\label{omega-shift}
\end{align}
Indeed, the variation (\ref{gauge-variation-2}) is localized at the simple poles $z=z_{+,\alpha}\,,z_{-,\beta}$ of $\omega_{\text{eff}}$ as
\begin{align}
    &\frac{1}{2\pi \hbar}\int_{\cM\times \mathbb{CP}^1} d\omega_{\text{eff}}\wedge \Tr\left( d\epsilon \wedge A\right)\no\\
    &= - \frac{1}{4\pi}\sum_{\alpha=1}^{n_{+}}k_{+,\alpha}\int_{\cM\times \{z_{+,\alpha}\}}  \Tr\left( d\epsilon\wedge A\right)+ \frac{1}{4\pi}\sum_{\beta=1}^{n_{-}}k_{-,\beta}\int_{\cM\times \{z_{-,\beta}\}} \Tr\left( d\epsilon\wedge A\right)\,,\label{gaugetx}
\end{align}
and these terms precisely cancel out the gauge-dependent terms of the chiral anomalies which take the form given in (\eqref{c_var2} and \eqref{a_var2}). While we studied only infinitesimal gauge transformation above, essentially the same cancellation mechanism works for finite gauge transformations, as described in Appendix \ref{Ano-canf}.

Note that the absence of an extra pole at infinity requires the 
condition
\begin{align}
  \sum_{\alpha=1}^{n_+} k_{+, \alpha} =  \sum_{\beta=1}^{n_-} k_{-, \beta} \;. \label{inf-con}
\end{align}
This is the condition needed for the cancellation of the total anomalies of the theory. When this condition is not satisfied, we expect that the classical integrability of the system will be broken by the quantum effects of the theory.

Although the consistent coupling of chiral and anti-chiral defects to the 4d CS theory requires a quantum correction to $\omega$, it is natural to expect that this does not affect the computation of R-matrices in \ref{sec:lattice-cs}, since the order $\hbar$ contribution to the R-matrix remains the same, and this uniquely determines the form of the R-matrix up to overall normalization.
Let us comment on a possible refinement to this statement that can be made. Note that the quantum-corrected meromorphic one-form can be rewritten as 
\ie 
\omega(z)dz =d\theta\,,
\fe 
where 
\ie 
\theta = z-\frac{\hbar}{4\pi i} \textrm{log}\frac{\prod_{\alpha=1}^{n_+}(z-z_{+,\alpha})^{k_{+,\alpha}}}{\prod_{\beta=1}^{n_-}(z-z_{-,\beta})^{k_{-,\beta}}}\,.
\fe
Since
\ie
d\zbar A_{\zbar}&=d\bar{\theta} A_{\bar{\theta}}\,,\\
d\zbar \partial_{\zbar}&=d\bar{\theta} \partial_{\bar{\theta}}\,,
\fe
the 4d CS action can be rewritten as 
\ie
\int d\theta \wedge {\rm CS}[A(\theta,\bar{\theta})]\,.
\fe
The boundary condition in the rational case, i.e., $A=0$ at $z=\infty$ maps to $A=0$ at $\theta =\infty $. 
We then observe that the R-matrices one computes from the 4d CS theory on the $\theta$-plane ought to be rational R-matrices, whose expressions require branch cuts when mapped to the $z$-plane.

Before closing this section, we would like to comment on the effect of the shift of the meromorphic one-form $\omega$ on the integrable system.
According to the results of \cite{Vicedo:2019dej}, the meromorphic one-form $\omega$ is closely related to the coefficients of the non-ultralocal terms appearing in the Poisson algebra of the Lax pair. This observation indicates that the 2d integrable field theories associated with the chiral surface defects supporting the affine Kac-Moody algebra are non-ultralocal at the quantum level.
Indeed, in \cite{deVega:1983gn}, the Schwinger term appears in the current algebra at the quantum level, so the quantum algebra formed by the Lax pair for the massless Thirring model is non-ultralocal. We expect this phenomenon to occur for other 2d integrable field theories arising from other chiral surface defects.
Further investigation of the relationship between the shift of $\omega$ at the quantum level and the non-ultralocal term can provide an important clue for understanding the quantization of the non-ultralocal integrable field theories.

\subsection{Emergence of Disorder Defects}

Before closing this section, 
let us comment further on the implications 
of the one-loop shift to the holomorphic one-form.

Let us consider the simplest example of one chiral and one anti-chiral defect.
In order to satisfy the anomaly cancellation condition \eqref{inf-con}
the levels for the two defects should be the same---we denote this level as $k$.

The one-loop corrected one-form is given by 
\begin{align}
    \omega_{\text{eff}}=\omega - \frac{\hbar}{4\pi i }\left(\frac{k}{z-z_{+}}-\frac{k}{z-z_{-}}\right)dz\,,
 \end{align}
By rewriting the one-form as 
\begin{align}
    \omega_{\text{eff}}=\frac{(z-\zeta_+)(z-\zeta_-)}{(z-z_+)(z-z_-)}dz\,, \label{oform}
\end{align}
we find that $\omega_{\rm eff}$ has the following poles and zeros :
\begin{align}
    \mathfrak{p}=\{z_{\pm},\infty\}\,,\qquad \mathfrak{z}=\{\zeta_{\pm}\}\,,
\end{align}
where $z=\infty$ is a double pole, $z=z_{\pm}$ are simple poles, and $z=\zeta_{\pm}$ are simple zeros
\begin{align}
    \zeta_{\pm}=\frac{1}{2}\left(z_++z_-\pm \sqrt{(z_+-z_-)\Bigl(z_+-z_{-}-\frac{i \hbar k}{\pi}\Bigr)}\right)\,.
\end{align}
Note that in the semiclassical limit $\hbar\to 0$ the poles $z_{\pm}$ and zeros $\zeta_{\pm}$ will cancel each other.
In other words, the quantum correction ``pair-creates'' a pair of poles and zeroes of the one-form.

As discussed in \cite{Costello:2019tri}, when we study the 4d CS theory with meromorphic one-form with poles and zeros as in \eqref{oform},
we need to impose appropriate boundary conditions on the gauge fields to fulfill the requirement of choosing elliptic boundary conditions. More concretely, at the poles $z_{\pm}$, for some choice of gauge,
we require Dirichlet boundary conditions where 
chiral and anti-chiral components of the gauge fields are required to vanish
\begin{align}
\begin{split}
A_+ | _{z=z_-} &= 0\,, \\
A_- | _{z=z_+} &= 0\,,
\end{split}
\end{align}
while on the simple zeros $\zeta_{\pm}$ we require 
singular boundary condition where chiral and anti-chiral components of the gauge fields have poles:
\begin{align}
\begin{split}
\label{A_singular}
A_+ \sim  \mathcal{O}\left( \frac{1}{z-\zeta_+} \right)\,, \\
A_- \sim  \mathcal{O}\left( \frac{1}{z-\zeta_-} \right)\,.
\end{split}
\end{align}
We can verify that the two types of boundary conditions cancel out in the limit $\zeta_{+}\to z_{+}$ or $\zeta_{-}\to z_{-}$.
We shall further investigate the implications of the emergence of disorder defects in Section \ref{sec:chiralWZW}. 

\section{Duality Web of Line and Surface Defects}
  \label{sec:duality}

\begin{figure}[htbp]
\centering
\scalebox{0.85}{
\begin{tikzpicture}
    \node[rectangle, draw, align=center](2B) at (6,6) {4d CS \\+ 2d defects $\mathcal{B}$};
    \node[rectangle, draw, align=center](2B') at (6,0) {4d CS \\+ 2d defects $\mathcal{B}'$};
    \node[rectangle, draw, align=center](2eB) at (11.5, 5) {2d Integrable\\ Field Theory $T[\mathcal{B}]$};
    \node[rectangle, draw, align=center](2eB') at (11.5, 1) {2d Integrable\\ Field Theory $T[\mathcal{B}']$};
    \node[rectangle, draw, align=center](1B) at (0,6) {4d CS\\ + 1d defects $B$};
     \node[rectangle, draw, align=center](1B') at (0,0) {4d CS\\ + 1d defects $B'$};
    \node[rectangle, draw, align=center](1) at (-2,3) {4d CS\\ + Wilson Lines}; 
    \draw [<->, >={Latex[round]}, shift=({5pt, 5pt})] (1B)--(1) node[midway, left] {Dualize};
    \draw [<->, >={Latex[round]}, shift=({5pt, 5pt})] (1B')--(1) node[midway, left] {Dualize};
    \draw [->, >={Latex[round]}, shift=({5pt, 5pt})]  (1B)--(2B) node[midway, above, align=center] {Thermodynamic\\ limit};
    \draw [->, >={Latex[round]}, shift=({5pt, 5pt})]  (1B')--(2B') node[midway, above, align=center] {Thermodynamic\\ limit};
    \draw [->, >={Latex[round]}, shift=({5pt, 5pt})] (2B)--(2eB) node[midway, above, align=center] {Integration\\along $C$};
    \draw [->, >={Latex[round]}, shift=({5pt, 5pt})] (2B')--(2eB') node[midway, below, align=center] {Integration\\along $C$};
    \draw [<->, >={Latex[round]}, shift=({5pt, 5pt})] (2eB)--(2eB') node[midway, right, very thick] {\textbf{2d Duality}};
    \draw[dashed] (3,-1)--(3,7);
    \draw [thick, decorate, decoration={brace, mirror, amplitude=10pt}] (-3, -1)--(2.5, -1) node[midway, below, align=center, yshift=-5pt] {1d Integrable\\ Lattice Model};
    \draw [thick, decorate, decoration={brace, mirror, amplitude=10pt}] (3.5, -1)--(13.5, -1) node[midway, below, align=center, yshift=-5pt] {2d Integrable \\ Field Theory};
\end{tikzpicture}
} 
\caption{By starting with two sets of 1d defects which are dual to the Wilson lines in the same representation, we obtain a duality between 2d integrable field theories.}
\label{duality_web}
\end{figure}
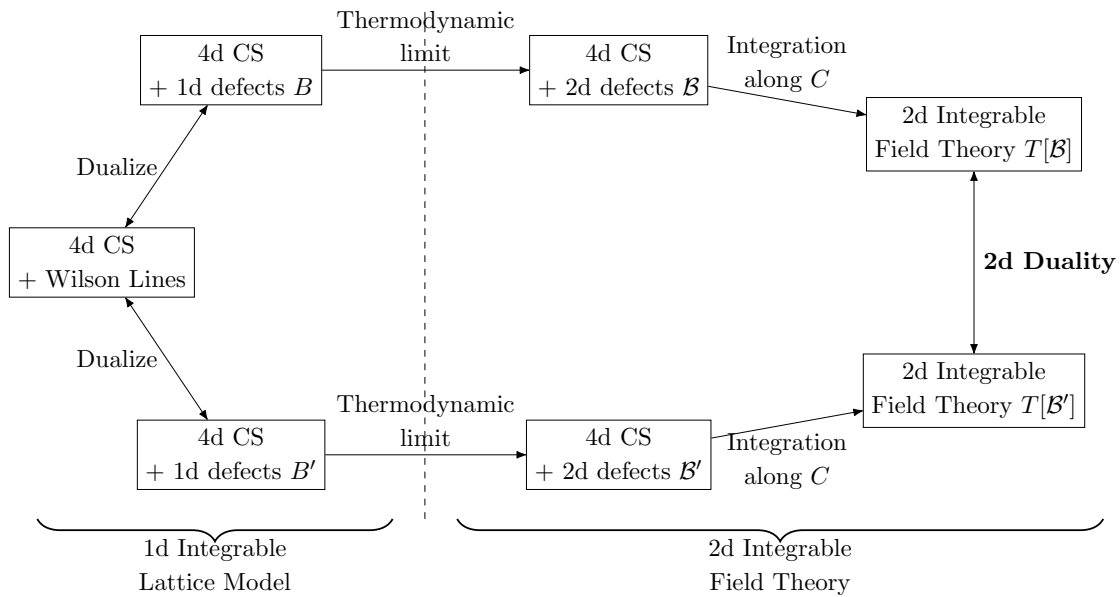

In Sections \ref{sec:disc_defect} and \ref{sec:VA} we discussed discretizations of 2d order surface defects
into 1d line defects, and we have seen that in many examples the resulting 1d defect can be identified with (i.e., dualized into)
the Wilson lines of the 4d CS theory. This in itself might not be too much of a surprise from the viewpoint of the 4d CS theory,
whose only natural order line defects are Wilson lines. We obtain, however, a rather surprising statement once we 
go back to integrable field theories by taking the thermodynamic limit (following the procedure in Section \ref{sec:backwards})---the equivalence of the 
line defects imply the equivalence of surface defects after the thermodynamic limit, which then 
imply the equivalence (duality) between 2d integrable field theories! Since we have many different defects that are equivalent to the Wilson lines as shown in Figure \ref{1d_as_WL},
we will obtain a huge duality web of integrable field theories. We can summarize this logic as in Figure \ref{duality_web}.

In this section, we study duality webs among integrable field theories that arise as a consequence of their lattice discretizations via the 4d CS theory.
While we leave the exhaustive analysis of this duality web for future work, 
we will make steps in this direction by 
first studying the triality of coadjoint orbit defects, free fermion defects, and free $\beta\gamma$ defects (Section \ref{sec:triality}).

\subsection{\texorpdfstring{Triality of Defects: Coadjoint Orbit, Free $\beta\gamma$ and Free Chiral Fermion}{Coadjoint Orbit, Free Beta-Gamma and Free Chiral Fermion}} \label{sec:triality}

\begin{figure}[!htbp]
\centering
\begin{tikzpicture}
    \node[rectangle, draw, align=center](CA) at (0,2) {Coadjoint Orbit \\ Defect};
    \node[rectangle, draw, align=center](FF) at (-2.5, -1.5) {Free Fermion\\ Defect \\with Constraints};
    \node[rectangle, draw, align=center](BG) at (2.5,-1.5){Free \\ $\beta\gamma$ Defect\\with Constraints}; 
    \draw [<->]  (CA)--(BG); 
    \draw [<->]  (FF)--(BG); 
    \draw [<->]  (CA)--(FF);
    \end{tikzpicture}
    \caption{Seed triality among 1d defects.}
    \label{triality}
\end{figure}
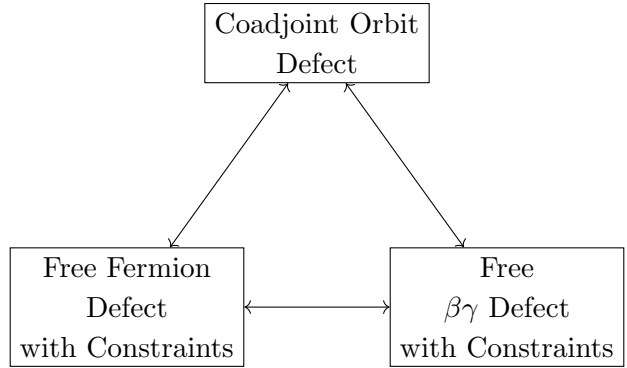

We will study as examples the triality of the
three defects in Figure \ref{triality}.\footnote{This was already included in Figure \ref{1d_as_WL}.
In this figure we emphasized that free fermion defects and free $\beta\gamma$ defects are constrained by suitable projection operators, so that we can match coadjoint orbit defects in a suitable irreducible representation.}

\begin{figure}[!htbp]
\centering
\begin{tikzpicture}
    \node[rectangle, draw, align=center](CA) at (0,2) {Generalized \\Faddeev-Reshetikhin\\ Model};
    \node[rectangle, draw, align=center](FF) at (-2.5, -1.5) {Constrained \\Fermionic Massless\\ Thirring Model};
    \node[rectangle, draw, align=center](BG) at (2.5,-1.5){Constrained \\ Bosonic Massless\\ Thirring Model}; 
    \draw [<->]  (CA)--(BG); 
    \draw [<->]  (FF)--(BG); 
    \draw [<->]  (CA)--(FF);
    \end{tikzpicture}
        \caption{An example of the triality statement as generated from the seed triality of the defects shown in Figure \ref{triality}.}
        \label{triality_FT}
\end{figure}
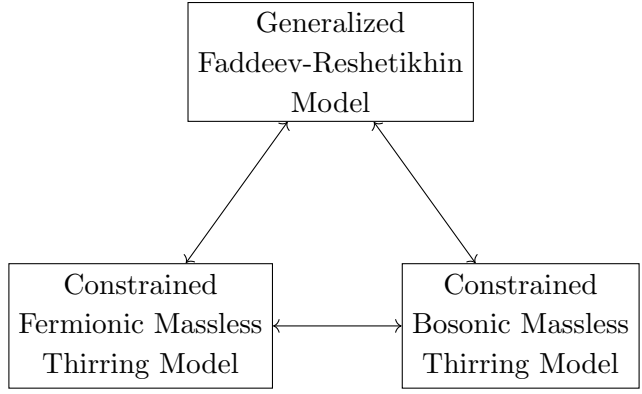

By studying the effective 2d theories obtained by incorporating each of these defects (one chiral and one anti-chiral),
we will obtain the triality of interest, depicted in Figure \ref{triality_FT}.
This includes novel instances of bosonization, and involves the generalized Faddeev-Reshetikhin model that arises from a pair of coadjoint surface operators defined with respect to \textit{different} highest weights associated with the two defects, as well as their bosonic and fermionic counterparts. Each of these models is associated with a pair of fixed, irreducible representations of $\SU(N)$, and these are the representations of the Wilson lines in the two lightcone directions upon discretization.

In what follows, we shall provide further evidence for this triality via explicit computations of characters of Wilson loop operators. In the next section, we shall discuss another instance of bosonization involving the massless Thirring model, which, in principle, extends a special case of the above triality (involving a pair of fundamental representations) with an additional duality.

As we have observed in previous sections, free $\beta\gamma$ systems are discretized to 1d quantum mechanical systems, whose partition functions, upon imposition of a projection condition, can be interpreted as Wilson lines in general irreducible representations.
As we will see below, these two path integral representations of the Wilson loop operators are related via the bosonization procedure.

To make the situation more specific, let us consider an irreducible representation $\cR$ 
 of $\SU(N)$ (or $\U(N)$) which is characterised by a Young tableau with $k_I$ boxes in the $I$-th column ($1\leq I\leq K$) and $l_{\alpha}$ boxes in the $\alpha$-th row ($1\leq \alpha\leq L$), respectively:
\begin{center}
\begin{ytableau}
    \none  &  &  & & &  & \none & \none[l_1] \\
    \none  &  &  & & \none & \none & \none & \none[l_2] \\
    \none [\cR:\qquad] &  &  & \none & \none & \none & \none & \none[\vdots]\\
    \none  &  &  & \none & \none & \none & \none &\none[\vdots]\\
    \none  &  & \none & \none & \none & \none & \none & \none[\vdots]\\
    \none  &  & \none & \none & \none & \none & \none &\none[l_L]\\
     \none & \none[k_1] & \none[k_2]  & \none[$\dots$]& \none[$\dots$] & \none[k_K]  & \none
\end{ytableau}
\end{center}
We have by definition $N \geq k_1\geq k_2 \geq \dots \geq k_{K}$ and $l_1\geq l_2 \geq \dots \geq l_{L}$\,.
As discussed in Section \ref{genrep} and \ref{sec:fbg-sec}, the Wilson loop operators in the representations $\cR$ can be described by the path integral representation of either the 1d fermion systems or the $\beta\gamma$ systems.
To remind us, we present the expressions again:
\begin{align}
    W_{\cR}^{f}[A]&=\int \cD\psi\cD \psi^{*}\cD \tilde{A}^f\,\exp \left(i S^f\right)\exp\left(-i\int dt\,\sum_{I=1}^{K}k_{\text{eff},I}\,(\tilde{A}_{t}^{f})_{II}\right)\,, \label{f-part1}\\
    S^f&=\int_{t_0}^{t_f} dt\,\sum_{I,J=1}^{K}\sum_{i,j=1}^{N}\psi^{I*}_i(i\delta^{IJ} \delta^{i}_{j}\partial_t+\delta^{IJ}A_t^a\rho(t_a)^{i}{}_{j}+\delta^{ij}(\tilde{A}_t^f)_{IJ})\psi_{J}^j\,,\label{1d-f-action-g2}
\end{align}
and
\begin{align}
   W_{\cR}^b[A]&=\int \cD{\bf z} \cD {\bf z}^*\cD \tilde{A}^b\,\exp \left(iS^b\right)\exp\left(-i\int dt\,\sum_{\alpha=1}^{L}l_{\text{eff},\alpha}\,(\tilde{A}_{t}^{b})_{\alpha\alpha}\right)\,,\label{b-part-ge1}\\
S^b&=\int_{t_0}^{t_f}dt\sum_{\alpha,\beta=1}^{L}\sum_{i,j=1}^{N}z_i[\alpha]^*\left(\delta^{\alpha\beta}\delta^{i}_{j}\partial_t +\delta^{\alpha\beta}A^a_t  \rho(t_a)^{i}{}_{j}+\delta^{ij}(\tilde{A}_t^b)_{\alpha\beta}\right)z^j[\beta]\,,\label{1d-b-action-g2}
\end{align}
where the only non-zero components of $(\tilde{A}_t^f)_{IJ}$ and $(\tilde{A}_t^b)_{\alpha\beta}$ are the $I\geq J$ and $\alpha \geq \beta$ components respectively.
The (quantum corrected) 1d CS coefficients $k_{\text{eff},I}\,, l_{\text{eff},\alpha}$ are
\begin{align}
 k_{\text{eff},I}=k_I-\frac{N}{2}\,, \qquad l_{\text{eff},\alpha}=l_{\alpha}+\frac{N}{2}\,.\label{k-l-eff}
\end{align}
Here, we give a comment on the case where $\U(N)$ is considered instead of $\SU(N)$. In this case, the Wilson loop operators above would have global anomalies under the large gauge transformations for $\U(1)\subset \U(N)$ which would change the overall sign of the Wilson loop operators. To cancel such anomaly, we need to add additional 1d CS terms  (For Wilson loop defect actions in other theories, the same global anomaly cancellation has been considered in \cite{Assel:2015oxa,Lozano:2020txg})
\begin{align}
    &\text{Fermionic:}\quad \frac{K}{2}\int dt\,\Tr(A_t)\,,\qquad \text{Bosonic:}\quad \frac{L}{2}\int dt\,\Tr(A_t)\label{U(1)-ano}
\end{align}
to the defect actions in (\ref{f-part1}) and (\ref{b-part-ge1}), respectively.\footnote{This anomaly is a one-dimensional counterpart of the parity anomaly in 3d massive Dirac fermions \cite{Niemi:1983rq,Redlich:1983kn,Redlich:1983dv}.}

Note that our path integral expressions (\ref{f-part1}), (\ref{b-part-ge1}) for the Wilson loop operators differ from those considered in \cite{Gomis:2006sb}, while we have the same results (except for the overall factor) as those obtained in \cite{Gomis:2006sb}.
Use of the path integral representations (\ref{f-part1}), (\ref{b-part-ge1}) has an advantage compared with the ones considered in \cite{Gomis:2006sb}. As we will see below, it is easy to consider the thermodynamic limit of multiple copies of a Wilson line in a general irreducible representation of $\SU(N)$, and hence we can immediately write down the corresponding 2d surface defect action. 

Now, as in \cite{Gomis:2006sb}, we will show that the two Wilson loop operators above are related to each other by bosonization.
To see this, we explicitly perform both path integrals under certain gauge fixings according to \cite{Gomis:2006sb} and see that they are indeed equivalent.
In the following, we only give an overview of this computation, the details will be presented in Appendix \ref{sec:fermion_betagamma}.
To this end, we take a gauge choice such that the time components of the gauge fields $A_t$ and $\tilde{A}_t^f$ are diagonalized to have constant eigenvalues $w_i$\,, $\theta_{I}$ using complexified gauge symmetry $\SL(N, \mathbb{C})\times B$ \cite{Corradini:2016czo} (see also \cite{Gomis:2006sb}), where $B$ is the Borel subgroup of $G_{\mathbb{C}}=\SL(N, \mathbb{C})$:
\begin{align}
    A_t=\text{diag}(w_1,\dots,w_{N})\,,\qquad \tilde{A}_t^f=\text{diag}(\theta_1,\dots, \theta_{K})\,,\label{gauge-fix-f}
\end{align}
and then the 1d fermion action (\ref{1d-f-action-g2}) becomes
\begin{align}
   S_{f}= \int_{t_0}^{t_f} dt\,\sum_{I=1}^{K}\sum_{j=1}^{N}\psi^{I*}_{j}(i\partial_t+w_j+\theta_I)\psi^{I}_j\,.\label{f-part-fix}
\end{align}
Note that the off-diagonal entries of the gauge fields do not contribute to the partition function (see Appendix A in \cite{Corradini:2016czo}).
For the 1d bosonic case, a similar gauge fixing of the gauge symmetry $\SL(N, \mathbb{C})\times B$ can be chosen, and the bosonic action (\ref{1d-b-action-g2}) becomes
\begin{align}
    S_b=\int_{t_0}^{t_f}dt\,\sum_{\alpha=1}^{L}\sum_{i=1}^{N}z_i^*[\alpha](\partial_t+w_i+\theta_{\alpha})z_i[\alpha]\,,
\end{align}
where $\theta_{\alpha}(\alpha=1,\dots, L)$ are eigenvalues of $\tilde{A}_t^b$\,.

In this gauge choice, we can perform the path integrals over the fermionic and bosonic fields, and then we obtain
\begin{align}
    W_{\cR}^{f}
    &=\prod_{I=1}^{K}\oint\frac{dz_I}{2\pi i}\frac{1}{z^{k_{I}+1}_I} \prod_{1\leq I<J \leq K}\left(1-\frac{z_J}{z_I}\right)\prod_{j=1}^{N}\prod_{J=1}^{K}(1+x_jz_{J})\,,\\
    W_{\cR}^{b}&=\prod_{\alpha=1}^{L}\oint\frac{dz_\alpha}{2\pi i}\frac{1}{z^{l_{\alpha}+1}_{\alpha}}\prod_{1\leq \alpha<\beta\leq L}\left(1-\frac{z_{\beta}}{z_{\alpha}}\right)
    \prod_{j=1}^{N}\prod_{\beta=1}^{L}\frac{1}{1-x_{j} z_{\beta}}\,,
\end{align}
where we introduced the new variables $z_I=e^{i \theta_I}\,, z_{\alpha}=e^{i\theta_{\alpha}}$ and $x_j=e^{i w_j}$\,.
Note that we have chosen $w_j$ complex so that $x_j$ is outside the unit-circle contour for $z_\alpha$'s.
By evaluating $z_{I}\,, z_{\alpha}$ integrals, we can see the partition functions are described by the Schur functions for the representation $\cR$
and its transpose:\footnote{By definition of the Wilson loop operator, the result (\ref{part-schur}), where the gauge field vanishes, i.e., $x_i=1$ for all $i$, gives the correct dimension of the representation $\cR$ with the Young tableaux $(l_1,l_2,\dots,l_L)$ i\,.e.\,,
\begin{align}
    \Tr_{\cH_{(l_1,\dots, l_{L})}}(1) = s_{(l_1,\dots, l_{L})}(1^{N})&=\prod_{1\leq \alpha<\beta\leq L}\frac{l_\alpha-l_\beta+\beta-\alpha}{\beta-\alpha}\,.
\end{align}}:
\begin{align}
    W^f_{\cR}[A]&=s_{(k_1,\dots, k_{K})^t}({\bf x})\,,\qquad
    W^b_{\cR}[A]=s_{(l_1,\dots ,l_{L}) }({\bf x}) \,,\label{part-schur}
\end{align}
where $(k_1,\dots, k_{K})^t$ is the transpose of the representation $(k_1,\dots, k_{K})$\,.
Note that the two expressions in (\ref{part-schur}) are equivalent, as are expected since we started with the same Young tableau.
This fact means that a 1d chiral fermion defect in an irreducible representation $\cR$ of $\SU(N)$ is dual to $\beta\gamma$ defect in the representation $\cR^t$ which is conjugate of $\cR$.

Although we have argued for a bosonization duality via the discretization of surface defects, we would like to know if the duality continues to hold without discretizing the field theories, i.e., we would like to know if the 2d defects themselves are dual to each other prior to discretization. To achieve this we would like to show that the levels of the affine Kac-Moody algebras generated by each defect are equal.

Firstly, we recall that the coadjoint orbit defects can be interpreted as curved $\beta\gamma$ defects, at least classically, as shown in Appendix \ref{betagammacoadjoint}. The latter give rise to affine Kac-Moody algebras at the critical level, i.e., $k=-\sh^{\vee}$ (see \eqref{critakm}). Moreover, as explained in Appendix \ref{sec:equivalence}, the coadjoint orbit defect can be understood to arise from a reduced chiral WZW defect, where the conserved current of the latter obeys $\textrm{Tr}(J^2)=\textrm{constant}$. As explained below, the latter behaviour also arises from the constrained 2d boson and fermion defects, and also suggests that the level of the associated affine Kac-Moody algebra is critical. 

Notably, the constraints on the 2d boson and fermion defects fix the particle number in these systems, which implies that the rescaled Sugawara energy-momentum tensor built out of the currents ought to be constant.
The upshot is that this energy-momentum tensor, $\tilde{T}(z)=(J^aJ_a)$, is proportional to the identity operator on Fock space. Therefore the OPE between $\tilde{T}(z)$ and any other operator, including $J^a(z)$, should be regular.
This in turn means that the modes of $\tilde{T}(z)$ commute with each other, as well as with all the modes of the affine Kac-Moody current. This can be understood to arise from the $k\rightarrow -\sh^{\vee}$ limit of the conventional Sugawara energy-momentum tensor,
defined as 
\ie 
T(z) = \frac{1}{2(k+\sh^{\vee})}(J^aJ_a)\,, 
\fe 
whose modes satisfy 
\ie 
\left[L_n, J_m^a\right]=-m J_{n+m}^a\,.
\fe
If we denote the modes of $\tilde{T}(z)$ as $\tilde{L}_n$, we would have
\ie 
\left[\tilde{L}_n, J_m^a\right]=-2 (k+\sh^{\vee}) m J_{n+m}^a\,,
\fe
and in the critical level limit, the RHS goes to zero. This suggests that all three defects that were used to derive the bosonization in this section can be understood as constrained systems, where the constraints fix the level of the affine Kac-Moody algebra to be critical.

\subsection{Triality of Field Theories in Thermodynamic Limit}

We may take the thermodynamic limit of multiple copies of a Wilson line in a general irreducible representation of $\SU(N)$. Since it has two descriptions in terms of bosonic or fermionic degrees of freedom, one ends up with two descriptions in the thermodynamic limit at surface operators with bosonic or fermionic degrees of freedom. For example, for a collection of Wilson lines in an irreducible representation of $\SU(N)$, supported along the $\sigma^+$ direction, the two descriptions of the surface operators one obtains in the thermodynamic limit are  

\begin{align}
    Z_{\rho}^{f}[\psi,\psi^{*},A]&=\int \cD\psi\cD \psi^{*}\cD a\,\exp \left(i S^f\right)\exp\left(-i\int d\sigma^+d\sigma^-\,\sum_{I=1}^{K}k_{\text{eff},I}\,a_{+}^{I}\right)\,, \label{f-part}\\
S^f&=\int d\sigma^+d\sigma^-\,\sum_{I,J=1}^{K}\sum_{i,j=1}^{N}\psi^{I*}_i(i\delta^{IJ} \delta^{i}_{j}\partial_++\delta^{IJ}A_+^a\rho(t_a)^{i}{}_{j}+\delta^{ij}(\tilde{A}_+)_{IJ})\psi_{J}^j\,,\label{1d-f-action-g22}
\end{align}
and
\begin{align}
   Z_{\rho^t}^b[z,z^*,A]&=\int \cD{\bf z} \cD {\bf z}^*\cD a\,\exp \left(iS^b\right)\exp\left(-i\int d\sigma^+d\sigma^-\,\sum_{I=1}^{L}l_{\text{eff},I}\,a_{+}^{I}\right)\,,\label{b-part-ge2}\\
S^b&=\int d\sigma^+d\sigma^-\sum_{I,J=1}^{L}\sum_{i,j=1}^{N}z_i[I]^*\left(\delta^{IJ}\delta^{i}_{j}\partial_+ +\delta^{IJ}A^a_+  \rho(t_a)^{i}{}_{j}+\delta^{ij}(\tilde{A}_+)_{IJ}\right)z^j[J]\,.\label{1d-b-action-g22}
\end{align}
In \eqref{1d-f-action-g22} and \eqref{1d-b-action-g22}, we have
\begin{align}
    (\tilde{A}_+)_{IJ}=\begin{cases}
        a_+\qquad & (I=J) \,,\\
        a_{+,IJ}\qquad & (I<J)\,,\\
        0\qquad & (I>J)\,.
    \end{cases}\
\end{align}
Thus, if we were to take the thermodynamic limit of a collection of Wilson lines each in the same irreducible representation of $\SU(N)$, we would arrive at two equivalent surface operator descriptions, that is, either as a free $\beta\gamma$ system, or as a chiral free fermion system, with projection conditions imposed. Hence, a version of bosonization holds for such a  surface operator, given its equivalent descriptions in terms of bosonic or fermionic degrees of freedom.

Let the aforementioned surface operator be located at $z_+$ on the holomorphic plane $\mathbb{CP}^1$ of the 4d CS. Then, considering another surface operator of opposite chirality at position $z_-\in \mathbb{CP}^1$, with the same dual descriptions, would lead us to dual integrable field theories that are related via bosonization.  Suppressing the projection conditions, the dual integrable field theories take the form 
\ie  
S^f
    =\int d^2\sigma&\Bigg(
        \sum_{I=1}^Ki\psi^{I*}_i \partial_+\psi^i_I 
        + \sum_{I'=1}^{K'}i\psi^{{I'}*}_i \partial_-\psi ^i_{I'} 
    \\
    & \qquad \qquad +\frac{i}{4}\frac{\hbar}{z_+-z_-}\sum_{I=1}^K\sum_{J'=1}^{K'} \left(\psi_i^{I*}(t^a)^i_{\textrm{ }j}\psi^j_I \right)\left(\psi_k^{J'*}(t_a)^k_{\textrm{ }l}\psi^l_{J'}\right)
    \Bigg)\,,
\fe
and
\begin{align}
\begin{split}
S^b=\int d^2\sigma
&\Bigg(
\sum_{I=1}^L z_i[I]^* \partial_+z^i[I] + \sum_{J=1}^{L'} z_i[J]^* \partial_-z^i[J] \\
&\qquad \qquad +\frac{i}{4}\frac{\hbar}{z_+-z_-} \sum_{I=1}^L\sum_{J'=1}^{L'} \left(z_i[I]^*(t^a)^i_{\textrm{ }j}z^j[I]\right) \left(z_k[J']^*(t_a)^k_{\textrm{ }l}z^l[J']\right)\Bigg)\,.
\end{split}
\end{align}
The chiral and anti-chiral surface operators considered here can each be in an arbitrary irreducible representation of $\SU(N)$, which we denote $\cR$ and $\cR'$ respectively. 

Thus, we have successfully employed discretization of surface operators in the 4d CS theory to prove the bosonization duality between constrained fermionic and constrained bosonic massless Thirring models, that are each associated with a pair of fixed irreducible representations $\cR$ and $\cR'$  of $\SU(N)$. Since the constrained bosonic massless Thirring model is further related via discretization
to the generalized Faddeev-Reshetikhin model associated with the highest weights $\cR$ and $\cR'$, denoted respectively as $\Lambda_{\cR}$ and $\Lambda_{\cR'}$, with the action
\begin{align}
\begin{split}
&S_{\mathrm{FR}}\left[g_{(\pm)}\right] \\
&\quad =\int_{\Sigma}  d^2\si \, \operatorname{Tr}\left(\Lambda_{\cR} g_{(+)}^{-1} \partial_{-} g_{(+)}+\Lambda_{\cR'} g_{(-)}^{-1} \partial_{+} g_{(-)}-\frac{i}{4}\frac{\hbar}{z_+-z_-} g_{(+)} \Lambda_{\cR} g_{(+)}^{-1} g_{(-)} \Lambda_{\cR'} g_{(-)}^{-1}\right),
\end{split}
\end{align}
we then arrive at a triality relationship between the constrained bosonic massless Thirring model, its fermionic counterpart, and the Faddeev-Reshetikhin model. It would be interesting to see the relations between the models more explicitly.

\section{Summary of the Discretization Setups}
  \label{sec:summary}

Let us summarize the setups of our discretization discussions.

We consider the 4d CS theory on $\mathbb{C}\times C$,
where the spectral curve is 
\begin{itemize}
    \item $C=\mathbb{C}$ (rational model)
    \item $C=\mathbb{C}^{\times}$ (trigonometric model)
    \item $C=\mathbb{C}/(\mathbb{Z} + \mathbb{Z} \tau) $ (elliptic model)
\end{itemize}

We consider multiple chiral or anti-chiral surface defects located at a point on $C$, as depicted in Figure \ref{torus}. 

Some examples of the defects include 
\begin{itemize}
  \item Coadjoint orbit defect (Section \ref{sec:coadjoint})
  \item Free fermion defect (Section \ref{sec:free_fermion})
  \item Free $\beta\gamma$ system (Sections \ref{sec:fbg-sec})
   \item Curved $\beta\gamma$ system (Sections \ref{sec:curved_betagamma} and \ref{sec:curved-bg})
  \item Vertex algebra defect (Section \ref{sec:VA})
\end{itemize}
Each of these defects are two-dimensional chiral (or anti-chiral) theories, which can be discretized into their 1d dimensionally-reduced counterparts. 
Note that these examples are not mutually exclusive, and for example the vertex algebra defect essentially includes 
all the examples discussed in this paper (for curved $\beta\gamma$ systems we need to consider a sheaf of vertex algebras, as we 
discussed in Section \ref{sec:VA}).

In all the cases we discussed, we find that the 1d defect after the discretization can be recast as Wilson lines, and hence are dual to each other.
By going back to the field theories by taking the thermodynamic limit, this immediately implies that 
there will be a gigantic zoo of dualities between 
two-dimensional field theories.

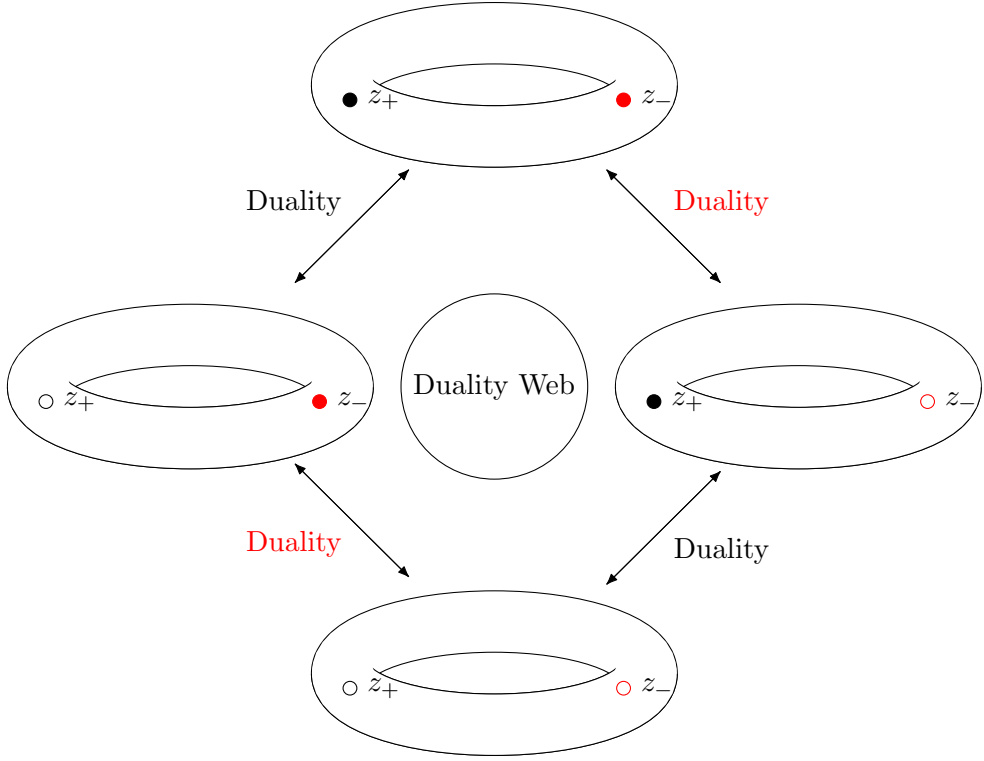
\begin{figure}[!ht]
\begin{center}
  \begin{tikzpicture}[fill=white, circ/.style={shape=circle, inner sep=2pt, draw, node contents=}]
    \pic at (0,4) {torus={2cm}{4mm}{70}} 
      node[above=7mm,font=\tt,blue]{};
    \pic at (4,0) {torus={2cm}{4mm}{70}}
      node[above=7mm,font=\tt,blue]{};
    \pic at (-4,0) {torus={2cm}{4mm}{70}}
      node[above=7mm,font=\tt,blue]{};
    \draw[  ->, >={Latex[round]}, shift=({5pt, 5pt})]  (2.8,1.2)--(1.3,2.7) node[text=red,midway,above right] {Duality};;
    \draw[ ->, >={Latex[round]}, shift=({5pt, 5pt})]  (1.3,2.7)--(2.8,1.2) ;
    \draw[ ->, >={Latex[round]}, shift=({5pt, 5pt})]  (-2.8,1.2)--(-1.3,2.7) ;
    \draw[ ->, >={Latex[round]}, shift=({5pt, 5pt})]  (-1.3,2.7)--(-2.8,1.2) node[midway,above left] {Duality};     
    \draw[ ->, >={Latex[round]}, shift=({5pt, 5pt})]  (2.8,-1.3)--(1.3,-2.8) ;
     \draw[ ->, >={Latex[round]}, shift=({5pt, 5pt})]  (1.3,-2.8)--(2.8,-1.3) node[midway,below right] {Duality};
     \draw[ ->, >={Latex[round]}, shift=({5pt, 5pt})]  (-2.8,-1.2)--(-1.3,-2.7) ;
     \draw[ ->, >={Latex[round]}, shift=({5pt, 5pt})]  (-1.3,-2.7)--(-2.8,-1.2) node[text=red,midway,below left] {Duality};
     \pic at (0,-3.8) {torus={2cm}{4mm}{70}} node[above=7mm,font=\tt,blue]{};
      
    \node[label=right:{$z_-$},draw=red,shape=circle,fill=red, scale=0.5] at (-2.3,-0.2) {};
    \node[label=right:{$z_+$},draw=black,shape=circle,fill=white, scale=0.5] at (-5.9,-0.2) {};        
    \node[label=right:{$z_+$},draw=black,shape=circle,fill=black, scale=0.5] at (2.1,-0.2) {};
    \node[label=right:{$z_-$},draw=red,shape=circle,fill=white, scale=0.5] at (5.7,-0.2) {};
    \node[label=right:{$z_+$},draw=black,shape=circle,fill=white, scale=0.5] at (-1.9,-4) {};
    \node[label=right:{$z_-$},draw=red,shape=circle,fill=white, scale=0.5] at (1.7,-4) {};
    \node[label=right:{$z_+$},draw=black,shape=circle,fill=black, scale=0.5] at (-1.9,3.8) {};
    \node[label=right:{$z_-$},draw=red,shape=circle,fill=red, scale=0.5] at (1.7,3.8) {};
    \node[circle, draw, align=center](42) at (0,0) {Duality Web};   
 
     \end{tikzpicture}
     \label{torus4}
     \caption{By using the dualities among surface defects as ``seed'' dualities, we have a huge web of dualities among integrable field theories. Here we consider one chiral and one anti-chiral defect, where we consider two possibilities (say, free fermion and free $\beta\gamma$ defect) for each defect.}
\end{center}
\end{figure}

Note that the dualization discussed in Section \ref{sec:disc_defect} works for each individual defect. For example, consider a special class of theories described by $n$ chiral and $n$ anti-chiral defects, where each defect is described by either a coadjoint orbit defect or a free fermion defect in the adjoint representation. Since the defects are dual, we can use either type for each of the four defects, giving us $2^{2n}$ possibilities, and we expect all these theories to be dual to each other. By pushing this idea further for multiple defects of different types as listed above,
we will expect a huge web of dualities among integrable field theories
(see Figure \ref{torus4} for the case of $n=1$). It is an exciting problem to study these dualities further.

\section{Quartet of Defects and Non-Abelian Bosonizations}\label{sec:duality-boson}

In Section \ref{sec:duality} we have seen that the discretization of integrable field theories
leads to duality statements between them. In this section, we consider 
dualities between integrable field theories more directly,
without relying on integrable discretizations. 
The idea is to employ bosonization for each order surface defect; when we combine 
many such defects, we have a rich web of dualities among integrable field theories,
including a new infinite class of non-Abelian bosonizations between massless Thirring-type models
and coupled WZW models,
where the latter were discussed previously in the context of affine Gaudin models \cite{Vicedo:2017cge}. 

\subsection{Quartet of Defects  from Bosonization of Defects}\label{sec:chiralWZW}

\subsubsection{Summary of Statements}\label{summary-d}

In this section we are going to discuss the equivalence between the following four different types of defects.
We will list both chiral and anti-chiral defects.

\begin{enumerate}

\item \textbf{Free Fermion Defect}

The free fermion defects have Lagrangians
\begin{align}
    S_+^{f}[A_-]=\sum_{I=1}^{N_F} \int_{\cM\times \{z_{+}\}} d^2 \si \, \bar{\Psi}_{I}i\hat{\cancel{D}}_L\Psi_I \,,  \label{Zhat-f-R0}\\
    S_-^{f}[A_+]=\sum_{I=1}^{N_F} \int_{\cM\times \{z_{-}\}} d^2 \si \, \bar{\Psi}_{I}i\hat{\cancel{D}}_R\Psi_I \,,
    \label{Zhat-f-R}
\end{align}
where the left/right-handed covariant derivatives $\hat{\cancel{D}}_L,  \hat{\cancel{D}}_R$ are
\begin{align}
  \hat{\cancel{D}}_L=\cancel{\partial}+\cancel{A}\frac{1+\gamma_5}{2} = \cancel{\partial}+\cancel{A}P_+\,,\\
   \hat{\cancel{D}}_R=\cancel{\partial}+\cancel{A}\frac{1-\gamma_5}{2}=\cancel{\partial}+\cancel{A}P_-\,.
\end{align}

\item \textbf{Edge Mode Defect}

The Lagrangians of the defects are given by 
\ie \label{swzwua}
{S}^{\textrm{edge}}_+[g,A]&=S_{\rm WZW}[g]  +\frac{1}{2\pi}\int_{\cM\times \{z_{+}\}} d^2 \si \, \textrm{Tr} (A_+A_-)
\fe
with the constraint 
\ie \label{edgcon1}
A_- |_{z_+}&=  g^{-1} \partial_{-} g\,,
\fe 
and
\ie
S^{\mathrm{edge}}_-[g,A]&=\tilde{S}_{\rm WZW}[g]  +\frac{1}{2\pi}\int_{\cM\times \{z_{-}\}} d^2 \si \, \textrm{Tr} (A_+A_-)
\fe
with the constraint 
\ie \label{edgcon2}
A_+|_{z_-}&= g^{-1} \partial_{+} g\,,
\fe
where $S_{\rm WZW}[g]$ is the WZW action 
\ie \label{wzw-action}
S_{\rm WZW}[g] &=-\frac{1}{2 \pi} \int_{\Sigma} d^2 \sigma \, \textrm{Tr}\left( g^{-1}\partial_+ g g^{-1}\partial_- g\right)+S_{\rm WZ}[g]\,,\\
S_{\rm WZ}[g]&=\frac{1}{12 \pi} \int_{\Sigma \times \R_+} \textrm{Tr} \left(g^{-1} dg\wedge g^{-1} d g \wedge g^{-1} d g\right)\,,
\fe
and $\tilde{S}_{\rm WZW}[g]=S_{\rm WZW}[g^{-1}]$ is its anti-chiral counterpart.
Note that here $g$ is not a separate dynamical degrees of freedom, 
but rather an ``edge mode'' of the bulk gauge field $A$: $g$ is determined by 
solving the constraints \eqref{edgcon1} and \eqref{edgcon2}
at the locations of the defect. Note that this is a 4d counterpart (cf.\ Section 8 of  \cite{Costello:2019tri}) of the 
holomorphic boundary condition for the 3d CS theory, which is well-known to 
generate the 2d WZW model as an edge mode \cite{Elitzur:1989nr}.

\item \textbf{WZW Defect}
The Lagrangians for the WZW defects are given by
\ie
S_{+}^{b}[A,\cG_{(+)};N_F]
    &= N_F S_{\rm WZW}[\cG_{(+)}]-\int_{\cM\times \{z_{+}\}} d^2\sigma\,\Tr(A_- \mathcal{J}_{+})\\&
 -\frac{N_F}{2\pi}\int_{\cM\times \{z_{+}\}} d^2\sigma\,\Tr(A_+ A_-)\,,\\
S_{-}^{b}[A,\cG_{(-)};N_F]
    &= N_F S_{\rm WZW}[\cG^{-1}_{(-)}]+ \int_{\cM\times \{z_{-}\}} d^2\sigma\,\Tr(A_+ \cJ_- )\\
    &-\frac{N_F}{2\pi}\int_{\cM\times \{z_{-}\}} d^2\sigma\,\Tr(A_+ A_-)
\,,\label{bo-rdefect}
\fe
where the bulk gauge fields couple to the current of the WZW model
\begin{align}
\mathcal{J}_{+}&=\frac{N_F}{\pi} \partial_+ \cG_{(+)}\cG_{(+)}^{-1}\,,
\\
\mathcal{J}_{-}&=\frac{N_F}{\pi} \partial_- \cG_{(-)}\cG_{(-)}^{-1}\,.
\end{align}
We emphasize that the fields $\cG_{(+)}$ and $\cG_{(-)}$ are dynamical degrees of freedom,
which are integrated in the path integral.\footnote{The WZW defects play a crucial role in showing the holomorphic factorization of the WZW model path integral \cite{Witten:1991mm}.}
Note that the fields in the WZW defects are not coupled to the 4d CS gauge field in such a way that the whole defect actions are gauge invariant. The interaction terms with the gauge field are chosen to have the same gauge transformation as the chiral anomaly of the free fermion systems, which are necessary to discuss the equivalence of the four different defects. Of course, there are various gauged versions of the WZW defects, but they are not in the interest of this section, so we will not discuss them further in this paper.

\item \textbf{Dirichlet Defects}

For Dirichlet defects we have no separate dynamical degrees of freedom localized at the surface defects.
We instead simply impose the Dirichlet boundary conditions
for chiral / anti-chiral components of the gauge fields:
\ie \label{gauge_pm_0}
A_-|_{z_+}=0\,, \\
A_+|_{z_-}=0\,.
\fe 

\end{enumerate}

In each of these cases, the holomorphic one-form is given by 
\ie 
\omega=  \left(1 - \frac{\hbar}{4\pi i}\left(\sum_{\alpha=1}^{n_{+}}\frac{k_{+,\alpha}}{z-z_{+,\alpha}}-\sum_{\beta=1}^{n_{-}}\frac{k_{-,\beta}}{z-z_{-,\beta}}\right) \right) dz\,.
\fe 
For the first 
three types of defects, the terms proportional to $\hbar$ arise from quantum corrections as discussed in Section \ref{sec:anomaly}.
By contrast, for the Dirichlet defect the poles of the one-form
are required already at the classical level,
and the boundary conditions for the gauge fields \eqref{gauge_pm_0}
are then required in order to choose elliptic boundary conditions.

It is important to note that since we consider $n_+\geq 1$ and $n_-\geq 1$, $\omega$ has zeroes, and therefore, in order to ensure a finite action, we ought to consider gauge field configurations that are singular close to these zeroes, i.e., disorder defects. In other words, we shall study all four of the defects listed above in the presence of disorder defects. 

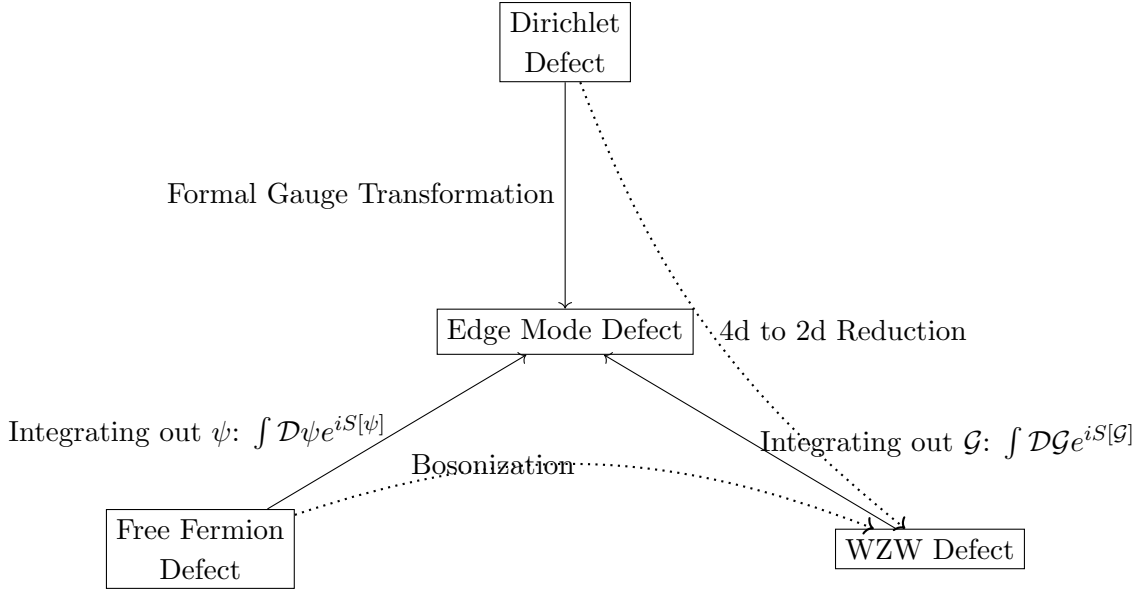
\begin{figure}[htbp]
\centering
\begin{tikzpicture}[scale=0.96]
    \node[rectangle, draw, align=center](DI) at (0,4) {Dirichlet\\ Defect};
    \node[rectangle, draw, align=center](ID) at (0, 0){Edge Mode Defect};
    \node[rectangle, draw, align=center](FF) at (-5, -3) {Free Fermion\\ Defect};
    \node[rectangle, draw, align=center](WZ) at (5,-3){WZW Defect}; 
    \draw [->]  (DI)--(ID) node[midway, left] {Formal Gauge Transformation };
    \draw [->]  (FF)--(ID) node[midway, left] {Integrating out $\psi$: $\int \cD \psi e^{iS[\psi]}$ };
    \draw [->]  (WZ)--(ID) node[midway, right] {Integrating out $\cG$: $\int \cD \cG e^{iS[\cG]}$ };
    \draw [->, dotted, style=thick] (DI) to [bend right=15] node[midway, right] {4d to 2d Reduction } (WZ)
    ;
    \draw [->, dotted, style=thick] (FF) to [bend left=20] node[midway, left]{ Bosonization} (WZ)
    ;
\end{tikzpicture}
\caption{The relationship between Dirichlet defects that realize coupled WZW models, edge mode defects obtained from formal gauge transformations thereof, free fermion defects, and WZW defects.
Here, $\int \cD \psi e^{iS[\psi]}$ and $\int \cD \cG e^{iS[\cG]}$ indicate the introduction of appropriate measures by multiplying the 4d CS path integral by constant factors, which can be extracted from its overall normalization.}
    \label{dis}
\end{figure}

Figure \ref{dis} summarizes the results of this subsection.

In the following we will derive the equivalence of the four types of defects.
Since the discussions are essentially the same for chiral (left-handed) and anti-chiral (right-handed) defects,
we will discuss the case of chiral (left-handed) defects only and omit the counterparts for anti-chiral (right-handed) defects.

\subsubsection{Free Fermion Defect into Edge Mode Defects}\label{fermtoedge}

Let us consider the nonabelian bosonization of the gauged fermionic system. While the bosonization has been discussed extensively in the literature \cite{Mandelstam:1975hb, Coleman:1974bu, Witten:1983ar, DiVecchia:1984df, Gonzales:1984zw, DiVecchia:1984ksr} (for details, see e.g.\ \cite{Fujikawa:2003az}), we characterize the bosonization as a two-step process, where the 
intermediate step will play a crucial role when making contact with pure 4d CS theory with only Dirichlet and disorder defects, as we shall see.

We consider the left-handed chiral free fermion defect with the partition function 
\begin{align}
    Z_+^f[A_-]
    =\int \cD\psi_L \cD \psi^*_L\,\exp\left({\frac{i}{\hbar_{\rm 2d}}\int_{\cM\times \{z_{+}\}}d^2\sigma\,\sum_{I=1}^{N_F}\psi_{L,I}^{*}i\cancel{D}_L\psi_{L,I}}\right)\,.\label{fermaction-p}
\end{align}
The bosonization of the gauged Weyl fermions system can be performed according to the results of the chiral anomaly for this system.
As explained in Appendix \ref{chiralanomaly}, one way to compute chiral anomalies in a well-defined way is to consider a chiral gauged fermion system by adding a free right-handed fermion system \cite{Zumino:1983rz,Alvarez-Gaume:1983ict} to (\ref{fermaction-p}). 
Here, we follow this prescription and consider bosonization of the chiral gauged fermion system
\begin{align}
     \hat{Z}_+^f[A_-]
     =\int \cD\Psi \cD\bar{\Psi}\,\exp\left({\frac{i}{\hbar_{\rm 2d}}\int_{\cM\times \{z_{+}\}} d^2\sigma\,\sum_{I=1}^{N_F}\bar{\Psi}_{I}i\hat{\cancel{D}}_L\Psi_I}\right)\,,\label{meff-chiralL}
\end{align}
where the modified covariant $\hat{\cancel{D}}_{L}$ is defined as 
\begin{align}
   \hat{\cancel{D}}_L=\cancel{\partial}+\cancel{A}\frac{1+\gamma_5}{2}\,.
\end{align}

Employing the formula \eqref{W_anomaly} of chiral anomaly, we can rewrite the partition function (\ref{meff-chiralL}) as
\begin{align}
   \hat{Z}_+^f[A_{-}|_{z_+}]= \hat{Z}_+^f[A^g_{-}|_{z_+}]\, \exp(-i N_F \Gamma_+[A_{-}^g,g^{-1}] )\;,
   \label{Z_f_chiral_anomay}
\end{align}
where we have replaced $A_{-}\to  A^g_{-}$ with $A^{g}: = g A g^{-1} - dg g^{-1}$ in \eqref{W_anomaly},
and we have (see \eqref{Gamma_def})
\begin{align}
    \Gamma_+[A_-,g^{-1}]:=S_{\rm WZW}[g]-\frac{1}{\pi}\int d^2\sigma\,
    \Tr\bigl(\partial_{+}gg^{-1}A_-\bigr) \,.
    \label{finite-G-L}
\end{align}

We can use this gauge degrees of freedom to set $A_{-}^{g}|_{z_+}=0$
to decouple the 4d CS gauge field from the 2d fermions. This is achieved by choosing 
$g=g_{(+)}$  with 
\begin{align}
    A_-\lvert_{z_+}=g_{(+)}^{-1}\partial_-g_{(+)}\,.\label{Ad-ch}
\end{align}
We will then have 
\begin{align}
\hat{Z}_{+}^{f}[A_{-}]
  &= \exp\left(-iN_F\Gamma_{+}[0 ,g_{(+)}^{-1}]\right) \hat{Z}_{+}^{f}[A=0]\no \\
  &= \exp\left(-iN_F S_{\rm WZW}[g_{(+)}]\right) \hat{Z}_{+}^{f}[A=0]\;.
\end{align}
Let us now add a local counter-term $\Tr(A_+A_-)$ to the bosonized action, with the overall factor in front of the counter term chosen so that the
we have the Wess-Zumino consistency condition \cite{Wess:1971yu}. We then reproduce the edge mode defect as
\begin{align}
\hat{Z}_{+}^{f}[A_{-}]
  &= \exp\left(-iN_F S_{\rm WZW}[g_{(+)}]-\frac{iN_F}{2\pi}\int d^2\sigma\,\Tr(A_+A_-)\right) \hat{Z}_{+}^{f}[A=0] \no\\
  &= \exp\left(-iN_F {S}^{\textrm{edge}}_+[g_{(+)},A]\right) \hat{Z}_{+}^{f}[A=0]  \,.\label{lbod-th}
\end{align}

A similar calculation can be performed for the right-hand system to obtain the corresponding edge mode defects.\begin{align}
    \hat{Z}_{-}^{f}[A]
    &=\exp\left(-iN_F S^{\mathrm{edge}}_-[g_{(-)},A]\right)\hat{Z}_{-}^{f}[A=0]\,.\label{rbod-th}
\end{align}
Here, the plus component, $A_+$, of the gauge field is taken as
\begin{align}
    A_+\lvert_{z_-}=g_{(-)}^{-1}\partial_+g_{(-)}\,.\label{Ap-ch}
\end{align}

We may also derive the equivalence between free-fermion defects and edge mode defects using an argument similar to that of \cite{Polyakov:1983tt,Polyakov:1984et}.
We have shown in Section \ref{sec:anomaly} that the anomalies associated with chiral and anti-chiral defects supporting affine Kac-Moody algebras (prior to the inclusion of Pauli-Villars counter-terms) are 
\ie \label{edgese}
    \delta W_{+}=\frac{k_{+}}{2 \pi} \int_{\Sigma} d^2 w \partial_w \epsilon_a(w, \bar{w}) A_{\bar{w}}^a(w, \bar{w}) \,,
\fe
and
\ie 
    \delta W_{-}=\frac{k_{-}}{2 \pi} \int_{\Sigma} d^2 w \partial_{\bar{w}} \epsilon_a(w, \bar{w}) A_w^a(w, \bar{w}) \,.
\fe
In Minkowski signature these are 
\ie
    \delta W_{+}=-\frac{k_{+}}{2 \pi} \int_{\Sigma} d\sigma^+ d\sigma^- \epsilon_a\partial_+ A_{-}^a \,,
\fe
and
\ie 
    \delta W_{-}=-\frac{k_{-}}{2 \pi} \int_{\Sigma} d\sigma^+ d\sigma^- \epsilon_a\partial_- A_{+}^a \,.
\fe
Let us focus on the chiral defect. The variation of the effective action can also be described as 
\ie 
\begin{aligned}
    W_+[{A}_-] & \rightarrow W_+[{A}_{-}-D_- \epsilon] \\
    & =W_+[{A}]-\int d\sigma^+ d\sigma^- \operatorname{tr}\left(D_- \epsilon \frac{\delta}{\delta A_-} W_+[A_-]\right) \\
    & =W_+[{A}]+\int d\sigma^+ d\sigma^- \operatorname{tr}\left( \epsilon D_- \frac{\delta}{\delta A_-} W_+[A_-]\right) .
\end{aligned}
\fe
We then have 
\ie \label{wrdid}
    \partial_-\frac{\delta W_+}{\delta A_-} + \left[A_-,\frac{\delta W_+}{\delta A_-}\right] = -\frac{k_+}{2\pi } \partial_+ A_- \,.
\fe
If we express $A_-$ as $A_-=g^{-1} \partial_- g$, then we can obtain the following expression for $\frac{\delta W_+}{\delta A_-}$ from \eqref{wrdid}
\ie 
\frac{\delta W_+}{\delta A_-}= -\frac{k_+}{2\pi} g^{-1} \partial_+ g \,.
\fe
Assuming that the chiral defect has no modes that decouple from the 4d CS gauge field, an arbitrary variation of the effective action then takes the form 
\ie 
    \delta W_+ = &\int d\si^+ d \si^-\, \textrm{Tr} \left(\frac{\delta W_+}{\delta A_-}\delta A_- \right)\\
    =& -\frac{k_+}{2\pi}\int d\si^+ d \si^- \textrm{Tr} (g^{-1}\partial_+g \delta (g^{-1}\partial_-g ) )\\
    =& \frac{k_+}{2\pi}\int d\si^+ d \si^- \textrm{Tr} ( ( [g^{-1}\partial_-g, g^{-1}\partial_+g]+\partial_- (g^{-1}\partial_+g))  g^{-1}\delta g  )\\
    =& \frac{k_+}{2\pi}\int d\si^+ d \si^- \textrm{Tr} ( \partial_+ (g^{-1}\partial_-g)  g^{-1}\delta g  ) \,.
\fe
Then, we find that 
\ie 
    W_+=-\frac{k_+}{4\pi } \int_{\Sigma} d\si^+ d\si^-\textrm{Tr}(g^{-1}\partial_+g g^{-1}\partial_-g)-\frac{k_+}{12\pi }\int_{\Sigma \times \mathbb{R}_+} \textrm{Tr} (g^{-1}dg\wedge g^{-1}dg\wedge g^{-1}dg ) \,,
\fe
which is the chiral edge mode action before the inclusion of Pauli-Villars counter-term, multiplied by $-k_+$, as expected.
An analogous computation relates anti-chiral defects and the anti-chiral edge mode action.

Although we have emphasized the duality between free fermion defects and edge mode defects, it is clear from the derivations above that the duality to edge mode defects also holds more generally for other chiral and anti-chiral defects.

\subsubsection{WZW Defects into Edge Mode Defects}

Let us next explain how to convert WZW defect into the edge mode defect,
thus completing the derivation of defect bosonization. The equivalence of edge mode defects and WZW defects is nothing but a classical result of \cite{Polyakov:1984et}.

We focus on the left-handed side of the WZW defect with the path integral
\begin{align}
    \hat{Z}_{+}^b[A]
    &= \int \cD \cG_{(+)}\exp\bigl(iS_{+}^{b}[A_-,\cG_{(+)}]\bigr)\,,\label{WZWd-p}\\
    S_{+}^{b}[A,\cG_{(+)}]
    &=N_F S_{\rm WZW}[\cG_{(+)}] -\frac{N_F}{\pi}\int_{\cM\times \{z_{+}\}} d^2\sigma\,\Tr(A_- \partial_+\cG_{(+)}\cG_{(+)}^{-1})\no\\
    &\quad -\frac{N_F}{2\pi}\int_{\cM\times \{z_{+}\}} d^2\sigma\,\Tr(A_+ A_-)\,.
\end{align}
For our purpose, we parametrize the minus component $A_-$ of the gauge field by (\ref{Ad-ch}) as in the case of free fermion defects, and use the Polyakov-Wiegmann identity \cite{Polyakov:1984et}
{\normalsize
\begin{align}
    S_{\rm WZW}[g_{(+)}\cG_{(+)}]\!=\!S_{\rm WZW}[g_{(+)}]+S_{\rm WZW}[\cG_{(+)}]-\frac{1}{\pi}\int_{\cM\times \{z_{+}\}} \!d^2\sigma\Tr\left(\!A_-\partial_+\cG_{(+)}\cG_{(+)}^{-1}\!\right)\,. \label{PW-id}
\end{align}
}
We can then rewrite (\ref{WZWd-p}) as
\begin{align}
    \hat{Z}_{+}^b[A]
    &= \int \cD \cG_{(+)}\exp\biggl(iN_F\,S_{\rm WZW}[g_{(+)}\cG_{(+)}]-iN_F\,S_{\rm WZW}[g_{(+)}]\no\\
    &\qquad\qquad\qquad\qquad-\frac{N_F}{2\pi}\int_{\cM\times \{z_{+}\}} d^2\sigma\,\Tr(A_+ A_-)\biggr)\no\\
    &=\exp\biggl(-iN_F\,{S}^{\textrm{edge}}_+[g_{(+)},A]\biggr)\int \cD \cG_{(+)}\exp\left(iN_F\,S_{\rm WZW}[\cG_{(+)}]\right)
\,,
\end{align}
where in the second equality we used the invariance of the integration measure with respect to the left action, i.e.\ $\cD(g_{(+)}\cG_{(+)}) =\cD \cG_{(+)}$\,.

We can also consider the anti-chiral edge mode defect action, and relate it to a WZW defect in an analogous manner. 

\subsubsection{Dirichlet Defects into Edge Mode Defects}
Let us next discuss yet another type of defect, a Dirichlet defect.
We consider a chiral and anti-chiral Dirichlet defects located at $z=z_{+,\alpha}$ and $z=z_{-,\beta}$, i.e., 
\ie \label{dirchi}
A_-|_{z_{+,\alpha}}=0\,,\qquad A_+|_{z_{-,\beta}}=0\,.
\fe 

Contrary to the case of an order defect, a Dirichlet defect 
has no additional degrees of freedom, 
but instead appropriate boundary conditions for the gauge fields as in \eqref{gauge_pm_0}.
In the following, we explain how to convert this defect into the edge mode defect discussed previously,
thus completing the task of connecting the quartet of defects.

One of the features of the Dirichlet defect is that the boundary condition \eqref{gauge_pm_0}
seems to break the gauge symmetry at the boundary: in order to preserve the boundary condition,
the gauge transformation $A\to -d\hat{g}\hat{g}^{-1}+\hat{g}A \hat{g}^{-1}$ is constrained to be 
of the form
 \ie  \label{g_defect}
g\to 1 \quad \textrm{as} \quad z\to z_+ \;.
\fe
This is in contrast with the case of some other defects, which have manifest $G$ global symmetries.

However, we can be more 
explicit about the $G$-symmetry of the defect by 
allowing for  gauge transformations 
which do not obey \eqref{g_defect}.
Since the path integral measure of the bulk 4d CS theory is invariant under this transformation,
it is legitimate to apply this gauge transformation
as long as one keeps track of the changes of both the Lagrangian and the boundary conditions (Such a gauge transformation was called a formal gauge transformation in \cite{Delduc:2019whp,Benini:2020skc}).

Let us now apply the gauge transformation
\ie\label{A_gauge}
A=-d\hat{g}\hat{g}^{-1}+\hat{g}\cL \hat{g}^{-1}.
\fe
By using this gauge transformation, 
one can set $\cL_{\zbar}=0$, as before.
This is possible since we can write the $\bar{z}$-component of the 4d CS gauge field as 
\ie \label{cpureg}
A_{\zbar}= -\partial_{\zbar} \hat{g}\hat{g}^{-1},
\fe
where $\hat{g}$ is determined by a map from $\Sigma$ to the moduli space of ${G}$-bundles on $\mathbb{CP}^1$ where the boundary conditions \eqref{dirchi} are imposed.
This is because in defining the 4d CS theory, we consider a topologically trivial ${G}$-bundle on $\Sigma \times \mathbb{CP}^1$, and a topologically trivial complex bundle on $\mathbb{CP}^1$ is generically holomorphically trivial as well, allowing us to express $A_{\zbar}$ as in \eqref{cpureg}.
The question of when this can be done for general twist functions is discussed in Remark 5.1 of \cite{Benini:2020skc}. 

Let us choose the gauge $\cL_{\zbar}=0$ and list the effect of the gauge transformation \eqref{A_gauge}.
The 4d CS action is transformed into  \ie \label{formaltx}
S_{\rm CS}[A]=\frac{1}{2\pi \hbar } \int_{\Sigma \times C} \omega_{\textrm{eff}}\wedge  \left(\textrm{CS}(\cL) +d\left(\textrm{Tr} ({\hat{g}}^{-1} d {\hat{g}} \wedge \cL)\right) +I_{\rm WZ}[\hat{g}]
\right)\,,
 \fe
 where $I_{\rm WZ}[\hat{g}]$ is the Wess-Zumino three-form defined as 
 \ie 
     I_{\rm WZ}[\hat{g}]=\frac{1}{3}\textrm{Tr}\left( {\hat{g}}^{-1} d {\hat{g}}\wedge {\hat{g}}^{-1} d {\hat{g}} \wedge {\hat{g}}^{-1} d {\hat{g}}\right),
\fe
while the boundary conditions are transformed to 
\ie 
    \label{genconn2}
    \cL_-|_{z_{+,\alpha}}
    &=g_{+,\alpha}^{-1}\partial_-g_{+,\alpha}|_{z_{+,\alpha}}\,,\\ \cL_+|_{z_{-,\beta}}
    &=g_{-,\beta}^{-1}\partial_+g_{-,\beta}|_{z_{-,\beta}}\,,
\fe 
where we have denoted the value of $\hat{g}$ at $z_{+,\alpha}$ and $z_{-,\beta}$ to be $g_{+,\alpha}$ and $g_{-,\beta}$, respectively. 

The second and third terms in \eqref{formaltx} can be shown, using the Cauchy-Pompeiu integral formula, to take the form 
 \ie  
-&\sum_{\alpha=1}^{n_+} \frac{k_{+,\alpha}}{4\pi} \bigg( \int_{\Sigma \times \{z_{+,\alpha}\}} d\sigma^+d\sigma^-\textrm{Tr} (g_{+,\alpha}^{-1} \partial_+g_{+,\alpha} \cL_--g_{+,\alpha}^{-1} \partial_-g_{+,\alpha} \cL_+)
\\ & \qquad \qquad \quad +\int_{\Sigma  \times \mathbb{R}_+ \times \{z_{+,\alpha}\}}I_{\rm WZ}[g_{+,\alpha}] 
\bigg)\\
+ &\sum_{\beta=1}^{n_-} \frac{k_{-,\beta}}{4\pi} \bigg(\int_{\Sigma \times \{z_{-,\beta}\}} d\sigma^+d\sigma^- \textrm{Tr} (g_{-,\beta}^{-1} \partial_+g_{-,\beta} \cL_--g_{-,\beta}^{-1} \partial_-g_{-,\beta} \cL_+)
\\ & \qquad \qquad \quad +\int_{\Sigma  \times \mathbb{R}_+\times \{z_{-,\beta}\}}I_{\rm WZ}[g_{-,\beta}]
\bigg)\,.
\label{pCS-d2}
\fe 
At this point, one may go on to derive an integrable field theory which takes the form of integrable coupled WZW models following the techniques of \cite{Costello:2019tri,Delduc:2019whp,Benini:2020skc}, that is, by varying $\cL$, $g_+$, and $g_-$ in the bulk action, equation \eqref{pCS-d2}, and the boundary constraints \eqref{genconn2}, and solving the resulting equations of motion. This integrable field theory will be studied further in the next subsection.

To relate to edge mode defects, we instead use the boundary constraints \eqref{genconn2}, to rewrite \eqref{pCS-d2} into
\ie \label{pCS-d3}
&-\sum_{\alpha=1}^{n_+}\frac{k_{+,\alpha}}{4\pi} \bigg(\int_{\Sigma \times \{z_{+,\alpha}\}} d\sigma^+d\sigma^-\textrm{Tr} (g_{+,\alpha}^{-1} \partial_+g_{+,\alpha} g_{+,\alpha}^{-1}\partial_-g_{+,\alpha} -\cL_+ \cL_-) \\ & \qquad \qquad \qquad + \int_{\Sigma  \times \mathbb{R}_+ \times \{z_{+,\alpha}\}}I_{\rm WZ}[g_{+,\alpha}]  \bigg)
\\
&- \sum_{\beta=1}^{n_-}\frac{k_{-,\beta}}{4\pi}\bigg( \int_{\Sigma \times \{z_{-,\beta}\}} d\sigma^+d\sigma^- \textrm{Tr} (g_{-,\beta}^{-1} \partial_+g_{-,\beta} g_{-,\beta}^{-1} \partial_-g_{-,\beta}-\cL_+\cL_-) \\ & \qquad \qquad \qquad - \int_{\Sigma  \times \mathbb{R}_+ \times \{z_{-,\beta}\}}I_{\rm WZ}[g_{-,\beta}] \bigg)\,.
\fe 
Thus, the path integral has the form 
\ie \label{dlpi}
\int \mathcal{D} \cL\, e^{iS_{\textrm{CS}}[\cL]} \prod_{\alpha=1}^{n_+} \textrm{exp} (-ik_{+,\alpha}{S}^{\textrm{edge}}_+[g_{+,\alpha},\cL]) \prod_{\beta=1}^{n_-} \textrm{exp} (-ik_{-,\beta}{S}^{\textrm{edge}}_-[g_{-,\beta},\cL]) \,,
\fe 
where ${S}^{\textrm{edge}}_+[g_{+,\alpha},\cL]$ and ${S}^{\textrm{edge}}_-[g_{-,\beta},\cL]$ are the Lagrangians for the edge mode defects given in \eqref{swzwua}.

\subsubsection{Comparison of Anomalies}

Before closing this subsection, let us comment on the anomalies of the defects.
As we have seen, the chiral anomaly of the fermion defects plays a crucial role 
when we discuss the equivalence of the defects. The natural question is then what are the counterparts of the anomaly
for other defects.

Firstly, for the WZW defect \eqref{WZWd-p}  with the action $S_+^b[\cG_{(+)},A]$, under the gauge transformation 
\ie 
\mathcal{G}_{(+)}^h=h \mathcal{G}_{(+)}\,, \quad A^h=h A h^{-1}-d h h^{-1}\,,
\fe
the Polyakov-Wiegmann formula can be used to show that the path integral changes by a factor of 
\ie \label{clasano1}
\exp \bigg(\frac{i k_+}{4 \pi} \int_{\Sigma \times\left\{z_{+}\right\}} \operatorname{Tr}\left(h^{-1} d h \wedge A\right)  +i k_+ S_{\mathrm{WZ}}[h] \bigg) \,,
\fe 
which cancels the chiral part of the bulk gauge transformation. 

Likewise, for the WZW defect (\ref{bo-rdefect}) with the action $S_-^b[\cG_{(-)},A]$, under the gauge transformation 
\ie 
\mathcal{G}_{(-)}^h= \mathcal{G}_{(-)}h^{-1}, \quad A^h=h A h^{-1}-d h h^{-1}\,,
\fe
the Polyakov-Wiegmann formula can be used to show that the path integral changes by a factor of 
\ie \label{clasano2}
\exp \bigg(\frac{-i k_-}{4 \pi} \int_{\Sigma \times\left\{z_{-}\right\}} \operatorname{Tr}\left(h^{-1} d h \wedge A\right)  - i k_- S_{\mathrm{WZ}}[h] \bigg) \,,
\fe 
which cancels the anti-chiral part of the bulk gauge transformation. 

As expected, the gauge transformations are compatible with the consistent chiral anomaly of the chiral gauged fermionic system.
Note that the anomaly in the dual description arises not from the measure of the path integral, but rather from the 
classical Lagrangian, and in this sense transformations \eqref{clasano1} and \eqref{clasano2} can be regarded as
classical avatars of the chiral anomalies for the fermion systems.

The edge mode defects also demonstrate a counterpart of the anomaly. 
To observe this, we note the following property of formal gauge transformations. 
Under $A=-dg g^{-1}+g\tilde{A}g^{-1}$, we have
\ie \label{forchk}
\textrm{CS}(A)=\textrm{CS}(\tilde{A})+d\left(\textrm{Tr} ({g}^{-1} d {g} \wedge \tilde{A})\right) +I_{\rm WZ}[g]
\,.
\fe
Now, upon a further formal gauge transformation $\tilde{A}=-dh h^{-1}+hA'h^{-1}$, \eqref{forchk} becomes 
\begin{align}
  &\textrm{CS}(A') +I_{\rm WZ}[g]+I_{\rm WZ}[h]\no
\\ 
&- \textrm{Tr}\left( g^{-1}dg \wedge dhh^{-1}\right) +d\left( \textrm{Tr} \left(h^{-1}dh \wedge A'\right)\right)  +d\left(\textrm{Tr} ( g^{-1}dg \wedge h A'h^{-1 } )
\right)\,.  
\end{align}
Using the Polyakov-Wiegmann relation, this can be put in the form 
\ie 
    \textrm{CS}(A)= \textrm{ } & \textrm{CS}(A')+d\left(\textrm{Tr} \left( {(gh)}^{-1} d {(gh)} \wedge A'\right)\right) +I_{\rm WZ}[gh] \,.
\fe
Moreover, if we have a boundary condition, e.g., $\tilde{A}_-=g^{-1}\partial_-g|_{z_+}$,
this is transformed to $A'_-=(gh)^{-1}\partial_-(gh)|_{z_+}$. Hence, the edge mode defects in \eqref{pCS-d3} (that can be obtained from a formal gauge transformation) also admit a notion of anomaly inflow,
up to a change of choice of pure gauge used to define the effective action of each free fermion surface defect.

\subsection{New Bosonizations from Multiple Defects}\label{Boson-multi}

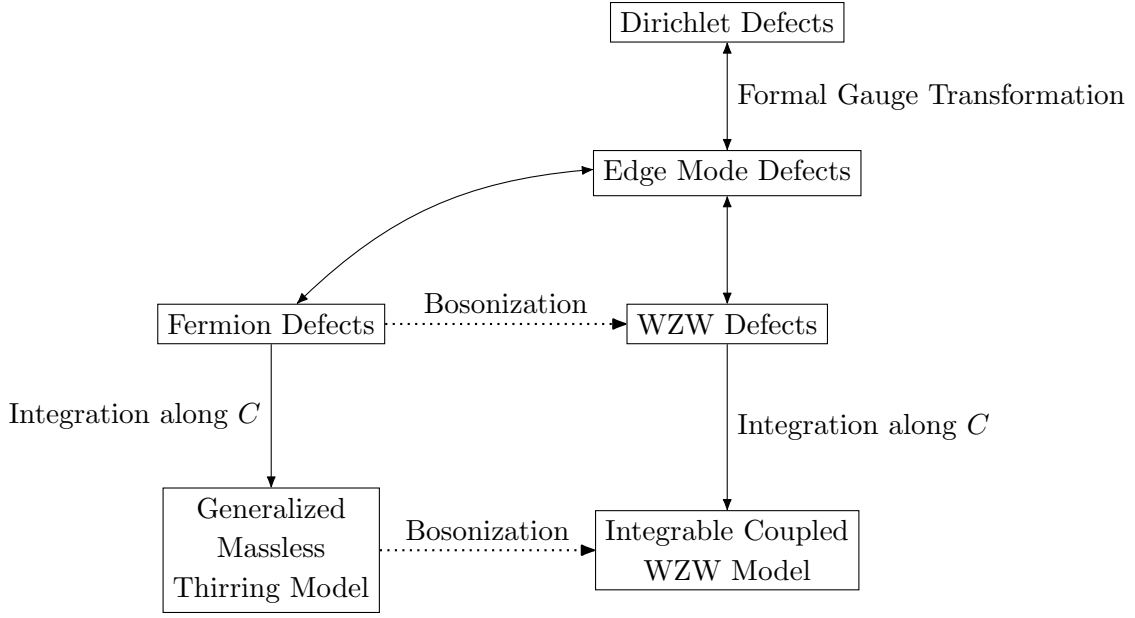
\begin{figure}[htbp]\label{duality-map}
\centering
\begin{tikzpicture}
    \node[rectangle, draw, align=center](31) at (0, 3) {Fermion Defects};
    \node[rectangle, draw, align=center](1) at (0, 0) {Generalized\\ Massless \\ Thirring Model};
    \node[rectangle, draw, align=center](32) at (6,3){WZW Defects};
    \node[rectangle, draw, align=center](2) at (6,0){Integrable Coupled \\ WZW Model};
    \node[rectangle, draw, align=center](52) at (6,5){Edge Mode Defects};
    \node[rectangle, draw, align=center](72) at (6,7){Dirichlet Defects};
    \draw[ ->, >={Latex[round]}, shift=({5pt, 5pt})]  (31)--(1) node[midway, left]{ Integration along $C$};  
    \draw[ ->, >={Latex[round]}, shift=({5pt, 5pt})]  (32)--(2)  node[midway, right]{ Integration along $C$};   
    \draw[ ->, dotted, style=thick, >={Latex[round]}, shift=({5pt, 5pt})]  (31)--(32) node[midway, above]{ Bosonization} ;  
    \draw[ ->, dotted, style=thick, >={Latex[round]}, shift=({5pt, 5pt})]  (1)--(2)  node[midway, above]{ Bosonization} ;  
    \draw[ <->, >={Latex[round]}, shift=({5pt, 5pt})]  (72)--(52) node[midway, right]{Formal Gauge Transformation};
    \draw[ <->, >={Latex[round]}, shift=({5pt, 5pt})]  (52)--(32) ;  
    \draw[<->,  >={Latex[round]},  shift=({5pt, 5pt})]  (52) to [bend right=20] (31) ;    
\end{tikzpicture}
\caption{The new bosonization discussed in this subsection is derived by including multiple defects 
discussed in the previous subsection, and integrating the theory along $C$ to obtain the effective two-dimensional theory.}
\label{fig:multi_bosonization}
\end{figure}

Let us next discuss bosonizations of integrable field theories obtained by including multiple defects.
The strategy is depicted in Figure \ref{fig:multi_bosonization}: since we already know the bosonizations between the defects, we can simply include multiple defects at various points on the spectral curve, both before and after bosonizations.
When we integrate the theories along the spectral curve, we obtain two effective integrable field theories
related by bosonization.

Let us consider $n_+$ chiral and $n_-$ anti-chiral Dirichlet defects:
\ie
    \label{chidiri}
    A_-|_{z_+,\alpha}=0\,,\qquad A_+|_{z_-,\beta}=0\,.
\fe 
with the one-form given by
\ie \label{geneff}
   \omega_{\text{eff}}
   =\varphi(z)dz= dz - \frac{\hbar}{4\pi i}\left(\sum_{\alpha=1}^{n_{+}}\frac{k_{+,\alpha}}{z-z_{+,\alpha}}-\sum_{\beta=1}^{n_{-}}\frac{k_{-,\beta}}{z-z_{-,\beta}}\right)dz\,.
\fe
As we have seen, we can trade the Dirichlet defects into either edge mode of WZW models,
where the one-form is generated by anomalies.

Let us consider an equal number of chiral and anti-chiral defects, i.e., $n_+=n_-=N$.
We assume the anomaly cancellation condition \eqref{inf-con} so that there is no pole at infinity.
such that \eqref{geneff} can be written as 
\ie \label{septwi}
    \omega_{\text{eff}}&=\prod_{\alpha=1}^{N} \frac{(z-\zeta_{-,\alpha})(z-\zeta_{+,\alpha})}{(z-z_{-,\alpha})(z-z_{+,\alpha})}dz\,\\
    &= \tilde{\varphi}_+(z)\tilde{\varphi}_-(z)\,,
\fe
where
\ie \label{phitil}
\tilde{\varphi}_+(z)=&\prod_{\alpha=1}^{N} \frac{(z-\zeta_{-,\alpha})}{(z-z_{-,\alpha})}\,,\qquad \tilde{\varphi}_-(z)=\prod_{\alpha=1}^{N} \frac{(z-\zeta_{+,\alpha})}{(z-z_{+,\alpha})}\,.
\fe

It turns out that the resulting model is a special limit of the 
coupled WZW models studied in \cite{Delduc:2018hty, Costello:2019tri}.
The twist function of the latter corresponds to the meromorphic one-form
\ie \label{cycoup}
    \omega
    =\frac{\prod_{i=1}^{2N}\left(z-q_i^+\right) \prod_{j=1}^{2N}\left(z-q_j^{-}\right)}{\prod_{k=1}^{2N}\left(z-p_k\right)^2} \mathrm{~d} z\,,
\fe 
where non-chiral Dirichlet boundary conditions 
\ie
    A_{\pm} |_{z=p_k}=0 
\fe
are imposed at the double poles at $p_k$, while at the zeroes $q_i^+$ and $q_i^{-}$, $A_+$ and $A_-$ are respectively allowed to have poles, i.e., 
\ie 
    A_+ \sim \frac{1}{z-q_i^+}\,,\quad \quad  A_- \sim \frac{1}{z-q_i^-}\,.
\fe
Let us factorize the denominator of \eqref{cycoup}  following \cite{Costello:2019tri,Delduc:2018hty}, by expressing the meromorphic one-form \eqref{cycoup} as
\ie \label{factorizedtw}
    \omega&= \varphi_+(z)\varphi_-(z) dz \,,
\fe
where 
\ie 
\varphi_+(z)=\frac{ \prod_{j=1}^{2N}\left(z-q_j^{+}\right)}{\prod_{k=1}^{2N}(z-p_k)} \,,
\fe
and 
\ie 
\varphi_-(z)=\frac{ \prod_{j=1}^{2N}\left(z-q_j^{-}\right)}{\prod_{k=1}^{2N}(z-p_k)} \,.
\fe
We also make use of the expressions
\ie
\varphi_{+, i}(z) & =\left(z-p_i\right) \varphi_+(z)\,, \\
\varphi_{-, i}(z) & =\left(z-p_i\right) {\varphi_-}(z)\,,
\fe
 defined in \cite{Costello:2019tri, Delduc:2018hty}.
The action for these models, in the form derived by \cite{Costello:2019tri}
can be expressed as
\ie \label{sumaction}
S_{\text {total }}=S_{\text {kin }}+S_{\textrm{2-form }}+S_{\text {3-form }},
\fe 
is given by\footnote{In \cite{Costello:2019tri} the factor of $i/\hbar$ is written as $k/4\pi$.}  
\ie 
S_{\mathrm{kin}}=&\frac{i}{\hbar } \int_{\mathbb{R}^2}d\sigma^+d\sigma^- \bigg(\sum_i^{2N} \textrm{Tr}(J_{ -,i} J_{ +,i})\left(-\varphi_{ +,i}\left(p_i\right) \varphi_{-,i}^{\prime}\left(p_i\right)+\varphi_{ +,i}^{\prime}\left(p_i\right) \varphi_{-,i}\left(p_i\right)\right) \\& -\sum_{j \neq i}^{2N} \frac{\varphi_{ -,i}\left(p_i\right) \varphi_{+,j}\left(p_j\right)-\varphi_{-,j}\left(p_j\right) \varphi_{ +,i}\left(p_i\right)}{p_i-p_j} \textrm{Tr}(J_{ -,i} J_{ +,j})\bigg)\,,
\fe

\ie
S_{\text {2-form }}=-\frac{i}{\hbar}\int_{\mathbb{R}^2} d\sigma^+d\sigma^- \sum_{j \neq i}^{2N} \frac{\varphi_{ -,i}\left(p_i\right) \varphi_{ +,j}\left(p_j\right)+\varphi_{ -,j}\left(p_j\right) \varphi_{+,i}\left(p_i\right)}{p_i-p_j} \textrm{Tr}\left( J_{-,i} J_{+,j} \right)\,,
\fe 
and 
\ie
S_{\text {3-form }}=-\frac{2i}{\hbar } \sum_i^{2N} \int_{\mathbb{R}^2 \times \mathbb{R}_{\geq 0}} d\sigma^+d\sigma^- ds \operatorname{Tr}\left(J_{-, i} J_{+, i} J_{s, i}\right)\left(\varphi_{ +,i}\left(p_i\right) \varphi_{ -,i}^{\prime}\left(p_i\right)+\varphi_{ +,i}^{\prime}\left(p_i\right) \varphi_{ -,i}\left(p_i\right)\right)\,, \label{wzw}
\fe 
where $J_{ +,i}=\left(\partial_+ \si_i\right) \si_i^{-1}, J_{-,i}=\left(\partial_{-} \si_i\right) \si_i^{-1}$, with $\si_i$ replaced by $\widehat{\si}_i$ in the same expressions in \eqref{wzw}, and $J_{s, i}=\left(\partial_s \widehat{\si}_i\right) \widehat{\si}_i^{-1}$. In what follows, we shall set $\si_i^{-1}=g_i$.

Now, colliding a zero, $q^-_i$ ($q^+_i$), of the twist function, where $A_-$ ($A_+$) has a pole, with a double pole, $p_k$, of the twist function results in a simple pole where $A_+$ ($A_-$) satisfies a chiral Dirichlet boundary condition. 
We shall collide the zeroes labelled $q_j^+$ for $j=1,\ldots, N$ with the double poles $p_k$ for $k=N+1,\ldots 2N$, and collide the zeroes labelled $q_j^-$ for $j=1,\ldots, N$ with the double poles $p_k$ for $k=1,\ldots N$. 
The former results in $N$ simple poles which can be identified with $z_{+,\alpha}$ and the  latter results in $N$ simple poles which can be identified with $z_{-,\alpha}$.
Renaming the remaining $N$ zeroes labelled $q_j^-$ as $\zeta_{+,\alpha}$ and the remaining $N$ zeroes labelled $q_j^+$ as $\zeta_{-,\alpha}$,
we can retrieve a twist function of the form \eqref{septwi} with chiral and anti-chiral Dirichlet boundary conditions imposed as in \eqref{chidiri}. In particular, in this limit, we find that 
\ie 
\varphi_+(z)\rightarrow \tilde{\varphi}_+(z) 
\fe
and 
\ie 
\varphi_-(z)\rightarrow \tilde{\varphi}_-(z) \,.
\fe
In addition, in this limit,
 \ie \label{speclim}
\varphi_{+,\alpha}(z_{+,\alpha}) \rightarrow 0\,,\\
\varphi_{-,\alpha}(z_{-,\alpha}) \rightarrow 0\,,\\
 \fe
and 
\ie 
\varphi'_{+,\alpha}(z_{+,\alpha})&\rightarrow  \tilde{\varphi}_{+}(z_{+,\alpha})\,, \\
\varphi'_{-,\alpha}(z_{-,\alpha}) &\rightarrow  \tilde{\varphi}_{-}(z_{-,\alpha})\,.
\fe 

Once we understand the limit, we can write down the action of the boson integrable field theory as a limit of that of the coupled WZW models with the action \eqref{sumaction}:\ie 
     \label{genagm}
    S[g_{+,\alpha},g_{-,\beta}] & \\
 =-\frac{i}{ \hbar} \sum_{\alpha=1}^N \bigg( & - \int_{\Sigma}d\sigma^+ d\sigma^- \tilde{\varphi}_{-,\alpha} (z_{+,\alpha})\tilde{\varphi}_{+} (z_{+,\alpha}) \Tr\left(g^{-1}_{+,\alpha} \partial_+ g_{+,\alpha}  g^{-1}_{+,\alpha}\partial_-g_{+,\alpha}\right) \\
     &+ \int_{\Sigma}d\sigma^+ d\sigma^- \tilde{\varphi}_{+,\alpha} (z_{-,\alpha})\tilde{\varphi}_{-} (z_{-,\alpha}) \Tr\left( g^{-1}_{-,\alpha} \partial_+ g_{-,\alpha}  g^{-1}_{-,\alpha}\partial_-g_{-,\alpha}\right) \\
     &+\int_{\Sigma}d\sigma^+ d\sigma^- \sum_{\beta=1}^N \frac{\tilde{\varphi}_{+,\beta}(z_{-,\beta})}{z_{+,\alpha}-z_{-,\beta}} \tilde{\varphi}_{-,\alpha}(z_{+,\alpha}) \Tr\left(g^{-1}_{-,\beta}\partial_+g_{-,\beta} g_{+,\alpha}^{-1}\partial_-g_{+,\alpha}\right)\\
     &-\int_{\Sigma}d\sigma^+ d\sigma^- \sum_{\beta=1}^N \frac{\tilde{\varphi}_{-,\beta}(z_{+,\beta})}{z_{-,\alpha}-z_{+,\beta}} \tilde{\varphi}_{+,\alpha}(z_{-,\alpha}) \Tr\left( g^{-1}_{+,\beta}\partial_-g_{+,\beta} g_{-,\alpha}^{-1}\partial_+g_{-,\alpha}\right) \\
     &-\int_{\Sigma \times \mathbb{R}_+} \tilde{\varphi}_{-,\alpha}(z_{+,\alpha}) \tilde{\varphi}_{+}(z_{+,\alpha}) I_{\rm WZ}[\hat{g}_{+,\alpha}] -\int_{\Sigma \times \mathbb{R}_+} \tilde{\varphi}_{+,\alpha}(z_{-,\alpha}) \tilde{\varphi}_{-}(z_{-,\alpha}) I_{\rm WZ}[\hat{g}_{-,\alpha}]
 \bigg)\\
 =-\sum_{\alpha=1}^{N}\biggl[k_{+,\alpha}&S_{\textrm{WZW}}[g_{+,\alpha}]+k_{-,\alpha}S_{\textrm{WZW}}[g_{-,\alpha}^{-1}]+\sum_{\beta=1}^{N}\rho_{\alpha\beta}\int_{\Sigma}d^2\sigma\,\Tr\left(j_+^{(\alpha)} j_-^{(\beta)}\right)
 \biggr]\,,
\fe
where 
\begin{align}\label{rho-gene}
    \rho_{\alpha\beta}=\frac{2i}{\hbar}\frac{\tilde{\varphi}_{+,\alpha}(z_{-,\alpha})\tilde{\varphi}_{-,\beta}(z_{+,\beta})+ \tilde{\varphi}_{-,\beta}(z_{+,\beta}) \tilde{\varphi}_{+,\alpha}(z_{-,\alpha})}{z_{-,\alpha}-z_{+,\beta}}  \,,
\end{align}
and 
 \ie 
    \tilde{\varphi}_{\pm,\alpha}(z)=(z-z_{\mp,\alpha})\tilde{\varphi}_{\pm}(z)\,,\qquad j_{ \pm}^{(\alpha)}(\tau,\si)=g_{\mp,\alpha}^{-1} \partial_{ \pm} g_{\mp,\alpha}\,.
 \fe 
In addition, the Lax connection is given by
\ie\label{laxgenaff}
    \cL_{ \pm}(\tau,\si,z)
    =\sum_{\alpha=1}^N \frac{\tilde{\varphi}_{ \pm, \alpha}\left(z_{\mp,\alpha}\right)}{\tilde{\varphi}_{ \pm, \alpha}(z)} j_{ \pm}^{(\alpha)}(\tau,\si)\,,
\fe
and satisfies the boundary condition (\ref{genconn2}) since 
\ie 
\frac{1}{\varphi_{\pm,\alpha} (z_{\mp,\beta})}=\delta_{\alpha,\beta}\frac{1}{\varphi_{\pm,\alpha} (z_{\mp,\alpha})}\,.
\fe
 In particular, the action (\ref{genagm}) and Lax connection (\ref{laxgenaff}) arise in the limit described in the previous paragraph from the action and Lax connection derived in \cite{Costello:2019tri, Delduc:2018hty}.

On the other side of the bosonization,
we have the 4d CS theory coupled to multiple chiral free fermion surface operators at $z_{+,\alpha}$ and multiple anti-chiral free fermion surface operators at $z_{-,\alpha}$ :
\ie \label{dlpi3}
&\int \mathcal{D} A\, e^{\frac{i}{2\pi \hbar}\int \omega_{\textrm{eff}}\wedge \textrm{CS}({A)}} \prod_{\alpha=1}^{N}\int \cD\Psi_{\alpha} \cD\bar{\Psi}_{\alpha} \,e^{{\frac{i}{\hbar_{\rm 2d}}\int_{\cM\times \{z_{+,\alpha}\}} d^2\sigma\,\big(\sum_{I=1}^{k_{+,\alpha}}\bar{\Psi}_{\alpha}^{I}i\hat{\cancel{D}}_L\Psi_{\alpha}^I} - \frac{k_{+,\alpha} \hbar_{\rm 2d}}{2\pi}  \textrm{Tr} (A_+A_-)\big)}\\& \times \prod_{\beta=1}^{N} \int \cD\tilde{\Psi}_{\beta} \cD\bar{\tilde{\Psi}}_{\beta}\,e^{{\frac{i}{\hbar_{\rm 2d}}\int_{\cM\times \{z_{-,\beta}\}} d^2\sigma\, \big(\sum_{J=1}^{k_{-,\beta}}\bar{\tilde{\Psi}}_{\beta}^{J}i\hat{\cancel{D}}_R\tilde{\Psi}_{\beta}^J}  - \frac{k_{-,\beta} \hbar_{\rm 2d}}{2\pi}  \textrm{Tr} (A_+A_-)\big)}\,.
\fe 
Taking $\hbar$ to be small such that $\omega_{\textrm{eff}}$ can be approximated by $\omega =dz$, and integrating out the 4d CS gauge fields in the $A_{\zbar}=0$ gauge gives us an integrable field theory with the action
\ie\label{genmulthir2}
\frac{1}{\hbar_{2d}}\int_{\cM} d^2\sigma\bigg(&{\sum_{\alpha=1}^N\,\sum_{I=1}^{k_{+,\alpha}}\bar{\Psi}_{\alpha}^{I}iP_+{\cancel{\partial}}\Psi_{\alpha}^I} + {\sum_{\beta=1}^N \,\sum_{J=1}^{k_{-,\beta}}\bar{\Psi}_{\beta}^{J}iP_-{\cancel{\partial}}\Psi_{\beta}^J} \\ &+\frac{\hbar_{2d}}{4}\sum_{\alpha=1}^N\sum_{\beta=1}^N \frac{i\hbar}{z_{+,\alpha}-{z_{-,\beta}}}\big(\sum_{I=1}^{k_{+,\alpha}}\bar{\Psi}_{\alpha}^{I} i \rho(t^a)P_+\Psi_{\alpha}^I \big)\big(\sum_{J=1}^{k_{-,\beta}}\bar{\Psi}_{\beta}^{J} i \rho(t_a)P_-\Psi_{\beta}^J \big)\bigg).
\fe

Hence, the bosonization duality of integrable quantum field theories that we have derived is an equivalence between the path integrals for the generalized multi-flavor massless Thirring model \eqref{genmulthir2}
and the coupled WZW models \eqref{genagm}.   

Note that in the action \eqref{genmulthir2} we have suppressed decoupled modes that do not couple to the 4d CS gauge field.
These modes were included in \eqref{Zhat-f-R0} and \eqref{Zhat-f-R} only for the purpose of computing the chiral anomaly. 
We can suppress such decoupled modes as we are only interested in interacting degrees of freedom.
More generally, if we employ the infinitesimal anomaly to show the equivalence between defects (as in the derivation that starts from \eqref{edgese}), we observe that the equivalence between defects holds up to modes that decouple from the 4d CS gauge fields, and therefore the bosonization duality we derived holds up to decoupled free fields. 
 
\subsubsection{Bosonization from WZW Defects and Boson-Boson Duality}

Now, let us focus on the case where there are two WZW defects whose couplings to the gauge field are chiral and anti-chiral.
The coupled WZW model obtained from these defects is\footnote{The coupled WZW models (\ref{2d-baction-2}) can be rewritten to a single WZW model with the current-current interaction by using the Polyakov-Wiegmann identity \eqref{PW-id}. Furthermore, we can see that the limit
\begin{align}
    \left(\frac{k}{2\pi}\right)^2\frac{i \hbar}{z_+-z_-}=-\frac{1}{\pi}\frac{k^2}{(k+1)^2}
\end{align}
reduces the bosonized model (\ref{2d-baction-2}) to the conformal WZW model.
}
\begin{align}
    S_{\rm 2d}
    &=S_{\rm WZW}^{(k)}[g_{(+)}]+S_{\rm WZW}^{(k)}[g_{(-)}]
    -\frac{8i}{\hbar}\rho_{+-}\int_{\cM}d^2\sigma\,\Tr(\partial_+g_{(+)} g_{(+)}^{-1} g_{(-)}^{-1}\partial_-g_{(-)})\,,\label{2d-baction-2}
\end{align}
where $\rho_{+-}$ is given by
\begin{align}
    \rho_{+-}= -\frac{(z_+-\zeta_+)(z_--\zeta_-)}{2(z_+-z_-)}=\frac{(z_+-z_--(\zeta_+-\zeta_-))^2}{8(z_+-z_-)}\,.
\end{align}
By construction, this model should be a bosonized model of the non-abelian Thirring model presented in Section \ref{Th-4dCS}. 
The bosonization of this model has been extensively discussed in a number of works  \cite{Polyakov:1984et}. The bosonized model obtained from the WZW defects has levels with opposite signs compared to that given in \eqref{genagm} for $N=1$, and in fact there are (at least) two bosonization dualities for the massless Thirring model. see e.g., \cite{Santos:2023dax} for a recent discussion on this.
Indeed, for the $k=N_F$ flavor massless Thirring model,
we can find a bosonized model of the form \eqref{2d-baction-2} where the two level $k$ WZW models have the same current-current interaction term \cite{Santos:2023dax} (see Appendix \ref{pboson} for derivation). The bosonized path integral is 
\begin{align}
Z^{\rm Th}_b&=\frac{1}{\cN}\int  \cD g_{(+)}\cD g_{(-)}\exp\biggl(ikS_{\rm WZW}[g_{(+)}]+ikS_{\rm WZW}[g_{(-)}]\no\\
&\qquad +i\left(\frac{k}{\pi}\right)^2\left(a+G^{-1}\right)^{-1}\int_{\cM} d^2\sigma\,\Tr\left(\partial_+g_{(+)}g_{(+)}^{-1}g_{(-)}^{-1}\partial_-g_{(-)}\right)\biggr)\,,\label{bNTh}
\end{align}
where $a$ is a parameter characterizing the ambiguity of the effective action, and the coupling $G$ is given by
\begin{align}
    G=\frac{1}{4}\frac{i\hbar}{z_+-z_-}\,.
\end{align}

The other bosonized model has a form analogous to \eqref{bNTh}, 
but where the level and coupling constant are replaced by $k=-N_F-\sh^{\vee} $ and $\left(a+G^{-1}\right)$, respectively \cite{Polyakov:1984et} (see also  \cite{Bardakci:1994ij,Hull:1995gj}), that is, it takes the form
\ie
Z^{\rm Th}_{b}=\frac{1}{\cN}\int \cD u_{(+)} \cD u_{(-)}  \exp \bigg(& -i(N_{F} +\sh^{\vee})S_{\rm WZW}[u_{(+)}]-i (N_{F} +\sh^{\vee})S_{\rm WZW}[u_{(-)}^{-1}]
\\ & +i\int d^2\sigma\, \left(a+G^{-1}\right)\Tr(u_{(-)}^{-1}\partial_+u_{(-)}u_{(+)}^{-1}\partial_-u_{(+)})\bigg)\,.
\label{bosinv}
\fe 
(The derivation is reviewed in Appendix \ref{pboson}.) We have derived a bosonized model of this form from the 4d CS in the previous subsection as a special case of \eqref{genagm}; the correction of $-\sh^{\vee} $ to the levels can be understood to arise from the Jacobian factor incurred when transforming  the path integral measure of the 4d CS gauge fields to the measure of the edge mode fields. This follows since the path integral measure for the 4d CS gauge fields in \eqref{dlpi} can be split into measures for the gauge fields at the poles of $\omega$ and their bulk complement :
\ie 
\int \mathcal{D} \cL = \int \mathcal{D} \cL_{\rm bulk} \prod_{\alpha=1}^N\mathcal{D} \cL_-|_{z_{+,\alpha}}\prod_{\beta=1}^N\mathcal{D} \cL_+|_{z_{+,\beta}}.
\fe 
The imposition of the bulk equation of motion $\partial_{\zbar} \cL=0$ removes the 4d CS kinetic term from the path integral. The path integral over bulk components of the gauge field can then be evaluated to give a constant.
Transforming the remaining path integral measures to path integral measures for the edge modes using \eqref{genconn2} incurs a Jacobian, whose effect is to shift the levels of the WZW models by $-\sh^{\vee} $ \cite{Polyakov:1984et}.

Strictly speaking, the dimensional reductions of the 4d CS theory along $C$ to obtain our bosonized models \eqref{genagm} and \eqref{2d-baction-2}  are performed at the semi-classical level, and there could be quantum corrections to the coupling constant.\footnote{In order to incorporate quantum effects on the 4d CS side, it is necessary to consider higher orders for the gauge field $A$, but the perturbation theory for the case where the meromorphic one-form $\omega$ has zeros has not been well-understood yet.}
On the other hand, the bosonized models in the literature are derived at the quantum level.
Hence, a precise agreement of the two coupling constants of the two bosonized actions we have derived with those in the literature is not necessarily expected at this stage.
Furthermore, the coupling constant also depends on the choice of regularization, so the comparison of the bosonized actions needs to be made carefully. 

Nonetheless, we can study the coupling constants of \eqref{genagm} (for $N=1$) and \eqref{2d-baction-2} in the limit where $\hbar$ is small. In \eqref{genagm}, for the case of $N=1$, the coupling takes the form
\ie 
-\rho_{11}&=-\frac{4i}{\hbar}\frac{(z_--\zeta_-)(z_+-\zeta_+)}{z_--z_+}\\
&=-\frac{8i}{\hbar} \bigg( -\frac{(z_-+\sqrt{(z_--z_+)^2}-z_+)^2}{8(z_--z_+)} + \frac{(z_-+\sqrt{(z_--z_+)^2}-z_+)}{8\sqrt{(z_--z_+)^2}}\frac{i\hbar k}{\pi}\\
&\qquad \qquad +\frac{\left(\frac{(i\hbar k)}{\pi}\right)^2}{32 \sqrt{(z_--z_+)^2} }
+ O(\hbar^3)\bigg)\\
&=\frac{4}{i\hbar} (z_+-z_-) + O(\hbar^0)\,,
\fe 
if we choose the branch of the square root to be $\sqrt{(z_--z_+)^2} =z_--z_+$. The leading order term is thus $G^{-1}$, in agreement with \eqref{bosinv} for $a=0$. In \eqref{2d-baction-2}, the coupling takes the form 
\ie 
-\frac{8i}{\hbar}\rho_{+-} &=-\frac{8i}{\hbar} \left(-\frac{(z_+-\zeta_+)(z_--\zeta_-)}{2(z_+-z_-)}\right)\\&=-\frac{8i}{\hbar} \bigg( -\frac{(z_-+\sqrt{(z_--z_+)^2}-z_+)^2}{8(z_--z_+)} + \frac{(z_-+\sqrt{(z_--z_+)^2}-z_+)}{8\sqrt{(z_--z_+)^2}}\frac{i\hbar k}{\pi}\\
&\qquad \qquad +\frac{\left(\frac{(i\hbar k)}{\pi}\right)^2}{32 \sqrt{(z_--z_+)^2} }
+ O(\hbar^3)\bigg)\\
&=\frac{k^2}{\pi^2}\frac{1}{4}\frac{i\hbar}{z_+-z_-} + O(\hbar^2),
\fe
if we choose $\sqrt{(z_--z_+)^2} =z_+-z_-$. The leading order term is thus $(k^2/\pi^2)G$, in agreement with \eqref{bNTh} for $a=0$. Hence, we find agreement of our bosonization derivations with those in the literature (for $a=0$) in the small $\hbar$ (semi-classical) limit, but only after making a particular choice of branch for  $\sqrt{(z_--z_+)^2}$. Such a choice could be justified by demanding the convergence of the path integral of the bosonized theories, which ought to restrict the form of the coupling constants. 

Thus, our procedure of bosonization of defects in the 4d CS theory well describes both the bosonizations of the multi-flavor massless Thirring model in the semi-classical limit, and moreover furnishes the boson-boson duality between the two models.

In fact, the relationship between the two bosonized models indicates that these two models exhibit a strong-weak duality,
including the transformation of level, which has been investigated in some literature \cite{Georgiou:2015nka,Georgiou:2017oly,Santos:2023dax}. Since the coupling constant is characterized by the positions of defects on $C$ in the 4d CS theory, it would be interesting to further investigate this strong-weak duality in the context of the 4d CS theory.

\section{Thermodynamic Limit as Polarization of D-branes in String Theory}
  \label{sec:string}

In this section, we shall furnish a string theoretic interpretation of the process whereby a large number of Wilson lines gives rise to an order surface operator in the 4d CS theory.  The main ideas are described in Section \ref{sec:string_sym}, which deals with finite-dimensional symmetric and anti-symmetric representations. The remaining sections deal with 
generalizations to more general  finite-dimensional irreducible representations (Section \ref{sec:string_finite}) and infinite-dimensional representations (Section \ref{sec:string_infinite}).

We expect that the string theory realization of our setup, when combined with the power of string dualities,  will make it possible to connect our discussion to many other examples of discretizations of integrable field theories and thermodynamic limits of integrable lattice models.
For example, it would be interesting to discuss the duality chain to the 
Gauge/YBE correspondence of \cite{Yamazaki:2012cp,Terashima:2012cx,Yamazaki:2013nra,Yamazaki:2018xbx} and discuss the thermodynamic limit of integrable lattice models discussed therein. We also expect our discussion to shed light on the study of vertex algebras in the Schur sector of 4d $\mathcal{N}=2$ supersymmetric field theories \cite{Beem:2013sza,Beem:2014kka}. 

\subsection{Type IIB String Theory Embedding of 4d Chern-Simons Theory}

We shall utilize the embedding of the 4d CS theory with gauge group $\GL(N,\mathbb{C})$ in type IIB string theory derived in \cite{Costello:2018txb} utilizing a stack of $N$ D5-branes in an $\Omega$-background.\footnote{This is equivalent via T-duality to a type IIA string theory embedding of the 4d CS theory using the D4-NS5 brane system, derived in \cite{Ashwinkumar:2018tmm,Ashwinkumar:2019mtj}.} Let us briefly recall this type IIB string theory configuration. The 10d (Euclidean) spacetime is specified to take the form 
\ie 
    ds^2=ds^2_{T^*\Sigma} + ds^2_C + \mathrm{d} s_{\mathrm{TN}}^2 \,,
\fe 
where $\mathrm{d} s_{\mathrm{TN}}^2$ refers to the following Taub-NUT background,
 \ie 
    \mathrm{d} s_{\mathrm{TN}}^2
    =U \mathrm{~d} \vec{x} \cdot \mathrm{d} \vec{x}+\frac{1}{U}\, (\mathrm{d} \theta+\vec{\omega} \cdot \mathrm{d} \vec{x})^2.
 \fe
 Here, $\vec{x}$ is a coordinate of $\mathbb{R}^3$ and $\theta$ parametrizes a circle of radius $r$, which is referred to as the Taub-NUT circle. In addition, $U=1/r+1/\lambda^2$, while $\vec{\omega}$ is a vector on $\mathbb{R}^3$ that satisfies $\mathrm{d} U=\star_{\mathbb{R}^3}(\vec{\omega} \cdot \mathrm{d} \vec{x})$. Parametrizing $\mathbb{R}^3$ by a radial coordinate, $\rho=\sqrt{\vec{x}\cdot\vec{x}}$, and two angular coordinates, one finds that a 2d surface at fixed values of these angular coordinates is a cigar with coordinates $\rho$ and $\theta$. In addition, the background includes a nontrivial dilaton and RR 2-form.
 
 This Taub-NUT geometry preserves 16 of the 32 spacetime supersymmetries. 
The cotangent bundle $T^*\Sigma$, with $ds^2_{T^*\Sigma}$ specified to be a Calabi-Yau metric, preserves a quarter of these supersymmetries. 
 A stack of D-branes in this background would thus preserve two supercharges.\footnote{In the T-dual D4-NS5 system \cite{Ashwinkumar:2019mtj}, the Taub-NUT geometry is replaced by an NS5-brane, intersecting a stack of D4-branes. Four scalar supercharges are preserved in the corresponding worldvolume theory due to the partial-twisting induced by the $T^*\Sigma$ geometry, and the NS5-brane reduces this to two supercharges.  }
 
Let us be more explicit about the supercharges. Type IIB supersymmetry involves two Weyl spinors of identical chirality, a linear combination of which can be written as $\epsilon_L Q_L + \epsilon_R Q_R $. The chirality constraints are $i \Gamma_{0 \cdots 9} \epsilon_L=\epsilon_L, \quad i \Gamma_{0 \cdots 9} \epsilon_R=\epsilon_R$.
The D5 branes constrains the type IIB supersymmetries by:
\ie \label{branecon1}
    \epsilon_R
    =i \Gamma_{012345} \epsilon_L \,,
\fe
which halves the number of supercharges to 16. Since $\epsilon_L$ is determined entirely in terms of $\epsilon_R$ we can use the former to refer to the preserved supercharge, and we shall also refer to it as $\epsilon$.
The topological-holomorphic twist was performed on the D5-brane worldvolume theory by imposing 
\ie \label{twistcon}
\left(\Gamma_{\mu \nu}+\Gamma_{\mu+6, \nu+6}\right) \epsilon=0 \,, \quad \mu, \nu=0, \ldots, 3 \,,
\fe
or equivalently
\ie
    \label{gunum}
    \epsilon
    =\Gamma_{\mu \nu \mu+6, \nu+6} \epsilon \,,
\fe
which are three independent constraints on $\epsilon$. There are thus $16 \times(1 / 2)^3=2$ preserved supercharges.
A single supercharge $Q$ is picked by imposing either
\ie \label{topq1}
    \epsilon
    =-\mathrm{i} \Gamma_{\mu, \mu+6} \epsilon 
\fe
or
\ie \label{topq2}
    \epsilon
    =\mathrm{i} \Gamma_{\mu, \mu+6} \epsilon \,.
\fe
Either set of four equations is compatible with the chirality condition 
\ie\label{chirality}
\mathrm{i} \Gamma_{0123456789} \epsilon=\epsilon 
\fe
and \eqref{gunum}.
 
 In particular, a linear combination of these supercharges, denoted $Q$,  induces an $\Omega$-deformation of the worldvolume theory of any D-brane that wraps a cigar in the Taub-NUT geometry, such that the theory can be described as an $\Omega$-deformed B-model whose target space is a space of maps. This supercharge squares to a Lie derivative generating a rotation of the Taub-NUT circle.  To realize $\GL(N,\mathbb{C})$ 4d CS theory on $\Sigma \times C$, one needs to place a stack of $N$ D5-branes along $C$, a cigar in the Taub-NUT geometry, and $\Sigma \subset T^*\Sigma$.\footnote{As described in \cite{Bershadsky:1995qy}, the twist of the normal bundle to $\Sigma \subset T^*\Sigma$ realizes the R-symmetry twist that ensures topological invariance along $\Sigma$. } The derivation of the 4d CS theory and its associated defects from $\Omega$-deformed field theories is outlined in Appendix \ref{omegappendix}.
 
\subsection{Symmetric and Anti-Symmetric Representations}\label{sec:string_sym}
 
 We shall now utilize this type IIB string theory background with additional D3-branes, which will realize Wilson lines within
 the 4d CS theory, as shown in the figure below \cite{Ishtiaque:2021jan}:

\begin{equation} 
\parbox{0.89\linewidth}{
\begin{tikzpicture}[overlay]
\draw [decorate,decoration={brace,amplitude=4pt},xshift=36pt,yshift=0pt]
(2.65,-0.2) -- (3.79,-0.2) node [black,midway,yshift=9pt] 
{\footnotesize $\Sigma$};
\draw [decorate,decoration={brace,amplitude=4pt},xshift=36pt,yshift=0pt]
(4.05,-0.2) -- (5.19,-0.2) node [black,midway,yshift=9pt] 
{\footnotesize $\mathbb{R}_{\hbar}^2$};
\draw [decorate,decoration={brace,amplitude=4pt},xshift=36pt,yshift=0pt]
(5.52,-0.2) -- (6.52,-0.2) node [black,midway,yshift=9pt] 
{\footnotesize $C$};
\draw [decorate,decoration={brace,amplitude=4pt},xshift=36pt,yshift=0pt]
(7.03,-0.2) -- (8.04,-0.2) node [black,midway,yshift=9pt] 
{\footnotesize $N\Sigma\!\!\subset \!\!T^\ast \Sigma$};
\draw [decorate,decoration={brace,amplitude=4pt},xshift=36pt,yshift=0pt]
(8.33,-0.2) -- (9.34,-0.2) node [black,midway,yshift=9pt] 
{\footnotesize $\mathbb{R}_{-\hbar}^2$};
\end{tikzpicture}
\begin{center}
\begin{tabular}{@{}lllllllllll@{}}
\toprule
  & \textbf{0} & \textbf{1} & \textbf{2} & \textbf{3} & \textbf{4} & \textbf{5} & \textbf{6} & \textbf{7} & \textbf{8} & \textbf{9} \\
   \textbf{D5} & $\times$ & $\times$  & $\times$  & $\times$  &$\times$  &$\times$  & & & &\\
    $\textbf{D3}^+_b$ & $\times$   & & $\times$  & $\times$  &  &  & &  $\times$    & &\\
    $\textbf{D3}^-_b$ &   & $\times$ & $\times$  & $\times$  &  &  &    $\times$ &  & &\\
   $\textbf{D3}^+_f$ &  $\times$  & &  &  &  &  &   & $\times$  & $\times$  &$\times$  \\
    $\textbf{D3}^-_f$ &    &$\times$ &  &  &  &      & $\times$ & & $\times$  &$\times$  \\
\bottomrule
 \end{tabular}
\end{center}
}
\end{equation}

Here, we shall take the 0 and 1 directions to be the analytic continuation to Euclidean signature of the lightcone coordinates utilized in previous sections.
In addition, we have denoted the Taub-NUT geometry as $\mathbb{R}^2_{\hbar}\times \mathbb{R}^2_{-\hbar}$ for notational simplicity, where the two planes can be identified with two antipodal cigars in the Taub-NUT geometry. The opposite signs of the deformation parameters indicate that the supercharge $Q$ squares to a symmetry that rotates the two cigars in opposite directions. Here, we have included D3-branes that spread along $\sigma^{\pm}$ and wrap $\mathbb{R}^2_{\hbar}$ , denoted D3$^{\pm}_b$, and D3-branes that spread along $\sigma^{\pm}$ and wrap $\mathbb{R}^2_{-\hbar}$, denoted D3$^{\pm}_f$.

Following \cite{Ishtiaque:2018str}, let us put the ${\rm D3}^+_b$ brane along the 0237 directions, which imposes
\ie \label{branecon2}
\epsilon_R=i \Gamma_{0237} \epsilon_L \,.
\fe
Thus, we have 
\ie 
i\Gamma_{012345}\epsilon = i\Gamma_{0237}\epsilon \,. 
\fe
Noting that  \eqref{chirality} and \eqref{gunum} imply 
\ie \label{45con}
i\Gamma_{45}\epsilon=\epsilon\,,
\fe
we find via the Clifford algebra that 
\ie 
\epsilon = i\Gamma_{17}\epsilon \,,
\fe
which is one of the relations of \eqref{topq2}. So the ${\rm D3}^+_b$-brane preserves this topological supercharge. 
Next, let us place the D3$_b^-$-brane along the $1236$ directions. In order to preserve supersymmetry, this should actually be an anti-D3-brane.  Let us see why this is the case. The anti-D3-brane will impose the constraint
\ie 
\epsilon_R=-i\Gamma_{1236}\epsilon_L \,,
\fe
which implies 
\ie 
i\Gamma_{012345}\epsilon=-i\Gamma_{1236}\epsilon\,.
\fe
Using the Clifford algebra and \eqref{45con} we arrive at 
\ie 
\epsilon=i\Gamma_{06}\epsilon \,,
\fe
which is one of the relations of \eqref{topq2}. Thus the topological-holomorphic supercharge corresponding to  \eqref{topq2} is preserved in the D5-D3-anti D3-brane configuration.

The $\textrm{D3}_b$-$\textrm{D5}$ system shares a 3d worldvolume with topology $\mathbb{R}\times \mathbb{R}^2_{\hbar}$, where strings stretched between them give rise to a 3d $\mathcal{N}=4$ hypermultiplet, which further localizes to a 1d theory of bosons  \cite{Yagi:2014toa}, with the action 
\ie \label{phdpp}
\frac{1}{\hbar_{\rm 1d}} \int_{\mathbb{R}} \operatorname{Tr}_{\mathbb{C}^N}\left(\varphi \mathrm{~d}^A \tilde{\varphi}\right),
\fe
where $\varphi$ and $\tilde{\varphi}$ are bosonic fields in the fundamental representation of $\U(N)$, since we are considering a single D3$_b$ brane. Note that if we were to consider a stack of $k$ D3$_b$ branes instead, we would obtain bosonic fields in bifundamental representations of $\U(N)\times \U(k)$; in deriving \eqref{phdpp} we have frozen the D3-brane center-of-mass degree of freedom. This is precisely the line operator that we obtain when discretizing free $\beta\gamma$ systems.
The projection to a particular symmetric representation arises due to the gauge field supported by the single D3-brane. Explicitly, the total action takes the form 
\begin{align}
    S_{\text{defect}}&=S_{g{\rm F}1}+S_{\rm D3int}\,,\label{def-action-1}\\
    S_{g{\rm F}1}&=\frac{1}{\hbar_{\rm 1d}}\int_{\mathbb{R}}\Tr_{\mathbb{C}^{N}}\left(\varphi d_{A} \tilde{\varphi}\right)\,,\label{gQM-1}\\
    S_{\rm D3int}&=\frac{1}{\hbar_{\rm 1d}}\int_{\mathbb{R}} \cB  \left( \Tr_{\mathbb{C}^N}\left( \varphi\tilde{\varphi}\right)-l\right)\,,\label{D3int-2}
\end{align}
where the field $\mathcal{B}$ in \eqref{D3int-2} is the gauge field of the D3-brane worldvolume theory, and appears here because the quantum mechanics on $\mathbb{R}$ arises from the hypermultiplet which is coupled to D5- and D3-brane worldvolume theories. The term involving the number $l$ (which determines the associated symmetric representation of $\GL(N,\C)$) arises from a Chern-Simons coupling on the D3-brane worldvolume theory. 

The D3$_f$-D5 strings are described by the dimensional reduction of the D4-D6 I-brane system given by $N$ chiral free fermions \cite{Dijkgraaf:2007sw, Bachas:1997kn, Itzhaki:2005tu} (if we were to consider a stack of $k$ D3$_f$ branes instead, we would obtain bosonic fields in bifundamental representations of $\U(N)\times \U(k)$), with the action 
\ie
\frac{1}{\hbar_{\rm 1d}} \int_{\mathbb{R}} \operatorname{Tr}_{\mathbb{C}^N}\left(\psi \mathrm{~d}^A \tilde{\psi}\right).
\fe
This is the analytic continuation of the line operator that we obtain via discretization of chiral free fermion surface operators. The associated representation of the line operator is a direct sum of anti-symmetric representations, and the projection to a fixed anti-symmetric representation arises via a coupling of the form \eqref{D3int-2} (with bosonic fields replaced by their fermionic counterparts), where $\mathcal{B}$ is the gauge field of the $\textrm{D3}_f$-brane worldvolume theory. 

\subsubsection{Thermodynamic Limit of Multiple Line Defects}

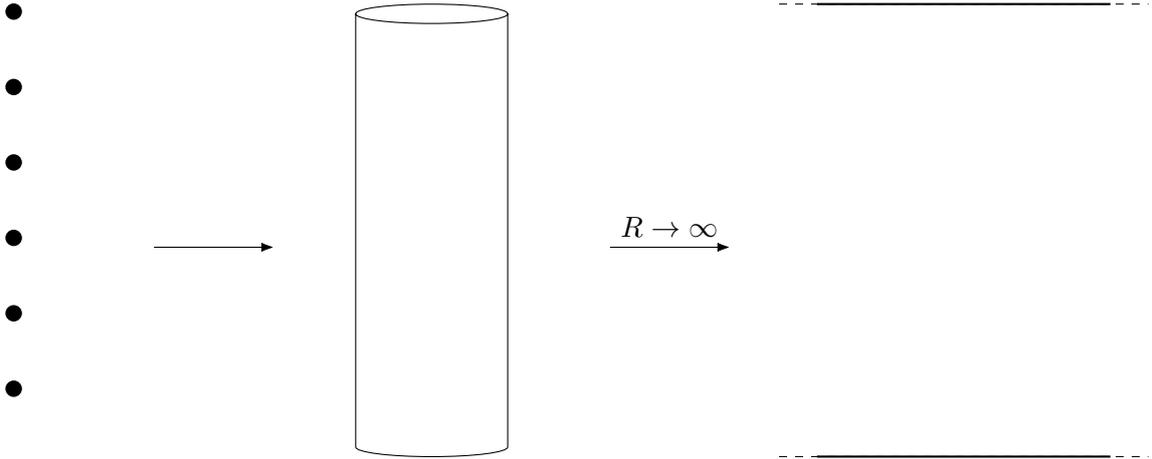
\begin{figure}
    \begin{tikzpicture}[>=latex,shorten >=2pt,shorten <=2pt,shape aspect=1]
    \node[shape=circle,fill=black, scale=0.6] at (1.5,0) {};
    \node[shape=circle,fill=black, scale=0.6] at (1.5,1) {};
    \node[shape=circle,fill=black, scale=0.6] at (1.5,-1) {};
    \node[shape=circle,fill=black, scale=0.6] at (1.5,2) {};
    \node[shape=circle,fill=black, scale=0.6] at (1.5,-2) {};
    \node[shape=circle,fill=black, scale=0.6] at (1.5,3) {};
    \node (A) [cylinder, shape border rotate=90, draw,minimum height=6cm,minimum width=2cm] at (7,0)
{};
    \draw[ ->, >={Latex[round]}, shift=({5pt, 5pt})]  (3.1,-0.3)--(4.8,-0.3) node[text=black,midway,below left] {};
    \draw[ ->, >={Latex[round]}, shift=({5pt, 5pt})]  (9.1,-0.3)--(10.8,-0.3) node[text=black,midway,above] {$R\rightarrow \infty $};       
    \draw[dashed] (11.5,3.1) -- (16.5,3.1);\\
    \draw[dashed] (11.5,-2.9) -- (16.5,-2.9);\\   
    \draw[thick] (12,3.1) -- (16,3.1);\\
    \draw[thick] (12,-2.9) -- (16,-2.9);
\end{tikzpicture}
 \caption{The polarization of a large number of D3-branes (depicted as points) to a D5-brane that realizes the thermodynamic limit of Wilson lines in the 4d CS theory. For example, for $\textrm{D3}_b^+$, the vertical direction is `1' while the horizontal direction is `6'.}
  \label{polarizationpicture}
\end{figure}
In what follows, we shall take the `0' and `1' directions to each be lightcone directions, since we are interested in taking the thermodynamic limit of Wilson lines along such lightcone directions.

Now, if we place a large number of D3$_b$ or D3$_f$ branes at various points along, say, the `1' direction, these branes 
can polarize via the Myers effect \cite{Emparan:1997rt,Myers:1999ps} into a single  D5$_b$ or D5$_f$ brane, wrapping the `1' and `6' directions. 
This is related to how a finite number of fundamental strings polarize to D3-branes realizing Wilson lines in fixed symmetric representations in 4d $\mathcal{N}=4$ SYM \cite{Gomis:2006im,Gomis:2006sb}, which is itself the low-energy worldvolume theory of a stack of D3-branes. Indeed, S-duality of said fundamental strings results in D1-branes, and two T-dualities takes the D1-D3 system to a D3-D5 system, which is precisely our starting point for the realization of line operators in the 4d CS theory. The D3-branes we are interested in can thus be understood to polarize to D5-branes via the Myers effect. 

The version of the Myers effect that is relevant is related to that of \cite{Mateos:2001qs}, where D-branes polarized into supertubes. For example, for $\textrm{D3}_b^+$, if modify the `6' direction, which is the fiber direction of the cotangent bundle $T^*\Sigma$ (with the topology of $\mathbb{R}$), to be a circle with radius $R$, a large number of D3-branes placed at points along the `1' direction can polarize into a D5-brane with supertube topology. The radius of the supertube depends on the strength of the magnetic field generated by the D3-branes, and considering an infinite number of them (since we are interested in the thermodynamic limit) allows for polarization into a D5-brane where $R\rightarrow \infty$, as in Figure \ref{polarizationpicture}.

The resulting configuration is shown in the following table:
\begin{equation} 
\parbox{0.89\linewidth}{
\begin{tikzpicture}[overlay]
\draw [decorate,decoration={brace,amplitude=4pt},xshift=36pt,yshift=0pt]
(2.65,-0.2) -- (3.79,-0.2) node [black,midway,yshift=9pt] 
{\footnotesize $\Sigma$};
\draw [decorate,decoration={brace,amplitude=4pt},xshift=36pt,yshift=0pt]
(4.05,-0.2) -- (5.19,-0.2) node [black,midway,yshift=9pt] 
{\footnotesize $\mathbb{R}_{\hbar}^2$};
\draw [decorate,decoration={brace,amplitude=4pt},xshift=36pt,yshift=0pt]
(5.52,-0.2) -- (6.52,-0.2) node [black,midway,yshift=9pt] 
{\footnotesize $C$};
\draw [decorate,decoration={brace,amplitude=4pt},xshift=36pt,yshift=0pt]
(7.03,-0.2) -- (8.04,-0.2) node [black,midway,yshift=9pt] 
{\footnotesize $N\Sigma\!\!\subset \!\!T^\ast \Sigma$};
\draw [decorate,decoration={brace,amplitude=4pt},xshift=36pt,yshift=0pt]
(8.33,-0.2) -- (9.34,-0.2) node [black,midway,yshift=9pt] 
{\footnotesize $\mathbb{R}_{-\hbar}^2$};
\end{tikzpicture}
\begin{center}
\begin{tabular}{@{}lllllllllll@{}}
\toprule
  & \textbf{0} & \textbf{1} & \textbf{2} & \textbf{3} & \textbf{4} & \textbf{5} & \textbf{6} & \textbf{7} & \textbf{8} & \textbf{9} \\
   \textbf{D5} & $\times$ & $\times$ & $\times$& $\times$ & $\times$ & $\times$ & & & &\\
    $\textbf{D5}^+_b$ & $\times$  & $\times$ & $\times$ & $\times$  &  &  & $\times$ & $\times$ & &\\
        $\textbf{D5}^-_b$ & $\times$  & $\times$ & $\times$ & $\times$  &  &  & $\times$ & $\times$ & &\\
   $\textbf{D5}^+_f$ &  $\times$ & $\times$ &  &  &  &  & $\times$  & $\times$ & $\times$ & $\times$ \\
   $\textbf{D5}^-_f$ &  $\times$ & $\times$ &  &  &  &  & $\times$  & $\times$ & $\times$ & $\times$ \\
\bottomrule
 \end{tabular}
\end{center}
}
\end{equation}

Let us try to understand the ${\rm D5}$-${\rm D5}^+_f$ system. 
The ${\rm D5}^+_f$ constraint is 
\ie 
\epsilon_R=i\Gamma_{016789}\epsilon_L \,.
\fe
Combining it with the constraint coming from the main D5-brane, we find
\ie 
i\Gamma_{012345}\epsilon=i\Gamma_{016789}\epsilon \,.
\fe
Using 
\ie 
i\Gamma_{0123456789}\epsilon=\epsilon \,,
\fe
we arrive at 
\ie 
i\Gamma_{012345}\epsilon=\Gamma_{2345}\epsilon \,,
\fe
or 
\ie \label{chi1}
i\Gamma_{01}\epsilon=\epsilon \,.
\fe
Likewise, for the anti ${\rm D5}_f$-brane denoted ${\rm D5}_f^-$ (obtained via polarization of the anti ${\rm D3}_f$-brane, ${\rm D3}_f^-$), we will get
\ie \label{chi2}
-i\Gamma_{01}\epsilon=\epsilon \,.
\fe
The conditions \eqref{chi1} and \eqref{chi2} indicate that the preserved supercharges at the two intersections of the ${\rm D5}$-${\rm D5}^+_f$ and ${\rm D5}$-${\rm D5}^-_f$ systems have opposite charges with regard to rotations along the $01$-plane.\footnote{Essentially the same D-brane setup involving a D5-brane intersecting at different hyperplanes with a D5-brane and an anti-D5-brane was considered in \cite{Antonyan:2006pg}, where the effective description of the system as a Gross-Neveu model was also derived.  } Moreover, using the supersymmetry algebra, these conditions can be used to show that at the intersections, the theories are either holomorphic or anti-holomorphic, i.e., either the $\partial_{\wbar}$ or $\partial_w$ operator is trivial in $Q$-cohomology. This is analogous to how the condition \eqref{45con} can be used to show that the twist of the D5-brane worldvolume theory that localizes it to the 4d CS theory has anti-holomorphic dependence (on $C$) that is trivial with regard to the cohomology of the supercharge used in the twist \cite{Costello:2018txb}. Notably, the conditions \eqref{chi1} and \eqref{chi2} are not compatible with the supercharge that respects topological invariance along $\Sigma$ and that results in the 4d CS theory. However, this is expected since the surface operators have holomorphic or antiholomorphic dependence on $\Sigma$.

 We can perform an analogous analysis for the bosonic case using the constraint coming from $\Omega$-deformation. $\Omega$-deformation requires topological twisting in the 23-direction, and thus requires
 \ie \label{omegadc}
(\Gamma_{23}+\Gamma_{89})\epsilon=0 \,.
 \fe
 The D5$^+_b$-brane constraint is 
 \ie 
\epsilon_R=i\Gamma_{012367}\epsilon_L \,.
 \fe
 Combining it with the constraint coming from the main D5-brane, we obtain 
 \ie 
i\Gamma_{012345}\epsilon=i\Gamma_{012367}\epsilon \,.
 \fe
Using 
\ie 
i\Gamma_{0123456789}\epsilon=\epsilon \,,
\fe
we arrive at 
\ie 
i\Gamma_{012345}\epsilon=\Gamma_{4589}\epsilon \,.
\fe
Using \eqref{omegadc} we find 
\ie 
i\Gamma_{012345}\epsilon=-\Gamma_{4523}\epsilon \,,
\fe
which is equivalent to 
\ie 
i\Gamma_{01}\epsilon=-\epsilon \,.
\fe
Likewise, an analogous analysis involving the anti-D5$_b$-brane, D5$_b^-$, will give
\ie 
i\Gamma_{01}\epsilon=\epsilon \,.
\fe

Here, the D5$_b$-D5 system shares a four-dimensional worldvolume with topology $\mathbb{C}\times \mathbb{R}^2_{\hbar}$, where strings stretched between them give rise to a 4d $\mathcal{N}=2$ hypermultiplet subject to the $\Omega$-deformation of Kapustin's topological-holomorphic twist \cite{Kapustin:2006hi}, which further localizes to the 2d theory of bosons with the form \cite{Oh:2019bgz}, which can be identified with a free $\beta\gamma$ surface operator in the 4d CS theory:
\ie
\int_{\Sigma} \operatorname{Tr}_{\mathbb{C}^N}\left(\varphi \mathrm{~\partial}_{\wbar}^A \tilde{\varphi}\right).
\fe 
The $\wbar$ derivative is a result of the dependence of the twist on the holomorphic structure of $\Sigma$.

The D5$_f$-D5 brane system is T-dual to
the D4-D6 brane system that can be described by a surface operator supporting $N$ chiral free fermions \cite{Hung:2006nn,Dijkgraaf:2007sw, Green:1996dd}, since it is related via multiple T-dualities to the D1-D9 brane system, and thus the strings stretched between the D5$_f$ and D5 branes give rise to a 2d theory of chiral fermions with the action
\ie
\int_{\Sigma} \operatorname{Tr}_{\mathbb{C}^N}\left(\psi \mathrm{~\partial}_{\wbar}^A \tilde{\psi}\right).
\fe 
This is a free-fermion surface operator in the 4d CS theory. 

\subsection{Irreducible Finite-Dimensional Representations}\label{sec:string_finite}

In this subsection, we shall furnish a string-theoretic interpretation of the thermodynamic limit for Wilson lines in the 4d CS theory associated with fixed, finite-dimensional irreducible representations of the gauge group. We shall focus mostly on Wilson lines realized by quantum mechanics with bosonic degrees of freedom, as a similar realization is expected for the case with fermionic degrees of freedom. 
This boson quantum mechanics can be equivalently described in terms of coadjoint orbit line defects or free $\beta\gamma$ systems with constraints; this equivalence is shown in Appendix \ref{sec:coadjointbeta}. 

We shall start with a brane configuration of Wilson lines for finite representations of $\mathfrak{gl}(N)$ in the 4d CS theory which is analogous to the construction of Wilson lines in 4d $\mathcal{N}=4$ SYM by Gomis and Passerini \cite{Gomis:2006sb,Gomis:2006im}, and was also recently employed in the context of 4d supergroup CS in \cite{Ishtiaque:2021jan}.
The construction involves a stack of multiple D3-branes ending on the stack of D5-branes realizing the 4d CS theory, instead of just a single D3-brane. Each of the D3-branes are understood to arise from polarization of a fixed number of fundamental strings, such that each D3-brane has a fixed amount of fundamental string charge dissolved in it. 

As shown in the table below, the brane configuration consists of the $K$ D3-branes D3$^{l_{\alpha}}_\alpha\,(1\leq \alpha \leq K )$, where $l_{\alpha}$ indicates the amount of fundamental string charge dissolved in the $\alpha$-th D3-brane, and the $N$ D5-branes D5$_i\,(1\leq i\leq N)$ that share the 3d spacetime $\mathbb{R}\times \mathbb{R}^2_{+\hbar}$:

\begin{equation} 
\parbox{0.89\linewidth}{
\begin{tikzpicture}[overlay]
\draw [decorate,decoration={brace,amplitude=4pt},xshift=36pt,yshift=0pt]
(2.65,-0.2) -- (3.79,-0.2) node [black,midway,yshift=9pt] 
{\footnotesize $\Sigma$};
\draw [decorate,decoration={brace,amplitude=4pt},xshift=36pt,yshift=0pt]
(4.05,-0.2) -- (5.19,-0.2) node [black,midway,yshift=9pt] 
{\footnotesize $\mathbb{R}_{\hbar}^2$};
\draw [decorate,decoration={brace,amplitude=4pt},xshift=36pt,yshift=0pt]
(5.52,-0.2) -- (6.52,-0.2) node [black,midway,yshift=9pt] 
{\footnotesize $C$};
\draw [decorate,decoration={brace,amplitude=4pt},xshift=36pt,yshift=0pt]
(7.03,-0.2) -- (8.04,-0.2) node [black,midway,yshift=9pt] 
{\footnotesize $N\Sigma\!\!\subset \!\!T^\ast \Sigma$};
\draw [decorate,decoration={brace,amplitude=4pt},xshift=36pt,yshift=0pt]
(8.33,-0.2) -- (9.34,-0.2) node [black,midway,yshift=9pt] 
{\footnotesize $\mathbb{R}_{-\hbar}^2$};
\end{tikzpicture}
\begin{center}
\begin{tabular}{@{}lllllllllll@{}}
\toprule
  & \textbf{0} & \textbf{1} & \textbf{2} & \textbf{3} & \textbf{4} & \textbf{5} & \textbf{6} & \textbf{7} & \textbf{8} & \textbf{9} \\
   \textbf{D5}$_i$ & $\times$ & $\times$  & $\times$  & $\times$  &$\times$  &$\times$  & & & &\\
    $\textbf{D3}^{l_{\alpha}}_{\alpha}$ & $\times$   & & $\times$  & $\times$  &  &  &  $\times$  &  & &\\
\bottomrule
 \end{tabular}
\end{center}
}
\end{equation}

Before topological twisting and turning on an $\Omega$-deformation, light excitations of fundamental strings stretched between D3- and D5-branes produce a 3d $\cN=4$ hypermultiplet in the bifundamental representation of $\U(K)\times \U(N)$ \cite{DeWolfe:2001pq}.\footnote{The $K$ coincident D3-branes give rise to 4d $\cN=4$ $\U(K)$ SYM, and these end on the stack of $N$ D5-branes, with a 3d $\cN=4$ hypermultiplet supported along the intersection of the two stacks.} In order to give rise to the bifundamental hypermultiplet, each of the D3-branes ought to intersect all the D5-branes, as described, for example, in Section 3.4 of 
\cite{Gaiotto:2008sa}. In this configuration, a D3-brane segment between two D5-branes can break and move independently of other segments, such that the value of a component of the bifundamental hypermultiplet corresponds to the position of a D3-brane that connects two D5-branes.
After twisting and turning on $\Omega$-deformation, the hypermultiplet localizes to the quantum mechanical system on $\mathbb{R}$ with the kinetic term \cite{Yagi:2014toa}
\begin{align}
    S_{{\rm F}1}=\frac{1}{\hbar_{\rm 1d}}\sum_{\alpha=1}^{K}\int_{\mathbb{R}} \Tr_{\mathbb{C}^{N}}\left(\varphi_{\alpha}d \tilde{\varphi}^{\alpha}\right)\,,
\end{align}
where the scalar fields $\tilde{\varphi}^{\alpha}_i$ and $\varphi_{\alpha}^i$ transform as vectors for $\mathbb{C}^N\otimes (\mathbb{C}^K)^*$ and $(\mathbb{C}^N)^*\otimes \mathbb{C}^K$ under $\GL(N)\times \GL(K)$\,, respectively.

Including the coupling to the gauge fields of the D3- and D5-branes, the action of the resulting line defect is given by 
\begin{align}
    S_{\text{defect}}&=S_{g{\rm F}1}+S_{\rm D3int}\,,\label{def-action-2}\\
    S_{g{\rm F}1}&=\frac{1}{\hbar_{\rm 1d}}\sum_{\alpha=1}^{K}\int_{\mathbb{R}}\Tr_{\mathbb{C}^{N}}\left(\varphi_{\alpha}d_{A} \tilde{\varphi}^{\alpha}\right)\,,\label{gQM-2}\\
    S_{\rm D3int}&=\frac{1}{\hbar_{\rm 1d}}\sum_{\alpha\,,\beta\,,\gamma\,,\delta=1}^{K}\int_{\mathbb{R}} \left(\cB^{\alpha \beta}\Tr_{\mathbb{C}^N}\left( \varphi_{\gamma}\rho(E_{\alpha \beta})^{\gamma}{}_{\delta}\tilde{\varphi}^{\delta}\right)-\cB^{\alpha\alpha}l_{\alpha}\right)\,,\label{D3int}
\end{align}
where $\rho(E_{\alpha \beta})\,(\alpha,\beta=1,\dots, K)$ are the fundamental representation of the $\mathfrak{gl}(K)$ generators $E_{\alpha\beta}$ satisfying the commutation relations 
\begin{align}
    [E_{\alpha\beta},E_{\gamma\delta}]=\delta_{\beta\gamma}E_{\alpha\delta}-\delta_{\delta\alpha}E_{\gamma\beta}\,.
\end{align}
The field $\mathcal{B}$ in \eqref{D3int} is the gauge field of the D3-brane worldvolume theory, and appears here because the quantum mechanics on $\mathbb{R}$ arises from the hypermultiplet which is coupled to D5- and D3-brane worldvolume theories.\footnote{The field $\mathcal{B}$ is a complex gauge field which takes values in the Borel subalgebra of the complexified gauge group $G_{\mathbb{C}}$;
for generic FI parameters $l_{\alpha}$ the gauge symmetry $G$ is broken to the Cartan subalgebra, and the complex counterpart of this is that the complex gauge symmetry takes values in the Borel subalgebra.}

The polarization process described in the previous subsection also takes place when considering a large number of stacks of D3-branes in the configuration that we are presently studying. To be precise, we can separate the D3-branes in the initial configuration along the `8' or `9' direction, consider a large number of copies of such a configuration, whereby the polarization process results in a D5-brane at the location of each D3-brane in the configuration prior to polarization, and then bringing together the resulting D5-branes to form a stack again. The polarization process results in a surface operator associated with a fixed, irreducible representation of $\GL(N,\C)$, which can be described either via a coadjoint orbit surface defect or a free $\beta \gamma$ system with constraints. The brane configuration that results from the polarization process is  :

\begin{equation} 
\parbox{0.89\linewidth}{
\begin{tikzpicture}[overlay]
\draw [decorate,decoration={brace,amplitude=4pt},xshift=36pt,yshift=0pt]
(2.65,-0.2) -- (3.79,-0.2) node [black,midway,yshift=9pt] 
{\footnotesize $\Sigma$};
\draw [decorate,decoration={brace,amplitude=4pt},xshift=36pt,yshift=0pt]
(4.05,-0.2) -- (5.19,-0.2) node [black,midway,yshift=9pt] 
{\footnotesize $\mathbb{R}_{\hbar}^2$};
\draw [decorate,decoration={brace,amplitude=4pt},xshift=36pt,yshift=0pt]
(5.52,-0.2) -- (6.52,-0.2) node [black,midway,yshift=9pt] 
{\footnotesize $C$};
\draw [decorate,decoration={brace,amplitude=4pt},xshift=36pt,yshift=0pt]
(7.03,-0.2) -- (8.04,-0.2) node [black,midway,yshift=9pt] 
{\footnotesize $N\Sigma\!\!\subset \!\!T^\ast \Sigma$};
\draw [decorate,decoration={brace,amplitude=4pt},xshift=36pt,yshift=0pt]
(8.33,-0.2) -- (9.34,-0.2) node [black,midway,yshift=9pt] 
{\footnotesize $\mathbb{R}_{-\hbar}^2$};
\end{tikzpicture}
\begin{center}
\begin{tabular}{@{}lllllllllll@{}}
\toprule
  & \textbf{0} & \textbf{1} & \textbf{2} & \textbf{3} & \textbf{4} & \textbf{5} & \textbf{6} & \textbf{7} & \textbf{8} & \textbf{9} \\
   \textbf{D5}$_i$ & $\times$ & $\times$  & $\times$  & $\times$  &$\times$  &$\times$  & & & &\\
    $\textbf{D5}^{l_{\alpha}}_{\alpha}$ & $\times$   & $\times$ & $\times$  & $\times$  &  &  &  $\times$  & $\times$ & &\\
\bottomrule
 \end{tabular}
\end{center}
}
\end{equation}

Analogous analysis holds for the thermodynamic limit of fermion quantum mechanics realizing irreducible finite-dimensional representations, with the resulting defect being realized by the following configuration : 
\begin{equation} 
\parbox{0.89\linewidth}{
\begin{tikzpicture}[overlay]
\draw [decorate,decoration={brace,amplitude=4pt},xshift=36pt,yshift=0pt]
(2.65,-0.2) -- (3.79,-0.2) node [black,midway,yshift=9pt] 
{\footnotesize $\Sigma$};
\draw [decorate,decoration={brace,amplitude=4pt},xshift=36pt,yshift=0pt]
(4.05,-0.2) -- (5.19,-0.2) node [black,midway,yshift=9pt] 
{\footnotesize $\mathbb{R}_{\hbar}^2$};
\draw [decorate,decoration={brace,amplitude=4pt},xshift=36pt,yshift=0pt]
(5.52,-0.2) -- (6.52,-0.2) node [black,midway,yshift=9pt] 
{\footnotesize $C$};
\draw [decorate,decoration={brace,amplitude=4pt},xshift=36pt,yshift=0pt]
(7.03,-0.2) -- (8.04,-0.2) node [black,midway,yshift=9pt] 
{\footnotesize $N\Sigma\!\!\subset \!\!T^\ast \Sigma$};
\draw [decorate,decoration={brace,amplitude=4pt},xshift=36pt,yshift=0pt]
(8.33,-0.2) -- (9.34,-0.2) node [black,midway,yshift=9pt] 
{\footnotesize $\mathbb{R}_{-\hbar}^2$};
\end{tikzpicture}
\begin{center}
\begin{tabular}{@{}lllllllllll@{}}
\toprule
  & \textbf{0} & \textbf{1} & \textbf{2} & \textbf{3} & \textbf{4} & \textbf{5} & \textbf{6} & \textbf{7} & \textbf{8} & \textbf{9} \\
   \textbf{D5}$_i$ & $\times$ & $\times$  & $\times$  & $\times$  &$\times$  &$\times$  & & & &\\
    $\textbf{D5}^{k_{\alpha}}_{\alpha}$ & $\times$   & &  &  &  &  &  $\times$  & &$\times$  & $\times$ \\
\bottomrule
 \end{tabular}
\end{center}
}\label{brane-F-f-2}
\end{equation}

\subsection{Infinite-Dimensional Representations}\label{sec:string_infinite}

In this subsection, we describe how the thermodynamic limit of line operators of the 4d CS theory in infinite-dimensional representations of the gauge group can be realized in string theory. These line operators were previously obtained via discretization of curved $\beta \gamma$ surface operators in Section \ref{sec:curved-bg}.

To realize these line operators, we use the D3-D5-NS5 system studied in \cite{Ishtiaque:2021jan,Ishtiaque:2022bck}, where one considers the \textit{same} number of D3, D5 and NS5-branes, in a configuration whereby each D3-brane ends on a single D5-brane and a single NS5-brane, which ensures that the s-rule is satisfied. 

\begin{figure}
    \centering
    \includegraphics[width=0.5\linewidth]{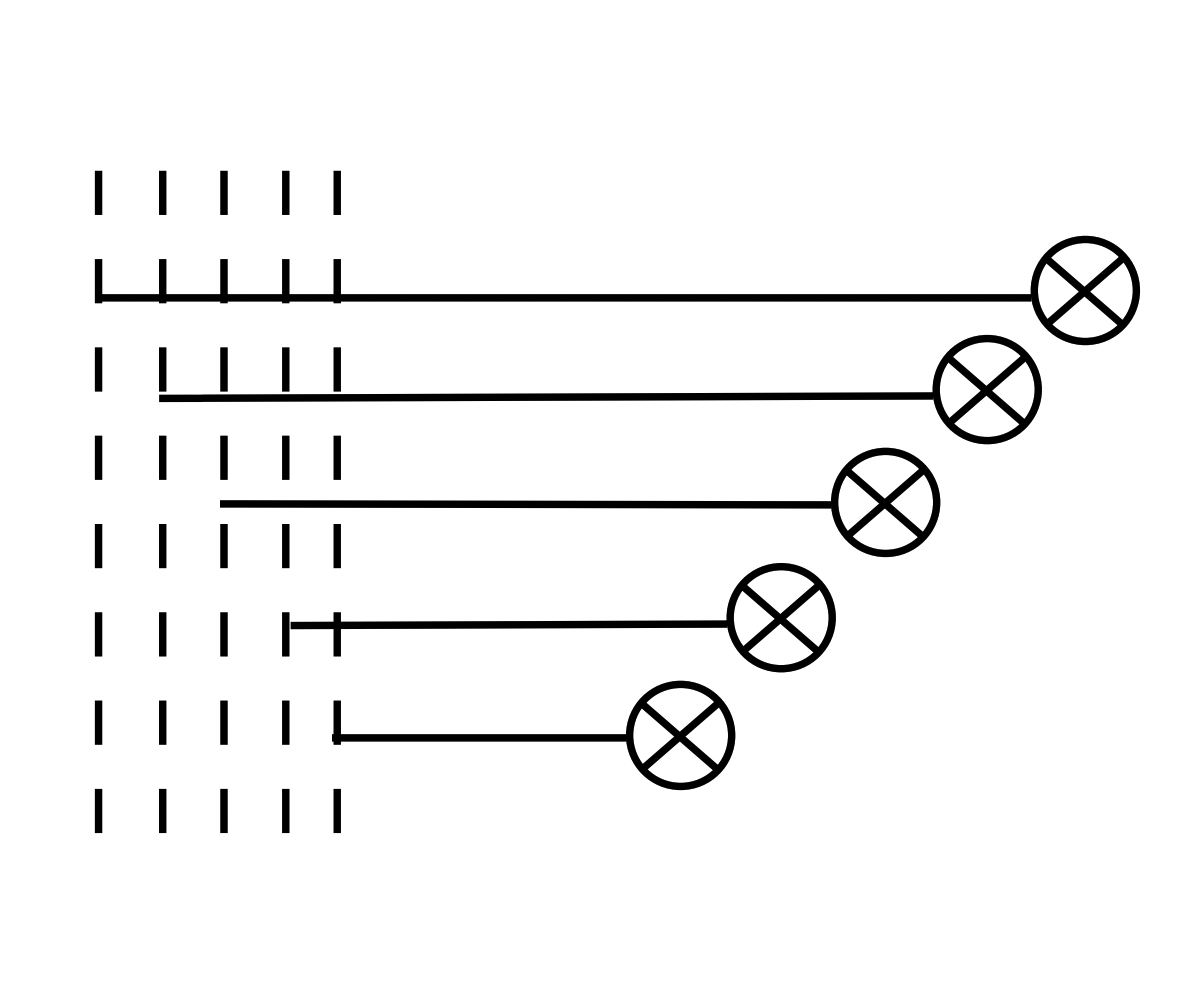}
    \caption{Configuration of D3-branes ending on D5-branes and NS5-branes, where the latter are depicted by crossed circles.}
    \label{D3D5NS5}
\end{figure}

The D3-D5-NS5  configuration is depicted in Figure \ref{D3D5NS5}. The support of each brane can be understood using the following table:

\begin{equation} 
\parbox{0.89\linewidth}{
\begin{tikzpicture}[overlay]
\draw [decorate,decoration={brace,amplitude=4pt},xshift=36pt,yshift=0pt]
(2.65,-0.2) -- (3.79,-0.2) node [black,midway,yshift=9pt] 
{\footnotesize $\Sigma$};
\draw [decorate,decoration={brace,amplitude=4pt},xshift=36pt,yshift=0pt]
(4.05,-0.2) -- (5.19,-0.2) node [black,midway,yshift=9pt] 
{\footnotesize $\mathbb{R}_{\hbar}^2$};
\draw [decorate,decoration={brace,amplitude=4pt},xshift=36pt,yshift=0pt]
(5.52,-0.2) -- (6.52,-0.2) node [black,midway,yshift=9pt] 
{\footnotesize $C$};
\draw [decorate,decoration={brace,amplitude=4pt},xshift=36pt,yshift=0pt]
(7.03,-0.2) -- (8.04,-0.2) node [black,midway,yshift=9pt] 
{\footnotesize $N\Sigma\!\!\subset \!\!T^\ast \Sigma$};
\draw [decorate,decoration={brace,amplitude=4pt},xshift=36pt,yshift=0pt]
(8.33,-0.2) -- (9.34,-0.2) node [black,midway,yshift=9pt] 
{\footnotesize $\mathbb{R}_{-\hbar}^2$};
\end{tikzpicture}
\begin{center}
\begin{tabular}{@{}lllllllllll@{}}
\toprule
    & \textbf{0} & \textbf{1} & \textbf{2} & \textbf{3} & \textbf{4} & \textbf{5} & \textbf{6} & \textbf{7} & \textbf{8} & \textbf{9} \\
    \textbf{D5} & $\times$ & $\times$  & $\times$  & $\times$  &$\times$  &$\times$  & $A$ & & &\\
    $\textbf{D3}$ & $\times$   & & $\times$  & $\times$  &  &  &  $\times$  &  & &\\
    $\textbf{NS5}$ &  $\times$  & & $\times$ & $\times$  &  &  &  $B$  &    $\times$ & $\times$  &$\times$  \\
    \bottomrule
 \end{tabular}
\end{center}
}
\end{equation}
where the support of the D3-brane is on a finite-sized interval along the `6' direction. To elucidate this, we have indicated the positions of the D5- and NS5-branes along the `6' direction as $A$ and $B$ respectively. 

One approach to understanding the line operator obtained via this construction is by noting that at low energy, the D3-brane worldvolume theory can be described by a 3d $\mathcal{N}=4$ supersymmetric gauge theory supported along the `0', `2' and `3' directions. At low energy, the theory can be described in terms of a 3d sigma model on the Higgs branch, and in a B-type $\Omega$-background, this localizes to topological quantum mechanics on the Higgs branch \cite{Yagi:2014toa}.  

\begin{figure}
    \centering
    \includegraphics[width=80mm]{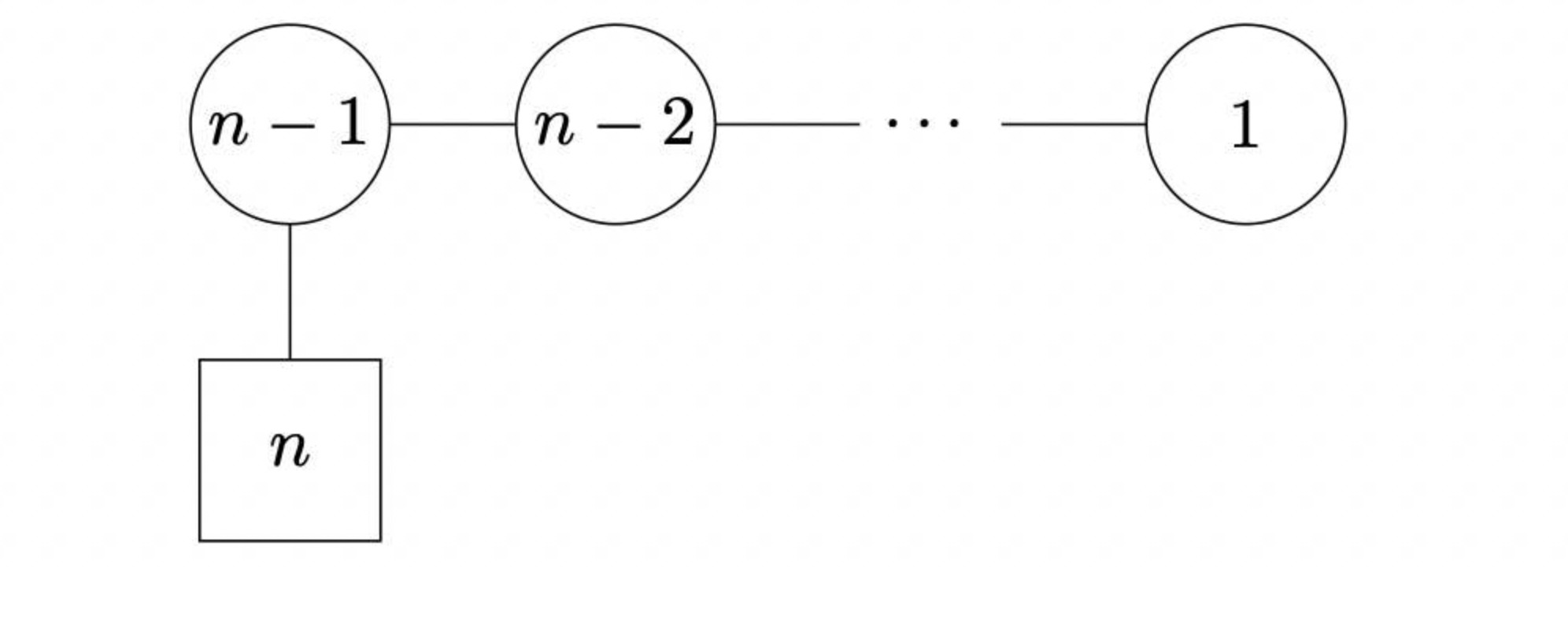}
    \caption{Quiver diagram associated with the 3d $\mathcal{N}=4$ theory $T[\U(N)]$.}
    \label{quiver}
\end{figure}

The D3-D5-NS5 system of interest is associated with the theory $T[\U(N)]$ and the quiver diagram shown in Figure \ref{quiver}.  The supersymmetric vacua that give rise to its Higgs branch can be described in terms of the nilpotent cone, $\mathcal{N}$, of $G_{\mathbb{C}}=\GL(N, \mathbb{C})$, which admits the Springer resolution (achieved by turning on certain FI parameters, as described on page 103 of \cite{Gaiotto:2008ak}) in terms of $T^*(G_{\mathbb{C}}/B)$, where $B$ is the  Borel subgroup of $\GL(N, \mathbb{C})$. 

Thus, the line operator can be described in terms topological quantum mechanics on $T^*(G_{\mathbb{C}}/B)$, that is, the discretization of a curved $\beta \gamma$ system with target space the flag manifold $G_{\mathbb{C}}/B$. 
The aforementioned FI parameters are related to the K\"ahler parameters of the Higgs branch, and are therefore related to the highest weights of the Verma modules that arise from quantization of the line operators of interest, as described towards the end of Section \ref{sec:DO_Zhu}.\footnote{The relationship between FI parameters and highest weights of Verma modules was also described in \cite{Ishtiaque:2021jan}.}

For our application, the crucial point is that considering a large number of stacks of D3-branes that were used to realize line operators associated to Verma modules results in polarization of these branes into D5-branes in the following brane configuration, which are expected to realize the curved $\beta \gamma$ surface operators : 

\begin{equation} 
\parbox{0.89\linewidth}{
\begin{tikzpicture}[overlay]
\draw [decorate,decoration={brace,amplitude=4pt},xshift=36pt,yshift=0pt]
(2.65,-0.2) -- (3.79,-0.2) node [black,midway,yshift=9pt] 
{\footnotesize $\Sigma$};
\draw [decorate,decoration={brace,amplitude=4pt},xshift=36pt,yshift=0pt]
(4.05,-0.2) -- (5.19,-0.2) node [black,midway,yshift=9pt] 
{\footnotesize $\mathbb{R}_{\hbar}^2$};
\draw [decorate,decoration={brace,amplitude=4pt},xshift=36pt,yshift=0pt]
(5.52,-0.2) -- (6.52,-0.2) node [black,midway,yshift=9pt] 
{\footnotesize $C$};
\draw [decorate,decoration={brace,amplitude=4pt},xshift=36pt,yshift=0pt]
(7.03,-0.2) -- (8.04,-0.2) node [black,midway,yshift=9pt] 
{\footnotesize $N\Sigma\!\subset \!T^\ast \Sigma$};
\draw [decorate,decoration={brace,amplitude=4pt},xshift=36pt,yshift=0pt]
(8.33,-0.2) -- (9.34,-0.2) node [black,midway,yshift=9pt] 
{\footnotesize $\mathbb{R}_{-\hbar}^2$};
\end{tikzpicture}
\begin{center}
\begin{tabular}{@{}lllllllllll@{}}
\toprule
  & \textbf{0} & \textbf{1} & \textbf{2} & \textbf{3} & \textbf{4} & \textbf{5} & \textbf{6} & \textbf{7} & \textbf{8} & \textbf{9} \\
   \textbf{D5} & $\times$ & $\times$  & $\times$  & $\times$  &$\times$  &$\times$  & A & & &\\
    $\textbf{D5}$ & $\times$   & $\times$ & $\times$  & $\times$  &   &  &  $\times$  & $\times$  & &\\
   $\textbf{NS5}$ &  $\times$  & & $\times$ & $\times$  &  &  &   B &    $\times$ & $\times$  &$\times$  \\
\bottomrule
 \end{tabular}
\end{center}
}
\end{equation}
The intersection of the two stacks of D5-branes is expected to be described by a 4d $\mathcal{N}=2$ theory, that localizes to the path integral description of these surface operators. Indeed, identifying the `7' direction with a circle of large radius, T-duality along this direction leads us to a system of D4-branes suspended between D6-branes and NS5-branes, which at low energy should be described by a 4d $\mathcal{N}=2$ theory that is related to the 3d $\mathcal{N}=4$ sigma model on $T^*(G_{\mathbb{C}}/B)$ via dimensional reduction.

Moreover, given that the discretization of these surface operators can be described in terms of Zhu algebras of affine Kac-Moody algebras as discussed in Section \ref{bg-anomaly}, the string theory interpretation of this discretization thus suggests a geometric interpretation of the Zhu algebra. This is because the polarization process naturally associates representations of the vertex algebras arising from the curved $\beta \gamma$ surface operators with the Higgs branch of the low energy effective 4d $\mathcal{N}=2$ theory described in the previous paragraph.\footnote{See \cite{Dedushenko:2019mzv} for closely related analyses of the Schur sector of 4d $\mathcal{N}$=2 theories.} 

\section{Discussion}
  \label{sec:discussion}

There are several future directions that can be pursued based on the results in this work. 

 In this paper we concentrated on the discretization of order defects. We leave it as a future problem to consider discretization of disorder defects. The natural expectation is that the disorder defects are discretized into 't Hooft loops.

In this paper we have only considered the discretization of integrable field theories associated with rational, trigonometric and elliptic lattice models. From the 4d CS perspective we may also couple order surface operators to the 4d CS holomorphic surface, $C$, of genus greater than one. The analysis of such discretizations will require the analysis of disorder defects.

When we consider a general integrable field theory known as the non-ultralocal type, the Poisson brackets of the spatial component of these Lax connections in general contain derivatives of the delta function \cite{Maillet:1985ek}.
This makes it a challenge to quantize the theory.
We expect non-ultralocal field theories to be realized via disorder surface operators \cite{Costello:2019tri,Vicedo:2019dej}. 
In Section \ref{sec:anomaly} we have seen that even when we start with classical order defects, the gauge anomaly cancellation of the chiral (or anti-chiral) defects requires the one-loop corrections
to modify the meromorphic one-form and creates  disorder defects. This suggests an interesting interplay between classical and quantum effects on the one hand, and order and disorder defects, on the other. 

In particular, the coupled WZW models obtained through bosonization in Section \ref{sec:duality-boson} are non-ultralocal and can be realized via disorder operators in the 4d CS theory, and we essentially derived a duality between a classically non-ultralocal field theory and a classically ultralocal one where non-ultralocality emerges at the quantum level. The latter can be discretized when $\hbar$ is small. This is reminiscent of the quantum inverse scattering approach to quantization of \cite{Faddeev:1985qu}, where non-ultralocality is understood as arising from excitations around a ``true" vacuum, which corresponds to the Dirac sea of excitations of a ``false" vacuum (where the excitations are described by the Bethe ansatz equations). 
We leave the further analysis of integrable discretizations of non-ultralocal integrable field theories for future work.

 In this paper, we discussed the discretizations of the physical systems themselves, and have refrained from studying the detailed properties of the theories.
For example, we expect that we can use the methods of this paper to discuss the physical quantities of the theory, such as the energy spectrum.

The type IIB string theory realization of line operators in the 4d CS theory via D5-D3-NS5 brane configurations in general can realize topological quantum mechanics on Cherkis Bow varieties \cite{Ishtiaque:2022bck}, and therefore it is expected that the thermodynamic limit of such line operators give rise to integrable sigma models on Cherkis Bow varieties. 

 The integrable discretizations of quantum field theories as discussed in this paper
have recently been revisited in the context of the integrable Trotterizations (integrable Floquet dynamics) needed for 
quantum simulations, see e.g.\ \cite{PhysRevLett.121.030606,Gritsev:2017zdm,Miao:2022dau,Maruyoshi:2022jnr}.
Some of our technical discussion of e.g., coherence states are also discussed in this context.
We expect our results to be of use when we wish to simulate integrable two-dimensional quantum field theories.

The realization of the 4d CS theory coupled with 2d WZW defects of non-ultralocal 2d integrable systems could be regarded as a straightforward extension of the order defect case. 
Recalling that the lightcone discretization of an ultralocal system is reduced to a simpler problem of discretization of order defects coupled to the 4d CS theory, one might expect that the problem of discretization of a non-ultralocal 2d integrable field theory is also reduced to that of discretization of 2d WZW defects.\footnote{
As discussed in \cite{Alekseev:1991wq,Alekseev:1992wn,Falceto:1992bf,Alekseev:1996jz} (see also Section 4 of \cite{Hollowood:2015dpa}), the construction of a lattice Kac-Moody algebra associated with the WZW model might be helpful in addressing this problem.
}
We leave these important subjects for future work.

\section*{Acknowledgements}

We would like to thank Kevin Costello, Simeon Hellerman, Nafiz Ishtiaque, Shigenori Nakatsuka, Yuji Tachikawa,  Meng-Chwan Tan, Benoit Vicedo, Junya Yagi and Yehao Zhou for helpful discussions. Results from this work were presented in several conferences, e.g.\ by M.A.\ at the Simons Center for Geometry and Physics program ``Geometric and Representation-Theoretic Aspects of Quantum Integrability" (October 2022), and we would like to thank the audience of this talk for valuable feedback. M.Y.\ would like to thank KITP Santa Barbara for hospitality, where part of this work was performed during the Integrable22 program.

This research was supported in part by WPI Research Center Initiative, MEXT, Japan. This research was also supported by the JSPS Grant-in-Aid for Scientific Research No.\ 19H00689,  20H05860, [M.A.\ and M.Y.], 19K03820, 23H01168, 23K17689 [M.Y.]. M.Y.\ is also supported in part by JST, Japan (PRESTO Grant No.\ JPMJPR225A, Moonshot R\&D Grant No.\ JPMJMS2061). This research was supported in part by the National Science Foundation under Grant No.\ NSF PHY-1748958. J.S.\ is supported by an INFN postdoctoral senior level 3 research grant in physics and by the INFN network project SFT.

\begin{appendix}

\section{Conventions and Useful Formulae}\label{convention}

In this appendix, we list the conventions as well as several useful formulae used in the main text.
\begin{itemize}
    \item The measure on $\Sigma$ is defined as \ie 
\int_{\Sigma} d^2\si = \int_{\Sigma} d \tau \wedge d \sigma = -\frac{1}{2}\int_{\Sigma} d\sigma^+ \wedge d \sigma^- .
\fe

\item The Minkowski metric on $\Sigma$ is $\eta = \textrm{diag} (-1,1)$. $\eta^{+-}=-2$  and $\epsilon^{+-}=-2$.

\item The anti-commuting relation of gamma matrices $\gamma^{\mu}\,(\mu=0,1)$ are normalized as 
\begin{align}
    \{\gamma^{\mu},\gamma^{\nu}\}=2\eta^{\mu\nu}\,.\label{anti-com}
\end{align}
The gamma matrices are 
\ie 
\gamma^0=
\left(\begin{array}{cc}
  0   & i \\
   i  & 0
\end{array} \right)\,,\quad 
\gamma^1=
\left(\begin{array}{cc}
  0   & i \\
   -i  & 0
\end{array} \right)\,,\quad
\gamma^5=
\left(\begin{array}{cc}
  1  & 0 \\
   0  & -1
\end{array} \right)\,.
\fe 

\item A Dirac fermion $\Psi$ can be decomposed into eigenstates $\Psi_L\,, \Psi_R$ of $\gamma^5$ with eigenvalues $\pm 1$\,,
\begin{align}
    \Psi_{L}=\frac{1+\gamma^5}{2}\Psi=\frac{1}{\sqrt{2}}\begin{pmatrix}
    \psi_L\\
    0
    \end{pmatrix}\,,\qquad \Psi_{R}=\frac{1-\gamma^5}{2}\Psi=\frac{1}{\sqrt{2}}\begin{pmatrix}
    0\\
    \psi_R
    \end{pmatrix}\,.
\end{align}

\item We often use the Cauchy-Pompeiu integral formula, which amounts in practice to the delta function identities 
\ie \label{deltident}
&\frac{1}{2\pi i}\partial_{\zbar}\frac{1}{z}= -\delta^{(2)}(z,\bar{z})\,,\\
&\frac{1}{2\pi i}\partial_{z}\frac{1}{\zbar}= -\delta^{(2)}(z,\bar{z})\,.
\fe 
\item We define the following WZW actions\,:
\begin{align}
    S_{\rm WZW}[u]&=\frac{1}{8\pi}\int_{\cM} d^2\sigma\,\Tr\left( u^{-1}\partial_{\mu}u\,u^{-1}\partial^{\mu}u\right)+S_{\rm WZ}[\tilde{u}]\,,\label{WZW-action-app}\\
    S_{\rm WZ}[\tilde{u}]&=\frac{1}{12\pi}\int_{0}^{1}ds\int_{\Sigma} d^2\sigma\,\epsilon^{\mu\nu\rho}\,\Tr\left( \tilde{u}^{-1}\partial_{\mu}\tilde{u}\,\tilde{u}^{-1}\partial_{\nu}\tilde{u}\,\tilde{u}^{-1}\partial_{\rho}\tilde{u}\right)\,,
\end{align}
and
\begin{align}
    \tilde{S}_{\rm WZW}[u]&=\frac{1}{8\pi}\int_{\cM} d^2\sigma\,\Tr\left( u^{-1}\partial_{\mu}u\,u^{-1}\partial^{\mu}u\right)-S_{\rm WZ}[\tilde{u}]\,.\label{tWZW-action-app}
\end{align}
Here $\epsilon^{\mu \nu s} =\epsilon^{\mu \nu}$, where $\epsilon^{01}=1$. Stokes' theorem for Wess-Zumino terms is defined with no minus sign incurred, i.e., $\tilde{u}$ is trivial at $s=0$. This convention is the opposite of that of Delduc et al.\ \cite{Delduc:2019whp}. 

\end{itemize}

\section{Framing Anomaly}\label{sec:framing_anomaly}

When we consider Wilson lines that do not cross at right angles, one needs to take into account the framing anomaly \cite{Costello:2017dso}.
The framing anomaly states that when one rotates the direction of the Wilson line by an angle $\theta$ along $\Sigma=\mathbb{C}$, then this is accompanied in the shift of the spectral parameter $z$ by $\hbar\, \sh^{\vee} \theta/\pi$, where $\sh^{\vee}$ is the dual Coxeter number of the Lie algebra $\mathfrak{g}$ for the gauge symmetry.

This anomaly is relevant when we consider crossing of the Wilson lines.
We explained in Section \ref{sec:lattice-cs} that
we obtain the R-matrix $R(z-z')$ when we have two Wilson lines 
located at positions $z, z'$ of the spectral curve.
Strictly speaking, this statement applies only when 
the two Wilson lines are almost parallel \cite{Costello:2017dso},
and we will instead obtain $R(z-z' + \hbar\, \sh^{\vee} \theta/\pi)$
when the two Wilson lines intersect at an angle $\theta$.

When we consider the total monodromy matrix in \eqref{T_A},
the Wilson lines in the horizontal direction (associated with the auxiliary Hilbert space)
intersects the lightcone Wilson lines at angles $\pm \pi/4$. This means that 
the monodromy matrix \eqref{T_A} is corrected to be 
\begin{align}
\begin{split}
    &\cdots
     R_{\rho_A, \rho_{-}}\left(z-\left(\nu + \frac{3\hbar\, \sh^{\vee}}{4} \right)\right)
    R_{\rho_A, \rho_{+}}\left(z+\left(\nu - \frac{\hbar\, \sh^{\vee}}{4} \right)\right) \\
    &\qquad \qquad\times R_{\rho_A, \rho_{-}}\left(z-\left(\nu + \frac{3\hbar\, \sh^{\vee}}{4} \right)\right)
    R_{\rho_A, \rho_{+}}\left(z+\left(\nu - \frac{\hbar\, \sh^{\vee}}{4} \right)\right)
    \cdots.
 \end{split} 
\end{align}
 The framing anomaly thus does not drastically affect the realization of the spin chain with alternating inhomogeneities and, and the shift can be absorbed into the redefinition $z'=z-\hbar\, \sh^{\vee}/2$ and $\nu'=\nu+\hbar\, \sh^{\vee}/4$. 

\section{Faddeev-Reshetikhin Model}\label{sec:frapp}

In this appendix we shall provide a brief review of the Faddeev-Reshetikhin (FR) model (also known as the linear chiral model).\footnote{A notable application of this model is to the study of a truncation of the $AdS_5 \times S^5$ superstring, in particular, its truncation to an $\mathbb{R} \times S^3$ submanifold \cite{Rivelles:2008rq}. } 

The Lagrangian for the FR model is given by
\begin{equation}
S_{\mathrm{FR}}\left[g_{(\pm)}\right]=2\int_{\Sigma}  d^2\sigma \, \operatorname{Tr}\left(\Lambda g_{(+)}^{-1} \partial_{-} g_{(+)}+\Lambda g_{(-)}^{-1} \partial_{+} g_{(-)}-\frac{1}{2 \nu} g_{(+)} \Lambda g_{(+)}^{-1} g_{(-)} \Lambda g_{(-)}^{-1}\right)\,.
\end{equation}
Here, both $g_{(+)}$ and $g_{(-)}$ are $\SU(2)$-valued fields on $\cM$, while $\Lambda$ is the Cartan generator of $\mathfrak{su}(2)$.
The Lax connection for this model takes the form
\ie
\Lax_\pm(z)=\frac1{\nu\mp z}\JJ_\pm\, ,\qquad \mathcal{J}_{(\pm)} := g_{(\pm)} \cdot \Lambda \cdot g_{(\pm)}^{-1}\,,
\label{leq}
\fe
where the equations of motion can be stated as
\ie
\partial_\mp\JJ_\pm=\mp\frac1{2\nu  }[\JJ_+,\JJ_-]\,.
\label{jeq}
\fe
The current $\JJ_{\pm}$ satisfies the following additional constraint
\begin{equation}
\operatorname{Tr}\left[\left(\mathcal{J}_{(\pm)}\right)^{n}\right]: \text { constant } \;. \label{J-con}
\end{equation}
The constraint (\ref{J-con}) is necessary to obtain a consistent dynamical system because the quantity $\operatorname{Tr}\left[\left(\mathcal{J}_{(\pm)}\right)^{n}\right]$ is a central element of the modified Poisson bracket.
Note that for an appropriate choice of $\nu$, the equations of motion (\ref{jeq}) take the same form as those of the principal chiral model.
Note that the Faddeev-Reshetikhin model with $\SU(2)$ symmetry can be straightforwardly extended to the $\SU(n)$ case (for more details, see \cite{Appadu:2017fff}); our discussion in Section \ref{sec:coadjoint} makes is clear how to generalize the model for an arbitrary Lie group $G$.

\section{Wilson Lines as Quantum Mechanics on Coadjoint Orbits}\label{sec:WL-CA}

In this appendix, we review relations between Wilson lines, quantum mechanics on coadjoint orbits, and coherent states.

\subsection{Coadjoint Orbits as Symplectic Manifolds}
Let $G$ be a semisimple Lie group and $\mathfrak{g}$ the corresponding Lie algebra of $G$ with generators $\{t_a\}\,(a=1\,,\dots\,, \text{dim}\,\mathfrak{g})$\,.
We denote $\rho(t_a)$ by a representative matrix of the generator of $\mathfrak{g}$ in an irreducible representation of $\mathfrak{g}$ which is characterized by a highest weight $\Lambda$\,.
Note that the highest weight $\Lambda$ is in the lattice of dominant weights of $\mathfrak{g}^*$ which is the dual space of $\mathfrak{g}$\,.

For a given $\Lambda\in \mathfrak{g}^{*}$\,, we can consider the associated coadjoint orbit
\begin{align}
    \cO_{\Lambda}=\{\text{Ad}^*_{g}(\Lambda)\,\lvert \, g\in G\}\,,
\end{align}
where $\text{Ad}^*_{g}$ is the coadjoint action of $g$ in the dual space $\mathfrak{g}^*$ defined as
\begin{align}
    \text{Ad}^*_{g}(X)=g\cdot X \cdot g^{-1}\,.
\end{align}
The codjoint orbit is diffeomorphic to a homogeneous space $G/H_{\Lambda}$\,, where $H_{\Lambda}$ is an isotropy subgroup 
\begin{align}
    H_{\Lambda} :=\{\text{Ad}^*_{h}(\Lambda)=\Lambda~\lvert~ h\in G\}\,.
\end{align}
It is known that for any Lie group $G$\,, the coadjoint orbit $\cO_{\Lambda}$ is a sympletic manifold equipped with a sympletic 2-form, so called Kirillov-Kostant-Souriau symplectic form
\begin{align}
    \omega_{\Lambda}(K_{\xi}\,, K_{\eta})=-\langle \Lambda\,,[\xi, \eta] \rangle \,,\label{KKS}
\end{align}
where $K_{x}$ is the associated vector field for a given $x \in \mathfrak{g}$\,, and the non degenerated bilinear form $\langle \cdot ,\cdot \rangle : \mathfrak{g}^*\times \mathfrak{g}\to \mathbb{R}$\,. 
The symplectic potential $\vartheta_{\lambda}$ of $\omega_{\Lambda}$ is defined as
\begin{align}\label{omega}
   \omega_{\Lambda}=-d\vartheta_{\Lambda}\,.
\end{align}

For any $X\in\mathfrak{g}$\,, the corresponding vector field $K_{X}$ is called Hamiltonian when there exists a single-valued function $f_{X}$ such that
\begin{align}
    \iota_{K_{X}}\omega_{\Lambda}=df_{X}\,.
\end{align}
Then, a moment map $\mu:\cO_{\Lambda}\to \mathfrak{g}^*$ is defined so that for every $X \in \mathfrak{g}$, the function $\langle  \mu (\,\cdot\,), X\rangle : \cO_{\Lambda} \to \mathbb{R}$ is Hamiltonian i.e.\,,
\begin{align}
    d\langle \mu (x)\,, X\rangle=\iota_{K_{X}}\omega_{\Lambda}\,,\qquad x\in \cO_{\Lambda}\,.\label{H-mu}
\end{align}
From the definition (\ref{KKS}) of $\omega_{\Lambda}$\,, the moment map $\mu$ needs to satisfy $\mu(\Lambda)=\Lambda$\,.
We define
\begin{align}
    \mu_a:=\langle \mu\,, K_a \rangle \label{mua}
\end{align}
by using the Killing vector fields $K_a$ satisfying the commutation relations of $\mathfrak{g}$
\begin{align}
    [K_a,K_b]=f_{ab}{}^{c}K_c\,.
\end{align}
The functions $\mu_a$ on $\cO_{\Lambda}$ can realize a (classical) representation of $\mathfrak{g}$ as functions over the coadjoint orbit $\cO_{\Lambda}$ with Poisson bracket
\begin{align}
    \iota_{K_a}d\mu_b=\{\mu_a,\mu_b\}=f_{ab}{}^{c}\mu_c\,,
\end{align}
where we used (\ref{KKS}) and $\iota_{K_a}\iota_{K_b}\omega_{\Lambda}=-\omega_{\Lambda}(K_a,K_b)$\,.

\subsection{Wilson Lines from the Geometric Quantization of Coadjoint Orbits}

We shall next review how geometric quantization of the action of a particle on the coadjoint orbit $G/H_{\Lambda}$ gives rise to a Wilson line in the highest weight representation determined by $\Lambda$. We shall first consider the case of $G=\SU(2)$, which has the action
\ie 
S= \frac{{n}}{2} \int dt \,\, \Tr\left(\sigma_3 {g}^{-1} (\partial_t+A_t)g \right)\,,
\fe 
and describe how it it equivalent to a Wilson line in the spin-$n/2$ representation of $\SU(2)$. We shall mostly follow the 
review of \cite{Nair:2016ufy}. From the action, we observe that the canonical one-form on the phase space is
\begin{equation}\label{canone}
\mathcal{A}= \frac{{n}}{2} \operatorname{Tr}\left(\sigma_3 g^{-1} dg\right)\,.
\end{equation}
A convenient parametrization of $g$ is  
\ie
g=\frac{1}{\sqrt{1+Z \bar{Z}}}\left(\begin{array}{cc}
1 & Z \\
-\bar{Z} & 1
\end{array}\right)\left[\begin{array}{cc}
e^{i \theta} & 0 \\
0 & e^{-i \theta}
\end{array}\right]\,,
\fe
where $Z$ and $\bar{Z}$ are local complex coordinates on $\mathbb{CP}^1$. 

In these coordinates the action takes the form 
\ie\label{zzbaction}
{S}= \int d t \bigg( i \frac{n}{2} \frac{ Z \dot{\bar{Z}}-\bar{Z} \dot{Z}}{1+Z \bar{Z}} + A_t^a \mu_a(Z,\bar{Z})\bigg)\,,
\fe
where 
\ie \label{gfunctions}
\mu_{+}=-n \frac{Z}{1+Z \bar{Z}}\,, \quad \mu_{-}=-n \frac{\bar{Z}}{1+Z \bar{Z}}\,, \quad \mu_3=-\frac{n}{2}\left(\frac{1-Z \bar{Z}}{1+Z \bar{Z}}\right)\,.
\fe 
The $\theta$ dependence of (\ref{zzbaction}) only contributes a boundary term which vanishes via Stokes' theorem and a suitable boundary condition. The action (\ref{zzbaction}) is invariant under the $\U(1)$ gauge transformation with a parameter $\exp(\varphi(t)\sigma_3)$,\footnote{More precisely, the action (\ref{zzbaction}) is invariant up to a boundary term $\int dt \dot{\varphi}(t)$. We assume that such a boundary term vanishes by imposing appropriate boundary conditions.} and the dynamics is restricted to $\SU(2)/\U(1)\simeq \mathbb{CP}^1$. For this $\U(1)$ symmetry, we can take a gauge choice $\theta=0$, and then the canonical one-form \eqref{canone} takes the form \ie 
\mathcal{A}=\cA_{Z}\,dZ+\cA_{\bar{Z}}\,d\bar{Z}=\frac{i n }{2}\left[\frac{Z d \bar{Z}-\bar{Z} d Z}{1+Z \bar{Z}}\right]\,.
\fe

Now, we define a prequantum line bundle on $\mathbb{CP}^1$ with curvature 
\ie 
\omega = d \mathcal{A}=\frac{in dZ\wedge d\bar{Z}}{(1+Z\bar{Z})^2}\,.
\fe 
This is precisely Kirillov-Kostant-Souriau symplectic form \eqref{omega} with the highest wight $\Lambda=\frac{n}{2}\sigma_3$\,, and $\mu_a$ is the corresponding moment map.
To obtain a Hilbert space of states, we impose the holomorphic polarization condition on sections of this bundle, denoted collectively as $\Psi$, i.e.,
\ie 
D_{\bar{Z}}\Psi=\left(\partial_{\bar{Z}}-i \mathcal{A}_{\bar{Z}}\right) \Psi=\left[\partial_{\bar{Z}}+\frac{n}{2} \frac{Z}{(1+Z \bar{Z})}\right] \Psi=0\,.
\fe 
The solution to this condition is 
\ie 
\Psi=\exp \left(-\frac{{n}}{2} \log (1+Z \bar{Z})\right) f(Z)
\fe 
for some holomorphic function $f(Z)$. Note that the single-valuedness of these wavefunctions requires that $\Omega$ satisfies the Bohr-Sommerfeld quantization condition, i.e.,\footnote{In general, the Bohr-Sommerfeld quantization condition amounts to the statement that the integral of $\omega$ on closed noncontractible two-surfaces ought to be be quantized as $2\pi$ times an integer. }
\ie 
    \int_{\mathbb{CP}^1} \omega = 2\pi n\,,\qquad n\in \mathbb{Z}\,.
\fe

The requirement of being normalizable with respect to the canonical inner-product 
\ie
\langle 1 \mid 2\rangle =\int \mathrm{d} \rho(\mathbb{CP}^1) \Psi_1^* \Psi_2\,,
\fe 
where $\rho(\mathbb{CP}^1)$ is the Liouville measure on $\mathbb{CP}^1$ defined by $\omega$,  restricts the size of the Hilbert space according to the integer $n$. That is, a basis of non-singular wavefunctions is defined as $f(Z)=1, Z, Z^2, \ldots, Z^n$, as higher powers of $Z$ give rise to wavefunctions with non-finite inner product. 

Now, the prequantum operators corresponding to the functions \eqref{gfunctions} can be defined via $-i \xi^{\mu}_a D_{\mu}+\mu_a$, where $\xi_a=\xi^{\mu}_a\partial_{\mu}$ is the Hamiltonian vector field corresponding to $\mu_a$, which gives 
\ie\label{j11}
    \mathsf{P}\left(J_{+}\right)&=\left(Z^2 \partial_Z-\frac{n Z}{2} \frac{2+\bar{Z} Z}{1+\bar{Z} Z}\right)-i \xi_{+}^{\bar{Z}} {D}_{\bar{Z}}\,,\\
    \mathsf{P}\left(J_{-}\right)&=\left(-\partial_Z-\frac{n}{2} \frac{\bar{Z}}{1+Z \bar{Z}}\right)-i \xi_{-}^{\bar{Z}} {D}_{\bar{Z}}\,, \\
    \mathsf{P}\left(J_3\right)&=\left(Z \partial_Z-\frac{n}{2} \frac{1}{1+Z \bar{Z}}\right)-i \xi_3^{\bar{Z}} {D}_{\bar{Z}}\,.
\fe 
The second terms in these expressions act on polarized wave functions to give zero. The action of these operators on the holomorphic factor $f(Z)$ of the wavefunction can be shown to be 
\ie\label{j22}
    \hat{J}_{+} f &=\left(Z^2 \partial_Z-n Z\right) f\,, \\
    \hat{J}_{-} f &=\left(-\partial_Z\right) f\,, \\
    \hat{J}_3 f &=\left(Z \partial_Z-\frac{1}{2} n\right) f\,,
\fe
which is just the monomial representation of the spin-$n/2$ irreducible representation of $\mathfrak{su}(2)$. Hence, the partition function of the quantum mechanical system with the Hamiltonian $\hat{H}=A^a_t\hat{\mu}_a$ on a circle, which can be defined as
\ie 
    \int \mathcal{D} g\,e^{i S[g]}=\operatorname{Tr}_{\text {Hilbert space }} \cP \exp \left(\int_0^{2 \pi} i \hat{H}(t) d t\right),
\fe
turns out to be precisely a Wilson loop in the spin-$n/2$ irreducible representation of $\mathfrak{su}(2)$.

It is useful, for generalizations to higher rank gauge groups, to also consider the geometric quantization outlined above in terms of Wigner D-functions.  Firstly, recall that given a function on $\SU(2)$, we can express it in terms of Wigner D-functions as
$$
    \Psi(g)=\sum_j \sum_{a, b} C_{a b}^{(j)} \mathcal{D}_{a b}^{(j)}(g)=\sum_j \sum_{a, b} C_{a b}^{(j)}\left\langle a\left|e^{ \hat{J}_{i}^j \theta^i}\right| b\right\rangle\,.
$$
Now, under the right action $g \rightarrow g h, h=\exp \left(- \frac{\sigma_3}{2} \theta\right)$, the canonical one-form changes as $\mathcal{A} \rightarrow \mathcal{A}+({n} / 2) \mathrm{d} \theta$, which implies that the wave functions must obey
\ie
\Psi\left(g e^{-\hat{J}_3 \theta}\right)=\Psi(g) \exp \left(\frac{i {n}}{2} \theta\right)\,.
\fe
Hence, we observe that the state $|{b}\rangle=$ $\left|{j},-\frac{n}{2}\right\rangle$.

Next, we consider the right action with respect to $R_{-}=R_1 - i R_2$, where $R_a$ is defined by 
\ie
R_a g=-ig \frac{\sigma_a}{2}\,.
\fe 
In particular, the polarization condition $\mathcal{D}_{\bar{z}} \Psi=0$ can be shown to be identical to 
\ie 
R_{-} \Psi=-i\sum_j \sum_{a, b} C_{a b}^{(j)}\left\langle a\left|e^{ \hat{J}^j_i \theta^i} \hat{J}^j_{-}\right| b\right\rangle=0\,.
\fe 
In other words,  $|b\rangle$ can be identified with the lowest weight state of the module. Moreover, since we know that $|{b}\rangle=$ $\left|{j},-\frac{n}{2}\right\rangle$, we find that $j=n/2$, and that the wavefunction ought to take the form 
\ie
\Psi=\sum_a C_{a,-\frac{n}{2}}^{\left(\frac{n}{2}\right)} \mathcal{D}_{a,-\frac{n}{2}}^{\left(\frac{n}{2}\right)}(g)\,.
\fe 
Hence, given that the index $a$ takes $2j + 1$ values, the Hilbert space corresponds to the unitary spin $j=n / 2$ irreducible representation of $\mathrm{SU}(2)$. 

Moreover, the operators $J_i$ given in \eqref{j11} and \eqref{j22} correspond to the left action on $g$, whereby
\ie 
J_i \Psi(g)=\sum_a C_{a,-\frac{n}{2}}^{\left(\frac{n}{2}\right)} \mathcal{D}_{a,-\frac{n}{2}}^{\left(\frac{n}{2}\right)}\left(\frac{\sigma_i}{2} g\right)=-i\sum_{a, c} C_{a,-\frac{n}{2}}^{\left(\frac{n}{2}\right)}\left(J_i\right)_{a c} \mathcal{D}_{c,-\frac{n}{2}}^{\left(\frac{n}{2}\right)}(g)\,,
\fe
which is expected from the results reviewed above. Hence, we find that the language of group theory can be used to show that quantum mechanics on the coadjoint orbit $\SU(2)/\U(1)$ is equivalent to a Wilson line as well. 

Moreover, in this formalism, the results can be straightforwardly generalized to gauge groups of higher rank. As an example, we shall consider flag manifolds of the form $G/H$, where $H$ is a maximal torus, which are always K{\"a}hler. For this purpose, we first introduce generators of $\mathfrak{g}$ in the Cartan-Weyl basis $\{H_j, E_{\alpha}\,, E_{-\alpha}\}$\,,
where $H_j\,(j=1,\dots ,r)$ are generators of the Cartan subalgebra $H$\,.
The commutation relations of these generators take the following forms:
\begin{align}\label{cartanweyl}
\begin{split}
    &[H_i\,, H_j]=0\,,\qquad [H_j, E_{\alpha}]=\alpha_j\,E_{\alpha}\,,\qquad [E_{\alpha}\,, E_{-\alpha}]=\alpha^i\,H_i\,,\\
    &[E_{\alpha}\,, E_{\beta}]=N_{\alpha,\beta}E_{\alpha+\beta}\,,
\end{split}
\end{align}
where $N_{\alpha,\beta}$ is a constant, and where $R$ is the root system with simple roots $\alpha_{j}\,(j=1\,,\dots\,, r)$\,.

On $G/H$, the canonical one-form takes the form 
\ie
\mathcal{A}(g)= \sum_j \Lambda_j \operatorname{Tr}\left(H_j g^{-1} d g\right)\,.
\fe
Functions on $G$ can be expressed in terms of Wigner D-functions as 
\ie 
\Psi=\sum_{\rho} \sum_{a, b} C_{a b}^{(\rho)} \mathcal{D}_{a b}^{(\rho)}(g)=\sum_{\rho} \sum_{a, b} C_{a b}^{(\rho)}\left\langle a\left|e^{ \hat{J}_{j}^\rho \theta^j}\right| b\right\rangle\,,
\fe
where $\rho$ denotes irreducible representations of $G$.
Under the right action, $g \rightarrow g h$, where $h=\exp \left(- \frac{H_j}{2} \theta^j\right)$, the canonical one-form changes as $\mathcal{A} \rightarrow \mathcal{A}+({\Lambda_j} / 2) \mathrm{d} \theta^j$, which implies that the wave functions must obey
\ie
\Psi\left(g e^{- \hat{H}_j \theta^j}\right)=\Psi(g) \exp \left(\frac{i {\Lambda_j}}{2} \theta^j\right)\,.
\fe
Thus, the state $|{b}\rangle=$ $\left|\rho,-\frac{\Lambda}{2}\right\rangle$. The polarization condition can be chosen to be the requirement that 
\ie 
E_{\alpha} \Psi=0
\fe
for all positive roots, i.e., it is the highest weight state. Thus we find that the Hilbert space corresponds to an irreducible unitary representation defined by the highest weight $\Lambda=(\Lambda_1,\ldots, \Lambda_r)$.

\subsection{Coherent States and Coadjoint Orbits}\label{bgroupch}

We can naturally introduce coherent states associated with symplectic manifolds $(\cM,\omega)$ (more specifically, coadjoint orbits) \cite{Perelomov:1971bd}.
The use of such a coherent state enables us to relate the classical phase space to the Hilbert space of quantum mechanics.
For a nice review on this state, see for example \cite{Zhang:1990fy}.

Following the conventions below \eqref{cartanweyl}, let $\Lambda_j\,(j=1,\dots ,r)$ be a $r$-dimensional vector.
The highest weight state $\lvert \Lambda \rangle$ of the irreducible representation $R$ is constructed by imposing conditions
\begin{align}
    H_j\lvert \Lambda \rangle=\Lambda_j\lvert \Lambda \rangle\,,\qquad E_{\alpha}\lvert \Lambda \rangle=0\qquad (\alpha \in R_+)\,,
\end{align}
where $R_+(R_-)$ is a subsystem of positive (negative) roots, and the state $\lvert \Lambda \rangle$ is normalized as
\begin{align}
\langle \Lambda   \lvert \Lambda \rangle=1\,.
\end{align}
For this reference state, we can find the subgroup $\tilde{H}\subset G$ whose elements leave $\lvert \Lambda \rangle$ invariant up to a phase factor i.e.
\begin{align}
    h\,\lvert \Lambda \rangle=e^{i \phi(h)} \lvert \Lambda \rangle\,,\qquad h \in \tilde{H}\,.
\end{align}
This condition means that the projection operator $\lvert \Lambda \rangle \langle \Lambda\lvert$ under the adjoint action of $h \in \tilde{H}$ is invariant, 
\begin{align}
    \text{Ad}_g\left(\lvert \Lambda \rangle \langle \Lambda \lvert\right)=\lvert \Lambda \rangle \langle \Lambda \lvert\,,
\end{align}
and so we can regard $\tilde{H}$ as $H_{\Lambda}$ by identifying $\lvert \Lambda \rangle \langle \Lambda \lvert$ with $\Lambda$.

Now, by using a representative $\xi \in G/H_{\Lambda}$\,, we can define the associated coherent state as \cite{Perelomov:1971bd}
\begin{align}
    \lvert \xi, \Lambda \rangle =\xi\,\lvert \Lambda \rangle\,,
\end{align}
or equivalently, 
\begin{align}
    \lvert \cG, \Lambda \rangle = \cG\lvert \Lambda\rangle\,,\qquad \cG=\xi\,h\in G\,,\quad h\in H_{\Lambda}\,.
\end{align}
By definition, the space generated by this coherent state is in one-to-one correspondence with the coset $G/H_{\Lambda}$ i.e. the coadjoint orbit $\cO_{\Lambda}$\,.
In general, the coherent state is non-orthogonal i.e. 
\begin{align}
    \langle \cG', \Lambda \lvert \cG, \Lambda \rangle \neq 0
\end{align}
but is normalized to unity
\begin{align}
    \langle \cG, \Lambda \lvert \cG, \Lambda \rangle=1 \,.\label{n-c}
\end{align}
In this normalization, the completeness relation is
\begin{align}
    \int d\mu_{G}(\cG) \, \lvert \cG, \Lambda \rangle \langle \cG, \Lambda\lvert \, =1\,,\label{g-compl}
\end{align}
where $d\mu_{G}(\cG)$ is the group invariant measure of $G$\,.

As we have seen in the previous subsection, the coadjoint orbit $\cO_{\Lambda}$ naturally equips a symplectic structure with the symplectic 2-form $\omega$ (\ref{KKS}), and we can consider a classical dynamical system for this symplectic manifold $(\cM\,, \omega)$ and its quantization.
By employing the coherent state defined above, we can connect the Hilbert space $\cH$ for $\cO_{\Lambda}$ to the irreducible representation $R$ of $\mathfrak{g}$ with the highest weight $\Lambda$ as in examples shown later.
In the mathematical literature, the identification is performed by the Borel-Weil-Bott theorem \cite{MR89473}.

\subsection{Wilson Lines from Coherent States}\label{Fd-Wilson}

In this Section, we will explain how the trace of a given Wilson loop is described by the path integral of a 1d quantum mechanical systems \cite{Diakonov:1989fc}.

Let $C$ be a loop on 2d Minkowski spacetime $\cM$\,, and a point $\sigma$ on $C$ is parameterized by a parameter $s$\,.
We will denote $R$ by a representation of a Lie group $G$\,.
The Wilson loop in $R$ is given by
\begin{align}
    W_{R}(C)=\Tr_{R}\left(\cP\exp\left(-\oint_{C}ds\,A(s)\right) \right)\,,\label{Wilson}
\end{align}
where the symbol $\cP$ denotes path-ordering, and the trace $\Tr_{R}$ is taken over the vector space for the representation $R$\,.
The Wilson line (\ref{Wilson}) satisfies the Schr\"{o}dinger-like differential equation
\begin{align}
    i\frac{d}{ds}\cP\exp\left(-\oint_{s_0}^{s}ds'\,A(s')\right)=-i\,A(s)\,\cP\exp\left(-\oint_{s_0}^{s}ds'\,A(s')\right)\,.
\end{align}
Then, if we identify $s$ with the time $t$, (\ref{Wilson}) can be interpreted as a wave function of a quantum mechanical system with the Hamiltonian $H(t)=-i\,A(t)$\,.
As we will show, this enables us to rewrite it as path integral representations.
 
To do this, we first divide the Wilson line (\ref{Wilson}) into $N$ segments as
\begin{align}
   U\left(t_f, t_0\right)&= \cP \exp\left[-i\int_{t_0}^{t_{f}}dt\, H(t)\right]
   =\cP\prod_{n=0}^{N-1}\exp\left\{-i\,\epsilon\,H(t_n)\right\}\,,
\end{align}
where $t_f:=t_{N-1}$\,, and the length $\epsilon$ of each segment and each point $t_n$ are given by
\begin{align}
    \epsilon=\frac{t-t_0}{N}\,,\qquad t_n=n\,\epsilon\,.
\end{align}
We then insert the completeness relation (\ref{g-compl}) in between each segment.
Then, $U\left(t_f, t_0\right)$ becomes
\ie
U\left(t_f, t_0\right)= \lim _{N \rightarrow \infty} \int\left(\prod_{n=1}^{N-1}d\mu(\cG_{n})\right) \prod_{n=1}^N\left\langle\cG_{n}\left|\exp \left\{-i \varepsilon H\left(t_n\right)\right\}\right|\cG_{n-1}\right\rangle\,.
\fe
Up to first order in $\varepsilon$\,, the expectation value $\left\langle\cG_{n}\left|\exp \left\{-i \varepsilon H\left(t_n\right)\right\}\right|\cG_{n-1}\right\rangle$ is
\ie
\left\langle\cG_{n}\left|\exp \left\{-i \varepsilon H\left(t_n\right)\right\}\right|\cG_{n-1}\right\rangle &\approx\left\langle\cG_{n} \mid \cG_{n-1}\right\rangle \exp \left(-i \varepsilon \frac{\left\langle\cG_n\left|H\left(t_n\right)\right| \cG_{n-1}\right\rangle}{   \left\langle \cG_{n} \mid \cG_{n-1}    \right\rangle}\right) \no\\
&\approx \left\langle\cG_{n} \mid \cG_{n-1}\right\rangle \exp \left(-i \varepsilon \left\langle\cG_n\left|H\left(t_n\right)\right| \cG_{n}\right\rangle\right)\,,
\fe
where we used the normalization (\ref{n-c}) of the coherent state.  Taking the limit $\varepsilon \rightarrow 0$ (i.e. $N \rightarrow \infty$ ), we find
\begin{align}
&\langle\cG_{n} \mid \cG_{n-1} \rangle =1-\langle\cG_{ n } |(|\cG_n\rangle-|\cG_{n-1}\rangle)\simeq \exp \bigg\{i \varepsilon\langle\cG_n| i\,\left(\frac{|\cG_{n}\rangle-|\cG_{n-1}\rangle}{\varepsilon}\right)\bigg\}\,.
\end{align}
Therefore, the transition amplitude takes the following path integral representation: 
\begin{align}
&\langle\cG(t_f)|U(t_f, t_0 )| \cG(t_0) \rangle\no\\
=& \lim _{N \rightarrow \infty} \left(\prod_{n=1}^{N-1}d\mu(\cG_{n})\right) \exp i \sum_{i=1}^N \varepsilon \biggl[\langle\cG_n| i\,\left(\frac{|\cG_{n}\rangle-|\cG_{n-1}\rangle}{\varepsilon}\right)
+i\left\langle\cG_n\left|H\left(t_n\right)\right| \cG_{n-1}\right\rangle\biggr] \no\\
=& \int \mathcal{D}\mu(\cG) \exp \left[i \int_{t_0}^{t_f} d t\Bigl(\langle\cG|i \frac{d}{d t}| \cG\rangle+i\langle\cG |H| \cG \rangle \Bigl)\right]\,,\label{tr-path}
\end{align}
where the measure $\mathcal{D}\mu(\cG)$ is defined as the product of $d\mu(\cG_{n})$ at each segment point
\begin{align}
   \mathcal{D}\mu(\cG)=\lim_{\substack{N \to \infty\\ \epsilon\to 0}}\prod_{n=1}^{N}d\mu(\cG_n)\,.
\end{align}
The trace of a given operator $\cO$ is rewritten as
\begin{align}
    \frac{1}{N}\Tr_{R}\left(\cO\right)=\int d\mu(\cG_N)\,\langle \cG_N, \Lambda   \lvert  \,\cO\,\lvert\cG_N, \Lambda \rangle \,.
\end{align}
Then, after doing a small computation, we can obtain a path integral representation of $W_{R}(C)$
\begin{align}
    W_{R}(C)=\int \mathcal{D}\mu(\cG)\exp\left[-\oint_{C} dt\,\langle \Lambda\lvert \cG(t)^{-1}\left(\frac{d}{dt}+A(t) \right)\cG(t)\lvert \Lambda \rangle\right]\,,\label{WL-path1}
\end{align}
where we introduced a group element $\cG$ of $G$ taking the form
\begin{align}
    \cG=\xi\,h \in G\,,\qquad h\in H_{\Lambda}\,. 
\end{align}
If we introduce the projection operator
\begin{align}
    \Lambda=\lvert \Lambda \rangle\langle \Lambda\lvert\,,
\end{align}
the expression (\ref{WL-path1}) is rewritten as
\begin{align}
    W_{R}(C)=\int \mathcal{D}\mu(\cG)\exp\left[ -\oint_{C} dt\,\Tr_{R}\left(\Lambda\, \cG(t)^{-1}\left(\frac{d}{dt}+A(t) \right)\cG(t)\right)\right]\,.\label{WL-coh}
\end{align}
Note that the kinetic term $\Tr_{R}(\Lambda\cG^{-1}d\cG)$ is a symplectic potential of Kirillov-Kostant-Souriau symplectic form (\ref{KKS}) for the coadjoint orbit $\cO_{\Lambda}$\,.
Therefore, the trace of the Wilson loop (\ref{Wilson}) can be expressed as the path integral of quantum mechanics with the symplectic structure on the coadjoint orbit $\cO_{\Lambda}$.

\subsection{Relationship between Coherent State and Geometric Quantization}

The use of the above coherent states allows us naturally to describe the geometric quantization of a given classical dynamical system.

For a given symplectic manifold $(\cM\,, \omega)$, the procedure of geometric quantization aims to construct the Hilbert space of the corresponding quantum mechanical system \cite{Kostant:1970,Souriau:1970}.
In this procedure, we first quantize the classical Poisson algebra of functions on a given phase space i.e. for each classical observable $f_i$ we consider a map $f_i \mapsto \hat{f}_i$ such that 
\begin{align}
    \{f_1\,, f_2\}=f_3 \qquad \to \qquad  [\hat{f}_1\,, \hat{f}_2]=i\,\hat{f}_3\,.
\end{align}
In general, the representation of the operator algebras obtained from the above procedure is not irreducible.
In order to obtain an irreducible representation, we need to do a certain kind of reduction, i.e., take a polarization.

Let us consider the geometric quantization of the coadjoint orbit $\cO_{\la}$\,.
For a given coherent state $\lvert \psi\rangle\,, \lvert \varphi \rangle$\,, we define a wave function 
\begin{align}
    \psi(g)=\langle \varphi\lvert g \lvert \psi \rangle\,,\qquad g \in G\,.
\end{align}
We can introduce the coherent state representation of the Lie algebra $\mathfrak{g}$ defined by
\begin{align}
    [\rho(A)\psi](g)=\langle \varphi \lvert g\,\hat{A}\,\lvert \psi \rangle\,,\qquad A\in \mathfrak{g}\,.
\end{align}
The prequantization of the Hamiltonian function $Q_a$ leads to a map
\begin{align}
 Q_a \to  \hat{Q}_a=i \nabla_{Q_a}+Q_a\,.
\end{align}
Here, the operator $\nabla_{Q_a}$ is expressed as
\begin{align}
    \nabla_{Q_a}=X_{Q_a}+i\vartheta(X_{Q_a})\,,
\end{align}
where $X_{Q_a}$ is a Hamiltonian vector field for $Q_a$\,, and $\vartheta$ is a symplectic potential for $\cO_{\Lambda}$ satisfying $\omega=-d\vartheta$\,.
If $G$ is compact, the Hilbert space $\cH$ for the coherent states is the space of the square-integrable sections of the prequantum line bundle (holomorphic sections of a complex line bundle $\cL(\Lambda)$ over $\cO_{\Lambda}$)
\begin{align}
    \cH=H_{\bar{\partial}_{\vartheta}}^{0}\left(\cO_{\Lambda}, \cL(\Lambda)\right)\,,\qquad \Lambda \geq 0\,,
\end{align}
where $\bar{\partial}_{\vartheta}$ is the Dolbeault operator twisted with respect to $\vartheta_{\Lambda}$.
However in general, the representation $\hat{Q}_a$ is reducible.
Therefore, we need to reduce the number of degrees of freedom by taking a polarization that corresponds to taking the highest-weight state for the coherent state.

In this way, the use of coherent states enables us to connect the Hilbert space $\cH$ for $\cO_{\Lambda}$ to the irreducible representation $R$ with the highest weight $\Lambda$\,, i.e.
\begin{align}
    V_R\simeq H_{\bar{\partial}_{\vartheta}}^{0}\left(\cO_{\Lambda}, \cL(\Lambda)\right)\,,
\end{align}
where $V_R$ is the vector space of the representation $R$\,.
In mathematical literature, the identification is performed by the Borel-Weil-Bott theorem \cite{MR89473}.

\section{Wilson Lines from Discretized Free Fermion Defects}
    \label{sec:WL-FF}

In this Appendix we discuss equivalence between discretized free fermion defects and Wilson lines.

\subsection{Coherent State for Fermion Systems}

Here, we shall describe coherent states for fermionic fields, instead of bosonic fields.

To do this, we introduce the coherent states for fermionic operators which transform in the fundamental representation of $\SU(N)$\,.
Let $\psi^j$ be $N$ pairs of Grassmann fields, and $\psi^*_j$ is the complex conjugate of $\psi^j$.
Under a global transformation of $\SU(N)$, these fermions transform as
\begin{align}
    \psi^{'j}=U^j{}_k\, \psi^k\,,\qquad \psi^{*'j}= \psi^{*k}\,(U^{\dagger})_{k}{}^{j}\,,\qquad U \in \SU(N)\,.
\end{align}
After quantization, the Grassmann variables lead to fermionic creation and annihilation operators $\hat{c}^{j}$ and $\hat{c}_{j}^{\dagger}$ which satisfy the anticommutation relations
\begin{align}
    \{\hat{c}^j, \hat{c}^{\dagger}_{k}\}=\delta^j_k\,,\qquad 
    \{\hat{c}^j, \hat{c}^k\}=0\,,\qquad 
    \{\hat{c}^{\dagger}_{j}, \hat{c}^{\dagger}_{k}\}=0\,.
\end{align}
Suppose that the creation\,(annihilation) operators $\hat{c}^{\dagger}_{j}(\hat{c}^{j})$ and the Grassmann variables $\psi^j\,, \psi^*_j$ are mutually anti-commuting.
Let $\cH_{f}$ be the Hilbert space generated by the fermionic creation operators $\hat{c}^{\dagger\,j}$\,.

\medskip

Then, the fermionic coherent state is defined as an eigenstate of the annihilation operator $\hat{c}_j$ with the eigenvalue $\psi_j$ i.e.
\begin{align}
\hat{c}^j \lvert \psi \rangle=\psi^j\,\lvert \psi \rangle\,.
\end{align}
The explicit expression of $\lvert \psi \rangle$ is given by
\begin{align}
    \lvert \psi \rangle =e^{-\psi^j\,\hat{c}_{j}^{\dagger}}\,\lvert 0\rangle_{f}=\prod_{j=1}^{N}\left(1 -\psi^j\,\hat{c}_{j}^{\dagger}\right)\lvert 0\rangle_{f}\,,\label{ch-s-f}
\end{align}
where $\lvert 0\rangle$ is the vacuum state satisfying $\hat{c}^j\lvert 0\rangle=0$\,.
Note that the coherent states are overcomplete i.e.
\begin{align}
    \langle \psi^* \lvert \psi' \rangle=\prod_{j=1}^{N}\left(1+\psi^*_j\psi^{'j}\right)=e^{\psi^*_j\psi^{'j}}\,.
\end{align}
The relation leads to the following completeness condition:
\begin{align}
    1=\int d\psi^*d\psi\,e^{-\psi^*\psi}\,\lvert \psi\rangle \langle \psi^*\lvert\,,\label{compl-f}
\end{align}
where the integration over the Grassmann variables is 
\begin{align}
    &\int d\psi^*d\psi:=  \int d\psi^*_1d\psi^*_2\cdots d\psi^*_{N}d\psi_1d\psi_2\dots d\psi_{N}\,,\\
    &\int \psi_j \,d \psi_j=1\,,\qquad \int 1\,d\psi_j=0\,.
\end{align}
Note that the integration measure $d\psi^* d\psi$ is invariant under $\SU(N)$ rotations of the Grassmann variables $\psi\,, \psi^*$\,. 
For a bosonic operator $\cO$\,, the trace over the coherent state is 
\begin{align}
    \Tr_{\cH_{f}}\,\cO:=  \int d\psi^*d\psi\,\langle -\psi^*\lvert \cO \lvert \psi \rangle\,.\label{Tr-f}
\end{align}

\subsection{Wilson Loop as a 1d Fermionic System}\label{f-Wilson}

Now let us give the path integral representation of the Wilson loop operator defined by
\begin{align}
    W(A)=\Tr_{\cH_{f}}\,\cP\exp\left(-\int_{t_0}^{t_f} dt\,A_{t}^{a} \hat{\rho}_f(t_a)\right)\,,\label{Wilson-f-r}
\end{align}
where $\hat{\rho}_f(t_a)$ is the Schwinger representation of $t_a \in \SU(N)$ in the Weyl ordering.
By using the coherent state (\ref{ch-s-f}), we can show that the Wilson loop (\ref{Wilson-f-r}) can be rewritten as 
\begin{align}
    W(A)&=\int_{\text{APB}} \cD\psi^*\cD\psi\,\exp\left(  i\, S^{f}[\psi^*,\psi,A]\right)\,,\label{Wilson-f-r-path}\\
    S^{f}[\psi^*,\psi,A]&= \int_{t_0}^{t_f} dt\,\psi^*_{i}i\left(\delta^{i}_{j}\partial_{t}+A_{t}^a\rho(t_{a})^{i}{}_{j}\right)\psi^{j}\,.\label{Wilson-f-r-action}
\end{align}
The symbol $\text{APB}$ denotes the anti-periodic boundary condition for the fermion 
\begin{align}
    \psi(t_0)=-\psi(t_f)\,.
\end{align}

We note that the associated 1d action can be regarded as a discretization of 2d order defect action for the 2d massless Thirring model.
To see this, we decompose the path ordered integral (\ref{Wilson-f-r}) into $N$ segments with the length $\epsilon$ along the integration curve as in
\begin{align}
   \lim_{N\to \infty} \int \prod_{l=1}^{N}d\psi^*_l d\psi_l\,e^{-\psi^*_l\psi_l}\,\langle \psi^*_{l}\lvert (1-\epsilon\,A_{t}^{a}(t_l) \hat{\rho}_f(t_a))\lvert \psi_{l-1}\rangle\,,\label{Wf-dec}
\end{align}
where $\psi_l:= \psi(t_l)$.
Since the $-\psi^*$ appears in the definition (\ref{Tr-f}) of the trace $\Tr_{\cH_{{\bf f}}}$, the boundary condition is taken as the anti-periodic boundary condition
\begin{align}
    \psi^*_{N}=-\psi^*_0\,,\qquad \psi_{N}=-\psi_0\,.
\end{align}
Then, we insert the completeness relation (\ref{compl-f}) into each segment of (\ref{Wf-dec}), the Wilson loop is rewritten as
\begin{align}
    &\lim_{N\to \infty}\int \prod_{l=0}^{N-1}d\psi^*_l d\psi_l\,e^{-\psi^*_l \psi_l}e^{\psi^*_{l}\psi_{l-1}}e^{-\epsilon\, \,\psi^*_{l,i}A_{t}^a(t_l)\rho(t_{a})^{i}{}_{j}\frac{\psi_l^j+\psi_{l-1}^j}{2}}\no\\
    &=\lim_{N\to \infty}\int \prod_{l=0}^{N-1}d\psi^*_l d\psi_l\,
    \exp\left(-\epsilon\,\left(\psi_{l}^*\frac{\psi_{l}-\psi_{l-1}}{\epsilon}+\psi^*_{l,i}A_{t}^a(t_l)\rho(t_{a})^{i}{}_{j}\frac{\psi_l^j+\psi_{l-1}^j}{2}\right)\right)\no\\
    &=\int \cD\psi^*\cD\psi\,\exp\left(-\int_{t_0}^{t_f}dt\,\left(\psi^*\partial_{t}\psi+\psi^*_iA_{t}^a\rho(t_{a})^{i}{}_{j}\psi^j\right)\right)\,.
\end{align}
In this way, we obtain the path integral representation (\ref{Wilson-f-r-path}) of the Wilson loop (\ref{Wilson-f-r}).

\subsection{Wilson Loop in Irreducible Representations as a 1d Fermionic System}\label{f-Wilson-irep}

As briefly explained in the main body, the extension to the Wilson loop operator for general irreducible representations can be achieved by adding the projection operator 
\begin{align}
     W_{\rho}[A]=\Tr_{\cH_f}\left(\hat{P}_{\rho}\cP \exp\left(-\oint A^a \hat{\rho}_f(t_a)\right) \right)\,.
\end{align}
The projection operator $\hat{P}_{\rho}$ projects to the Hilbert space for the irreducible representation $\rho$ consisting of states $\lvert \Psi \rangle\in \cH_{f}$ satisfying the following constraints \cite{Corradini:2016czo} : 
\begin{align}
    \hat{L}^I{}_{I}\lvert \Psi \rangle=k_{I}\lvert \Psi \rangle\,,  \qquad   \hat{L}^{J}{}_{K}\lvert\Psi\rangle=0\,,\qquad J>K\,,\label{rep-const}
\end{align}
where $\hat{L}^{J}{}_{K}$ is defined by (\ref{Lf-def}), and the sum for the subscript $I$ in $\hat{L}^{I}{}_{I}$ is not taken. The explicit form of the projection operator $\hat{P}_{\rho}$ is (\ref{proj-f}).
The corresponding path integral representation (\ref{1d-f-action-g00}) can be obtained by straightforwardly extending the above method using coherent states to take into account flavor degrees of freedom and projection operators.

\subsubsection*{More Explanation on Realization of Irreducible Representations}

The remainder of this Appendix will show, through concrete examples, that the constraint (\ref{rep-const}) to realize an irreducible representation actually works well.

For wave functions, the first constraint (\ref{rep-const}) projects onto the wavefunction component with $k_I$ antisymmetric indices associated to each family,
\begin{align}
    \lvert\Psi\rangle\sim  \Psi_{i_{1}\dots i_{k_{1}},j_{1}\dots j_{k_{2}},\cdots,l_{1}\dots l_{k_{K}}}\hat{c}_{K}^{\dagger l_1} \dots \hat{c}_{K}^{\dagger l_{k_{K}}}\cdots \hat{c}_{2}^{\dagger j_1} \dots \hat{c}_{2}^{\dagger j_{k_{2}}}\hat{c}_{1}^{\dagger i_1} \dots \hat{c}_{1}^{\dagger i_{k_{1}}}\lvert 0\rangle_f\,.
\end{align}
This state is still in a reducible representation.
The second constraint of (\ref{rep-const}) allows this reducible representation to be projected onto an irreducible representation. Indeed, each $\hat{L}^{I}{}_{J}$ replaces $\hat{c}_J^{\dagger i}$ with $\hat{c}_{I}^{\dagger i}$, so imposing the constraint only leaves the states which are symmetrized for the $\SU(N)$ indices involved.
This is certainly a state to realize a $\SU(N)$ irreducible representation.

For example, consider the second symmetric representation of $\SU(N)$. If we impose the constraint
\begin{align} \label{12conf}
 \hat{c}_1^{i\dagger}  \hat{c}^2_i\lvert\Psi\rangle=0\,, \qquad \lvert \Psi\rangle=\Psi_{i,j}\hat{c}_2^{\dagger j}\hat{c}_1^{\dagger i}\lvert 0\rangle_{f}\,,
\end{align}
we can see $\Psi_{[ij]}=0$\,. That is, we find that the state corresponding to the second symmetric representation, described explicitly as 
\begin{align}
\lvert \Psi\rangle_{\left[k_{1}=1, k_{2}=1\right]}=\frac{1}{2}\Psi^{i,j}\left(\hat{c}_i^{1\dagger} \hat{c}_j^{2\dagger}+\hat{c}_j^{1\dagger} \hat{c}_i^{2\dagger}\right)|0\rangle_f\,,
\end{align}
remains. This follows since, for fixed $i$ and $j$,
\begin{align}
\begin{split}
\hat{c}_1^{k\dagger}  \hat{c}^2_k (\hat{c}_i^{1\dagger} \hat{c}_j^{2\dagger}+\hat{c}_j^{1\dagger} \hat{c}_i^{2\dagger})|0\rangle_f &= (\hat{c}_1^{k\dagger}  \hat{c}^2_k \hat{c}_i^{1\dagger} \hat{c}_j^{2\dagger}+\hat{c}_1^{k\dagger}  \hat{c}^2_k \hat{c}_j^{1\dagger} \hat{c}_i^{2\dagger} )|0\rangle_f\\
&=-(\hat{c}_1^{k\dagger}   \hat{c}_i^{1\dagger} \hat{c}^2_k\hat{c}_j^{2\dagger}+\hat{c}_1^{k\dagger}   \hat{c}_j^{1\dagger} \hat{c}^2_k\hat{c}_i^{2\dagger} )|0\rangle_f\\
&=-(\hat{c}_{1j}^{\dagger}   \hat{c}_i^{1\dagger} +\hat{c}_{1i}^{\dagger}   \hat{c}_j^{1\dagger}  )|0\rangle_f
\\
&=0 \,.
\end{split}
\end{align}
In this way, only the symmetrized family of indices representing the $k_I$-th antisymmetric representations remains, and we obtain an irreducible representation for the Young diagram specified by $\{k_I\}$.

\section{\texorpdfstring{Wilson Lines from Discretized $\beta\gamma$ Defects}
{Wilson Lines from Discretized Beta-Gamma Defects}}
    \label{sec:WL-BG}

\subsection{\texorpdfstring{Free $\beta\gamma$ Systems}{Free beta-gamma Systems}} \label{beta-gamma-Wilson} 

Let us first present the path integral representations of the Wilson loops obtained from discretization of $\beta\gamma$-systems with a target space $X$\,.
Here, we assume $X$ is a $N$-dimensional complex manifold with local coordinates $\{\gamma^i\}\,(i=1\,,2\,,\dots \,,N)$\,.

We start to see the 1d free $\beta\gamma$-system coupled with the gauge field $A_t \in \mathfrak{g}$ that we considered in Section \ref{sec:fbg-sec}.
In this case, the target space is taken as $X=\mathbb{C}^N$\,, and the action of the 1d system is given by \footnote{If $\beta$ and $\gamma$ have the same conformal dimensions 1/2\,, the $\beta\gamma$-system is equivalent to the chiral symplectic boson system.}
\begin{align}
    S_{\beta\gamma}=\oint_{S^1} dt\,\left( \beta_i \partial_t\gamma^i+A_{t}^a\mu_a\right)\,.\label{bega-action}
\end{align}
The fields $\beta_i(t)$ and $\gamma^i(t)$ are transformed as the (anti-)fundamental representations $R\,(R^*)$ of $\mathfrak{g}$\,.
The moment map $\mu_a$ (in the Lie algebra basis) is specified by taking the symplectic potential $\vartheta$ and the vector fields $K_a$ for $\mathfrak{g}$ on $X$\,, which are expressed as
\begin{align}
    K_a=\sum_{i}\rho(t_a)^i{}_j\gamma^j\frac{\partial}{\partial \gamma^i}\,,\qquad [K_a,K_b]=f_{ab}{}^{c}K_c\,.
\end{align}
Here, $\rho(t_a)$ are the fundamental representations of the generators $t_a$ of $\mathfrak{g}$\,.
The 1d system (\ref{bega-action}) has the holomorphic symplectic form $\omega$ 
\begin{align}
    \omega=\sum_{i=1}^{N}d\beta_{i}\wedge d\gamma^{i}\,,\label{KKS-bg}
\end{align}
and we take a symplectic potential as 
\begin{align}
    \vartheta=-\sum_{i}\beta_i d\gamma^i\,.
\end{align}
In this choice of $\vartheta$, by using (\ref{H-mu}) and (\ref{mua}), the moment map $\mu_a$ is given by
\begin{align}
    \mu_a=\langle \mu, K_{a} \rangle=-\iota_{K_a}\vartheta=\beta_{i}\rho(t_a)^i{}_j\gamma^j\,,\label{mu-flat}
\end{align}
and then we have the expression (\ref{1dbg-action}) considered in Section \ref{sec:fbg-sec}.
We can see that the 1d action (\ref{bega-action}) is invariant under the gauge transformation 
\begin{align}
    \delta \beta_i=\delta \epsilon^{a}\beta_j\rho(t_{a})^j{}_i\,,\qquad \delta\gamma^i=-\delta \epsilon^a \rho(t_{a})^i{}_j \gamma^j\,,\qquad \delta A_t^{a}=D_t\delta \epsilon^{a}\,.
\end{align}
We can check the gauge transformation of the $\beta\gamma$ action, which is  
\ie 
    &\int dt  \Big(\epsilon^a \beta_i \rho(t_a)^{i}{}_{j} \partial_t \gamma^j + \beta_i \partial_t (-\epsilon^a \rho(t_a))^{i}{}_{j}\gamma^j + \partial_t\epsilon^a \beta_i \rho(t_a)^{i}{}_{j}\gamma^j  \\& +A^a_t\epsilon^b \beta_i \rho(t_b)^i_{\textrm{ }j}\rho(t_a)^j_{\textrm{ }k}\gamma^k  - A^a_t\epsilon^b \beta_i \rho(t_a)^i_{\textrm{ }j}\rho(t_b)^j_{\textrm{ }k}\gamma^k  - f^{bac}A^a_t\epsilon_b \beta_i \rho(t_c)^i_{\textrm{ }j} \gamma^j\Big)=0\,.
\fe 
Here, we have used $\rho(t_b)\rho(t_a)-\rho(t_a)\rho(t_b)=f_{ba}{}^c\rho(t_c)$.

The expression of the action (\ref{bega-action}) is useful when extending to $\beta\gamma$-systems with other target spaces $X$.
In fact, the action of the curved $\beta\gamma$-system on flag manifold presented in (\ref{bg-flag}) can be obtained simply by replacing (\ref{mu-flat}) with the corresponding moment map.

\subsubsection{Coherent State}

To rewrite the path integral of (\ref{bega-action}) to the Wilson loop, let us introduce the coherent state for $\beta\gamma$-system.
To do so, we perform the canonical quantization of the quantum-mechanical system (\ref{bega-action}) and then introduce the operators $\{\hat{\beta}_i\,,\hat{\gamma}_i \,(i=1\,,\dots\,,N)\}$ satisfying the commutation relations
\begin{equation}
    [\hgamma^i,\hbeta_j]=i\delta^i_{j}\hat{I}\,,\qquad [\hgamma^i,\hgamma^j]=0\,,\qquad [\hbeta_i,\hbeta_j]=0\,,
\end{equation}
where $\hat{I}$ is the identity operator commuting with $\hat{\gamma}_i\,,\hat{\beta}^i$\,.
The algebra is known as the Heisenberg-Weyl algebra.
Note that the operators $\hat{\gamma}_i\,,\hat{\beta}^i$ are expressed in terms of the oscillator creation and annihilation operators like
\begin{align}
    \hat{a}_j=\frac{1}{\sqrt{2}}(\hat{\gamma}^j+i\hat{\beta}_j)\,,\qquad \hat{a}^{\dagger}_j=\frac{1}{\sqrt{2}}\left(\hat{\gamma}^j-i\hat{\beta}_j\right)\,,
\end{align}
and they satisfy 
\begin{equation}
     [\ha_i,\ha^{\dagger}_j]=\delta_{ij}\,,\qquad [\ha_i,\ha_j]=0\,,\qquad [\ha^{\dagger}_i,\ha^{\dagger}_j]=0\,.
\end{equation}
The operators can define the Fock space $\cH_{\beta\gamma}$ with the vacuum state $\lvert 0\rangle_b$ (i.e. highest weight state) defined as
\begin{align}
    \hat{a}_i\,\lvert 0\rangle_b= \frac{1}{\sqrt{2}}(\hat{\gamma}^j+i\hat{\beta}_j)\,\lvert 0\rangle_b =0\,,
\end{align}
and the identity operator $\hat{I}$ is normalized as $ {}_b\langle 0\lvert \hat{I} \lvert 0 \rangle_b =1$\,.

Let us suppose an element of the Heisenberg-Weyl group which is parameterized as
\begin{align}
    \hat{\rho}(g(\gamma,\beta,\theta)) =e^{-i\theta \hat{I}} e^{i(\beta_k\hat{\gamma}^k -\gamma_k\hat{\beta}^k)}\,.\label{group-HW}
\end{align}
Then, the coherent state is given by\footnote{This coherent state is the same as the one for the standard harmonic oscillator algebra 
\ie 
\begin{aligned}
    | \beta\gamma \rangle &=\exp \left\{i\sum^n_{k=1}\left(\beta_k\hat{\gamma}^k -\gamma_k\hat{\beta}^k\right)\right\}|0\rangle_b
    =\exp \left\{-\frac{1}{2} \sum_k^n\left|z_k\right|^2\right\} \exp \left\{\sum_k^n\left(z_k \ha_k^{\dagger}\right)\right\}|0\rangle_b\,.
\end{aligned}
\fe
}
\begin{align}
    \lvert \beta \gamma \rangle=\hat{\rho}(g(\gamma,\beta,\theta=0)\lvert 0\rangle_b= e^{i(\beta_k\hat{\gamma}^k -\gamma_k\hat{\beta}^k)}\lvert 0 \rangle_b\,,\label{bg-cs}
\end{align}
which satisfies 
\begin{align}
    \hat{\beta}\lvert \beta\gamma \rangle= \beta \lvert \beta\gamma \rangle\,,\qquad 
     \hat{\gamma}\lvert \beta\gamma \rangle= \gamma \lvert \beta\gamma \rangle\,. \label{HW-ch}
\end{align}
The coherent state (\ref{bg-cs}) satisfies the completeness relation 
\begin{align}
       \hat{I}=\int \frac{d \beta d\gamma}{2\pi}\,\lvert \beta\gamma \rangle \langle \beta\gamma  \lvert\,,\label{bg-comple}
\end{align}
and is normalized to unity.

\subsubsection{Path Integral Representation of Wilson Loop}

Now, we can rewrite the path integral of the free $\beta\gamma$-system
\begin{align}
    Z_{\beta\gamma}=\int \mathcal{D}\beta \mathcal{D}\gamma \exp \left[i\oint_{S^1} dt(\beta_i \partial_t \gamma^i+A_a  \mu^a)\right]
\end{align}
as the trace of the Wilson loop.
By using the results in Section \ref{Fd-Wilson},
we can easily see 
\begin{align}
   W_{R}= \Tr_{\cH_{\beta\gamma}}\left(\cP \exp\left(i\oint_{S^1} dt A_t^a \hat{\rho}(t_a) \right)\right)=Z_{\beta\gamma}\,,\label{Wilson-Fock}
\end{align}
where the trace is taken over the Fock space $\cH_{\beta\gamma}$\,, and $\hat{\rho}(t_a)$ is the Schwinger representation of $t_a\in \mathfrak{g}$ defined as
\begin{align}
    \hat{\rho}(t_a)=\hat{\gamma}_i\rho(t_a)^i{}_j\hat{\beta}^j\,.\label{osc-rep}
\end{align}
In fact, the transition amplitude (\ref{tr-path}) of the time evolution operator $\hat{H}(t)=-A_{t}^a\hat{\rho}(t_a)$ acting on the coherent state (\ref{bg-cs}) is given by
\ie
\left\langle \bt^{'}\gamma^{'}\left|U\left(t_f, t_0\right)\right|  \bt\gm \right\rangle
=& \int \mathcal{D}\beta \mathcal{D}\gamma\,\exp \left(i \int_{t_0}^{t_f} d t\Bigl[\langle \bt\gm|i \frac{d}{d t}| \bt \gm\rangle-\langle\bt\gm |\hat{H}|  \bt\gm\rangle\Bigr]\right) \\
=& \int \mathcal{D}\beta \mathcal{D}\gamma\,\exp \left(i \int_{t_0}^{t_f} d t \Bigl[\frac{1}{2}(\bt_i \dot{\gm}^i-\gm_i \dot{\bt}^i)+A_a  \mu^a\Bigr]\right)\,,
\fe
where in the second equality, we used the expectation value $\langle \beta\gamma  \lvert\hat{\rho}(t_a)\lvert \beta \gamma \rangle$ evaluated as
\begin{align}
     \langle \beta \gamma  \lvert\hat{\rho}(t_a)\lvert \beta \gamma \rangle =\beta_i   \rho(t_a)^i{}_j\gamma^j=\mu_a\,.
\end{align}
If we take a trace over all initial and final states, the relation (\ref{Wilson-Fock}) holds up to a total derivative term.
Note that the symplectic 2-form (\ref{KKS-bg}) of the free $\beta\gamma$-system can be reproduced from the Kirillov-Kostant-Souriau form (\ref{KKS}) by substituting the left-invariant current $\hat{\rho}(g)^{-1}d\hat{\rho}(g)$ for (\ref{group-HW}) into (\ref{KKS}) and identifying $\la$ with the projection operator $\hat{\la}=\lvert 0\rangle_b{}_b\langle 0\lvert$\,.

\subsection{\texorpdfstring{Curved $\beta\gamma$ Systems on Flag Manifolds}{Curved Beta-Gamma Systems on Flag manifolds}}\label{bg-flag}

Finally, we will consider the curved $\beta\gamma$-system on Flag manifold as introduced in subsection \ref{sec:curved-bg}.
In this case, we take $X$ as the flag manifold $G_{\mathbb{C}}/B$\,.

For simplicity, we focus on the case $\mathfrak{g}=\mathfrak{su}(2)$ which corresponds to consider $X=\mathbb{CP}^1$\,.
The 1d action is given by
\begin{align}
    S_{\beta\gamma}^{\text{flag}}=\oint_{S^1} dt\,\left( \beta \partial_t\gamma+A_{t}^a\mu_a\right)\,.\label{bega-flag-action}
\end{align}
Here, the moment map $\mu_a$ is given by
\begin{align}
    \mu_a=\beta\,K_{a}^{\gamma}\,,
\end{align}
where $K_{a}=K_a^{\gamma}\partial_{\gamma}$ $(a=\pm \,, 3)$ are the Killing vector fields of $\mathfrak{su}(2)$ given by
\begin{align}
    K_{+}=\frac{\partial}{\partial \gamma}\,,\qquad K_-=-\gamma^2\frac{\partial}{\partial \gamma}\,,\qquad K_3=\gamma\frac{\partial}{\partial \gamma}\,.
\end{align}
In this case, the corresponding Wilson loop can be obtained by replaying the representation $\hat{\rho}(t_a) $ in (\ref{Wilson-Fock}) with the Dyson-Maleev representation of $\mathfrak{su}(2)$ generators \cite{Dyson:1956zza,Maleev:1958}
\footnote{Here, we assume the curved $\beta\gamma$-system does not have any anomaly described, for example, in subsection \ref{bg-anomaly}.}
\begin{align}
    \hat{\rho}(t_+)=\hat{\beta}\,,\qquad  \hat{\rho}(t_-)=-\hat{\beta}\hat{\gamma}^2\,,\qquad  \hat{\rho}(t_3)=\hat{\beta}\hat{\gamma}\,.
\end{align}
We can check that $\hat{\rho}(t_a)$ satisfy the $\mathfrak{su}(2)$ commutation relations
\begin{align}
    [\hat{\rho}(t_{\pm}),\hat{\rho}(t_3)]=\mp i \hat{\rho}(t_{\pm})\,,\qquad [\hat{\rho}(t_+), \hat{\rho}(t_-)]=+2i \hat{\rho}(t_3)\,.
\end{align}
As in the previous case, we work on the coherent state (\ref{bg-cs}), and can obtain the same representation of the Wilson loop
\begin{align}
    W_{R}=\int \mathcal{D}\beta \mathcal{D}\gamma \exp\left[i\oint_{S^1} dt(\beta_i \partial_t \gamma^i+A_a  \langle \beta\gamma  \lvert\hat{\rho}(t_a)\lvert \beta\gamma \rangle )\right]\,.
\end{align}
The expectation values $\langle \beta\gamma  \lvert\hat{\rho}(t_a)\lvert \beta \gamma \rangle$ are evaluated as
\begin{align}
\begin{split}
  \mu_+=\langle \beta \gamma  \lvert\hat{\rho}(t_+)\lvert \beta \gamma \rangle&=\beta\,,\\ 
  \mu_-=\langle \beta \gamma  \lvert\hat{\rho}(t_-)\lvert \beta\gamma \rangle&=-\beta \gamma^2\,,\\
 \mu_3=\langle \beta\gamma  \lvert  \hat{\rho}(t_3)\lvert \beta \gamma \rangle&=\beta \gamma\,.
\end{split}
\end{align}
In this way, we can obtain the path integral representations of the Wilson loop in terms of curved $\beta\gamma$-systems.

\section{Derivation of Chiral Anomaly for Finite Gauge Transformations}\label{ano-con}

In this Appendix, let us give a derivation of the chiral anomaly for the 2d gauged Weyl fermion systems.

\subsection{Chiral Anomaly for Weyl Fermions}\label{chiralanomaly}

To see this, let us first consider the chiral anomaly of the left-handed Weyl fermion system coupled with the gauge field.

The effective action for this gauged fermion system is given by
\begin{align}
    e^{\frac{i}{\hbar_{\rm 2d}}W_{+}[A]}=\frac{1}{\cN}\int \cD\psi_L\cD\psi_L^{*}\,e^{\frac{i}{\hbar_{\rm 2d}}\int d^2\sigma\,\bar{\Psi}i\cancel{D}_{L}\Psi}\,,\label{effa-LW}
\end{align}
where we introduced the covariant derivative 
\begin{align}
    \cancel{D}_{L}=\left(\cancel{\partial}+\cancel{A}\right)\frac{1+\gamma^5}{2}\,.
\end{align}
The classical action in (\ref{effa-LW}) is invariant under the chiral transformation
\begin{align}
    \psi^{u}_{L}=u\,\psi_{L}\,,\qquad A^{u}=uAu^{-1}-du u^{-1}\,,\qquad u=e^{i\alpha^a_L(\sigma)T_a}\in \SU(N)\,, \label{ch-tr}
\end{align}
but the effective action $W_L[A]$ is not invariant.
Our purpose is to give the explicit expression of the anomalous term with a finite parameter $\alpha_L(\sigma)$\,, which is useful for the computation of the bosonization of order defect actions.

\medskip

Naively, integrating out the fermionic fields in (\ref{effa-LW}) gives the determinant of the operator $\cancel{D}_{L}$\,.
However, since $\{\cancel{D}_L,\gamma^5\}=0$\,, $\cancel{D}_L$ maps positive chirality spinors to negative ones, and so the determinant $\text{det}\,\cancel{D}_L$ or equivalently the eigenvalue problem "$\cancel{D}_L\phi_n=\la_n\phi_n$" is not well defined (For more details, see for example Section 5.2.2 in \cite{Bilal:2008qx}).
To resolve this issue, we add a free right-handed operator $\cancel{\partial}\frac{1-\gamma^5}{2}$ to $\cancel{D}_L$ in (\ref{effa-LW}) by following the procedure described in \cite{Zumino:1983rz,Alvarez-Gaume:1983ict}.
Namely, let us suppose the path integral
\begin{align}
    \hat{Z}^f_{+}[A_-]= e^{\frac{i}{\hbar_{\rm 2d}}W_+[A_-]}=\frac{1}{\cN}\int \cD\Psi \cD\bar{\Psi}\,e^{\frac{i}{\hbar_{\rm 2d}}\int_{\cM} d^2\sigma\,\bar{\Psi}i\hat{\cancel{D}}_L\Psi}\,,\label{effa-LWr}
\end{align}
where $\hat{\cancel{D}}_{L}$ is defined as
\begin{align}
    \hat{\cancel{D}}_L=\cancel{\partial}+\cancel{A}\frac{1+\gamma_5}{2}\,.
\end{align}
Note that adding these fermionic degrees of freedom only affects the overall normalization factor because the gauge field only couples with the negative chirality spinors. 
Hence, we expect the effective action defined in (\ref{effa-LWr}) to have the same chiral anomaly of (\ref{effa-LW}).

\subsubsection*{Chiral Anomaly Computation for Finite Gauge Transformations}

For the chiral gauged fermion system (\ref{effa-LWr}), we can apply Fujikawa's method \cite{Fujikawa:1979ay,Fujikawa:1980eg} to obtain the chiral anomaly which is derived from a Jacobian for the integral measure of the path integral. 

For later convenience, we present a chiral anomaly of the massless fermion system coupled to more general vector fields \cite{Bardeen:1969md} (see also \cite{Fujikawa:2003az})
\footnote{Our notation of the gamma matrix here is different from that in Fujikawa-Suzuki's paper \cite{Fujikawa:2003az}: 
$\gamma^\mu\lvert_{\rm Fujikawa-Suzuki} = -i\gamma^{\mu}\lvert_{\rm here}$\,. 
Also, the spacetime metric is $\eta_{\mu\nu}=\text{diag}(-1,1)$, and the antisymmetric tensor $\epsilon^{\mu\nu}$ is normalized as $\epsilon^{01}=1$\,.}: 
\begin{align}
\frac{1}{\cN}\int \cD\Psi \cD\bar{\Psi}\,\exp\left({\frac{i}{\hbar_{\rm 2d}}\int d^2\sigma\,\bar{\Psi}\gamma^{\mu}(\partial_{\mu}-i\mathcal{V}_{\mu}-i \cA_{\mu}\gamma_5)\Psi}\right)\,.
\end{align}
The chiral anomaly can be derived by using Fujikawa's method \cite{Fujikawa:1979ay,Fujikawa:1980eg}, which describes the anomaly as a Jacobian of the path integral.
To do this, let us suppose the chiral transformation 
\begin{align}
\Psi(\sigma)\to \Psi'(\sigma)=e^{i\alpha_L(\sigma)\frac{1+\gamma_5}{2}}\Psi(\sigma)\,,\quad 
\bar{\Psi}(\sigma)\to \bar{\Psi}'(\sigma)=\bar{\Psi}(\sigma)e^{i\,\alpha_L(\sigma)\frac{-1+\gamma_5}{2}}\,.
\end{align}
The measure of the fermionic fields is transformed as
\begin{align}
\cD \Psi' \cD \bar{\Psi}'=\cJ_5(\alpha_L)\cD \Psi \cD \bar{\Psi}\,,
\end{align}
where the Jacobian is given by 
\begin{align}
\ln \cJ_5(\alpha_L)=-\frac{i}{2\pi}\int d^2\sigma\,
\Tr\left[\alpha_L(\sigma)\left((\partial^{\mu}\cA_{\mu}-i[\mathcal{V}^{\mu},\cA_{\mu}])+\frac{1}{2}\epsilon^{\mu\nu}(F_{\mu\nu}(\mathcal{V})+i[\cA_{\mu},\cA_{\nu}])\right)\right]\,,\label{Jacobian-ge}
\end{align}
where $F_{\mu\nu}(\mathcal{V})$ is defined as
\begin{align}
F_{\mu\nu}(\mathcal{V})=\partial_{\mu}\mathcal{V}_{\nu}-\partial_{\nu}\mathcal{V}_{\mu}-i[\mathcal{V}_{\mu},\mathcal{V}_{\nu}]\,.
\end{align}
For the left-handed case, the vector fields $\mathcal{V}_{\mu}\,, \cA_{\mu}$ are taken as
\begin{align}
\cA_{\mu}=\mathcal{V}_{\mu}=\frac{i}{2}A_{\mu}\,,
\end{align}
and then the Jacobian (\ref{Jacobian-ge}) becomes
\begin{align}
 \ln \cJ_5(\alpha_L)&=\frac{1}{4\pi}\int_{\cM}d^2\sigma\Tr\left[\alpha_L(\sigma)(\eta^{\mu\nu}+\epsilon^{\mu\nu})\partial_{\mu}A_{\nu}\right]\no\\
 &=-\frac{1}{\pi}\int_{\cM} d^2\sigma\Tr\left(\alpha_L(\sigma)\partial_+A_-\right)\,.\label{inf-chan}
\end{align}

Next, we derive an expression for the chiral anomaly with a finite chiral transformation.
To this end, we repeatedly perform an infinitesimal chiral transformation
\begin{align}
\Psi(\sigma)\to \Psi'(\sigma)=e^{i\delta s\,\alpha_L(\sigma)\frac{1+\gamma_5}{2}}\Psi(\sigma)\,,\quad 
\bar{\Psi}(\sigma)\to \bar{\Psi}'(\sigma)=\bar{\Psi}(\sigma)e^{i\delta s\,\alpha_L(\sigma)\frac{-1+\gamma_5}{2}}\,,
\end{align}
where $\delta s \ll 1$. This infinitesimal transformation induces a transformation of (the minus component of) the gauge field: $A_-\to A'_-=e^{-i\delta s\,\alpha_L(\sigma)}A_-e^{i\delta s\,\alpha_L(\sigma)}+e^{-i\delta s\,\alpha_L(\sigma)}\partial_-e^{i\delta s\,\alpha_L(\sigma)}$\,. To see the variation of the gauge field by repeated infinitesimal transformations, it is convenient to formally extend the 2d basis space to a 3d space like
we extend 
\begin{align}
    \tilde{A}(\sigma,s)=\tilde{u}^{-1}(\sigma,s)A(\sigma)\tilde{u}(\sigma,s)+\tilde{u}(\sigma,s)^{-1}d\tilde{u}(\sigma, s)\,,\qquad \tilde{u}(\sigma,s)=e^{is\,\alpha_L(\sigma)}\,.
\end{align}
By definition, the extended gauge field $\tilde{A}_-(\sigma,s)$ satisfies the boundary conditions
\begin{align}
    \tilde{A}_-(\sigma,s=1)=A^{u^{-1}}_-(\sigma)\,,\qquad \tilde{A}_-(\sigma,s=0)=A_-(\sigma)\,,
\end{align}
where $A^{u^{-1}}(\sigma)$ is a transformed gauge field given by
\begin{align}
    A^{u^{-1}}(\sigma)=u(\sigma)^{-1}A(\sigma)u(\sigma)+u(\sigma)^{-1}du(\sigma)\,,\qquad u(\sigma)=e^{i\alpha_L(\sigma)}\,.\label{gauge-tr-app}
\end{align}
Using the result of the chiral anomaly (\ref{inf-chan}) with an infinitesimal parameter, we can obtain the flow equation of the effective action $W_+[\tilde{A}_-(\sigma,s)]$ along the extended space direction of $s$\,,
\begin{align}
    e^{\frac{i}{\hbar_{\rm 2d}}W_+[\tilde{A}_-(\sigma,s)]}&=\frac{1}{\cN}\int \cD\Psi \cD\bar{\Psi}\,e^{\frac{i}{\hbar_{\rm 2d}}\int_{\cM} d^2\sigma\,\bar{\Psi}i\left(\cancel{\partial}+\tilde{\cancel{A}}\frac{1+\gamma_5}{2}\right)\Psi}\no\\
    &=\frac{1}{\cN}\int \cD\Psi' \cD\bar{\Psi}'\,e^{\frac{i}{\hbar_{\rm 2d}}\int_{\cM} d^2\sigma\,\bar{\Psi}'i\left(\cancel{\partial}+\tilde{\cancel{A}}\frac{1+\gamma_5}{2}\right)\Psi'}\no\\
    &=e^{i\,\delta\Gamma_L[\tilde{A}_-(\sigma,s)]\delta s+\frac{i}{\hbar_{\rm 2d}}W_+[\tilde{A}_-(\sigma,s+\delta s)]}\,,
\end{align}
or equivalently,
\begin{align}
    \frac{\partial}{\partial s}W_+[\tilde{A}_-(\sigma,s)]=-\hbar_{\rm 2d}\,\delta\Gamma_+[\tilde{A}_-(\sigma,s)]\,.\label{diff-chirl-ano}
\end{align}
From (\ref{inf-chan}), the anomalous term $\delta\Gamma_+[\tilde{A}_-]=-i \ln \cJ_5(\alpha_L)$ is
\begin{align}
\delta\Gamma_L[\tilde{A}_-]&=-\frac{1}{4\pi}\int_{\cM} d^2\sigma\,
\Tr\left[\tilde{u}(\sigma,s)^{-1}\partial_s\tilde{u}(\sigma,s)\left(\partial^{\mu}\tilde{A}_{\mu}+\epsilon^{\mu\nu}\partial_{\mu}\tilde{A}_{\nu}\right)\right]\,.
\end{align}
Performing the integral with respect to $s$ on both sides of (\ref{diff-chirl-ano}), we obtain 
\begin{align}
    W_+[A^{u^{-1}}_-(\sigma)]=W_+[A_-(\sigma)]-\hbar_{\rm 2d}\,\int_0^1ds\,\delta\Gamma_+[\tilde{A}_-(\sigma,s)]\,.\label{chano-int}
\end{align}
The anomalous term in (\ref{chano-int}) can be computed as follows:
\begin{align}
   \Gamma_+[A_-,u^{-1}]&:= \int_0^1ds\,\delta\Gamma_+[\tilde{A}_-(\sigma,s)]\no\\
   &=-\frac{1}{4\pi}\int_0^1ds\int_{\cM} d^2\sigma\,
\Tr\left[\tilde{u}^{-1}\partial_s\tilde{u}\left(\partial^{\mu}\tilde{A}_{\mu}+\epsilon^{\mu\nu}\partial_{\mu}\tilde{A}_{\nu}\right)\right]\no\\
&=\frac{1}{4\pi}\int_0^1ds\int_{\cM} d^2\sigma\,
\Tr\biggl[\partial_{s}(\partial_{\mu}\tilde{u}\tilde{u}^{-1})(\eta^{\mu\nu}+\epsilon^{\mu\nu})A_{\nu}\no\\
&\qquad \qquad +\partial_{\mu}(\tilde{u}^{-1}\partial_{s}\tilde{u})(\tilde{u}^{-1}\partial^{\mu}\tilde{u}+\epsilon^{\mu\nu}\tilde{u}^{-1}\partial_{\nu}\tilde{u})\biggr]\no\\
&=S_{\rm WZW}[u]-\frac{1}{\pi}\int_{\cM} d^2\sigma\,
\Tr\bigl(\partial_{+}uu^{-1}A_-\bigr) \,,\label{finite-G}
\end{align}
where in the third equality, we used $\tilde{u}\partial_{\mu}(\tilde{u}^{-1}\partial_s\tilde{u})\tilde{u}^{-1}=\partial_s(\partial_{\mu}\tilde{u}\tilde{u}^{-1})$\,.
Here, $S_{\rm WZW}[u]$ is the WZW action with the level one given by
\begin{align}
    S_{\rm WZW}[u]&=\frac{1}{8\pi}\int_{\cM} d^2\sigma\,\Tr\left( u^{-1}\partial_{\mu}u\,u^{-1}\partial^{\mu}u\right)+S_{\rm WZ}[\tilde{u}]\,,\label{WZW-action}\\
    S_{\rm WZ}[\tilde{u}]&=\frac{1}{12\pi}\int_{0}^{1}ds\int_{\Sigma} d^2\sigma\,\epsilon^{\mu\nu\rho}\,\Tr\left( \tilde{u}^{-1}\partial_{\mu}\tilde{u}\,\tilde{u}^{-1}\partial_{\nu}\tilde{u}\,\tilde{u}^{-1}\partial_{\rho}\tilde{u}\right)\,,\label{WZ-term}
\end{align}
where $\epsilon^{\mu\nu \rho}\,(\mu,\nu,\rho=0,1,s)$ is an antisymmetric tensor normalized as $\epsilon^{\mu\nu s}=\epsilon^{\mu\nu}$\,.
In the final equality of (\ref{finite-G}), we used the identity
\begin{align}
  S_{\rm WZW}[u]&=\frac{1}{4\pi}\int_0^1ds\int d^2\sigma\,
\Tr\biggl[\partial_{\mu}(\tilde{u}^{-1}\partial_{s}\tilde{u})(\tilde{u}^{-1}\partial^{\mu}\tilde{u}+\epsilon^{\mu\nu}\tilde{u}^{-1}\partial_{\nu}\tilde{u})\biggr]\,.
\end{align}
In this way, we obtain the chiral anomaly for a finite chiral transformation
\begin{align}
    W_+[A^{u^{-1}}_-]&=W_+[A_-]-\hbar_{\rm 2d}\Gamma_+[A_-,u^{-1}] \label{W_anomaly_pre}\,, \\
    \Gamma_+[A_-,u^{-1}]&=S_{\rm WZW}[u]-\frac{1}{\pi}\int d^2\sigma \,
        \Tr\bigl(\partial_{+}uu^{-1}A_-\bigr) \label{Gamma_def}\,.
\end{align}
Note that the consistency of \eqref{W_anomaly} requires
\begin{align}
\Gamma_+[A_-,u^{-1}] + \Gamma_+[A^{u^{-1}}_-, u]=0 \;,
\end{align}
which can be verified from the definition of $\Gamma_+$.
For the use in the main text, it is useful to replace $A_-$ by $A_-^u$ in \eqref{W_anomaly_pre},
so that we have 
\begin{align}
    W_+[A_-]&=W_+[A^u_-]-\hbar_{\rm 2d}\Gamma_+[A^u_-,u^{-1}] \,. \label{W_anomaly}
\end{align}

\subsubsection*{Right-handed Case}

Next, we briefly see the right-hand side.
As explained on the left-hand side, we consider the chiral anomaly of the gauged right-handed fermion system with the additional free left-hand fermion
\begin{align}
    \hat{Z}^f_-[A_+]=e^{\frac{i}{\hbar_{\rm 2d}}W_-[A]}=\frac{1}{\cN}\int \cD\Psi\cD\bar{\Psi}\,e^{\frac{i}{\hbar_{\rm 2d}}\int d^2\sigma\,\bar{\Psi}i\hat{\cancel{D}}_{R}\Psi}\,,\label{effa-RWr0}
\end{align}
where the modified covariant derivative $\hat{\cancel{D}}_R$ is defined as
\begin{align}
    \cancel{D}_{R}=\cancel{\partial}+\cancel{A}\frac{1-\gamma_5}{2}\,.\label{mrCD}
\end{align}
The corresponding chiral anomaly can be obtained by performing repeated infinitesimal chiral transformations
\begin{align}
\Psi(\sigma)\to \Psi'(\sigma)=e^{i \delta s \alpha_R(\sigma)\frac{1-\gamma_5}{2}}\Psi(\sigma)\,,\quad 
\bar{\Psi}(\sigma)\to \bar{\Psi}'(\sigma)=\bar{\Psi}(\sigma)e^{i \delta s\, \alpha_R(\sigma)\frac{-1-\gamma_5}{2}}\,,
\end{align}
and we use the formula (\ref{Jacobian-ge}) by replacing $\alpha_L$ with $-\alpha_R$ and taking the vector fields $\mathcal{V}_{\mu}\,, \cA_{\mu}$ as
\begin{align}
-\cA_{\mu}=\mathcal{V}_{\mu}=\frac{i}{2}A_{\mu}\,.
\end{align}
After doing a small calculation, we can see the effective action $W_-[A_+]$ has the chiral anomaly 
\begin{align}
    W_-[A^{u^{-1}}_+(\sigma)]=W_-[A_+(\sigma)]-\hbar_{\rm 2d}\,\Gamma_-[A_+,u^{-1}]\,,
    \label{chano-int-R}
\end{align}
where the anomalous term $\Gamma_-[A_+,u^{-1}]$ is
\begin{align}
   \Gamma_-[A_+,u^{-1}]&= S_{\rm WZW}[u^{-1}]-\frac{1}{\pi}\int d^2\sigma\,
\Tr\bigl(\partial_{-}uu^{-1}A_+\bigr) \,.\label{finite-G-R}
\end{align}

\subsubsection*{Consistent Anomaly}

It is known that different regularizations used in the computation of chiral anomaly give different results.
As explained in Section \ref{akccalc}, we can add a non-gauge invariant local counter term to the effective action so that the anomaly satisfies the Wess-Zumino consistency condition \cite{Wess:1971yu} i.e. 
\begin{align}
    W_+^{(a)}[A]&=W_+[A_-]+\frac{a \hbar_{\rm 2d}}{8\pi}\int_{\cM} d^2\sigma\,\Tr(A_{\mu}A^{\mu})\,,\\
    W_-^{(a)}[A]&=W_-[A_+]+\frac{a \hbar_{\rm 2d}}{8\pi}\int_{\cM} d^2\sigma\,\Tr(A_{\mu}A^{\mu})\,.
\end{align}
This counter term is often called the Jackiw-Rajaraman term \cite{Jackiw:1984zi} (For more details, see for example \cite{Fujikawa:2003az}).
When $a=1$ is chosen, chiral anomalies (\ref{finite-G-L}), (\ref{finite-G-R}) take simpler forms:
\begin{align}
\Gamma_+^{(a=1)}[A,u^{-1}]&=\frac{1}{4\pi}\int_{\cM}
\Tr\bigl(duu^{-1}\wedge A\bigr)+S_{\rm WZ}[\tilde{u}]\,,\\
\Gamma_-^{(a=1)}[A,u^{-1}]&=-\frac{1}{4\pi}\int_{\cM}
\Tr\bigl(duu^{-1}\wedge A\bigr)-S_{\rm WZ}[\tilde{u}]\,.
\end{align}
The expressions of chiral anomalies are useful to see how chiral anomalies arising from fermion defects can be canceled out by considering the coupling with the 4d CS theory, as discussed in Section \ref{akccalc}.

\subsection{Anomaly Cancellation in 4d CS Theory for Finite Gauge Transformations}\label{Ano-canf}

Here we will discuss how the chiral anomalies for the order defect actions (\ref{fermaction1})\,, (\ref{fermaction2}) are canceled in the 4d--2d system. While this was already discussed in Section \ref{sec:anomaly}, here we discuss a finite gauge transformation which is not necessarily connected to the identity.

As shown in the previous subsection, the chiral anomalies for the order defect systems are
\begin{align}
\begin{split}
    \hat{Z}_{+}^f[A_-^{\hat{u}}\lvert_{z_+}]&= \hat{Z}_+^f[A_-\lvert_{z_+}]\exp\left(-i\Gamma_+^{(a=1)}[A,\hat{u};z_+]\right)\,,\\
   \hat{Z}_{-}^f[A_+^{\hat{u}}\lvert_{z_-}]&=\hat{Z}_-^f[A_+\lvert_{z_-}]\exp\left(-i\Gamma_-^{(a=1)}[A,\hat{u};z_-]\right)\,,\label{ch-ano}
\end{split}
\end{align}
where we explicitly labelled the positions $z=z_{\pm}$ on $\mathbb{CP}^1$ of each chiral anomaly for clarity, with each chiral anomaly being given in terms of the expressions
\begin{align}
\Gamma_+^{(a=1)}[A,\hat{u};z_+]&=-\frac{1}{4\pi}\int_{\cM \times \{ z_+ \}}
\Tr\bigl(\hat{u}^{-1}d\hat{u}\wedge A\bigr)-S_{\rm WZ}[\tilde{u};z_+]\,,\\
\Gamma_-^{(a=1)}[A,\hat{u};z_-]&=\frac{1}{4\pi}\int_{\cM \times \{z_- \} }
\Tr\bigl(\hat{u}^{-1}d\hat{u} \wedge A\bigr)+S_{\rm WZ}[\tilde{u};z_-]\,.
\end{align}
The chiral anomalies can be cancelled by considering a coupling to the 4d CS theory with a quantum-corrected meromorphic one-form $\omega$.

To see this, we consider a gauge transformation
\begin{align}
A\mapsto A^{\hat{u}}:=  \hat{u}\,A\,\hat{u}^{-1}-d\hat{u}\hat{u}^{-1}\,, 
\label{gauge transformation}
\end{align}
where $\hat{u}$ is a $G$-valued function defined on 
$\cM\times\mathbb{CP}^1$\,. 
Under the transformation, the 4d CS action is transformed as
\begin{align}
S_{\rm CS}[A^{\hat{u}}]=S_{\rm CS}[A]+\frac{1}{2\pi \hbar}\int_{\cM\times \mathbb{CP}^1} \omega \wedge I_{\rm WZ}[\hat{u}]+\frac{1}{2\pi \hbar}\int_{\cM\times \mathbb{CP}^1} d\omega\wedge \Tr\left( \hat{u}^{-1}d\hat{u}\wedge A\right)\,, 
\label{gauge-variation-1}
\end{align}
where $I_{\rm WZ}[\hat{u}]$ is the Wess-Zumino (WZ) 3-form defined as 
\begin{align}
I_{\rm WZ}[\hat{u}]:= \frac{1}{3}\Tr( \hat{u}^{-1}d\hat{u}\wedge \hat{u}^{-1}d\hat{u}\wedge \hat{u}^{-1}d\hat{u})\,.
\end{align}
The transformation rule (\ref{gauge-variation-1}) means that the chiral anomalies (\ref{ch-ano}) can be canceled out by considering the shift in $\omega$\,, 
\begin{align}
    \omega \mapsto \omega_{\text{eff}}=\omega-\frac{\hbar}{4\pi i}\left(\frac{1}{z-z_+}-\frac{1}{z-z_-}\right)dz\,.\label{omega-shift-1}
\end{align}
Indeed, the last term in (\ref{gauge-variation-1}) is localized at the simple poles $z=z_{\pm}$ of $\omega_{\text{eff}}$, using the formula \eqref{deltident}:
\begin{align}
    &\frac{1}{2\pi \hbar}\int_{\cM\times \mathbb{CP}^1} d\omega_{\text{eff}}\wedge \Tr\left( \hat{u}^{-1}d\hat{u}\wedge A\right)\no\\
    &= -\frac{1}{4\pi}\int_{\cM\times \{z_+\}} \Tr\left( \hat{u}^{-1}d\hat{u}\wedge A\right) +\frac{1}{4\pi}\int_{\cM\times \{z_-\}} \Tr\left( \hat{u}^{-1}d\hat{u}\wedge A\right)\,,
\end{align}
these terms precisely cancel out the gauge-dependent terms of the chiral anomalies (\ref{ch-ano}).
For the first term in (\ref{gauge-variation-1}), we use the Cauchy-Pompeiu integral formula as in (Proposition 2.10 in \cite{Benini:2020skc})
\begin{align}
    &\frac{1}{2\pi \hbar}\int_{\cM\times \mathbb{CP}^1} \omega_{\text{eff}} \wedge I_{\rm WZ}[\hat{u}]\no\\
    &=-\frac{1}{4\pi}\int_0^1 ds\int_{\Sigma} d^2\sigma I_{\rm WZ}[\hat{u}\lvert_{z=z_{+}}]+\frac{1}{4\pi}\int_0^1 ds\int_{\Sigma} d^2\sigma I_{\rm WZ}[\hat{u}\lvert_{z=z_{-}}]\,.
\end{align}
The last two terms cancel the WZ terms that arise from the chiral anomaly.

\section{Equivalence of Different Defects}
    \label{sec:equivalence}

\subsection{\texorpdfstring{Coadjoint Orbit Defects and $\beta\gamma$ Defects}{Coadjoint Orbit Defects From Constrained Beta-Gamma Defects}}
    \label{sec:coadjointbeta}
In this appendix, we shall describe coadjoint orbit defects in terms of constrained free $\beta\gamma$-defects as well as curved $\beta\gamma$-defects. Although we shall focus on 1d defects, the proofs generalize straightforwardly to surface defects. 

\subsubsection{Coadjoint Orbit Defects From Constrained $\beta \gamma$ Defects}
Let us derive the equivalence between coadjoint orbit line defects and free $\beta\gamma$ line defects with constraints, which can each be shown to be equivalent to Wilson lines associated with fixed irreducible representations of $\SU(N)$. The derivation is a generalization of the case of the fundamental representation, which was shown in \cite{Diakonov:2000kw}.

 We shall use the representation of $\mathfrak{su}(N)$ given by $N \times N$ traceless matrices. The Cartan subalgebra in this representation is spanned by the set of all traceless, diagonal matrices.  Denote the matrix with 0 everywhere, except for a single 1 at the $i$-th row and $j$-th column as $e_{i j}$. 
 We may then write elements of the Cartan subalgebra in the form \ie 
 H=\sum_{i=1}^N \epsilon_i e_{i i}\,,
 \fe
 where $\epsilon_i$ form an overcomplete basis of vectors, with $N-1$ components for each $i\in \{1,\ldots N\}$, with the constraint $\sum_{i=1}^N \epsilon_i=0$.  
 The vectors 
 satisfy $\epsilon_i \cdot \epsilon_j = \delta_{ij}-\frac{1}{N}$.

Given an $\mathfrak{su}(N)$ integrable highest weight $\lambda$,
we can specify it by its partition
\ie 
\lambda=\left\{l_1 ; l_2 ; \cdots ; l_{N-1}\right\}\,,
\fe
where 
\ie 
\lambda= \sum_{i=1}^{N-1} l_i  \epsilon_i\,. 
\fe 
The quantity $\Lambda$ that appears in the coadjoint orbit action can be written as 
\ie 
\label{weightt}
\Lambda &= \lambda \cdot H 
 =\sum_{i=1}^{N-1} l_i \epsilon_i  \cdot \sum_{j=1}^N \epsilon_j e_{jj} 
 = \sum_{i=1}^{N-1} l_i e_{ii}  -\sum_{i=1}^{N-1}\frac{l_i}{N} I\,.
\fe 
Given that $\textrm{Tr}( \cG \del_t{\cG}^{\dagger})=0$ and $\textrm{Tr}( \cG A_t {\cG}^{\dagger})=0$, the term in \eqref{weightt} that is proportional to the identity matrix drops out of the coadjoint orbit action, which can thus be written as\footnote{As explained in \cite{Diakonov:2000kw}, for generic values of $\Lambda$ the path integral for this action is over the full flag manifold, but setting some values of $l_i$ to zero corresponds to an integral over partial flag manifolds, such as $\mathbb{CP}^{N-1}$ in the case of the fundamental representation. }
\ie 
\int dt \, \textrm{Tr} \bigg[ 
    \bigg(\sum_{i=1}^{N-1} l_i e_{ii}\bigg)   
    \bigg(\cG \del_t {\cG}^{\dagger} +\cG A_t{\cG}^{\dagger} \bigg) 
    \bigg]\,.
\fe 

Writing the $\alpha$-th row of $\cG$ as $z^*_i[\alpha ]$ and $A_t=A^a_t t_a$ where $t_a$ are traceless matrices that generate $\SU(N)$, we find that the action can be reexpressed as 
\ie 
\int dt \sum_{\alpha=1}^{N-1} l_{\alpha} \bigg(\sum_{i=1}^N z^*_i [\alpha ] \del_t z_i[\alpha] + \sum_{i=1}^N z^*_i [\alpha ] A^a_t t^{ij}_a z_j[\alpha] \bigg)\,.
\fe 
In addition, there are several constraints that arise from the unitarity of $\cG$, that is, $\cG\cG^{\dagger}=I$. These can be obtained by studying the quantity 
\ie 
\textrm{Tr}(\cG^{\dagger} e_{\alpha \beta } \cG)\,,
\fe 
for $\alpha, \beta = 1,\ldots , N-1$.
For $\alpha =\beta$, we obtain
\ie \label{constrrainteq1}
\sum_{i=1}^N z^*_i[\alpha] z_i[\alpha] = 1\,, 
\fe 
while the cases where $\alpha \neq \beta$ give us 
\ie \label{constrrainteq0}
\sum_{i=1}^N z^*_i[\alpha] z_i[\beta] = 0\,.
\fe 

To obtain the familiar form of the action and constraints studied in the main text, we need to perform the rescaling 
\ie 
z_i[\alpha]\rightarrow \frac{z_i[\alpha]}{\sqrt{l_{\alpha}}}\,.
\fe 
This results in the action taking the form 
\ie 
\int dt \sum_{\alpha=1}^{N-1}  \bigg(\sum_{i=1}^N z^*_i [\alpha ] \del_t z_i[\alpha] + \sum_{i=1}^N z^*_i [\alpha ] A^a_t t^{ij}_a z_j[\alpha] \bigg)\,,
\fe 
and the constraints \eqref{constrrainteq1} taking the form
\ie \label{constrrainteq1res}
\sum_{i=1}^N z^*_i[\alpha] z_i[\alpha] = l_{\alpha}\,,
\fe 
while the constraints \eqref{constrrainteq0} remain unchanged.

The equations \eqref{constrrainteq0} and 
\eqref{constrrainteq1res} mean that $z[\alpha]$
specify mutually orthogonal copies of 
$\mathbb{CP}^1$ inside $\mathbb{CP}^{N-1}$,
and hence (assuming $l_{\alpha}\ne 0$ for all $\alpha$) defines a coordinate system for the full flag
$\GL(N, \mathbb{C})/ B$:
the constraints  \eqref{constrrainteq0} define  a direct product of $N $ copies of $\mathbb{CP}^N$ which define the flag variety as a multi-projective variety. 
Solving these constraints in terms of locally-defined inhomogeneous coordinates on the flag variety, and substituting the solution into the defect action,
we obtain a 1d action of the form \eqref{zzbaction}. To see this in the case of $\mathbb{CP}^1$, to solve the constraint $\sum_i^2 z_i^*z_i=l$ we substitute 
\ie 
z_1&=\frac{Z\sqrt{l}e^{i\varphi}}{\sqrt{1+|Z|^2}}\;, \quad
z_2=\frac{\sqrt{l}e^{i\varphi}}{\sqrt{1+|Z|^2}} \,,
\fe
whereby the kinetic term becomes 
\ie 
\int dt \sum_i^2 z_i^* \partial_t z_i= %
-\int dt \, \frac{1}{2}\frac{l (Z\del_t Z^*-Z^*\del_tZ)}{(1+|Z|^2)},
\fe 
where we neglected a total derivative term $\int dt (il\del_t\varphi )$,
which vanishes due to the Stokes' theorem.


\subsubsection{Chiral WZW Model as a Bosonized Model of the Weyl Fermion System}\label{rCWZW}

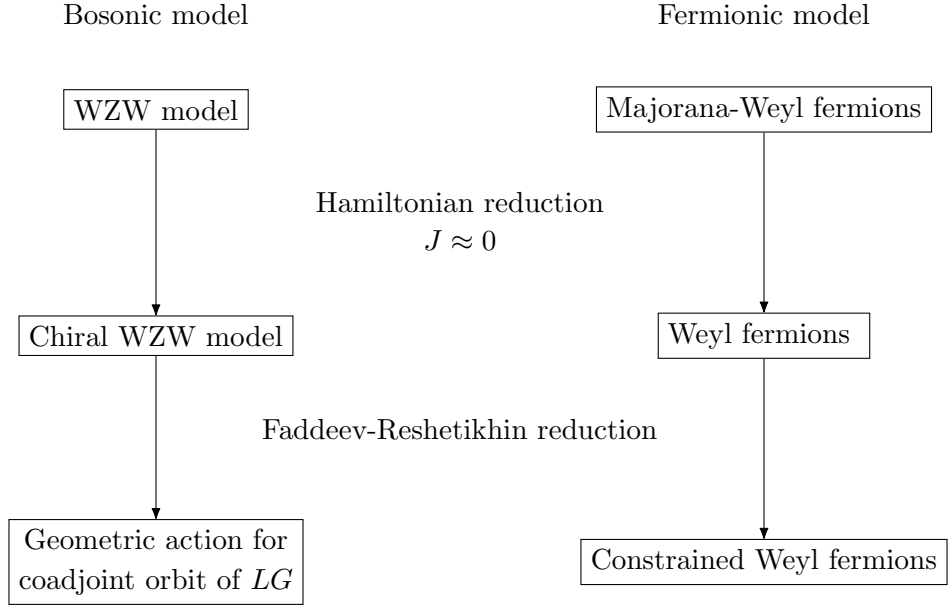
\begin{figure}[htbp]
\centering
\begin{tikzpicture}
    \node[rectangle, draw, align=center](91) at (0,9) {
   WZW model};
    \node[rectangle, draw, align=center](97) at (8,9) {Majorana-Weyl fermions};
    \node[rectangle, draw, align=center](61) at (0,6){
   Chiral WZW model};
   \node[rectangle, draw, align=center](68) at (8,6){ Weyl fermions };
    \node[rectangle, draw, align=center](31) at (0, 3) {Geometric action for\\
   coadjoint orbit of $LG$};
   \node[rectangle, draw, align=center](38) at (8, 3) {Constrained Weyl fermions};
    \draw[ ->, >={Latex[round]}, shift=({5pt, 5pt})]  (91)--(61) ;  
    \draw[ ->, >={Latex[round]}, shift=({5pt, 5pt})]  (61)--(31) ;  
    \draw[ ->, >={Latex[round]}, shift=({5pt, 5pt})]  (97)--(68) ;  
    \draw[ ->, >={Latex[round]}, shift=({5pt, 5pt})]  (68)--(38) ;  
    \draw (0,10)node[above]{Bosonic model};
    \draw (8,10)node[above]{Fermionic model};
    \draw (4.0,7.5)node[above]{Hamiltonian reduction};
    \draw (4.0,7.0)node[above]{ $J\approx 0$};
    \draw (4.0,4.5)node[above]{Faddeev-Reshetikhin reduction};
\end{tikzpicture}
\caption{This figure shows a series of reductions of the WZW model discussed in Section \ref{rCWZW}, and we have displayed the fermionic models dual to each of the reduced models.}
\end{figure}

In Section \ref{sec:chiralWZW}, in order to obtain a bosonized model of the (gauged) Weyl fermion, we considered a bosonization of the fermionic system that additionally incorporates a free Weyl fermion with opposite chirality.
As a result, the bosonized model is the WZW model, which has both left and right Kac-Moody currents, unlike the Weyl fermion with a single Kac-Moody current.
To obtain the bosonized model of the Weyl fermion system, one of the chiral currents, e.g., the "right-handed" current $\tilde{J}$, should be constrained to be exactly zero, leaving only the "left-handed" current $J$ (For the details, see \cite{Tsutsui:1992ca}) \footnote{Here, $J$ is one of the variables in the phase space in Hamiltonian analysis of the WZW model and is independent of the group element $g$ of a Lie group $G$. Also, $\tilde{J}$ is defined as $\tilde{J}=-g^{-1}Jg+ g^{-1}\partial_{\sigma}g$.}.
This reduced model, the so-called chiral WZW model \cite{Alekseev:1988ce,Wiegmann:1989hn,Perret:1990bc}, can be obtained by a way of a Hamiltonian reduction of the WZW theory and might be regarded as a (non-Abelian) chiral boson theory.
In \cite{Sonnenschein:1988ug,Frishman:1987se}, the chiral WZW model is proposed as a non-Abelian bosonized model of the Weyl fermion system. Here we will briefly review the chiral WZW model.

The chiral WZW model for the left-handed Kac-Moody algebra is given by
\begin{align}
    S_{\rm CWZW}^+[g_+]&=-\frac{1}{2\pi}\int_{\cM} d^2\sigma\,\Tr\left( g_+^{-1}\partial_{\sigma}g_+\,g_+^{-1}\partial_{-}g_+\right)+S_{\rm WZ}[\tilde{g}_+]\,,\label{cWZWp}
\end{align}
The equation of motion is
\begin{align}
    \partial_-(\partial_{\sigma}g_+g_+^{-1})=0\,.
\end{align}
The current $\partial_{\sigma}g_+g_+^{-1}$ generates the left-handed Kac-Moody algebra.
For the right-handed side, the action takes
\begin{align}
    S_{\rm CWZW}^-[g_-]&=\frac{1}{2\pi}\int_{\cM} d^2\sigma\,\Tr\left( g_-^{-1}\partial_{\sigma}g_-\,g_-^{-1}\partial_{+}g_-\right)-S_{\rm WZ}[\tilde{g}_-]\,,\label{cWZWm}
\end{align}
and the equation of motion is
\begin{align}
    \partial_+(\partial_{\sigma}g_-g_-^{-1})=0\,.
\end{align}
This current $\partial_{\sigma}g_-g_-^{-1}$ also generates a right-handed Kac-Moody algebra.
These chiral WZW models also satisfy the PW like identity:
\begin{align}
     S_{\rm CWZW}^{\pm}[g_{\pm}h_{\pm}]=S_{\rm CWZW}^{\pm}[g_{\pm}]+S_{\rm CWZW}^{\pm}[h_{\pm}]\mp \frac{1}{\pi}\int_{\cM}d^2\sigma\Tr\left[\partial_{\sigma}g_{\pm}g_{\pm}^{-1}\,\partial_{\mp}h_{\pm}h_{\pm}^{-1}\right]\,.
\end{align}
The canonical quantization of this chiral bosonic model is performed in \cite{Salomonson:1988mk}, and the model endows only one affine Kac-Moody algebra. Thus, the chiral WZW model is considered as a good candidate for a bosonized model of Weyl fermion.

We can write down a ``gauged" chiral WZW action that has the same transformation law as the chiral gauged fermion system under gauge transformation:
\begin{align}
  S_{+,{\rm chiral}}^{b}[g_+,A]&=S_{\rm CWZW}^+[g_+]
    -\frac{1}{\pi}\int_{\cM} d^2\sigma \Tr(A_{-} \partial_{\sigma}g_+g_+^{-1})\no\\
    &\quad+\frac{1}{2\pi}\int_{\cM} d^2\sigma\Tr(A_-^2)-\frac{1}{2\pi}\int_{\cM} d^2\sigma\Tr(A_+A_-)
    \,,\label{gcWZWp}
\end{align}
and
\begin{align}
    S_{-,{\rm chiral}}^{b}[g_+,A]&=S_{\rm CWZW}^-[g_-]
    +\frac{1}{\pi}\int_{\cM} d^2\sigma \Tr(A_{+} \partial_{\sigma}g_-g_-^{-1})\no\\
    &\quad+\frac{1}{2\pi}\int_{\cM} d^2\sigma\Tr(A_+^2)-\frac{1}{2\pi}\int_{\cM} d^2\sigma\Tr(A_+A_-)
    \,.\label{gcWZWm}
\end{align}
Indeed, we can see that under the transformation 
\begin{align}
    g^{h}_{\pm}=hg_{\pm}\,,\qquad A^h=hAh^{-1}-dhh^{-1}\,,
\end{align}
these actions transform as
\begin{align}
S_{+,{\rm chiral}}^{b}[g_+^h,A^h]&=S_{\rm CWZW}^+[g_+]+S_{\rm WZ}[\tilde{h}]
     +\frac{1}{4\pi}\int d^2\sigma \Tr(h^{-1}dh\wedge A)\,,\\
   S_{-,{\rm chiral}}^{b}[g_-^h,A^h]&=S_{\rm CWZW}^-[g_-]-S_{\rm WZ}[\tilde{h}]
     -\frac{1}{4\pi}\int d^2\sigma \Tr(h^{-1}dh\wedge A)
    \,.
\end{align}
These actions, for example, have been discussed in \cite{Harada:1989fn}.
Since they have the same transformation as the WZW defects, the gauge anomalies in the defects can be cancelled out by considering the same shift in the meromorphic 1-form $\omega$.
To obtain the bosonized model without the decoupled degrees of freedom of the multi-flavor massless Thirring model, it seems necessary to solve the equations of motion of the 4d-2d system of the 4d CS theory coupled with the chiral WZW model. However due to the existence of the $\Tr(A_{\pm }^2)$ terms in (\ref{gcWZWp}), (\ref{gcWZWm}), it seems to be difficult to solve the equations of motion. This problem would be left as a future work.

\subsection*{Coadjoint orbit defect from Faddeev-Reshetikhin reducton of chiral WZW model} 

The Faddeev-Reshetikhin procedure \cite{Faddeev:1985qu} or the alleviation procedure \cite{Delduc:2012qb} which was originally applied to PCM on a generic Lie group, is a way to reduce the original model by removing the non-ultralocality of the Poisson bracket while preserving the equations of motion.
In the following discussion, we will apply the reduction procedure to the chiral WZW model, and observe that the geometric action on the coadjoint orbit of the loop group $LG$ is obtained. 
This observation indirectly implies that the coadjoint orbit defect can be regarded as a bosonized model of the constraint fermion system.
Here, we focus on the right-handed case.

To do this, we modify the Poisson structure of the "right-handed" current $\tilde{J}$ that forms the Kac-Moody algebra by removing the non-ultralocal term, according to \cite{Faddeev:1985qu,Delduc:2012qb}. 
We denote the modified Poisson structure by
\begin{align}
    \{ \cJ^a_{(-)}(\sigma),\cJ^b_{(-)}(\sigma')\}_{\rm FR}=f^{ab}{}_{c}\cJ_{(-)}^c(\sigma)\delta(\sigma-\sigma')\,,
\end{align}
where we replaced $\tilde{J}$ and $\{ ,\}$ with  $\cJ_{(-)}$ and $\{ ,\}_{\rm FR}$, respectively.
By construction, the current $\cJ_{(-)}$ has to satisfy the equation of motion
\begin{align}
    \partial_+\cJ_{(-)}=0\,.\label{eom-cFR}
\end{align}
Note that chiral conserved currents $\Tr(\cJ_{(-)}^n)$ are centers of the modified Poisson bracket.
This means that the conserved quantity $\Tr(\cJ_{(-)}^2)$ corresponding to the original chiral Hamiltonian
\begin{align}
    H_{\rm CWZW}\propto \int d\sigma \Tr(\tilde{J}\tilde{J}) 
\end{align}
cannot generate a time evolution, so to obtain a consistent phase space we impose the following constraints:
\begin{align}
    \Tr(\cJ_{(-)}^n)=\textrm{constant}\,.
\end{align}
In particular, we take  the (chiral) energy-momentum tensor $\tilde{T}$ as
\begin{align}
\tilde{T}=\Tr(\cJ_{(-)}\cJ_{(-)})=\mu^2\,.\label{Tmm-const}
\end{align}
The constraint (\ref{Tmm-const}) enable us to parameterize the current $\cJ_{(-)}$ as
\begin{align}
    \cJ_{(-)}=\cG_{(-)}\Lambda \cG^{-1}_{(-)}\,,
\end{align}
where $\Lambda$ is the constant element of $\mathfrak{g}$ satisfying $\Tr[\Lambda^2]=\mu^2$\,.
By using this parametrization, we can find the first-order action, that derives the equations of motion (\ref{eom-cFR}),
\begin{align}
    S_-=\int d^2\sigma\,\Tr[\Lambda \cG_{(-)}^{-1}\partial_{+}\cG_{(-)}]\,.
\end{align}
This is precisely the geometric action on the coadjoint orbit of the loop group $LG$ with the highest weight $\Lambda$ i.e. the action is specified by
the symplectic 2-form is a Kirillov-Kostant-Souriau symplectic form 
\begin{align}
    \omega_{\textrm{FR}}=\int d\sigma^- \Tr(\Lambda \cG_{(-)}^{-1} \delta \cG_{(-)} \wedge \cG_{(-)}^{-1} \delta \cG_{(-})\,.\label{FR-KKS}
\end{align}

In terms of Hamiltonian analysis, the above derivation of the geometric action might be interpreted as follows.
Let us first recall the phase space action with ta symplectic form $\omega$ can be described by
\begin{align}
S=\int (d^{-1}\omega-H)dt\,.\label{phase-action}
\end{align}
In the case of the chiral WZW model, the symplectic form lives on the coadjoint orbit of the Kac-Moody group, and is proportional to the level \cite{Alekseev:1988ce} (see also \cite{Rai:1989js}). Here, the highest weight is taken as the vacuum.
Hence, removing the non-ultralocal term of the Poisson algebra of the Kac-Moody current would correspond to taking the symplectic form of the chiral WZW model to zero. 
Next, the condition  (\ref{Tmm-const})  then corresponds to changing the ground state energy from a vacuum to a state with nontrivial energy.
In the coadjoint orbit perspective, this means choosing a different orbit (with the highest weight $\Lambda$), which induces the Kirillov-Kostant-Souriau symplectic form (\ref{FR-KKS}) on the coadjoint orbit of the loop group $LG$ with the highest weight $\Lambda$. 
As a result, the reduced phase space action of the chiral WZW model is the one (\ref{phase-action}) with (\ref{FR-KKS}) which is a geometric action on the coadjoint orbit of $LG$.

\subsubsection{Curved $\beta\gamma$ and Coadjoint Orbit Defects }\label{betagammacoadjoint}

When no reality condition is imposed on the coadjoint orbit defects, they can be related to curved $\beta\gamma$ defects with flag manifold target spaces. This is expected since coadjoint orbits are affine deformations of cotangent bundles of flag manifolds, $T^*(G_{\mathbb{C}}/B)$.

The relationship between the defects can be derived as follows.\footnote{We shall consider 2d (chiral) defects, but the derivation works for 1d line defects as well.} Consider the coadjoint orbit action 
\ie \label{coadqm}
\int_{\Sigma} dw d \wbar \,\, \textrm{Tr} \Lambda (G^{-1}\partial_{\wbar} G+G^{-1}A^a_{\wbar}  t_a  G) \,.
\fe 
We then consider an Iwasawa decomposition of $G=\SL(2,\mathbb{C})$,
such that
\ie 
G=\begin{pmatrix}
1 & 0\\
\gamma & 1
\end{pmatrix} \begin{pmatrix}
e^{i\phi} & 0 \\
0 & e^{-i\phi}
\end{pmatrix}\begin{pmatrix}
1 & \beta \\
0 & 1 
\end{pmatrix}= \begin{pmatrix}
e^{i\phi}  &  e^{i\phi}\gamma \\
 e^{i\phi}\beta &  e^{i\phi}\beta \gamma + e^{-i\phi}
\end{pmatrix} \;,
\fe
where $\beta$, $\gamma$ and $\phi$ are maps from $\Sigma$ to $\mathbb{C}$.
We also fix the weight $\Lambda$ to be 
\ie 
\Lambda = a\begin{pmatrix}
1 & 0\\
0 & -1
\end{pmatrix}=2a t_{H} \,,
\fe
where $a\in \mathbb{C}$.

We can then show that 
\ie 
\textrm{Tr} (\Lambda G^{-1}\partial_{\wbar}  G)
=&  -2a e^{2i\phi}\beta \partial_{\wbar}  \gamma + 2a i \partial_{\wbar}\phi \,,\\
\textrm{Tr} (\Lambda G^{-1}t_H G)=& 2a e^{2i\phi}\beta \gamma +a  \,,\\
\textrm{Tr} (\Lambda G^{-1}t_E G)=& 2 a e^{2i\phi}\beta \gamma^2 +2 a \gamma\ \,,\
\textrm{Tr} (\Lambda G^{-1}t_F G)=& -2ae^{2i\phi}\beta  \,.
\fe
The term proportional to $\partial_{\wbar}\phi$ does not contribute to the action due to Stokes' theorem. Redefining $\beta$ as $-2ae^{2i\phi}\beta \rightarrow \beta$,
we find that the action \eqref{coadqm} becomes
\ie 
    &\int_{\Sigma} dw d\wbar ( \beta \partial_{\wbar}  \gamma + A_{\wbar}^H (-\beta \gamma +a) +A_{\wbar}^F \beta  +A_{\wbar}^E( -2\beta\gamma^2 +2 a\gamma )) \;,
\fe 
which is a curved $\beta \gamma$ defect with target space $\mathbb{CP}^1$, with the generators of the $\SL(2,\mathbb{C})$ isometry already taking the form of twisted differential operators, as discussed in Section \ref{sec:curved_betagamma}. Moreover, we can generalize this computation for coadjoint orbit defects associated with any flag variety. 

\subsection{\texorpdfstring{Constrained Free Fermions and Constrained $\beta\gamma$ Systems}{Constrained Free Fermions from Constrained Beta-GammaSystems}}\label{sec:fermion_betagamma}

Here, let us show that two Wilson loop operators (\ref{f-part}) and (\ref{b-part-ge}) are equivalent to the Schur functions associated with the Young tableau with $k_I$ boxes in the $I$-th column ($1\leq I\leq K$) and $l_{J}$ boxes in the $J$-th row ($1\leq J\leq L$).

\subsubsection{Fermionic Case}

We first evaluate the Wilson loop operator (\ref{f-part}) for the fermionic description. 

Before doing this, we briefly give a comment on the gauge fixing (\ref{gauge-fix-f}) of $\tilde{A}^f$.
The action (\ref{1d-f-action-g2}) is invariant under
\begin{align}
   \psi_{I}^{'j}=U_{I}{}^{J}\psi_{J}^{j}\,,\quad  \psi'{}^{I\dagger}_{j}=\psi^{J\dagger}_{j}(U^{-1})_{J}{}^{I}\,,\quad \tilde{A}^{f'}=U\tilde{A}^fU^{-1}-iU^{-1}\partial_t U\,,\label{gauge-f-W}
\end{align}
where $U=e^{i\varphi}\in B_{K}$\,.
The gauge transformation (\ref{gauge-f-W}) compatible with anti-periodic boundary conditions $\psi_{J}^{j}(2\pi)=-\psi_{J}^{j}(0)$ should satisfy
\begin{align}
    \varphi_{IJ}(2\pi)=\varphi_{IJ}(0)+2\pi n\,,\qquad n\in \mathbb{Z}\,,
\end{align}
or equivalently
\begin{align}
    \varphi_{IJ}(t)=\varphi_{0,IJ}(t)+n\,t\,,\qquad \varphi_{0,IJ}(2\pi)=\varphi_{0,IJ}(0)\,.
\end{align}
Therefore, we can take a gauge choice such that the gauge field $\tilde{A}^f_t$ takes values in a constant matrix $\Theta$,
\begin{align}
    \tilde{A}^f_t=\Theta\,,\qquad \Theta=\int^{2\pi}_{0}\frac{dt}{2\pi}\tilde{A}^f_t\,.
\end{align}
The matrix $\Theta$ takes the constant upper-triangular matrix.
Since eigenvalues of the upper-triangular matrix are the diagonal entries, the functional determinant for fermions only depends on the diagonal part of $\tilde{A}^f$\,.
Hence, after fixing a gauge choice, it is enough to focus on the constant diagonal entries $\theta_I$ of $\tilde{A}_t^f$ as in (\ref{gauge-fix-f}). 

We insert the identity relation (up to irrelevant factors that can be absorbed into the normalization factor of the path integral)
\begin{align}
    1&=\int \cD U\,\delta(\tilde{A}^f-\Theta)\text{Det}\left(\partial_t-i[\tilde{A}_t^f,~\cdot~]\right)e^{-i\sum_{I=1}^{K}\frac{2I-K-1}{2}\theta_I}\no\\
    &=\int \cD U\,\delta(\tilde{A}^f-\Theta)\prod_{I=1}^{K}\text{Det}_{\text{PB}}(i\partial_t)\prod_{I<J}\text{Det}_{\text{PB}}(\partial_t+(\theta_I-\theta_J))e^{-i\sum_{I=1}^{K}\frac{2I-K-1}{2}\theta_I}\,.\label{1FP}
\end{align}
Unlike the identity relation inserted in the standard Faddeev-Popov ghost procedure, there is an extra exponential factor in the first line.
This is because exponential factors with half-integer powers arise from the functional determinant:
\begin{align}
    \mu_f\left({\bf \theta}\right)&:=\prod_{1\leq I<J \leq K}\text{Det}_{\text{PB}}(i\partial_t+(\theta_I-\theta_J))=\prod_{1\leq I<J \leq K}2i\sin\left(\frac{\theta_I-\theta_J}{2}\right)\no\\
    &=\prod_{I=1}^{K}e^{-i\frac{2I-K-1}{2}\theta_I}\prod_{1\leq I<J \leq K}\left(1-e^{-i(\theta_I-\theta_J)}\right)\,.\label{FP-det}
\end{align}
Hence, the functional determinant alone is in general not invariant under large gauge transformations, and so we need to insert such an extra exponential factor to cancel the anomaly.

We now return to the computation of the path integral (\ref{f-part}).
By using the identity (\ref{1FP}), we can replace the integration over $\tilde{A}^{f}$ with \cite{Corradini:2016czo}
\begin{align}
    \int \cD \tilde{A}^f&\quad \to \quad \prod_{I=1}^{K}\int_0^{2\pi}\frac{d\theta_I}{2\pi}\mu_f\left({\bf \theta}\right)e^{-i\sum_{I=1}^{K}\frac{2I-K-1}{2}\theta_I} \,.\label{A-FP}
\end{align}
where the Faddeev-Popov measure $\mu_f\left({\bf \theta}\right)$ is given in (\ref{FP-det}).
We then perform the path integral with gaged fixed action (\ref{f-part-fix}) over the fermionic fields $\psi\,, \psi^{*}$ and obtain
\begin{align}
    \int \cD\psi \cD \psi^{*}\exp\left(i S^f\right)&=\prod_{i=1}^{N}\prod_{I=1}^{K}\text{Det}_{\text{APB}}\left(i\frac{d}{dt}+w_i+\theta_I\right)\no\\
    &=\prod_{i=1}^{N}\prod_{I=1}^{K} \prod_{n\in \mathbb{Z}} \left(2\pi n+w_i+\theta_I\right)\no\\
    &=\prod_{j=1}^{N}\prod_{I=1}^{K}2\cos\left(\frac{\theta_I+w_j}{2}\right)\no\\
    &=\prod_{j=1}^{N}\prod_{I=1}^{K}z_I^{-\frac{N}{2}}x_j^{-\frac{K}{2}}(1+x_jz_I)\,.\label{fermi-int}
\end{align}
Here we made changes $z_I=e^{i \theta_I}, x_j=e^{i w_j}$\,.
The functional determinant in (\ref{fermi-int}) is defined by performing is defined by the zeta function prescription (for the details, see e.g.\ Appendix B in \cite{Corradini:2016czo}).
Hence, combining (\ref{A-FP}) and (\ref{fermi-int}) yields the partition function (\ref{f-part}) to
\begin{align}
    W_{\cR}^{f}&=\prod_{I=1}^{K}\int_{0}^{2\pi}\frac{d\theta_I}{2\pi} e^{-i \theta_I k_{\text{eff},I}+\frac{i}{2}\sum_{j=1}^{N}w_j}\mu_f\left(\theta\right)\prod_{i=1}^{N}\left(2\cos\left(\frac{\theta_I+w_i}{2}\right)\right)\no\\
    &=\prod_{I=1}^{K}\oint\frac{dz_I}{2\pi i}\frac{1}{z^{k_{I}+1}_I} \prod_{1\leq I<J \leq K}\left(1-\frac{z_J}{z_I}\right)\prod_{j=1}^{N}\prod_{J=1}^{K}(1+x_jz_{J})\,. \label{part-f-z}
\end{align}
Note that the effect of the shift in the coefficients $k_{\text{eff},I}$ of the 1d CS terms $\int dt (\tilde{A}^f_t)_{II}$ precisely cancel with the products of the overall factors $z^{-(2I-K-1)/2}$ and $z^{-N/2}$ arising from (\ref{FP-det}) and (\ref{fermi-int}).
Similarly, $x_j^{-\frac{K}{2}}$ in (\ref{fermi-int}) cancels with the 1d CS term (\ref{U(1)-ano}) for $A$.
In this way, the integrand of (\ref{part-f-z}) is invariant under a large gauge transformation
\begin{align}
    \theta_I \to \theta_I+\partial_t\varphi_{I}\,,\qquad \varphi_{I}=n\, t\,,\qquad n\in \mathbb{Z}\,.
\end{align}

Finally, we integrate (\ref{part-f-z}) with respect to $z_I$ whose contour is the unit circle in the complex plane for $z_I$.
Before doing this, we rewrite the integrand in (\ref{part-f-z}) by using the following formulas:
\begin{align}
    \prod_{1\leq I<J \leq K}\left(1-\frac{z_{J}}{z_{I}}\right)&=\sum_{\sigma \in S_K}\text{sgn}(\sigma)\prod_{I=1}^{K}z_{I}^{I-\sigma(I)}\,, \label{V-ex}\\
    \prod_{i=1}^{N}\prod_{I=1}^{K}(1+x_{i} z_{I})&=\prod_{I=1}^{K}\sum_{m_{I}\geq 0}e_{m_{I}}({\bf x})\,z_{I}^{m_{I}}\,, \label{elsy-gen}
\end{align}
where the polynomial $e_{m}({\bf x})$ is the $m$-th elementary symmetric polynomial
\begin{align}
e_m({\bf x})=e_{m}(x_1,x_2,\dots, x_{n})=\sum_{1\leq j_1 <\dots <j_{n} \leq  m} x_{j_1}x_{j_2}\dots x_{j_n}\,.
\end{align}
Then, by using (\ref{V-ex}) and (\ref{elsy-gen}), we can rewrite (\ref{part-f-z}) as
\begin{align}
W_{\cR}^f
    &=\prod_{1\leq I<J \leq K}\left(1-\frac{z_{J}}{z_{I}}\right)
    \prod_{i=1}^{N}\prod_{I=1}^{K}(1+x_iz_I)\biggl|_{z_1^{k_1} z_2^{k_2}\cdots z_{K}^{k_{K}}}\no\\
    &=\sum_{\sigma \in S_K}\text{sgn}(\sigma)\prod_{I=1}^{K}\sum_{m_{I}\geq 0}e_{m_{I}}({\bf x})\,z_{I}^{m_{I}+I-\sigma(I)}\biggl|_{z_1^{k_1} z_2^{k_2}\cdots z_{K}^{k_{K}}}\no\\
    &=\sum_{\sigma \in S_K}\text{sgn}(\sigma)\prod_{I=1}^{K}e_{k_{I}-I+\sigma(I)}({\bf x})\no\\
    &=\text{det}(e_{k_{I}-I+J})_{I,J=1}^{K}=s_{(k_1,\dots, k_{K})^{t}}({\bf x})\,.\label{f-result}
\end{align}
In the final equality, we used the second Jacobi-Trudi formula \cite[Appendix A]{MR1153249}.

\subsubsection{Bosonic Case}

Next, we consider the partition function (\ref{b-part-ge}) for the 1d bosonic system with the same representation.

By the same argument as for the fermions, we can use the gauge symmetry $B_{\SL(L, \mathbb{C})}$ to choose a gauge fixing where the gauge field $\tilde{A}^b$ takes on a constant value. 
In particular, since the functional determinant does not depend on the off-diagonal components, we also focus on only the diagonal elements of $\tilde{A}^b$. 
The symmetry of the large gauge transformations of the Cartan part $\{\theta_{\alpha}\}$ restricts the range of the values to 0 to $2\pi$.
Then, evaluating the Faddeev-Popov determinant for the auxiliary gauge symmetry $B_{\SL(L, \mathbb{C})}$, the measure of $\tilde{A}^b$ can be rewritten as follows:
\begin{align}
   \int \cD \tilde{A}^b\quad \to \quad \prod_{\alpha=1}^{L}\int_0^{2\pi}\frac{d\theta_\alpha}{2\pi}\mu_b\left(\{\theta_{\beta}\}\right)e^{i\sum_{\alpha=1}^{L}\frac{2\alpha-L-1}{2}\theta_\alpha}\,.\label{Ab-fp}
\end{align}
The Faddeev-Popov measure $\mu_b\left(\{\theta_{J}\}\right)$ is given by
\begin{align}
    \mu_b(\{\theta_\beta\})=\prod_{0\leq \alpha<\beta \leq L}2i\,\sin\left(\frac{\theta_\alpha-\theta_\beta}{2}\right)=\prod_{\alpha=1}^{L}z_\alpha^{-\frac{2\alpha-L-1}{2}}\prod_{1\leq \alpha<\beta \leq L}\left(1-\frac{z_{\beta}}{z_\alpha}\right)\,,
\end{align}
where $z_\alpha=e^{i \theta_\alpha}$\,. As in the fermion case, an additional exponential factor is inserted in (\ref{Ab-fp}) to cancel out the global anomaly arising from the Faddeev-Popov determinant.
The integration over the bosonic fields $z[\alpha], z[\alpha]^*$ gives
\begin{align}
    \int \cD z \cD z^*\exp\left(i S^b\right)&=\prod_{i=1}^{N}\prod_{I=1}^{L}\text{Det}_{\text{PB}}\left(\frac{d}{dt}+w_i+\theta_\alpha\right)^{-1}\no\\
    &=\prod_{\alpha=1}^{L}\prod_{i=1}^{N}\left(2i\sin\left(\frac{\theta_\alpha+w_i}{2}\right)\right)^{-1}\no\\
    &=\prod_{j=1}^{N}\prod_{\alpha=1}^{L}z_\alpha^{\frac{N}{2}}x_j^{\frac{L}{2}}\frac{1}{1-x_jz_\alpha}\,,
\end{align}
Then, the partition function (\ref{b-part-ge}) becomes
\begin{align}
    W_{\cR}^b&=\prod_{\alpha=1}^{L}\oint\frac{dz_\alpha}{2\pi i}\frac{1}{z^{l_{\alpha}+1}_{\alpha}}\prod_{1\leq \alpha<\beta \leq L}\left(1-\frac{z_{\beta}}{z_{\alpha}}\right)
    \prod_{i=1}^{N}\prod_{\alpha=1}^{L}\frac{1}{1-x_{i} z_{\alpha}}\no\\
    &=\prod_{1\leq \alpha<\beta \leq L}\left(1-\frac{z_{\beta}}{z_{\alpha}}\right)
    \prod_{i=1}^{N}\prod_{\alpha=1}^{L}\frac{1}{1-x_{i} z_{\alpha}}\biggl|_{z_1^{l_1} z_2^{l_2}\cdots z_{L}^{l_{L}}}\,.\label{b-part-ge3}
\end{align}
Here, the value of the complex number $w_j$ is chosen so that $x_j$ is outside the unit-circle contour for $z_I$'s.
As in the fermionic case, we use the following formulae:
\begin{align}
    \prod_{1\leq \alpha<\beta \leq L}\left(1-\frac{z_{\beta}}{z_{\alpha}}\right)&=\sum_{\sigma \in S_{+}}\text{sgn}(\sigma)\prod_{\alpha=1}^{L}z_{\alpha}^{\alpha-\sigma(\alpha)}\,,\label{V-det -ex}\\
    \prod_{i=1}^{N}\prod_{\alpha=1}^{L}\frac{1}{1-x_{i} z_{\alpha}}&=\prod_{\alpha=1}^{L}\sum_{m_{\alpha}\geq 0}h_{m_{\alpha}}({\bf x})\,z_{\alpha}^{m_{\alpha}}\,,\label{h-gen}
\end{align}
where $h_{m}$ is the $m$-th complete homogeneous symmetric function
\begin{align}
h_m({\bf x})=h_{m}(x_1,x_2,\dots, x_{n})=\sum_{1\leq j_1 \leq \dots \leq j_{n} \leq  m} x_{j_1}x_{j_2}\dots x_{j_n}\,.
\end{align}
By using (\ref{V-det -ex}) and (\ref{h-gen}), we can rewrite the partition function (\ref{b-part-ge3}) as
\begin{align}
W_{\cR}^b&=\prod_{1\leq \alpha<\beta \leq L}\left(1-\frac{z_{\beta}}{z_{\alpha}}\right)
    \prod_{i=1}^{N}\prod_{\alpha=1}^{L}\frac{1}{1-x_{i} z_{\alpha}}\biggl|_{z_1^{l_1} z_2^{l_2}\cdots z_{L}^{l_{L}}}\no\\
&=\sum_{\sigma \in S_{+}}\text{sgn}(\sigma)\prod_{\alpha=1}^{L}\sum_{m_{\alpha}\geq 0}h_{m_{\alpha}}({\bf x})\,z_{\alpha}^{m_{\alpha}+\alpha-\sigma(\alpha)}\biggl|_{z_1^{l_1} z_2^{l_2}\cdots z_{L}^{l_{L}}}\no\\
&=\sum_{\sigma \in S_{+}}\text{sgn}(\sigma)\prod_{\alpha=1}^{L}h_{l_{\alpha}-\alpha+\sigma(\alpha)}({\bf x})\no\\
&=\text{det}(h_{l_{\alpha}-\alpha+\beta})=s_{(l_1,\dots ,l_{L}) }({\bf x})\,.
\end{align}
In the final equality, we used the first Jacobi-Trudi formula \cite[Appendix A]{MR1153249}.

\section{Non-Abelian Bosonization of Multiflavor Massless Thirring Model}\label{pboson}

In this Appendix, we will give a derivation of the non-abelian bosonization of the multi-flavor massless Thirring model \cite{Polyakov:1984et}.

\subsection{Standard Derivation}

To this end, we rewrite the action (\ref{NTh-ac}) of the multi-flavor massless Thirring model by introducing an auxiliary field $B$ as follows :
\begin{align}
   \tilde{S}^{\text{Th}}_{f}&=\frac{1}{\hbar_{2d}}\int_{\cM} d^2\sigma\biggl(\sum_{I=1}^{N_F}\psi_{L,I}^{*}i\partial_- \psi_{L,I}+\psi_{R,I}^{*}i\partial_+\psi_{R,I}+iB_-^aJ_{+,a}+iB_+^aJ_{-,a}\no\\
&\qquad\qquad \qquad    +\hbar_{2d}G^{-1}\Tr\left(B_{+}B_{-}\right)\biggr)\no\\
    &=S_{+}^{f}[B_-,\psi_L]+S_{-}^{f}[B_+,\psi_R]+\int_{\cM} d^2\sigma\,G^{-1}\Tr(B_{+}B_{-})\,,\label{gTh-action}
\end{align}
where the coupling $G$ is given by
\begin{align}
    G=\frac{1}{4}\frac{i\hbar}{z_+-z_-}\,.
\end{align}
In two-dimensional spacetime, the two independent components of the auxiliary field $B$ can be parametrized as
\begin{align}
B_+=u_{(-)}^{-1}\partial_+u_{(-)}\,,\qquad B_-=u_{(+)}^{-1}\partial_-u_{(+)}\,,\label{A-ch}
\end{align}
where $u_{(\pm)}\in G$\,.
As in the bosonization of chiral fermion defects described in Appendix \ref{chiralanomaly}, the partition functions for the chiral fermion parts can be rewritten as
\begin{align}
\hat{Z}^f_+[B_-]
&=e^{-iN_F S_{\rm WZW}[u_{(+)}]}\,,\qquad
\hat{Z}^f_-[B_+]
=e^{-iN_F S_{\rm WZW}[u_{(-)}^{-1}]}\,.
\end{align}
Hence, the partition function $Z^{\rm Th}$ with the action (\ref{NTh-ac}) becomes
\begin{align}
Z^{\rm Th}&=\int \cD \Psi\cD\bar{\Psi}\,
e^{i S^{\text{Th}}_{f}[\Psi,\bar{\Psi}]}\no\\
&=\int \cD \Psi\cD\bar{\Psi}\,\cD B\,
e^{i \tilde{S}^{\text{Th}}_{f}[\Psi,\bar{\Psi},B]}\no\\
&=\int \cD Be^{-iN_{F}S_{\rm WZW}[u_{(+)}]-i N_{F}S_{\rm WZW}[u_{(-)}^{-1}]
 +i\int d^2\sigma\, \left(a+G^{-1}\right)\Tr(B_{+}B_-)}\,,\label{lb-th}
\end{align}
where we added a local counter term $\Tr(B_+B_-)$ with a parameter $a$\,.
Similar to the bosonization of the order defect actions described in Section \ref{sec:chiralWZW}, we rewrite the path integral by multiplying the right-hand side by a constant
\begin{align}
    \int_{\cM} \cD g_{(+)} \cD g_{(-)}e^{-iN_FS_{\rm WZW}[g_{(+)}]-iN_FS_{\rm WZW}[g_{(-)}]}\,,\label{cWZW}
\end{align}
and by using the Polyakov-Wiegmann identity (\ref{PW-id}).
Then, we obtain
\begin{align}
Z^{\rm Th}
    &=\int \cD B \cD g_{(+)} \cD g_{(-)}\exp\biggl(iN_FS_{\rm WZW}[g_{(+)}]+iN_FS_{\rm WZW}[g_{(-)}]\no\\
    &\qquad -\frac{iN_F}{\pi}\int_{\cM} d^2\sigma\,\Tr(B_- \partial_+g_{(+)}g_{(+)}^{-1})+\frac{iN_F}{\pi}\int_{\cM} d^2\sigma\,\Tr(B_+ g_{(-)}^{-1}\partial_-g_{(-)})\no\\
    &\qquad +i\int_{\cM} d^2\sigma\, \left(a+G^{-1}\right)\Tr(B_{+}B_-)\biggr)\,.\label{gbTh2}
\end{align}
Finally, integrating the path integral (\ref{gbTh2}) with respect to the auxiliary fields $B_{\pm}$ gives
\begin{align}
    B_+=\frac{N_F}{\pi}\left(a+G^{-1}\right)^{-1}\partial_+g_{(+)}g_{(+)}^{-1}\,,\qquad B_-=-\frac{N_F}{\pi}\left(a+G^{-1}\right)^{-1}g_{(-)}^{-1}\partial_-g_{(-)}\,,
\end{align}
and then we have the bosonized action of the massless Thirring model
\begin{align}
Z^{\rm Th}
    &=\int  \cD g_{(+)}\cD g_{(-)}\exp\biggl(iN_FS_{\rm WZW}[g_{(+)}]+iN_FS_{\rm WZW}[g_{(-)}]\no\\
    &\qquad +i\left(\frac{N_F}{\pi}\right)^2\left(a+G^{-1}\right)^{-1}\int_{\cM} d^2\sigma\,\Tr\left(\partial_+g_{(+)}g_{(+)}^{-1}g_{(-)}^{-1}\partial_-g_{(-)}\right)\biggr)\,.
    \label{bNTh-2}
\end{align}
The conditions for Gaussian integration to converge impose restrictions on the possible values of the parameters $a$ and the coupling constant $G$.

In general, there are several ways to do bosonization.
For example, instead of multiplying by a constant (\ref{cWZW}), we may regard $u_{(\pm)}$ in \eqref{lb-th} as the bosonized fields and make the change of variables from $B_{\pm} \to u_{(\pm)}$\,. 
This transformation generates a nontrivial Jacobian \cite{Alvarez:1994np}.
By taking care of this, we again obtain  two interacting WZW models, however, the level $N_F$ and the coupling constant for the current-current interaction are replaced by $-N_{F}-\sh^{\vee}$ and $\left(a+G^{-1}\right)$, respectively, i.e., we obtain 
\ie
Z^{\rm Th}=\int \cD u_{(+)} \cD u_{(-)}  \exp \bigg(& -i(N_{F} +\sh^{\vee})S_{\rm WZW}[u_{(+)}]-i (N_{F} +\sh^{\vee})S_{\rm WZW}[u_{(-)}^{-1}]
\\ & +i\int d^2\sigma\, \left(a+G^{-1}\right)\Tr(u_{(-)}^{-1}\partial_+u_{(-)}u_{(+)}^{-1}\partial_-u_{(+)})\bigg).
\fe 
Here, $\sh^{\vee}$ is the quadratic Casimir of $G$ in the adjoint representation.
For more details, see, for example, \cite{Santos:2023dax} and the references within it.
The relationship between these two bosonized models might be regarded as a strong-weak duality, and the duality has been discussed in various works \cite{Georgiou:2015nka,Georgiou:2017oly,Santos:2023dax}.

\subsection{Bosonization from WZW Defects}\label{boson-WZW}

In Section \ref{sec:duality-boson}, we derived the bosonized action (\ref{2d-baction-2}) of multi-flavor massless Thirring model from the 4d CS theory with Dirichlet defects.
As shown in the figure \ref{duality-map}, this bosonized model can also be derived from the 4d CS theory coupled to the WZW defects on $C=\mathbb{CP}^1$.
The resulting 4d--2d system is 
\begin{align}
\begin{split}
    Z^{\text{Th}}_{b}(z_{\pm})&=\int \mathcal{D}A \mathcal{D}\cG_{(\pm)}\exp\left(iS^{\textrm{4d--2d}}[A,\cG_{(\pm)}]\right)\,,\\
    S^{\textrm{4d--2d}}[A,\cG_{(\pm)}]&=S_{\rm CS}[A]+S^{b}_{+}[A,\cG_{(+)}]+S^{b}_{-}[A,\cG_{(-)}^{-1}]\,.\label{4d-2d-bTh}
\end{split}
\end{align}
Unlike the case of the standard order defects, the WZW defects themselves are not gauge invariant, and gauge invariance of the 4d--2d system is achieved by the mechanism of anomaly flow accompanied by the shift of the meromorphic one-form $\omega$,
\ie 
    \label{simtwist}
    \omega_{\textrm{eff}} =
      dz- \frac{\hbar}{4\pi i}\left(\frac{k_{+}}{z-z_{+}}-\frac{k_{-}}{z-z_{-}}\right)  dz=
    \frac{(z-\zeta_1)(z-\zeta_2)}{(z-z_+)(z-z_-)} dz\,,
\fe
where we set $k:=k_+=k_-=N_{F}$\,.
Such 4d--2d systems have not been considered in the previous literature, and here we will briefly describe the derivation of the corresponding 2d integrable field theory by extending the procedure developed in \cite{Delduc:2019whp}.

To do this, we need to give the bulk and the boundary equations of motion of (\ref{4d-2d-bTh}).
Taking a variation of the gauge field, we obtain the bulk equations of motion 
\begin{align}
    F_{+-}&=0\,,\qquad
   \omega_{\rm eff}\, F_{\bar{z}\pm}=0\,,\label{NB-Bbulkeom2}
\end{align}
and the boundary equations of motion localized at the poles $\{z_{\pm}, \infty\}$ of $\omega_{\rm eff}$
\begin{align}
\begin{split}
   & -\frac{i}{\hbar}\epsilon^{\alpha\beta}\xi_{\infty}\partial_{\xi_{\infty}}\Tr(A_{\alpha}\delta A_{\beta})\lvert_{\xi_{\infty}}
   -\frac{k}{\pi}\Tr(\delta A_-(A_++\partial_+\cG_{(+)} \cG_{(+)}^{-1}))\lvert_{z=z_+}\\
   &+\frac{k}{\pi}\Tr(\delta A_+\,(A_--\cG_{(-)}^{-1}\partial_-\cG_{(-)} ))\lvert_{z=z_-}=0\,,
\end{split}
\end{align}
where $F_{\mu\nu}=\partial_{\mu}A_{\nu}-\partial_{\nu}A_{\mu}+[A_{\mu},A_{\nu}]$ and $\xi_{\infty}=1/z$\,.
For our purpose, we take the following values of the gauge field $A$ at the poles:
\begin{align}
    &A\lvert_{z=\infty}=0\,,\qquad
    A_+\lvert_{z=z_+}=-\partial_+\cG_{(+)} \cG_{(+)}^{-1}\,,\qquad
    A_-\lvert_{z=z_-}=\cG_{(-)}^{-1}\partial_-\cG_{(-)} \,.\label{bc-b}
\end{align}

\paragraph{Lax Pair and the Associated 2d Action}

Next, we introduce the Lax pair by performing the gauge transformation,
\begin{align}
    A&=\hat{g}\cL\hat{g}^{-1}-d\hat{g}\hat{g}^{-1}\,,\qquad  \cG_{(+)}=\hat{g}\lvert_{z=z_+}\cdot g_{(+)}\,,\qquad
    \cG_{(-)}= g_{(-)} \cdot \hat{g}^{-1}\lvert_{z=z_-}\,,\label{gauge-tr}
\end{align}
where $\hat{g}$ is a $G_{\mathbb{C}}$-valued function on $\cM\times C$\,.
Here, we will take a choice $\cL_{\bar{z}}=0$\,.
The bulk equations of motion (\ref{NB-Bbulkeom2}) reduce to
\begin{align}
    \partial_{+}\cL_--\partial_{-}\cL_++[\cL_+,\cL_-]&=0\,,\qquad
   \omega_{\rm eff}\, \partial_{\bar{z}}\cL_{\pm}=0\,,\label{NB-Bbulkeom3}
\end{align}
and the boundary conditions (\ref{bc-b}) are translated to
\begin{align}
    &\cL\lvert_{z=\infty}=0\,,\qquad  \cL_+\lvert_{z=z_+}=-\partial_+g_{(+)} g_{(+)}^{-1}\,,\qquad
    \cL_-\lvert_{z=z_-}=g_{(-)}^{-1}\partial_-g_{(-)} \,.\label{bc-b1}
\end{align}
The second equation of (\ref{NB-Bbulkeom3}) indicates that $\cL_{\pm}$ are meromorphic functions with the poles at the zeros of $\omega_{\rm eff}$\,, and the order of its poles is less than or equal to the order of the corresponding zeros of $\omega_{\rm eff}$.
By taking care of this, we can find a solution to the second equation of (\ref{NB-Bbulkeom3}) with the condition (\ref{bc-b1}) given by
\begin{align}
    \cL&=-\frac{z_+-\zeta_+}{z-\zeta_+}\partial_+g_{(+)}g_{(+)}^{-1}d\sigma^++
    \frac{z_--\zeta_-}{z-\zeta_-}g_{(-)}^{-1}\partial_-g_{(-)}d\sigma^-\,.\label{lax-NB}
\end{align}
where $\zeta_{\pm}$ are simple zeros of $\omega_{\rm eff}$
\begin{align}
    \zeta_{\pm}=\frac{1}{2}\left(z_++z_-\pm \sqrt{(z_+-z_-)\Bigl(z_+-z_{-}+\frac{i \hbar k}{\pi}\Bigr)}\right)\,.
\end{align}

\medskip

Now we can derive the corresponding 2d action by taking the classical solution (\ref{lax-NB}).
By gauge invariance, the 4d-2d action (\ref{4d-2d-bTh}) under the transformation (\ref{gauge-tr}) becomes
\begin{align}
    S^{\textrm{4d--2d}}&=S_{\textrm{CS}}[\cL]+S_{+}^{b}[\cL,g_{(+)}]+ S_{-}^{b}[\cL,g_{(-)}]\,.\label{gauge-2d4d}
\end{align}
Note that the first term vanishes due to the bulk equation of motion (\ref{NB-Bbulkeom2}) i.e.
\begin{align}
    S_{\textrm{CS}}[\cL]=\frac{1}{2\pi\hbar}\int_{\cM\times \mathbb{C}P^1}\omega_{\rm eff} \wedge \Tr\left(\cL\wedge d\cL\right)=0\,.
\end{align}
Unlike the standard disorder defect case, there is no contribution from the 4d CS action to the 2d action.
Finally, by substituting (\ref{lax-NB}) into (\ref{gauge-2d4d}), we obtain the 2d action 
\begin{align}
    S_{\rm 2d}&=S_{+}^{b}[\cL,g_{(+)}] + S_{-}^{b}[\cL,g_{(-)}^{-1}]\no\\
    &=S_{\rm WZW}^{(k)}[g_{(+)}]+S_{\rm WZW}^{(k)}[g_{(-)}]\no\\
    &\quad-\frac{k}{\pi}\int_{\cM\times \{z_{+}\}} d^2\sigma\,\Tr(\cL_- \partial_+g_{(+)}g_{(+)}^{-1})
    +\frac{k}{\pi}\int_{\cM\times \{z_{-}\}} d^2\sigma\,\Tr(\cL_+ g_{(-)}^{-1}\partial_-g_{R})\no\\
    &\quad-\frac{k}{2\pi}\int_{\cM\times \{z_{+}\}} d^2\sigma\,\Tr(\cL_+ \cL_-)-\frac{k}{2\pi}\int_{\cM\times \{z_{-}\}} d^2\sigma\,\Tr(\cL_+ \cL_-)\no\\
    &=S_{\rm WZW}^{(k)}[g_{(+)}]+S_{\rm WZW}^{(k)}[g_{(-)}]
    -\frac{8i}{\hbar}\rho_{+-}\int_{\cM}d^2\sigma\,\Tr(\partial_+g_{(+)} g_{(+)}^{-1} g_{(-)}^{-1}\partial_-g_{(-)})\,,\label{2d-baction}
\end{align}
where $\rho_{+-}$ is given by
\begin{align}
    \rho_{+-}= -\frac{(z_+-\zeta_+)(z_--\zeta_-)}{2(z_+-z_-)}=\frac{(z_+-z_--(\zeta_+-\zeta_-))^2}{8(z_+-z_-)}\,.
\end{align}

The 2d integrable field theory in \eqref{2d-baction} is a special case of integrable coupled $\sigma$-models derived from the affine Gaudin model studied in \cite{Delduc:2018hty}, up to the conventions of lightcone coordinates.
This can be seen by taking the parameters $(z_1,z_2,\zeta_1^{\pm},\zeta^{\pm}_2)$ of the twist function considered in \cite{Delduc:2018hty} as
\begin{align}
   z_1=\zeta_1^-=z_{+}\,,\qquad z_2=\zeta_1^+=z_{-}\,,\qquad
  \zeta^{+}_2= \zeta_+\,,\qquad \zeta^{-}_2=\zeta_-\,,\qquad l^{\infty}=-1\,.
\end{align}
We can see that under the choices, the twist function (1) and the master formula (2) of the 2d action in \cite{Delduc:2018hty} reduces to (\ref{simtwist}) and (\ref{2d-baction}), respectively.
In addition, this model would also be a special case of the doubly $\la$-deformed sigma model \cite{Georgiou:2017jfi,Georgiou:2016urf,Georgiou:2017oly}.

\section{Vertex Algebras and Zhu Algebras} \label{vz}

Let us briefly review standard facts regarding vertex algebras and their corresponding Zhu algebras.

Define a {\em field} on a vector space $V$ to be  a formal series $$a(z) =
\sum_{n\in \mathbb{Z}} a_{(n)} z^{-n-1} \in ({\rm End} V)[[z, z^{-1}]]$$
such that for any $v\in V$ one has $a_{(n)}v = 0$ if $n\gg 0$. We shall use $\mathcal{F}(V)$ to denote the space of all fields on $V$.

A {\em vertex algebra}  is a vector space $V$ appended with the following
data:
\begin{itemize}
  \item a linear map (the state-operator correspondence) $Y: V \to \mathcal{F}(V)$, 
    $V\ni a \mapsto a(z) = \sum_{n\in \Z} a_{(n)} z^{-n-1}$,
  \item an even vector ${\bf 1}\in V$, known as the {\em vacuum vector},
  \item a linear operator $\partial: V \to V$, known as the {\em translation operator}.
\end{itemize}
These data satisfy the following axioms:
\begin{enumerate}
  \item (Translation Covariance)
     $(\partial a)(z) = \partial_z a(z)$.

  \item (Vacuum)
    ${\bf 1}(z) = \textrm{id}$;
    $a(z)\vac \in V[[z]]$ and $a_{(-1)}\vac = a$.

  \item (Locality) 
      For all $a\,,b \in V$ there exists a positive integer $N$ such that 
      \begin{align} 
         (z-w)^{N}[Y(a,z),Y(b,w)]=0 \;.
      \end{align}
\end{enumerate}

To each vertex algebra $V$, one can define an associative algebra known as the Zhu algebra $A(V)$. As a vector space, it is a quotient space of $V$. Let $O(V)$ denote the linear span of elements  $\operatorname{Res}_z\left(Y(a, z)\left((\hbar_{\star} z+1)^{\operatorname{deg}(a)} / z^2\right) b\right)$. One can define a multiplication in $V$ as follows: for $a, b \in V$
\ie\label{star}
a \star b: =\operatorname{Res}_z\left(Y(a, z) \frac{(\hbar_{\star} z+1)^{\operatorname{deg}(a)}}{z} b\right) \;.
\fe
The $\star$-product induces the multiplication on the quotient $A(V):=V/O(V)$, and can be shown to be associative in $A(V)$.

In physics terminology, the Zhu algebra corresponds to the zero modes of the CFT Hilbert space. 
This is due to a theorem by Zhu \cite{zhu1990vertex}, which can be stated as follows: for a graded module, $M$, over a vertex algebra $V$, the top component $M_0$ is a module over the Zhu algebra $A(V)$. The assignment $M \mapsto M_0$ gives rise to a one-to-one correspondence between isomorphism classes of graded irreducible $V$-modules and those of irreducible $A(V)$ modules.

\section{\texorpdfstring{2d Curved $\beta\gamma$ Systems and Twisted Chiral Differential Operators }{2d beta-gamma Systems and Twisted Chiral Differential Operators }}

In this appendix, we describe the obstruction to defining a global sheaf of chiral differential operators on a K\"ahler manifold, $X$. We first compute the commutator of Laurent modes of currents associated with holomorphic vector fields that generate the $G$ symmetry of $X$, whose OPE is given in equation \eqref{JJ-OPE}, which includes the Lie algebra extension \eqref{zeroal}. 
We then review how a global sheaf of chiral differential operators can be defined on a 
general K\"ahler target space, following \cite{Witten:2005px}, emphasizing how this can be obstructed by the extension to $\mathfrak{g}$, and why this obstruction vanishes when $p_1(X)=0$.

\subsection{\texorpdfstring{Current-Current OPE in Curved $\beta\gamma$ Systems}{Current-current OPE in beta-gamma Systems}}\label{opeal}

Let us first write the Laurent expansion of the currents as 
\ie
J_V(w)=\sum_{n\in \mathbb{Z}}w^{-n-1}J_{Vn}\;,\quad \quad J_{Vn}=\frac{1}{2\pi i}\oint w^{n}J_V(w)\;.
\fe
The contribution of the first term to the commutator $[J_{Vn},J_{Wm}]$ can be computed using the formula 
\begin{equation}
    [A,B]=\oint_0 dw' \oint_w \textrm{ }dw \textrm{ }a(w)b(w')
\end{equation}
to be 
\begin{equation}
\begin{aligned}
&\frac{1}{(2\pi i)^2}\oint_0 dw' \oint_{w'} dw w^n w'^m  \bigg(-\frac{\partial_jV^i\del_iW^j(w')}{(w-w')^2}\bigg)\\
    &=\frac{1}{2\pi i}\oint_0 dw' \lim_{w\rightarrow w'}\frac{d}{dw}w^n (w')^m (-\partial_jV^i\del_iW^j(w'))\\
    &=\frac{1}{2\pi i}\oint_0 dw' \quad n \quad (w')^{n-1 }(w')^m\sum_{p\in \mathbb{Z}}(-\del_jV^i\del_iW^j)_p (w')^p\\
    &=n\delta_{n+m+p,0}(-\del_jV^i\del_iW^j)_p \;.
    \end{aligned}
\end{equation}
This is nonzero, but it does not contribute to the commutator $[J_{V0},J_{W0}]$ that is our main object of interest.

Defining $\tilde{J}(w)=-(\partial_k\partial_jV^i)(\partial_iW^j\partial\gamma^k)$, the remaining terms in the commutator are computed to be 
\begin{equation}
\begin{aligned}
&\frac{1}{(2\pi i)^2}\oint_0 dw' \oint_w dw w^n (w')^m \bigg(\frac{J_{[V,W]}(w')}{w-w'}+\frac{\tilde{J}(w')}{w-w'}\bigg)\\
&=\frac{1}{2\pi i}\oint_0 dw' (w')^n (w')^m\bigg( \sum_{m+n\in \mathbb{Z}}(w')^{-m-n-1}J_{[V,W]m+n}+\sum_{m+n\in \mathbb{Z}}(w')^{-m-n-1}\tilde{J}_{m+n}\bigg)\\
&=J_{[V,W]m+n}+\tilde{J}_{m+n} \;,
\end{aligned}
\end{equation}
where $[V,W]^j=(V^i\partial_iW^j-W^i\partial_iV^j)$.
The full commutator arising from the current-current OPE is thus
\ie\relax
[J_{Vn},J_{Wm}]= n\delta_{n+m+p,0}(-\del_jV^i\del_iW^j)_p +J_{[V,W]m+n}+\tilde{J}_{m+n} \;.
\fe 
The zero mode algebra is thus a Lie algebra extension defined via the exact sequence
\ie
0 \rightarrow \mathfrak{c} \rightarrow \mathfrak{g} \rightarrow \mathfrak{v} \rightarrow 0\,.
\fe

\subsection{Defining Global Sheafs of Twisted Chiral Differential Operators}\label{cdoapp}

Firstly, cover the K\"ahler manifold by open sets $U_a$. For each
$a,b$,  choose an open set $U_{ab}\subset U_{a}$, and also an
open set $U_{ba}\subset U_b$, and a holomorphic diffeomorphism
$f_{ab}$ between these open sets $f_{ab}:U_{ab}\cong U_{ba}$ such that
$f_{ba}=f_{ab}^{-1}$. In order to identify a point $P\in U_{ab}$
with a point $Q\in U_{ba}$ when $Q=f_{ab}(P)$, we would require that for any $U_a$, $U_b$, and $U_c,$ we have
\begin{equation}\label{con1}
{f_{ca}f_{bc}f_{ab}=1}
\end{equation}
on the triple intersection $U_{abc}$.

Analogously, to obtain a consistent global sheaf of chiral algebras defined over the entire target space, we ought to glue the chiral algebras using symmetries of the associated conformal field theory. For example, given two open sets $U_a$ and $U_b$, we can use a CFT symmetry, denoted $\hat{f}_{ab}$, to glue them, with the consistency condition on triple overlaps given by 
\begin{equation}\label{con2}
    {\hat{f}_{ca} \hat{f}_{bc} \hat{f}_{ab}=1}\,.
\end{equation}

Now, since we have a Lie algebra extension and thus a group extension as shown in Section \ref{bg-anomaly}, given such a consistent gluing of the sheaf of CDOs, one can always map the CFT symmetry group elements to geometrical symmetry group elements, i.e., \eqref{con2} implies \eqref{con1}, thereby defining the complex manifold that is the target space of the CFT. 

However, the $\hat{f}_{ab}$ are not uniquely determined by the ${f}_{ab}$, as we can always pick an element $C_{ab} \in H^{0}\left(U_{a b}, \Omega^{2, c l}(X )\right)$ (where  $\Omega^{2,cl}(X)$ is the sheaf of $(2,0)$-forms on $X$ that are annihilated by $\partial$) representing an element of $\mathfrak{c}$, and perform the transformation $\widehat{f}_{a b} \rightarrow \widehat{f}_{a b}^{\prime}=\exp \left(C_{a b}\right) \widehat{f}_{a b}$. If the CFT gluing condition on triple intersections is to still hold, we would then require that 
$$
C_{a b}+C_{b c}+C_{c a}=0\,,
$$
and hence the $C$ 's define an element of the \u{C}ech cohomology group $H^{1}\left(X, \Omega^{2, c l}(X)\right)$.

Now, let us understand the anomaly that is the obstruction to gluing. If we wish to lift the gluing condition \eqref{con1}, in general we will not obtain \eqref{con2}, but rather 
\ie
\widehat{f}_{c a} \widehat{f}_{b c} \widehat{f}_{a b}=\exp \left(C_{a b c}\right)
\fe
for some $C_{a b c} \in H^{0}\left(U_{a b c},  \Omega^{2, c l}(X )\right)$ which represents an element of $\mathfrak{c}$. Now, transforming $\widehat{f}_{a b} \rightarrow \exp \left(C_{a b}\right) \widehat{f}_{a b}$, we get
\ie \label{crel1}
C_{a b c} \rightarrow C_{a b c}^{\prime}=C_{a b c}+C_{a b}+C_{b c}+C_{c a} \,.
\fe
In addition, in quadruple overlaps $U_{a} \cap U_{b} \cap U_{c} \cap U_{d}$, we have the relation
\ie\label{crel2}
C_{a b c}-C_{b c d}+C_{c d a}-C_{d a b}=0 \;.
\fe 
The relations \eqref{crel1} and \eqref{crel2} imply that the $C's$ define an element of the sheaf cohomology group $H^{2}\left(X, \Omega^{2, c l}(X)\right)$. There is no obstruction to obtaining a globally defined sheaf of chiral algebras 
if it is possible to pick the $C_{a b}$ to set all $C_{a b c}^{\prime}=0$. This can be identified with the vanishing of $p_1(X)$.

\section{Examples of Trigonometric and Elliptic Discretizations}\label{sec:trig_ell}

In this appendix we provide examples of lattice discretizations of trigonometric and elliptic integrable field theories.

By repeating the arguments of Sections \ref{sec:disc_FT} and \ref{sec:disc_defect}, we can relate trigonometric and elliptic deformations of integrable field theories respectively to trigonometric and elliptic integrable lattice models/spin chains. For example, by taking $C=\mathbb{C}^{\times}$, the trigonometric analogue of the massless Thirring/Gross-Neveu model derived in \cite{Costello:2019tri},
\begin{equation}
\begin{aligned}
S_{\rm 2d}^{\mathrm{eff}}&=\frac{1}{\hbar}\left[2 \int \psi_{e} \bar{\partial} \psi_{f}+2 \int \psi_{h} \bar{\partial} \psi_{h}+2 \int \bar{\psi}_{e} \partial \bar{\psi}_{f}+2 \int \bar{\psi}_{h} \partial \bar{\psi}_{h}\right.\\
&\quad\qquad +\frac{4}{1-z_{1} / z_{2}} \int \psi_{f} \psi_{h} \bar{\psi}_{f} \bar{\psi}_{h}+\frac{4}{1-z_{2} / z_{1}} \int \psi_{e} \psi_{h} \bar{\psi}_{e} \bar{\psi}_{h} \\
&\quad\qquad \left.+\frac{z_{2}+z_{1}}{z_{2}-z_{1}} \int \psi_{e} \psi_{f} \bar{\psi}_{e} \bar{\psi}_{f}\right]\,,
\end{aligned}
\end{equation}
can be discretized to a trigonometric lattice model. We can also derive a trigonometric generalization of the Faddeev-Reshetikhin model
\begin{equation}
\begin{aligned}
S_{\rm 2d}^{\mathrm{eff}}=-\int_{\Sigma} \textrm{Tr}\bigg(&\Lambda g_{(+)}^{-1} \partial_{-} g_{(+)}+\Lambda g_{(-)}^{-1} \partial_{+} g_{(-)} +\frac{4}{1-z_{+} / z_{-}}  \mathcal{J}_+^e\mathcal{J}_-^f
\\&-\frac{4}{1-z_{-} / z_{+}} \mathcal{J}_+^f\mathcal{J}_-^e 
+\frac{z_{-}+z_{+}}{z_{-}-z_{+}}\mathcal{J}_+^h\mathcal{J}_-^h  \bigg) d \sigma^{+} \wedge d \sigma^{-}  \;,
\end{aligned}
\end{equation}
which is dual to the trigonometric massless Thirring model since they can be discretized to identical lattice models. Here we have, for example, $\mathcal{J}_+^e$ denoting the $e$-component of $\mathcal{J}_+=g_{(+)}\Lambda g_{(+)}^{-1}$.

Likewise, by picking $C$ to be an elliptic curve $E$, the elliptic analogue of the massless Thirring model 
\begin{equation}
\begin{aligned}
S_{\mathrm{2d}}^{\mathrm{eff}}&=\frac{1}{\hbar}\left[ -\frac{1}{2} \sum_{i=1}^{3} \int \psi_{i} \bar{\partial} \psi_{i}-\frac{1}{2} \sum_{i=1}^{3} \int \bar{\psi}_{i} \partial \bar{\psi}_{i}\right.\\
&\quad\qquad \left.+\frac{1}{4} \sum_{i, j, k, l, m=1}^{3} w_{i}\left(z_{0}-z_{1}\right) \epsilon_{i j k} \epsilon_{i l m} \int \psi_{j} \psi_{k} \bar{\psi}_{l} \bar{\psi}_{m}\right]
\end{aligned}
\end{equation}
derived in \cite{Costello:2019tri}
can be discretized to an elliptic lattice model. 
We can also derive an elliptic generalization of the Faddeev-Reshetikhin model
\begin{equation}
S_{\rm 2d}^{\mathrm{eff}}=-\int_{\Sigma} \operatorname{Tr}\left(\Lambda g_{(+)}^{-1} \partial_{-} g_{(+)}+\Lambda g_{(-)}^{-1} \partial_{+} g_{(-)} + \sum_{i=1}^3 w_i(z_+-z-)\mathcal{J}^i_+\mathcal{J}^i_-  \right) d \sigma^{+} \wedge d \sigma^{-}  \;,
\end{equation}
which is dual to the elliptic massless Thirring model since they can be discretized to identical lattice models. Here we have, for example, $\mathcal{J}_+^1$ denoting the $1$-component of $\mathcal{J}_+=g_{(+)}\Lambda g_{(+)}^{-1}$.

\section{\texorpdfstring{$\Omega$-deformed B-models and 4d CS Coupled to Defects}{Omega-deformed B-models and 4d CS Coupled to Defects}\label{omegappendix}}

In this appendix, we shall provide some details of how the 4d CS theory coupled to (line or surface) defects can be realized via the localization of $\Omega$-deformed supersymmetric gauge theories that arise from the brane configurations studied in Section \ref{sec:string}, using the constructions of \cite{Yagi:2014toa,Costello:2018txb,Oh:2019bgz}. Although the realizations of the 4d CS theory and its line and surface defects all involve supersymmetric gauge theories of different dimensions, they can each be described as a 2d sigma model with $\Omega$-deformed B-type supersymmetry and gauge symmetry, with disc worldsheet and target space defined as a space of maps. 

The $\Omega$-deformation is defined with respect to a Killing vector field, $V$, that generates a $\U(1)$ isometry of the disk that the B-model is defined on. The B-model is defined in terms of both a vector multiplet and a hypermultiplet. The vector multiplet contains the gauge field $A$, an auxiliary field $\mathrm{D}$, and bosonic and fermionic fields (distinguished with the notation `$\Pi$') of various form degrees: 
\ie
\mathrm{D} \in \Omega^0\,, \quad A, \sigma \in \Omega^1\,, \quad \alpha \in \Pi \Omega^0\,, \quad \lambda \in \Pi \Omega^1\,, \quad \zeta \in \Pi \Omega^2\,,
\fe
where $\Pi$ indicates a fermionic field. 
The chiral multiplet consists of a scalar field, $\varphi$, a complex auxiliary 2-form field, $\mathrm{F}$, and bosonic and fermionic fields of various form degrees: 
\ie 
\varphi \in \Omega^0\,, \quad \mathrm{~F} \in \Omega^2\,, \quad \bar{\eta} \in \Pi \Omega^0\,, \quad \rho \in \Pi \Omega^1\,, \quad \bar{\mu} \in \Pi \Omega^2\,.
\fe
The action of the deformed supercharge, $Q_V$ is given by  
$$
\begin{aligned}
\delta_V \mathcal{A} & =\iota_V \zeta\,, & \delta_V \overline{\mathcal{A}} & =\lambda-\iota_V \zeta\,, \\
\delta_V \lambda & =2 \iota_V F-2 \mathrm{i} D \iota_V \sigma\,, & \delta_V \zeta & =\mathcal{F}\,, \\
\delta_V \alpha & =\mathrm{D}\,, & \delta_V \mathrm{D} & =\iota_V \mathcal{D} \alpha \,,
\end{aligned}
$$
(where $\mathcal{A}=A+\mathrm{i} \sigma$, $\cF$ its curvature 2-form, and $\mathcal{D} =d +\mathcal{A}$ is its associated covariant derivative, while $F$ and $D=d +A$ are the curvature and covariant derivative for $A$) for the vector multiplet and
$$
\begin{aligned}
    & \delta_V \varphi=\iota_V \rho\,, \quad \delta_V \bar{\varphi}=\bar{\eta}\,, \\
    & \delta_V \rho=\mathcal{D} \varphi+\iota_V \mathrm{F}\,, \quad \delta_V \bar{\eta}=\iota_V \mathcal{D} \bar{\varphi}\,, \\
    & \delta_V \mathrm{F}=\mathcal{D} \rho-\zeta \varphi\,, \quad \delta_V \overline{\mathrm{F}}=\mathcal{D} \iota_V \bar{\mu}\,, \\
    & \delta_V \bar{\mu}=\overline{\mathrm{F}} \,, 
\end{aligned}
$$
for the chiral multiplet. It is not nilpotent but rather squares to the covariant Lie derivative defined with respect to $\mathcal{A}$, i.e.,   :
\ie
\delta_V^2=\mathrm{d}_{\mathcal{A}} \iota_V+\iota_V \mathrm{~d}_{\mathcal{A}} \,.
\fe

The action containing the standard kinetic terms can be written as $S_V+S_C+S_W$, where
\begin{align}
S_{\mathrm{V}}&=\delta_V \int_D \operatorname{Tr}((-\star \mathrm{D}+2 \mathrm{i} D \star \sigma) \alpha-\zeta \star \overline{\mathcal{F}})\;, \\
S_{\mathrm{C}}&=\delta_V \int_D\left(\left(\overline{\mathcal{D}} \bar{\varphi}+\iota_{\bar{V}} \overline{\mathrm{F}}\right) \wedge \star \rho+\star \bar{\varphi} \alpha \varphi+\bar{\mu} \star \mathrm{F}\right)\;, \\
S_W&=\int_D\left(\mathrm{F} \frac{\partial W}{\partial \varphi}+\frac{1}{2} \rho \wedge \rho \frac{\partial^2 W}{\partial \varphi \partial \varphi}-\delta_V\left(\bar{\mu} \frac{\partial \bar{W}}{\partial \bar{\varphi}}\right)\right)-\int_{\partial D} W \frac{\mathrm{d} \theta}{V^\theta} \,.
\end{align}
The boundary term in $S_W$ is needed for unbroken $Q_V$-invariance at the boundary. 

Localization of the path integral of this deformed B-model involves expanding each field around $Q_V$-invariant configurations, and computing the bosonic and fermionic one-loop determinants, that can be shown to cancel. The $Q_V$-invariant configurations are such that the non-$Q_V$-exact terms other than the boundary term also vanish. The end result of the localization is a (gauge-fixed) 0d theory  with action proportional to the superpotential $W$, and integration cycle specified by the $Q_V$-invariant configurations. These include the gradient flow equations that ensure the convergence of the path integral. 

To realize the 4d CS theory, the chiral multiplet scalars of the theory are chosen to form the partial connection 
$$
\mathcal{A}=\mathcal{A}_m \mathrm{~d} x^m+A_{\bar{z}} \mathrm{~d} \bar{z} \,,
$$
which corresponds to the gauge field of the theory on $\Sigma \times C$, where $x^1$ and $x^2$ are Euclidean coordinates on $\Sigma$. The deformed B-model thus has a target space 
which is defined to be the space of such connections.
The gauge group, $\mathcal{G}$, of the B-model is the group of maps from $\Sigma \times C$ to $G$, where the latter corresponds to the gauge transformations that are independent of the coordinates on the disk worldsheet.
Endowing the target space with the $\mathcal{G}$-invariant K\"ahler metric
\ie
g_X=-\frac{1}{2 e^2} \int_{\Sigma \times C} \sqrt{g_{\Sigma}} \mathrm{d}^2 x \mathrm{~d}^2 z \operatorname{Tr}\left(\delta \mathcal{A}^m \otimes \delta \overline{\mathcal{A}}_m+\delta \overline{\mathcal{A}}^m \otimes \delta \mathcal{A}_m+\delta A_{\bar{z}} \otimes \delta A_z+\delta A_z \otimes \delta A_{\bar{z}}\right)\,,
\fe
and defining $W$ to be the 4d CS action (with suitable coefficient) reproduces the $\Omega$-deformed D5-brane worldvolume theory studied in the main text from the aforementioned B-model. 

The supersymmetric gauge theories that give rise to line and surface defects in the 4d CS theory can each be similarly written as an $\Omega$-deformed B-model with target space specified as a space of maps from the defect to the space of fields supported on the defect. A convenient approach to understand this for the defects that we are concerned with is as follows. Starting with superpotential which we defined to be proportional to the 4d CS action in the previous paragraph, perform successive dimensional reductions down to a one-dimensional action, i.e., 
\ie 
&\int_{T^4} dz \wedge \textrm{CS}(\mathcal{A})\\
\rightarrow & \int_{T^3}  \textrm{CS}(\mathcal{A}) \\
\rightarrow & \int_{T^2} \textrm{Tr }\phi \mathcal{F}
\\
\rightarrow & \int_{S^1} dt (\phi^a \del_t\tilde{\phi}_a +f^{abc}\phi_a \mathcal{A}_{tb}\tilde{\phi}_c)\,,
\fe
where $\phi$ and $\tilde{\phi}$ are complex-valued. 
The last action is what we would obtain if we start with an $\Omega$-deformed 3d $\mathcal{N}=4$ theory with a vector multiplet and adjoint hypermultiplet. Replacing the adjoint hypermultiplet by an arbitrary representation, the localization procedure should give us an action of the form 
\ie 
\int_{S^1} dt (\phi^a \del_t\tilde{\phi}_a +\phi_a (T^b)_{ac}\mathcal{A}_{tb}\tilde{\phi}_c)\,,
\fe
which is precisely the gauged quantum mechanical action that we require to realize Wilson lines.
Moreover, from this perspective $l_{\alpha}$ can also be regarded as an FI parameter that can be added to the moment map $\phi_a (T^b)_{ac}\tilde{\phi}_c$ that is contracted with the complexified gauge field, without breaking supersymmetry.
Thus, to realize the 4d CS theory coupled to line defects along lightcone directions from supersymmetric gauge theory, we need to twist and $\Omega$-deform a coupled 6d-3d-3d supersymmetric theory.

This realization of chiral and anti-chiral surface defects can be understood using dimensional reduction of the 4d CS action functional which is holomorphic along $\Sigma$ instead of $C$, and using a different dimensional reduction, that is, along the topological directions instead of the holomorphic directions, reducing the 4d action down to two-dimensions. 
In this way we obtain : 
\ie 
&\int_{T^2 \times \Sigma} dw \wedge \textrm{CS}(\mathcal{A}) \\ \rightarrow & \int_{S^1 \times \Sigma} dw \wedge \textrm{Tr} \phi \mathcal{F}  \\ \rightarrow &  \int_{\Sigma} dw d\wbar(\phi^a \del_{\wbar}\tilde{\phi}_a +f^{abc}\phi_a A_{\wbar b}\tilde{\phi}_c)
\,.
\fe
This action is what we would obtain from an $\Omega$-deformed 4d $\mathcal{N}=2$ theory with a vector and adjoint hypermultiplet. Replacing the adjoint hypermultiplet by e.g., a fundamental hypermultiplet, the localization procedure should give an action of the form 
\ie 
\int_{\Sigma} dw d\wbar(\phi^a \del_{\wbar}\tilde{\phi}_a +\phi_a (T^b)_{ac}A_{\wbar b}\tilde{\phi}_c) \,.
\fe
Likewise, the complex conjugate the 4d CS action gives rise to an anti-chiral defect via dimensional reduction, i.e., 
\ie 
&\int_{T^2 \times \Sigma} dw d\wbar \wedge \textrm{CS}(\mathcal{A}) \\ \rightarrow & \int_{S^1 \times \Sigma} d\wbar \wedge \textrm{Tr} \phi \mathcal{F}  \\ \rightarrow &  \int_{\Sigma} dw d\wbar (\phi^a \del_{w}\tilde{\phi}_a +f^{abc}\phi_a A_{w b}\tilde{\phi}_c)\,.
\fe

\end{appendix}


\newpage
\bibliographystyle{ytphys}
\bibliography{4dCS} 


\end{document}